\documentclass[
 reprint,
 amssymb, amsmath,
 aip,jcp,
]{revtex4-1}

\usepackage[T1]{fontenc}
\usepackage{textcomp}
\usepackage{newunicodechar}
\usepackage{multirow}
\usepackage{graphicx}
\usepackage{rotating}
\usepackage{makeidx}
\usepackage{threeparttable}
\usepackage{float}
\usepackage{amsmath}
\usepackage{rotating}
\usepackage{hyperref}
\usepackage{relsize}
\usepackage{etoolbox}
\usepackage[most]{tcolorbox}
\usepackage{hyperref}
\usepackage{dcolumn}
\usepackage{bm}
\usepackage{wrapfig}
\usepackage{float}
\usepackage[toc,page]{appendix}
\usepackage{epstopdf}
\usepackage{subfiles}
\usepackage{comment}
\usepackage[autostyle=true]{csquotes}
\usepackage{makecell}
\usepackage{braket}

\usepackage{subfigure}
\usepackage{siunitx}
\usepackage[all]{hypcap}
\usepackage[most]{tcolorbox}
\usepackage{hyperref}
\usepackage{xcolor}
\usepackage[version=4]{mhchem}
\hypersetup{
	colorlinks   = true, 
	urlcolor     = blue, 
	linkcolor    = blue, 
	citecolor   = blue 
}

\usepackage{etoolbox}
\usepackage{txfonts}
\expandafter\ifx\csname package@font\endcsname\relax\else
 \expandafter\expandafter
 \expandafter\usepackage
 \expandafter\expandafter
 \expandafter{\csname package@font\endcsname}%
\fi
\begin{document}

	\title{Real-time non-adiabatic dynamics in the one-dimensional Holstein model: Trajectory-based vs exact methods}

	\author{M. ten Brink}%
	\affiliation{Institut f\"ur Theoretische Physik, Georg-August-Universit\"at G\"ottingen, Friedrich-Hund-Platz 1, 37077 G\"ottingen, Germany}%
	\affiliation{Institut f\"ur Theoretische Physik, Technische Universit\"at Clausthal, Leibnizstr. 10, 38678 Clausthal-Zellerfeld, Germany}%

	\author{S. Gr\"aber}%
	\affiliation{Institut f\"ur Theoretische Physik, Georg-August-Universit\"at G\"ottingen, Friedrich-Hund-Platz 1, 37077 G\"ottingen, Germany}%

	\author{M. Hopjan}%
	\affiliation{Institut f\"ur Theoretische Physik, Georg-August-Universit\"at G\"ottingen, Friedrich-Hund-Platz 1, 37077 G\"ottingen, Germany}%
	\affiliation{Department of Theoretical Physics, J. Stefan Institute, Jamova cesta 39, SI-1000 Ljubljana, Slovenia}

	\author{D. Jansen}%
	\affiliation{Institut f\"ur Theoretische Physik, Georg-August-Universit\"at G\"ottingen, Friedrich-Hund-Platz 1, 37077 G\"ottingen, Germany}%

	\author{J. Stolpp}%
	\affiliation{Institut f\"ur Theoretische Physik, Georg-August-Universit\"at G\"ottingen, Friedrich-Hund-Platz 1, 37077 G\"ottingen, Germany}%

	\author{F. Heidrich-Meisner}%
	\affiliation{Institut f\"ur Theoretische Physik, Georg-August-Universit\"at G\"ottingen, Friedrich-Hund-Platz 1, 37077 G\"ottingen, Germany}%

	\author{P. E. Bl\"ochl}%
	\affiliation{Institut f\"ur Theoretische Physik, Georg-August-Universit\"at G\"ottingen, Friedrich-Hund-Platz 1, 37077 G\"ottingen, Germany}%
	\affiliation{Institut f\"ur Theoretische Physik, Technische Universit\"at Clausthal, Leibnizstr. 10, 38678 Clausthal-Zellerfeld, Germany}%

\begin{abstract}
We benchmark a set of quantum-chemistry methods, including multitrajectory Ehrenfest, fewest-switches surface-hopping, and multiconfigurational-Ehrenfest dynamics, against exact quantum-many-body techniques by studying real-time dynamics in the Holstein model. This is a paradigmatic model in condensed matter theory incorporating a local coupling of electrons to Einstein phonons. For the two-site and three-site Holstein model, we discuss the exact and quantum-chemistry methods in terms of the Born-Huang formalism, covering different initial states, which either start on a single Born-Oppenheimer surface, or with the electron localized to a single site. For extended systems with up to 51 sites, we address both the physics of single Holstein polarons and the dynamics of charge-density waves at finite electron densities. For these extended systems, we compare the quantum-chemistry methods to exact dynamics obtained from time-dependent density matrix renormalization group calculations with local basis optimization (DMRG-LBO). We observe that the multitrajectory Ehrenfest method, in general, only captures the ultrashort time dynamics accurately. In contrast, the surface-hopping method with suitable corrections provides a much better description of the long-time behavior but struggles with the short-time description of coherences between different Born-Oppenheimer states. We show that the multiconfigurational Ehrenfest method yields a significant improvement over the multitrajectory Ehrenfest method and can be converged to the exact results in small systems with moderate computational efforts. We further observe that for extended systems, this convergence is slower with respect to the number of configurations. Our benchmark study demonstrates that DMRG-LBO is a useful tool for assessing the quality of the quantum-chemistry methods.
\end{abstract}

	\date{\today}

	\maketitle

	\section{Introduction\label{Sec:Intro}}

In quantum chemistry, the joint dynamics of electronic excitations coupled to nuclear degrees of freedom can lead to complex non-adiabatic effects, which cannot be described within the Born-Oppenheimer (BO) approximation.\cite{Gonzalez20,Curchod18,Agostini19}
The correct description of non-adiabatic effects is essential to understand the resulting intricate processes, such as interference close to and transitions through avoided crossings and conical intersections,\cite{Domcke2011,Yarkony1996} intraband relaxation, or energy transfer.\cite{Nelson2011}
This coupled dynamics of an electron-bosonic composite system is often initiated by the excitation  with an electromagnetic field, and complexity can further increase if the quantum nature of the electromagnetic field and its bosonic excitations are taken into account explicitly.\cite{Coles2014,Hutchison2012,Orgiu2015,Thomas2016,Bienfait2016,Flick2017}

Bosonic excitations naturally emerge in condensed matter, e.g.,  plasmons,\cite{Nabok2021} phonons,\cite{Giustino17,Franchini21,Kang21} or magnons.\cite{Nabok2021} 
Here, such excitations can couple to electrons and influence the electronic properties. 
For example, exciton-phonon coupling\cite{Raja2018,Shree2018,Werdehausen2018,Chen2020,Mor2021,Li2021} in semiconductors affects exciton mobilities,
the phonon-bottleneck mechanism reduces energy loss of hot carriers,\cite{Frost2017,Raiser2017,Park18,Ghosh20,Kressdorf20} and
phonon-magnon scattering\cite{Berk2019,Akimov2020} induced by electron-phonon coupling can lead to ultrafast demagnetization,\cite{Frietsch2020}
to mention a few examples. Clearly, there is an increasing interest in the theoretical description of such multi-component systems involving bosons, 
also in view of recent ultrafast dynamics experiments,\cite{Orenstein2012,Giannetti2016,Lloyd_Hughes_2021} where, for example, light excitations in electron-phonon 
coupled systems have been claimed to enhance superconductivity\cite{Mankowsky2014,Hu2014,Mitrano2016,Sentef2016,Babadi2017,Paeckel2020,Buzzi2020}  or where
phase transitions to a charge-density-wave phase\cite{Vogelgesang18,Storeck20,Storeck21} or a metal-insulator structural phase transition \cite{Horstmann20} can be driven.

Exact solutions to the real-time dynamics of composite electron-nuclear systems are scarce. Among the few systems for which numerically exact solutions exist, we mention, e.g., the dissipative spin-boson model,\cite{Leggett87,Makarov94,Egger94,Kehrein96,Thompson99,Wang00,MacKernan02,Wang2003,Wang08,MacKernan08,Wang09,Kananenka16,Chen2016} Tully's set of problems,\cite{Tully1990,Tully1998,Agostini16,Agostini18,Ibele2020}  the Shin-Metiu model of proton-charge transfer,\cite{Shin95,Abedi13,Agostini15,Eich16,Gossel19,Martinez21}
and its variant for desorption.\cite{Bostrom2016a,Bostrom2016b} 
Recently, some exactly solvable models for cavity quantum electrodynamics (QED) have been presented.\cite{Flick2017} 
To describe more complex systems, a quite large palette of quantum-chemical approximate methods has been devised so far.\cite{Gonzalez20} 
Examples are Ehrenfest and surface-hopping methods, which are widely used and computationally 
favorable algorithms.\cite{Tully1998,Gonzalez20}  
They use independent classical trajectories for the nuclei and  restore some
quantum effects by averaging over the trajectories. Their low computational costs allow for treating rather large molecular systems.

In condensed matter physics, the description of coupled electron-bosonic systems is naturally done within the formalism of diagrammatic quantum field methods,\cite{Giustino17,Sakkinen15a,Sakkinen15b,Karlsson20} which has recently been extended to the description of non-equilibrium real-time dynamics.\cite{Schuler16,Karlsson21,Pavlyukh21a,Pavlyukh21b} A related method based on the Bogoliubov-Born-Green-Kirkwood-Yvon hierarchy for the correlation 
matrices has recently been applied to the bosonic excitation of the cavity field in cavity QED.\cite{Hoffmann19}
We also mention parallel efforts with two-component density-functional theory,\cite{Bostrom19} density-matrix embedding theory,\cite{Reinhard19} and quantum Monte-Carlo methods.\cite{weber2021field,weber2021realtime} Combinations of Ehrenfest and surface-hopping dynamics with time-dependent density functional theory are popular as well.\cite{Tavernelli2005,Curchod2013,Pela2022}

For condensed matter problems, numerically exact solutions for the real-time dynamics of electron-phonon coupled systems based on the matrix-product representation of many-body states 
have started to emerge.\cite{Brockt15,Kloss19,Stolpp2020} 
Such methods can efficiently handle large bosonic Hilbert spaces.
This effort to develop efficient methods for (quasi) one-dimensional (1D) coupled electron-nuclear models based on 
the density matrix renormalization group (DMRG)\cite{White92,Schollwock2005,Schollwock2011} is ongoing.\cite{Brockt15,wall_16,Kloss19,Stolpp2020,Kohler21,Stolppkoehler2020} 
Specific methods are the pseudosite DMRG method,\cite{Jeckelmann1998} DMRG with local basis optimization (DMRG-LBO),\cite{Zhang1998,zhang99,Guo2012} and the projected purified DMRG.\cite{Kohler21} 
DMRG methods use an efficient matrix-product state (MPS) representation of the truncated wave function and DMRG-LBO adds the determination of an optimal basis for the local degrees of freedom obtained from diagonalizing local reduced density matrices. This optimal basis can, in many cases, be truncated with a negligible error, thereby making many algorithms computationally more efficient.\cite{Zhang1998,zhang99,Friedman2000,Wong08,Guo2012,Brockt15,Stolpp2020,Jansen2020,Jansen_Jooss_2021}

Naturally, one can ask if the approximate methods devised for quantum chemistry could be applied in condensed matter\cite{Horsfield_2006,Wang2016,Chen2016,Smith_2019} and vice versa.\cite{Mardazad21} 
To answer such questions, benchmarks, such as the ones that have been carried out in the context of cavity QED\cite{Hoffmann19} or the spin-boson model,\cite{Chen2016,Stock2005} are desirable. We also mention a recent comparative study of several quantum-chemistry methods performed in large chromophores.\cite{Freixas2021}
Applications of surface-hopping algorithms in extended condensed matter systems have started to appear, see Ref.~\onlinecite{Wang2020} and references therein. 
However, there is a need for systematic studies comparing such independent-trajectory methods to unbiased numerically exact results in extended condensed matter systems. One recent effort along these lines has been presented in Ref.~\onlinecite{Krotz2022}.
Therefore, in our work, we study the real-time dynamics 
in a paradigmatic condensed matter system, the Holstein model,\cite{Holstein1959a} with methods of quantum chemistry, specifically the Ehrenfest and fewest-switches 
surface-hopping algorithms (FSSH),\cite{Tully1990,Tully1998,Gonzalez20}
and the multiconfigurational Ehrenfest (MCE) algorithm\cite{Shalashilin2009,Shalashilin2010} and we benchmark them against exact diagonalization (ED) for small systems and DMRG-LBO \cite{Brockt15,Stolpp2020}
for large systems. We note that the multiconfigurational Ehrenfest method has recently been compared against the hierarchy equation of motion method\cite{Tanimura1989,Chen15} and the multiple Davydov D2 ansatz\cite{Zhou15} in a similar model for up to 16 sites.\cite{Chen2019}

The Holstein model \cite{Holstein1959a} is one of the prototypical systems to describe the formation of polarons, \cite{Pekar1954,Feynman1955} 
which were originally thought of as electrons that cause distortions in their surrounding polar lattice, now  broadly understood as electronic quasi-particles. 
The key ingredients of the Holstein model are  the local interaction of electrons and Einstein phonons. The model consists of one  harmonic oscillator 
on every lattice site, which is bi-linearly coupled to the electronic density on that site. The only coupling between lattice sites and hence oscillators originates from the electronic
hopping between sites, which is often restricted to nearest-neighbor hopping.  Consequently, the oscillators interact only indirectly via electrons. 
Despite its simplicity, the Holstein model has been used to describe polaronic signatures in materials.\cite{Franchini21} 
Apart from the polarons, the Holstein model can also host a 
Peierls-type lattice instability,\cite{peirls_55} the so-called charge-density wave (CDW) phase at half filling. \cite{Hirsch83,Scalletar89,Noack91,Noack92,bursill_98,creffield_05,Bradley21,Araujo21}
By tuning parameters of the 1D Holstein model, one can predict a transition between the CDW and Luttinger-liquid metallic phases.\cite{bursill_98,creffield_05} 
The CDW-to-metallic transition has recently been observed in experiments.\cite{Kang21} 
The physics of polarons and the CDW-to-metallic transition described in the Holstein model
generally involve a strong coupling between oscillators and electrons, for whose description non-perturbative theoretical methods are necessary even in equilibrium, see
Ref.~\onlinecite{Franchini21} and references therein.

\begin{figure}[t]
	\includegraphics[width=8.4cm]{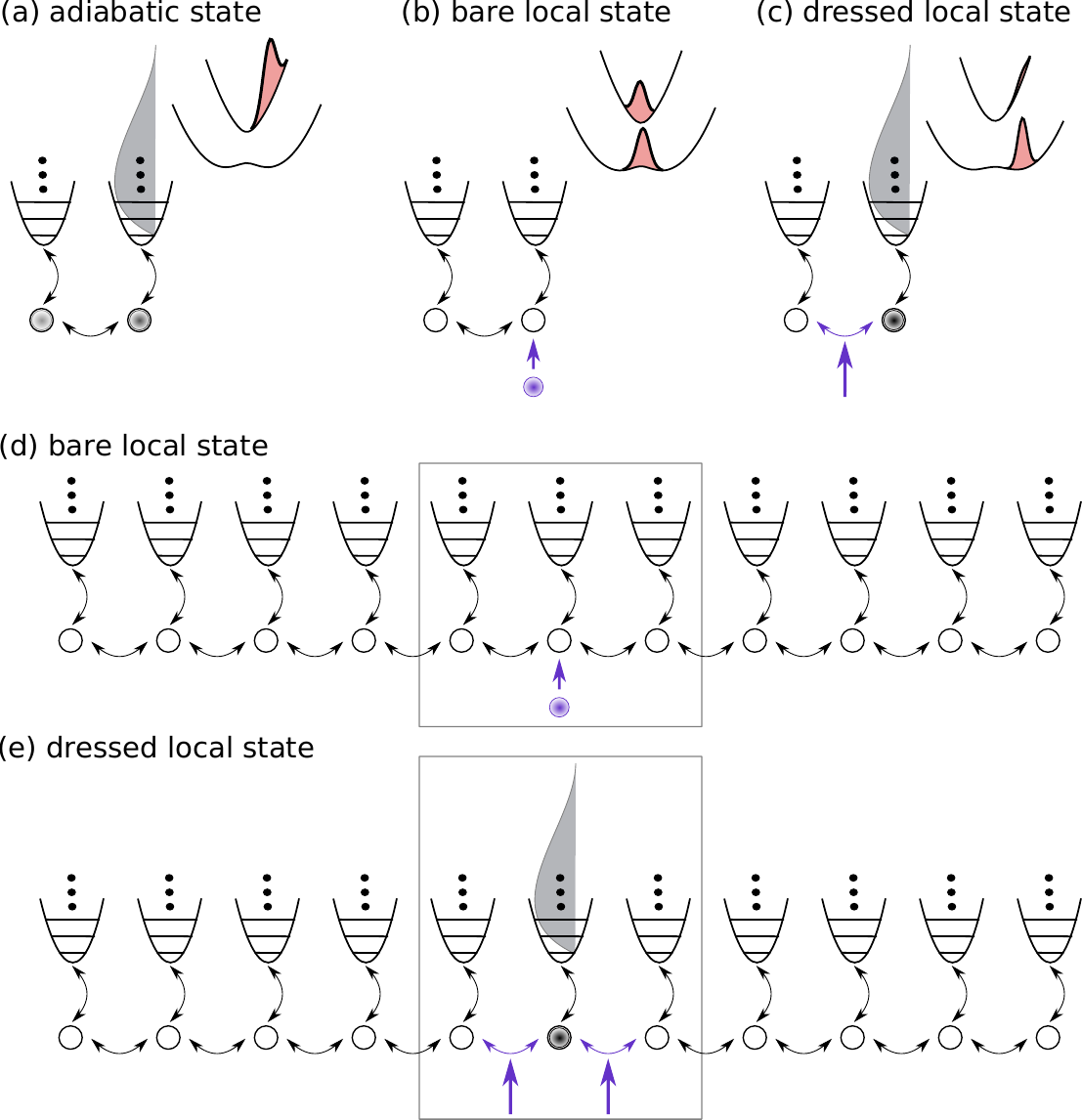}
	\caption{Examples of initial conditions considered in this work: (a)--(c) the two-site Holstein model and (d),(e) the extended Holstein model.
		For the two-site Holstein model, we also sketch the initial nuclear wave function densities on the potential energy (Born-Oppenheimer) surfaces in the Born-Huang formalism. (a) The adiabatic initial state has contributions only on the upper Born-Oppenheimer surface, (b) the bare local state represents the addition of an electron on one of the sites
		of an empty Holstein dimer, and (c) the dressed local state represents a quench of the hopping matrix element between the empty site and a local polaron on the other site. 
		(d) This bare local state represents an electron added to an empty Holstein lattice and addresses the formation of a polaron. 
		(e) The dressed local initial state represents a polaron localized to one site, which can be understood again as a quench of the hopping parameter. We note that conditions (d) and (e) 
		have been studied in Refs.$~$\protect\onlinecite{Kloss19,Pavlyukh21b}.
		The rectangles denote the reduction to the three-site Holstein model, for which the Born-Huang formalism in terms of potential energy surfaces is still practically tractable.}
	\label{Fig:Introduction:States_Sketch}
\end{figure}
To connect our model to the concepts of quantum chemistry, we first consider the two-site Holstein model,\cite{Ranninger92,Ranninger92,deMello97,Firsov97,Chatterjee00,Hakioglu00,Rongsheng02,QingBao05,Paganelli08,Paganelli08a,Zhang09}
see Figs.~\ref{Fig:Introduction:States_Sketch}(a)-(c), as 
it can be seen as a limit of both the spin-boson model and the Shin-Metiu model, i.e., the quantum-chemistry models mentioned above.
For the Shin-Metiu problem, one arrives at the two-site Holstein model by taking its first two Born-Oppenheimer surfaces into account, leading to a similar Born-Oppenheimer
Hamiltonian as for the Holstein model.\cite{Agostini19}  For the spin-boson model, the two-site Holstein model is obtained by limiting the number of bath oscillators to one, leading to the so-called one-mode spin-boson model.\cite{Sato18}  Moreover the two-site Holstein model can also be seen as the simplest appropriate model system in which two
diabatic states, representing reactants and products, are coupled through a single harmonic oscillator.\cite{McKemmish15,Reimers15}
It is thus natural to start with the two-site Holstein model and discuss the non-adiabatic dynamics from the perspective of the Born-Huang formalism
as it is common in quantum chemistry. 

For the two-site Holstein model, we consider both initial states that have contributions only on a single Born-Oppenheimer surface, see Fig.~\ref{Fig:Introduction:States_Sketch}(a), and local initial states, see Figs.~\ref{Fig:Introduction:States_Sketch}(b) and (c).
The former is typically used in quantum chemistry to study the dynamics of a wave
packet near an avoided crossing. Since the wave function at each nuclear position has contributions only in a single (adiabatic) Born-Oppenheimer state, we also call this an adiabatic initial state.
The latter two have recently been studied in condensed-matter-theory studies, for example, for the  Holstein model, see Refs.$~$\onlinecite{Kloss19,Pavlyukh21b}. 
The local initial conditions start in a coherent superposition of different adiabatic states, which poses a challenge for fewest-switches surface hopping.\cite{Subotnik2016}
We note that both, the adiabatic initial state which has contributions on a single Born-Oppenheimer surface and the local states, are idealized initial states and a generic initial state realized in an experiment, e.g., after an optical excitation, is probably in between the two initial conditions.
Even though idealized, the nuclear wave functions of all initial conditions can be exactly represented as well-defined probability distributions in phase space for the Ehrenfest and surface-hopping trajectory methods. In this way, we test the inherent approximations of the dynamics alone, and not approximations of the nuclear initial state.
The findings from the two-site Holstein model will help us to 
interpret the dynamics in larger Holstein chains, where the exact Born-Huang formalism is not practically tractable.

\begin{figure}[t]
	\includegraphics[width=8.4cm]{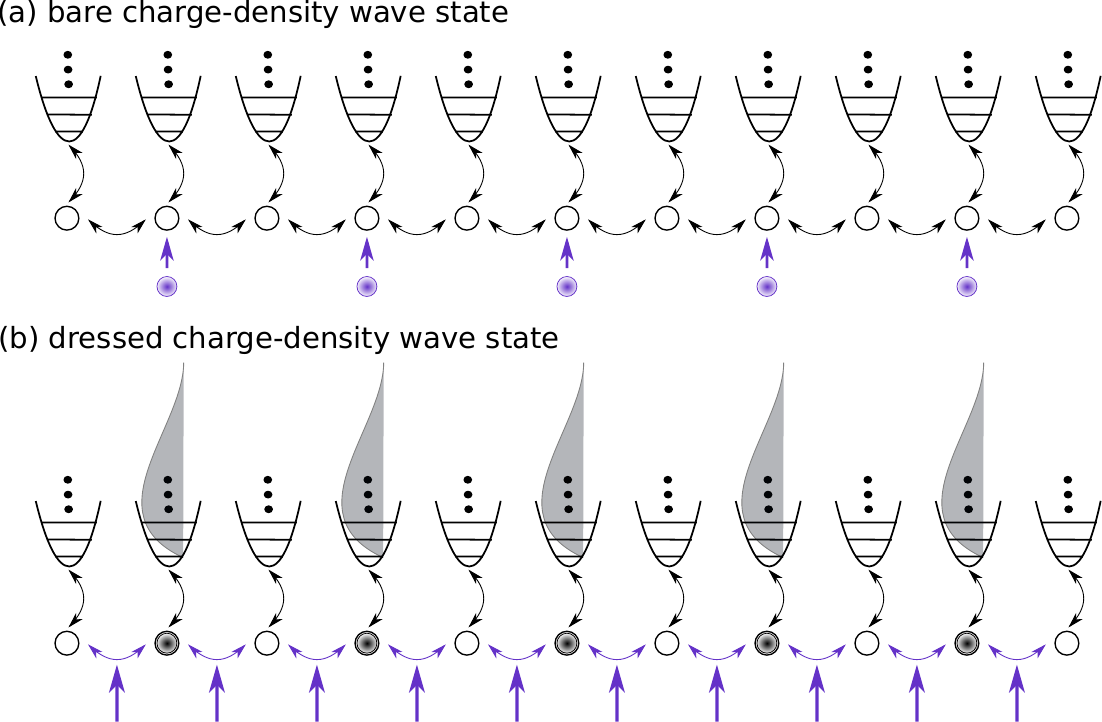}
	\caption{Examples of charge-density wave initial conditions considered in this work for the  Holstein chains mimicking ideal CDW orders. 
		(a) The bare CDW represents the addition of electrons on every second site. 
		(b) The dressed CDW represents  maximally localized  polarons placed on every second site. We note that conditions (a) and (b) 
		have been studied in Refs.$~$\protect\onlinecite{Stolpp2020} using DMRG-LBO.
	}
	\label{Fig:Introduction:States_Sketch2}
\end{figure}

In the Holstein chain, the coupling between bosons and electrons can lead to strong non-adiabatic effects, similar to those in the two-site Holstein model.
Thus,  out-of-equilibrium dynamics in the Holstein model constitute a challenging problem, for which several methods have been developed to describe various aspects of the model.
A straight-forward way to solve the problem is to use exact diagonalization,\cite{zhang99,capone_97,Ku07,Golez12,jansen19} however, one quickly reaches memory limits
due to the huge Hilbert space that needs to be considered. Therefore, one has to resort to efficient methods that seek to determine the relevant part of the Hilbert space such as diagonalization in a limited functional space,\cite{bonca99,golez12a,dorfner_vidmar_15} its recent extension,\cite{Kessing2021} and tailored implementations of the Lanczos method.\cite{wellein96,wellein98,Vidmar11,bonca2019} 
These methods can still describe only moderately large system sizes.  To increase the system sizes even further, matrix-product-state (e.g., the  density matrix renormalization group) 
methods\cite{Brockt15,shroeder_16,Brockt_17,Kloss19,Stolpp2020,Jansen2020} adapted to the large phononic Hilbert space can be used. 
These methods allow us to study both single polarons \cite{Kloss19,Jansen2020} and charge-density wave states \cite{Stolpp2020} in systems with a finite electronic density.
For the polaron problem,  alternative quantum-chemistry methods are the hierarchical equation-of-motion method \cite{Tanimura1989,Chen15,Jankovic21} or the Davydov D$_1$  or D$_2$ ansatz.\cite{Sun2010,Luo11,Zhou15,Chen2019}  Recently, the multiconfigurational Ehrenfest method has been  applied to the polaron problem with promising results.\cite{Chen2019} 
In this work, we consider some of these methods and apply them to initial conditions that probe the relevant physics of the Holstein model, i.e., polaron physics in chains with up to 51 sites, 
see Figs.~\ref{Fig:Introduction:States_Sketch}(d) and (e), and CDW physics, see Figs.~\ref{Fig:Introduction:States_Sketch2}(a) and (b), in chains with 13 sites.

We provide a brief account of our main observations.
Starting with the two-site system (the Holstein dimer) and the three-site system (the Holstein trimer), we show that the multitrajectory Ehrenfest (MTE) method, which is the cheapest method to
implement, has serious drawbacks and in general, only the ultrashort time dynamics is correctly described. The surface-hopping method improves the description of 
non-adiabatic effects such as wave-packet splitting. 
Even though the coherences included in the local initial states are not well suited for the surface-hopping method, the approach still gives a better qualitative description than the multitrajectory Ehrenfest method for later times in most cases.
For the multiconfigurational Ehrenfest method, we demonstrate convergence to exact dynamics with an increasing number of configurations, confirming that the method is in principle exact. For instance, it is able to
describe quantum tunneling processes. 

For the polaron problem in Holstein chains with up to 51 sites, the multitrajectory Ehrenfest method is incapable of sufficiently relaxing excitations back  to the lower energy surfaces and cannot describe the local trapping of the charge carrier. Contrarily, the surface-hopping method 
yields a significantly better agreement with DMRG than MTE and 
certain physics in the long-time behavior such as local trapping is captured qualitatively by this method.
We observe a slower  convergence for the multiconfigurational Ehrenfest (MCE) method than for the dimer or trimer. 
In practical calculations using MCE, we are able to converge most observables only in the short-time dynamics, while at later times, some of the observables deviate from the results of the exact DMRG-LBO method. 

Finally, for the charge-density-wave physics, for which we consider multi-electron systems, we demonstrate the failure of the multitrajectory Ehrenfest dynamics to describe energy transfer and the evolution of the total phonon number after 
ultrashort dynamics by comparison to the exact DMRG-LBO method. In contrast, in an adiabatic parameter regime, the charge-density wave order parameter is reasonably well  
described by MTE.

The plan of this work is as follows. We introduce the 1D Holstein model in Sec.$~$\ref{Sec:Holstein}. We review the Born-Huang formalism in Sec.~\ref{Sec:Born-Huang}.
In Sec.~\ref{Sec:Exact}, we discuss exact numerical methods, the exact diagonalization in second quantization and in the Born-Huang basis, and the density matrix renormalization 
group. 
The independent-trajectory methods are discussed in Sec.~\ref{Sec:Trajectory}.
In Sec.~\ref{Sec:Multi}, we follow up with the multiconfigurational Ehrenfest method. We then present our results for
the two-site model in Sec.~\ref{Sec:Dimer} and for the three-site Holstein model in Sec.~\ref{Sec:Results:Trimer}. The polaron physics for Holstein chains  is considered in Sec.~\ref{Sec:One-Electron_Extended}. Finally,  CDW physics is covered  in Sec.~\ref{Sec:Many-Electron_Extended}. We conclude and give prospects in Sec.~\ref{Sec:Conclusions}.

	\section{The Holstein model\label{Sec:Holstein}}

\subsection{The Holstein chain\label{Sec:Model_General}}

In this work, we consider the Holstein model\cite{Holstein1959a} for the special case of one dimension.
The Hamiltonian in the occupation number formalism with the phonon creation and annihilation operators $\hat b_i^\dagger$ and $\hat b_i$ for site $i$ and the corresponding spinless electronic operators $\hat c_i^\dagger$ and $\hat c_i$ can be written as:
\begin{align}
\hat H = \sum_{i} \Bigg[ &-t_{0} (\hat c_i^\dagger \hat c_{i+1} + \hat c_{i+1}^\dagger \hat c_i)
+ \hbar \omega_{0} (\hat b_i^\dagger \hat b_i +1/2) \nonumber\\
&- \gamma \hat n_i \left( \hat b_i^\dagger + \hat b_i  \right)\Bigg],\label{Eq:Holstein_model:Hamiltonian:2ndQ}
\end{align}
with $\hat n_i=\hat c_i^\dagger \hat c_i$.
Here, we can identify the three fundamental model parameters: the hopping matrix element $t_{0}$, the phonon frequency $\omega_{0}$, and the electron-phonon coupling $\gamma$.

It is instructive to transform the phonon operators to their position $\hat x_i$ and momentum $\hat p_i$ operators to get real-space representations for the solutions to this model. For this purpose, we insert the definition of the phonon ladder operators used above: $\hat b_i=\sqrt{\frac{m \omega_{0}}{2\hbar}} \left( \hat x_i + \frac{i}{m \omega_{0}} \hat p_i \right)$ (and $\hat b_i^\dagger$ accordingly), with the nuclear mass $m$, to write the Hamiltonian (\ref{Eq:Holstein_model:Hamiltonian:2ndQ}) equivalently as:
\begin{align}
\hat H= &\sum_{i} \Bigg[-t_{0} (\hat c_i^\dagger \hat c_{i+1} + \hat c_{i+1}^\dagger \hat c_i)
+ \frac{m \omega_{0}^2}{2} \hat x_i^2 + \frac{1}{2 m} \hat p_i^2\nonumber\\
&\phantom{\sum_i \Bigg[\,}- \sqrt{\frac{2 m \omega_{0}}{\hbar}} \gamma \hat x_i \hat n_i \Bigg]. \label{Eq:Holstein_model:Hamiltonian:1stQ}
\end{align}

Sometimes, in this work, we use a simplified dimensionless notation, which will always be indicated by a bar over a symbol: $\bar X$. The dimensionless variables are defined by setting $\hbar \omega_{0}$ as the energy unit: $\hat{\bar H}=\frac{\hat H}{\hbar \omega_{0}}$, $\bar{t}_{0}=\frac{t_{0}}{\hbar \omega_{0}}$ and  $\bar{\gamma}=\frac{\gamma}{\hbar \omega_{0}}$, and by using the natural length scale for the harmonic oscillators: $\hat{\bar x}_i=\hat x_i/\sqrt{\frac{\hbar}{m\omega_{0}}}$.

\subsection{Holstein dimer and trimer\label{Sec:Holstein_dimer_trimer}}

When visualizing potential energy surfaces later in this work, but also for the numerically exact calculation of eigenstates of the system, it is useful to reduce the number of phonon degrees of freedom for the small Holstein systems. 

For this, the phonon coordinates of the dimer Hamiltonian can be transformed into a relative and a center-of-mass coordinate: $\hat{\bar q} = (\hat{\bar x}_1-\hat{\bar x}_2)/\sqrt{2}$ and $\hat{\bar Q} = (\hat{\bar x}_1+\hat{\bar x}_2)/\sqrt{2}$ to obtain
\begin{align}
\hat{\bar H}_{Dimer} &= -\bar{t}_{0} (\hat c_1^\dagger \hat c_{2} + \hat c_{2}^\dagger \hat c_1)
+ \frac{ \hat{\bar Q}^2}{2} + \frac{\hat{\bar p}_Q^2}{2}  + \frac{ \hat{\bar q}^2}{2} + \frac{\hat{\bar p}_q^2}{2} \nonumber\\
&-\bar{\gamma} [\hat{\bar q} (\hat n_1 - \hat n_2) +\hat{\bar Q} (\hat n_1 + \hat n_2)]. \label{Eq:Model_Dimer_Trimer:Hdimer}
\end{align}
If we keep the total electron density constant at $\braket{\hat n_1}+\braket{\hat n_2}=const.$, the equation for the center-of-mass coordinate $\bar Q$ is a simple harmonic oscillator and independent of the rest of the system.
We are then left with a single phonon coordinate $\bar q$:
\begin{align}
\hat{\bar H}_{\bar{q}} = -\bar{t}_{0} (\hat c_1^\dagger \hat c_{2} + \hat c_{2}^\dagger \hat c_1) + \frac{ \hat{\bar q}^2}{2} + \frac{\hat{\bar p}_q^2}{2}-\bar{\gamma} \hat{\bar q} (\hat n_1 - \hat n_2).
\label{Eq:Model_Dimer_Trimer:Hdimer_reduced}
\end{align}

The center-of-mass coordinate can be removed for the Holstein model of any size, but we do this explicitly only for the dimer and trimer $(L=3)$. For the trimer, the phonon coordinates can be transformed into: 
\begin{align}
\hat{\bar X} &= \frac{1}{\sqrt{3}} \left(\hat{\bar x}_1 + \hat{\bar x}_2 + \hat{\bar x}_3\right) \nonumber \\
\hat{\bar x}_s &= \frac{2}{\sqrt{6}} \left( \hat{\bar x}_2 - \frac{\hat{\bar x}_1 + \hat{\bar x}_3}{2} \right)\label{Eq:Model_Dimer_Trimer:Trimer_phonon_transform} \\
\hat{\bar x}_a &= \frac{1}{\sqrt{2}} \left(\hat{\bar x}_3 - \hat{\bar x}_1\right). \nonumber
\end{align}
Here, $\hat{\bar X}$ is a center-of-mass coordinate, which, like in the Holstein dimer, does not couple to the rest of the system if we have a constant electron number. $\hat{\bar x}_s$ is a symmetric phonon mode around the central site and couples to the difference of the electron population on the central site and the average population on the edge sites. Finally, $\hat{\bar x}_a$ is an anti-symmetric phonon mode, which couples to the difference of the electronic populations on the edge Holstein sites ($\hat{\bar x}_1$ and $\hat{\bar x}_3$).

	\section{The Born-Huang formalism\label{Sec:Born-Huang}}

\subsection{General recapitulation of the Born-Huang approach\label{Sec:BH:recap}}

In this section, we provide a short recapitulation of the Born-Huang approach,\cite{BornHuang1954} which can be described as an expansion of the wave function in adiabatic electronic eigenstates. 
In this Born-Huang formalism we can understand the notion of non-adiabatic effects, which can lead to transitions between different adiabatic electronic states. 

We start by expanding the total state $\ket{\Psi}$ of the system in nuclear and electron basis states: $\ket{R}$ and $\ket{\phi_a(R)}$, where the nuclear state is defined by its (many-particle) position vector $R$, and the electron state, which might depend on the nuclear position, is labeled by a general (many-particle) index $a$:
\begin{align}
\ket{\Psi} = \int \textup{d}R\ \sum_a \ket{R,\phi_a(R)} \Psi_a(R) .\label{Eq:BH:WFexpansion}
\end{align}

The idea of the Born-Huang approach is to use a special set of electronic basis states: the eigenstates of the Born-Oppenheimer\cite{Born1927} (BO) Hamiltonian $\hat H^{BO}$, which is obtained from the total Hamiltonian $\hat{H}$ by removing the kinetic energy term of the phonons:
\begin{align}
&\hat{H}=\sum_k \frac{\hat P^2_k}{2 m_k} + \hat{H}^{BO}(\hat R)\nonumber\\
&\hat{H}^{BO}(R) \ket{\phi^{BO}_a(R)} = {E}^{BO}_a(R) \ket{\phi^{BO}_a(R)}.\label{Eq:BH:BO-eigenequation}
\end{align}
Here, $\hat{P}_{k}$ and $m_k$ are the nuclear momentum operators and nuclear masses. Note that in the second line of Eq.~\eqref{Eq:BH:BO-eigenequation} the Born-Oppenheimer Hamiltonian is a purely electronic operator: $\hat{H}^{BO}(R)=\braket{R|\hat H^{BO}(\hat R)|R}$. Its electronic eigenstates, called Born-Oppenheimer states, form a complete basis for every $R$.

Using the Born-Oppenheimer states $\ket{\phi_a^{BO}(R)}$ as electronic basis, the time-dependent Schr\"odinger equation for the wave function $\Psi_a(R)$ of the full system can be written as:
\begin{align}
i \hbar \frac{\partial}{\partial t}\Psi_a(R) = \sum_b \Bigg[\sum_k \frac{1}{2 m_k} \Bigg(\mathbf{1} \frac{\hbar}{i}\nabla_k + \mathbf{{A}}_{(k)}(R)\Bigg)^2& \nonumber\\
+ \mathbf{{E}}^{BO}(R)&\Bigg]_{a,b} \Psi_b(R), \label{Eq:BO:NucSchrEq}
\end{align}
where bold-face symbols represent matrices in the basis of Born-Oppenheimer states, with $\mathbf{{E}}^{BO}(R)$ being the diagonal matrix containing the Born-Oppenheimer energies, $\mathbf{1}$ the unit matrix and $\mathbf{{A}}_{(k)}(R)$ a new contribution called the derivative couplings:
\begin{align}
{A}_{a,b,(k)}(R)=\braket{\phi_a^{BO}({R})|\frac{\hbar}{i}\nabla_k|\phi_b^{BO}({R})}. \label{Eq:BO:DerivCoup}
\end{align}
We note that the derivative couplings are often defined without the factor $\hbar/i$.
The Schr\"odinger equation (\ref{Eq:BO:NucSchrEq}) takes the form of a wave function evolving in a set of potential energy surfaces ${E}_a^{BO}(R)$, also called Born-Oppenheimer surfaces, which are coupled via a vector potential $\mathbf{{A}}_{(k)}(R)$. 

The Born-Oppenheimer basis is also called the adiabatic basis. This is rooted in the adiabatic approximation, or sometimes called Born-Oppenheimer approximation, which amounts to neglecting the derivative couplings in Eq.~\eqref{Eq:BO:NucSchrEq}.\cite{Tully2000,Worth2004b} Within this approximation, the wave function evolves independently on all potential energy surfaces $E_a^{BO}(R)$ and the Born-Oppenheimer states $\ket{\phi^{BO}_a(R)}$ are treated like electronic eigenstates of the system. The methods studied in this work go beyond the adiabatic approximation to account for the influence of the derivative couplings on the wave function dynamics. 
Any effect induced by the derivative couplings is called a \textit{non-adiabatic} effect, and they often couple different adiabatic (Born-Oppenheimer) states. 

These non-adiabatic effects become especially important when two Born-Oppenheimer surfaces $E_a^{BO}(R)$ and $E_b^{BO}(R)$ come close in energy, as can be seen from a different representation of the derivative couplings: 
\begin{align}
{A}_{a,b,(k)}(R)=\frac{\braket{\phi_a^{BO}(R) | \left(\frac{\hbar}{i}\nabla_k \hat H^{BO}(R)\right) | \phi_b^{BO}(R)}}{E_a^{BO}(R) - E_b^{BO}(R)},\quad a\ne b. \label{Eq:BH:DerivCoup_Energy_expression}
\end{align}

Instead of an adiabatic electronic basis, which diagonalizes the Born-Oppenheimer Hamiltonian, one could also choose an electronic basis in which the derivative couplings defined in Eq.~\eqref{Eq:BO:DerivCoup} vanish. Such a basis is called a \textit{diabatic} electronic basis.\cite{Lichten1963,Smith1969,Baer1975} The obvious choice is a phonon-independent basis, in the following denoted as $\ket{\chi_a}$, which will also be used later in one of our implementations of exact diagonalization and the density matrix renormalization group. Note that such a phonon-independent, ''trivial diabatic`` basis is, in general, the only strictly diabatic basis.\cite{Mead1982}
In this work, when we refer to diabatic basis states, we always consider them as phonon-independent, or at least with a negligible $R$-dependence.

For later reference, the expansion of the state of the system in both the adiabatic Born-Oppenheimer and the diabatic phonon-independent basis is stated explicitly: 
\begin{align}
\ket{\Psi} 	&= \int \textup{d}R\ \sum_a \ket{R,\phi^{BO}_a(R)} \Psi^{(a)}_a(R)\label{Eq:BH:WFexpansion:adiabatic}\\
&= \int \textup{d}R\ \sum_a \ket{R,\chi_a} \Psi^{(d)}_a(R)\label{Eq:BH:WFexpansion:diabatic}.
\end{align}
Here, the superscripts $(a)$ and $(d)$ of the wave function refer to the adiabatic and diabatic basis, respectively, and the index $a$ refers to the Born-Oppenheimer and diabatic electronic states in the two cases. We can understand the $\Psi_a(R)$ as a multi-component nuclear wave function. 

Different diabatic states are coupled by the off-diagonal elements of the Born-Oppenheimer Hamiltonian: $V_{a,b}=\braket{\chi_a|\hat H^{BO}|\chi_b}$, which usually change smoother with respect to variations in the nuclear coordinates than the derivative couplings in an adiabatic basis,\cite{Baer1975,Smith1969,Van_voorhis2010,Koppel1984} thus in some cases leading to a more stable numerical integration, see also Sec.~\ref{Sec:FSSH:avoid_deriv_coup}. 
Then again, the peaked structure of the derivative couplings in the adiabatic basis allows the identification of strong non-adiabatic coupling regions in nuclear configuration space $R$ and other regions where the adiabatic surfaces are mostly isolated. This insight can be used to motivate approximations and restrictions to certain adiabatic states, which is not as easily obtained from a diabatic basis. 

\subsection{Born-Huang approach for the Holstein dimer\label{Sec:BH-Dimer}}

The Born-Oppenheimer Hamiltonian of the Holstein dimer $\hat {\bar{H}}^{BO}_{\bar q}$ (see Eq.~\eqref{Eq:Model_Dimer_Trimer:Hdimer_reduced}) can be diagonalized analytically (see, e.g., Ref.~\onlinecite{McKemmish15}), for which we will assume a system with exactly one electron. In the basis of the two site-local electronic states $\ket{\chi_{1}}$ and $\ket{\chi_2}$ (where $\ket{\chi_{i}}=c^\dagger_{i} \ket{0}$), it becomes:
\begin{align}
\hat{\bar{H}}^{BO}(\bar{q}) = \begin{pmatrix}
\frac{\bar{q}^2}{2} -\bar{\gamma}\bar{q} & -\bar{t}_0 \\
-\bar{t}_0  & \frac{\bar{q}^2}{2} +\bar{\gamma}\bar{q}
\end{pmatrix},
\end{align}
which we solve for the Born-Oppenheimer eigenenergies:
\begin{align}
\bar{E}^{BO}_\pm(\bar{q}) =\frac{\bar{q}^2}{2} \pm \sqrt{ \bar{q}^2 \bar{\gamma}^2 + \bar{t}_0^2} \label{Eq:BO:Energies}
\end{align}
and the corresponding eigenstates:
\begin{align}
\begin{pmatrix}
\phi^{BO}_{\pm,1}(\bar{q}) \\
\phi^{BO}_{\pm,2}(\bar{q})
\end{pmatrix}
=
\begin{pmatrix}
\left. \sqrt{1+\left(\frac{\bar{q} \bar{\gamma}}{\bar{t}_0} \pm \sqrt{\left(\frac{\bar{q} \bar{\gamma}}{\bar{t}_0}\right)^2+1}\right)^2} \right.^{-1} \\
\mp \left. \sqrt{1+\left(\frac{\bar{q} \bar{\gamma}}{\bar{t}_0} \mp \sqrt{\left(\frac{\bar{q} \bar{\gamma}}{\bar{t}_0}\right)^2+1}\right)^{2}} \right.^{-1}
\end{pmatrix}.\label{Eq:BH-Dimer:WF}
\end{align}

\begin{figure}[t]
	\includegraphics[width=8.5cm]{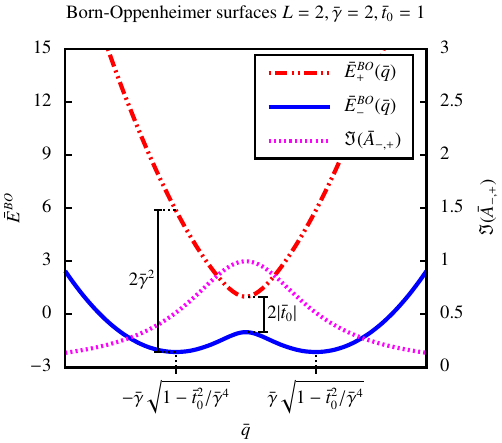}
	\caption{Born-Oppenheimer surfaces of the Holstein dimer according to Eq.\ (\ref{Eq:BO:Energies}) and the imaginary part of the derivative couplings from Eq.\ (\ref{Eq:BO:DimerDerivCoup}) for the parameters $\bar{\gamma}=2$ and $\bar{t}_0=1$. The splitting of the surfaces is $2\bar{t}_0$ at $\bar{q}=0$, while at the minimum of the lower surface ($\bar{q}=\pm\sqrt{\bar{\gamma}^2-\bar{t}_0^2/\bar{\gamma}^2}$), the difference in Born-Oppenheimer energies amounts to $2\bar{\gamma}^2$ (for $\frac{\bar{\gamma}^2}{\bar{t}_0}>1$).}\label{Fig:BO:BOSurfaces_and_DerivCoup}
\end{figure}

The Born-Oppenheimer surfaces $\bar{E}^{BO}(\bar{q})$ of the Holstein dimer are shown in Fig. \ref{Fig:BO:BOSurfaces_and_DerivCoup} for $\frac{\bar{\gamma}^2}{\bar{t}_0}>1$. 
For $\frac{\bar{\gamma}^2}{\bar{t}_0}<1$, the lower surface has only a single minimum at $\bar q=0$, a case studied, for example, in Refs.~\onlinecite{Spencer2016,Giannini2022}. This regime is not studied in this work, but is of interest for future investigations.

In the case of a slowly moving nuclear wave function, we can expect the system to evolve along these potential energy surfaces as long as they are sufficiently separated, i.e., for $q$ far away from zero. The non-adiabatic effects can be quantified by calculating the derivative couplings $\bar A$ via Eq.\ (\ref{Eq:BO:DerivCoup}). The diagonal elements vanish and we obtain for the off-diagonal elements:\cite{Landry2011,Landry2012}
\begin{align}
\bar{A}_{+,-} = \bar{A}^*_{-,+}=-i\frac{\bar{\gamma}}{2 \bar{t}_0} \frac{1}{1+\left(\bar{q}\bar{\gamma}/ \bar{t}_{0}\right)^2 }. \label{Eq:BO:DimerDerivCoup}
\end{align}
They are also depicted in Fig.~\ref{Fig:BO:BOSurfaces_and_DerivCoup}.
We see that the coupling is peaked with a Lorentzian curve around $\bar q{=}0$. The curve becomes more localized as $\bar{\gamma}/\bar{t}_0$ becomes large, i.e., a small hopping matrix element between the sites will lead to highly peaked derivative couplings around the avoided crossing point.\cite{Wang2020}
The easiest estimate for non-adiabatic transitions between the Born-Oppenheimer surfaces is the Landau-Zener formula,\cite{Landau1932,Zener1932} that assumes a predefined classical nuclear path $q(t)$ which evolves with constant velocity $\dot q$ through the avoided crossing region. Starting on the lower Born-Oppenheimer surface at $q\approx-\infty$, the transition probability to the upper Born-Oppenheimer surface at $q\approx\infty$ is then dominated by the avoided crossing point and can be approximated as:\cite{Zener1932} 
\begin{align}
P_{-\rightarrow+}^{LZ}=\exp\left(\frac{-\pi |\bar t_0|^2}{\dot{\bar{q}} \bar \gamma}\right).\label{Eq:BO:LZ}
\end{align}
The Landau-Zener formula (\ref{Eq:BO:LZ}) illustrates that we can expect non-adiabatic effects to become relevant for a small hopping matrix element, i.e., for large derivative couplings at the crossing region $q\approx0$.


\subsection{Born-Huang approach for the Holstein trimer\label{Sec:BH-Trimer}}

Similar to the dimer, we can write the Born-Oppenheimer Hamiltonian of the Holstein trimer with one electron in the basis of the three local electronic states as:
\begin{align}
\hat{\bar H}^{BO}_{Trimer}(\bar x_s,\bar x_a) &= \frac{\bar x_s^2}{2} + \frac{\bar x_a^2}{2} \label{Eq:BH-Trimer:BO_Hamiltonian}\\
&+ \begin{pmatrix}
\bar \gamma \left(\bar x_a + \frac{1}{\sqrt{3}} \bar x_s\right) & -\bar{t}_0 & 0 \\
-\bar{t}_0 & - \bar \gamma \frac{2}{\sqrt{3}} \bar x_s & -\bar{t}_0 \\
0 & -\bar{t}_0 & \bar \gamma \left(-\bar x_a + \frac{1}{\sqrt{3}} \bar x_s\right)
\end{pmatrix},\nonumber
\end{align}
where we again discarded the center-of-mass phonon coordinate. $\bar x_s$ and $\bar x_a$ are the symmetric and anti-symmetric phonon mode defined in Eq.~\eqref{Eq:Model_Dimer_Trimer:Trimer_phonon_transform}.

\begin{figure}[t]
	\includegraphics[width=8.5cm]{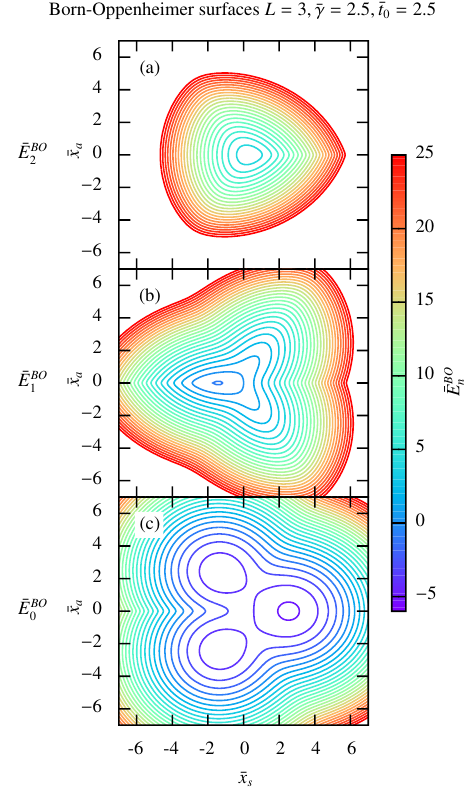}
	\caption{Contours of the Born-Oppenheimer surfaces of the Holstein trimer, obtained from the diagonalization of the Hamiltonian \eqref{Eq:BH-Trimer:BO_Hamiltonian} for $\bar \gamma=\bar t_0=2.5$ in the nuclear coordinate space of the symmetric $\bar x_s$ and anti-symmetric $\bar x_a$ phonon mode. We draw the contours for every $\Delta \bar E^{BO}_i=1$, for the (a) highest, (b) middle and (c) lowest Born-Oppenheimer surface.}\label{Fig:BH_Trimer:BO-contour}
\end{figure}

\begin{figure}[t]
	\includegraphics[width=8.5cm]{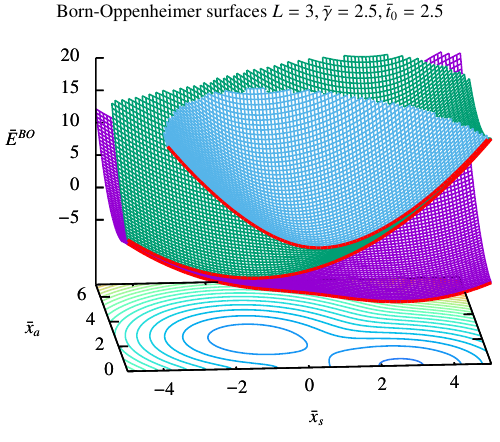}
	\caption{Born-Oppenheimer surfaces of the Holstein trimer, obtained from diagonalization of the Hamiltonian \eqref{Eq:BH-Trimer:BO_Hamiltonian} for $\bar \gamma=\bar t_0=2.5$ in the nuclear coordinate space of the symmetric $\bar x_s$ and anti-symmetric $\bar x_a$ phonon mode. Since the surfaces are invariant under the operation $\bar x_a \rightarrow -\bar x_a$, we show only non-negative values for the anti-symmetric mode. The lowest surface is drawn in magenta, the middle in green and the highest in blue. The Born-Oppenheimer energies along the symmetric phonon slice $\bar x_a=0$ are included as red lines. In the x-y-plane, a contour plot of the lowest (magenta) surface is included, corresponding to Fig.~\ref{Fig:BH_Trimer:BO-contour}(c).}\label{Fig:BH_Trimer:Trimer_Surf}
\end{figure}

The Born-Oppenheimer surfaces in the nuclear space spanned by $\bar x_s$ and $\bar x_a$ can be calculated analytically, or by numerical diagonalization of the Born-Oppenheimer Hamiltonian. We show contours of the Born-Oppenheimer surfaces calculated from diagonalization for $\bar \gamma=\bar t_0=2.5$ in Fig.~\ref{Fig:BH_Trimer:BO-contour}. 
The energy-separation of the different Born-Oppenheimer surfaces can be seen the best in a surface-plot, which is illustrated in Fig.~\ref{Fig:BH_Trimer:Trimer_Surf}. Since the surfaces are symmetric with respect to $\bar x_a$, we include only positive values for the anti-symmetric phonon mode in Fig.~\ref{Fig:BH_Trimer:Trimer_Surf}.

The lowest Born-Oppenheimer surface (Fig.~\ref{Fig:BH_Trimer:BO-contour}(c)) has three local minima, corresponding to the three different Holstein sites. The lowest (global) minimum is at positive $\bar x_s$ and at $\bar x_a=0$, i.e., here mainly the central site has a large phonon distortion. The other two minima correspond to large phonon distortions on the left and right edge sites of the Holstein trimer, respectively.
The higher Born-Oppenheimer surfaces (Figs.~\ref{Fig:BH_Trimer:BO-contour}(a),(b)) each have only one minimum, which for the highest surface, similar to the dimer case, is located at the symmetric point $\bar x_s=\bar x_a=0$.

Using the language of chemical bonds, we can associate the three Born-Oppenheimer states with a bonding, non-bonding, and antibonding state. Note that technically, the middle Born-Oppenheimer state is only truly non-bonding, with zero electronic population on the central site, for $\bar x_a=0$ (the red lines in Fig.~\ref{Fig:BH_Trimer:Trimer_Surf}). 
At the three minima of the lowest Born-Oppenheimer surface, the bonding state corresponds to the electron mostly localized to the site with a large nuclear coordinate $\bar x_i$, which one might consider as a small polaron state. 

In this work, we investigate symmetric initial states, with a phonon distribution around $\bar x_a =0$. The corresponding slice of the Born-Oppenheimer surfaces is included as red lines in Fig.~\ref{Fig:BH_Trimer:Trimer_Surf}. Here, the lowest (bonding) Born-Oppenheimer electronic wave function has the same sign on all three trimer sites, the middle (non-bonding) Born-Oppenheimer state has contributions only on the edge sites and changes sign between them, and the upper (antibonding) Born-Oppenheimer state changes its sign twice between the sites. The energy of the non-bonding state is near degenerate with the antibonding state for large positive $\bar x_s$, where the antibonding state has only little contribution on the central site. At the same time, for large negative values of the symmetric phonon mode, the non-bonding state becomes near degenerate with the bonding state.
These near degeneracies are lifted if the anti-symmetric phonon mode is shifted away from zero. 

The qualitative picture of these asymptotically approaching surfaces stays the same for any parameter choice with $\bar t_0$ and $\bar \gamma$ positive. In contrast to the Holstein dimer, we thus cannot achieve a complete separation of the Born-Oppenheimer surfaces by tuning the parameters of the system.

	\section{Exact diagonalization and density matrix renormalization group methods\label{Sec:Exact}}

\subsection{Exact diagonalization in second quantization\label{Sec:ED}}

In order to obtain exact results, we diagonalize the Hamiltonian in its second quantized form Eq.~\eqref{Eq:Holstein_model:Hamiltonian:2ndQ}. This is briefly reviewed here since the notation also appears in the description of the DMRG algorithm.  A thorough review on exact diagonalization and how it can be improved for larger systems can be found in Ref.~\onlinecite{Sandvik2010}.  
To diagonalize Eq.~\eqref{Eq:Holstein_model:Hamiltonian:2ndQ}, we first set up a basis for the Hilbert space.  For the dimer,  each state in the real-space occupation number basis takes the form $\ket{\alpha }=\ket{n^{e}_1,n^{ph}_{1},n^{e}_2, n^{ph}_2}$. We further truncate the local phonon number to $M$ so that $n_i^{ph}\in \{ 0, 1 ,\dots, M \}$.  With $n^e_i \in \{  0,1 \}$ and the constraint $n^{e}_1+n^{e}_2=1$, we have a total Hilbert-space dimension $D_{H}=L(M+1)^{L}$, where $L=2$ for the dimer. The complete sparse Hamiltonian can now easily be generated by iterating through the basis and determining which elements are nonzero $\bra{\alpha} \hat H \ket{\beta} \neq 0$. The Hamiltonian can then be diagonalized, and the time evolution of an initial state carried out exactly. In this work, exact diagonalization in second quantization is used for $L=2$ and $L=3$. 

\begin{figure}[t]
	\centering
	\includegraphics{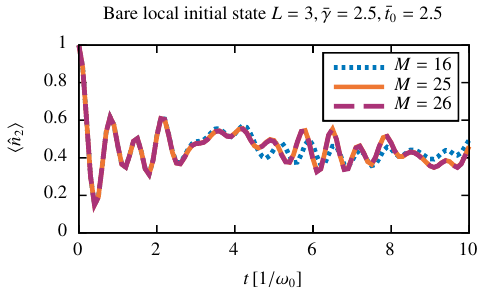}
	\caption{Convergence of exact diagonalization in second quantization with respect to the used local phonon cutoff $M$. We show the electron occupation on the central site for the Holstein trimer with $\bar\gamma=\bar t_0=2.5$, starting from the bare local initial state.\label{fig:trimer_conv_M}}
\end{figure}

Since we are working in the real-space phonon occupation-number basis, it is important to include enough phonons $M$ to capture the correct physics for a given set of parameters. In Fig.~\ref{fig:trimer_conv_M}, we illustrate this for the Holstein trimer. One sees that the observable deviates from its correct value if $M$ is chosen too small, as is the case for $M=16$. Once the data have been converged with respect to $M$, the expectation value should no longer change upon increasing $M$ further.

\subsection{Grid-based calculation of eigenstates in the Born-Oppenheimer basis\label{Sec:NucWFEigenstates}}

Another approach is to solve the eigenproblem of the Holstein dimer directly in the Born-Oppenheimer (adiabatic) electronic basis via the Schr\"odinger equation (\ref{Eq:BO:NucSchrEq}). This will lead to the same eigenenergies as in the exact diagonalization in second quantization method, but gives easier access to adiabatic observables, as for example the occupation of the different Born-Oppenheimer electronic states. 
Since the Holstein dimer has only one relevant nuclear coordinate, we can solve the Schr\"odinger equation  (\ref{Eq:BO:NucSchrEq}) directly on a grid and calculate the lowest $m$ eigenstates $\ket{\Psi^n},n=1,\dots, m$ of the system. For sufficiently large $m$, we can then represent any initial state in these eigenstates and time evolve the state to arbitrary long times.

Using the Born-Oppenheimer energies and derivative couplings of Eqs.~(\ref{Eq:BO:Energies}) and (\ref{Eq:BO:DimerDerivCoup}), we can write the time-independent version of the Schr\"odinger equation in the adiabatic electronic basis (\ref{Eq:BO:NucSchrEq}) as:
\begin{align}
\bar{E}^n \Psi^n_\pm(\bar{q},\bar{t})=&\phantom{+\ }\left[ -\frac{1}{2} \nabla_{\bar q}^2 + \frac{\bar{a}(\bar{q})^2}{2} + \bar{E}^{BO}_\pm(\bar{q}) \right] \Psi^n_\pm(\bar{q},\bar{t})\nonumber\\
&\mp \left[ \frac{1}{2} \frac{\partial a(\bar{q})}{\partial {\bar{q}}} +  \bar{a}(\bar{q}) \nabla_{\bar q} \right] \Psi^n_\mp(\bar{q},\bar{t}), \label{Eq:BO:DimerNucSchrEq}
\end{align}
where $\bar{a}(\bar{q})$ is the imaginary part of $\bar{A}_{-,+}$. 

There are many ways to solve the differential equations (\ref{Eq:BO:DimerNucSchrEq}). Here we use an iterative procedure to obtain the lowest $m$ eigenstates in a Car-Parrinello inspired scheme.\cite{Car1985} 

In this process, the two-component wave functions $\Psi^n_{\pm}(\bar q)$ time evolve with a fictitious kinetic energy on the instantaneous energy surfaces $\braket{\phi^{BO}_\pm(\bar q),\bar q|\hat{\bar{H}}|\Psi^n}$, discretized on a grid. For higher energy states, the wave functions and forces acting on them are orthogonalized to lower states, to ensure relaxation to eigenstates.
A friction term dampens the fictitious kinetic energy until the potential energy is minimized under the orthogonalization constraints. We use a Verlet-algorithm for the propagation in this relaxation process.\cite{Verlet1967} We finally end up with the lowest $m$ eigenstates, which can be used to represent the initial state and its time evolution at any later time.
One needs to ensure that enough eigenstates $m$ are used to accurately represent the desired initial state, which is the case if the norm of the projected state is close to 1. 

\subsection{Density matrix renormalization group methods\label{Sec:DMRG}}

\subsubsection{Brief review}
Exactly diagonalizing  Eq.~\eqref{Eq:Holstein_model:Hamiltonian:2ndQ} works well in many parameter regimes for the Holstein dimer and gives direct access to eigenstates and energies. However,  due to the exponential scaling of the Hilbert-space dimension with the system size ($\sim (M+1)^{L}$), the method quickly reaches its limits when larger systems and a large local phonon truncation $M$ are needed. Taking all symmetries of the model into consideration~\cite{Sandvik2010} only gives access to slightly larger systems. One type of methods, which has proven extremely powerful when dealing with one-dimensional systems, are density matrix renormalization group (DMRG) algorithms.  In this work,  we  describe the key ideas of DMRG, but there exist many comprehensive reviews, e.g., Refs.~\onlinecite{Schollwock2005, Schollwock2011}, and Ref.~\onlinecite{Paeckel2019} specifically for time evolution.

DMRG, though originally developed for the ground-state search,\cite{White92} has proven to be useful for a wide range of other applications such as time evolution\cite{Vidal2004,Daley2004,White04,Haegeman2011,Haegeman2016} and finite-temperature\cite{Verstrate2004,Sirker2005,Feiguin2005,Stoudenmire2010,Karrasch2012,Karrasch_2013} calculations.  Even though DMRG is primarily used in one dimension, there exist generalizations to two-dimensional systems,  see, e.g., Refs.~\onlinecite{Verstraete2004-2D,Stoudenmire2012,Orus2014,Zheng2017,Bruognolo2021}.

The key insight in DMRG is that in many situations,  the knowledge of the whole Hilbert space is not needed to capture the relevant physics. Rather, one  only  needs to focus on states in a small part of it which can be clearly identified based on the reduced-density matrices of the sub-system obtained by tracing out the remaining physical degrees of freedom.  In the eigenbasis of the reduced-density matrix of a system partitioned into subsystem $A$ and $B$,  the state $\ket{\psi}$ can be truncated to $\ket{\psi}_{\rm trunc}$ in a controlled way with a minimal distance ${||\ket{\psi}-\ket{\psi}_{\rm trunc} ||_2}$.
The reason why this approximation works is rooted in the sufficiently fast decay of reduced density-matrix eigenvalues in typical ground states of one-dimensional
systems. This, in turn, is related to the spatial entanglement encoded in many-body ground states, which under certain conditions obey an area law instead of a volume law, where the latter is typical in generic excited states.\cite{Schollwock2011,Eisert2010}

Such a bi-partitioned state on a one-dimensional lattice can be written as
\begin{equation}
\label{eq:bipartition_def}
\ket{\psi}=\sum\limits_{\vec{\sigma}_{i}, \vec{\sigma}_j} \psi^{\vec{\sigma}_{i} \vec{\sigma}_j}\ket{\vec{\sigma}_i}_A \ket{\vec{\sigma}_j}_B,
\end{equation}
where $\ket{\vec{\sigma}_i}_A=\ket{\sigma_1 ,  \sigma_2, \dots ,\sigma_w}_A$, $\ket{\vec{\sigma}_j}_B=\ket{\sigma_{w+1} ,  \sigma_{w+2}, \dots ,\sigma_L}_B$ and $\sigma_l$ corresponds to the physical state at site $l$.
Now,  one can perform a singular-value decomposition of $\psi^{\vec{\sigma_{i}}\vec{\sigma_{j}}}=\sum\limits_{\alpha,\beta }U^{\vec{\sigma_{i}}}_{\alpha} s_{\alpha\beta} V^{\dagger}{}_{\beta}^{ \vec{\sigma_{j}} }$.   $s_{\alpha \beta}$ are the $r$ non-zero singular values.  By rotating the states using $\ket{\alpha}_A=\sum\limits_{\vec{\sigma}_{i}}U^{\vec{\sigma_{i}}}_{\alpha} \ket{\vec{\sigma}_i}_A$ and  $\ket{\beta}_B=\sum\limits_{\vec{\sigma}_{j}}V^{\dagger}{}_{\beta}^{ \vec{\sigma_{j}}}\ket{\vec{\sigma}_j}_B$ and using the fact that  $s_{\alpha\beta}$ ($s_{\alpha}$ for brevity) is a diagonal matrix, one gets:
\begin{equation}
\ket{\psi}=\sum\limits_{\alpha=1}^{r} s_{\alpha } \ket{\alpha}_A \ket{\alpha}_B.
\end{equation}
After ordering the singular values $s_{\alpha}>0$ according to their magnitude such that $s_{1}\geq s_{2}\geq  s_3 \dots$, $\ket{\psi}$ can now be truncated by neglecting states, depending on their singular value $s_{\alpha}$.  For example, if there are $r$ singular values and one wants  to truncate at the dimension $m+1$, one cuts off all states belonging to $s_{\alpha}$ with $m<\alpha \leq r$.  The state can now be written as 
\begin{equation}
\label{def_psi_part_trunc}
\ket{\psi}_{\rm trunc}=\sum\limits_{\alpha=1}^{m} s_{\alpha}\ket{\alpha}_A \ket{\alpha}_B.
\end{equation}
The $s_{\alpha}^{2}$ are also the eigenvalues of the reduced density matrix of either subsystem and $||\ket{\psi}-\ket{\psi}_{\rm trunc} ||^2_2=\sum\limits_{\gamma=m+1}^{r}s_{\gamma}^{2}$.
We refer to this truncation as the standard DMRG truncation, affecting the so-called bond dimension, as opposed to the truncation of the local Hilbert space discussed next.

With this in mind, a state $\ket{\psi}$  can be represented as a matrix-product state (MPS) by iterating through the system and performing a series of singular-value decompositions. It  can then be written in its final form as a so-called mixed canonical MPS:
\begin{equation}
\label{MPS_def}
\ket{\psi}=\sum\limits_{\vec{\sigma}, \alpha_1,  \hdots, \alpha_{L-1} } A^{\sigma_1}_{\alpha_0 \alpha_1} \dots A^{\sigma_{i-1}}_{\alpha_{i-2} \alpha_{i-1}}M^{\sigma_i\sigma_{i+1}}_{\alpha_{i-1} \alpha_{i+1}}B^{\sigma_{i+2}}_{\alpha_{i+1} \alpha_{i+2}}\dots B^{\sigma_L}_{\alpha_{L-1} \alpha_L}\ket{\vec{\sigma}},
\end{equation}
where $A^{\sigma_j},B^{\sigma_l},M^{\sigma_i \sigma_{i+1}}$ are matrices with bond indices $\alpha_j$ and $\alpha_0=\alpha_L=1$,  $\ket{\vec{\sigma}}=\ket{\sigma_1 ,  \sigma_2, \dots ,\sigma_L}$ are the state vectors, and the $\sigma_i$ are the local degrees of freedom.  For the Holstein model,  we have $\ket{\vec{\sigma}}=\ket{n^{ \rm e}_1 n^{\rm ph}_1 , n^{ \rm e}_2 n^{\rm ph}_2, \dots ,n^{\rm e}_L n^{\rm ph}_L }$. The $A$ ($B$) matrices are left (right) normalized,  meaning that  $\sum\limits_{\sigma_i}A^{\dagger}{}^{\sigma^i}A^{\sigma^i}=\mathbb{I}$ ( $\sum\limits_{\sigma_i}B^{\sigma^i}B^{\dagger}{}^{\sigma^i}=\mathbb{I}$), where $\mathbb{I}$ is the identity matrix. This representation is exact if no truncation has taken place,  but is also valid for a truncated state.  Truncation of the MPS can be done variationally or through a series of singular-value decompositions.\cite{Schollwock2011} In practice, it is unfeasible to obtain the MPS from a general state and therefore, one often starts from one that can easily be written down in MPS form or is obtained efficiently by a ground-state search.  This representation now serves as the starting point for the time-evolution algorithm with local basis optimization.

\subsubsection{Time evolution with local basis optimization}
To time evolve the matrix-product state $\ket{\psi}$ with a Hamiltonian that only acts on neighboring sites, we first write it as a sum of local energy terms $\hat H= \sum\limits_i \hat h_i$ with $\hat h_i $ connecting two neighboring sites. We then carry out a Trotter-Suzuki decomposition to second order so that our time-evolution operator becomes
\begin{equation} \label{eq:trot}
e^{-i\Delta t\hat H/\hbar}=e^{-i\Delta t  \hat H_{\rm even} /(2\hbar)}e^{-i\Delta t \hat H_{\rm odd}/\hbar}e^{-i\Delta t \hat H_{\rm even} /(2\hbar)}+ O(\Delta t^3)\, ,
\end{equation}
where $\hat H_{\rm even (odd)}=\sum\limits_{l : even (odd)} \hat h_l$, and $\Delta t$ is the time step.  Since the even (odd) $\hat{h}_l $ commute we can write each exponential as a product of Trotter gates,  e.g.,  $e^{-i\Delta t \hat H_{\rm even} /(2\hbar)}=\prod\limits_{l : even }e^{-i\Delta t \hat h_l/(2\hbar)} $. To apply the time-evolution gate acting on sites $i$ and $i+1$,  we first bring the matrix-product state into the mixed canonical form of Eq.~\eqref{MPS_def} for those sites.
We then apply the corresponding time-evolution operator directly
\begin{equation} \label{eq:aoogate}
\Phi^{\sigma_i^{\prime}\sigma_{i+1}^{\prime} }_{\alpha_{i-1} \alpha_{i+1}}=\sum\limits_{{\sigma_i, \sigma_{i+1}}} U^{\sigma_i^{\prime}\sigma_{i+1}^{\prime}\sigma_{i}\sigma_{i+1}} M^{\sigma_i \sigma_{i+1}}_{\alpha_{i-1} \alpha_{i+1}},
\end{equation}
where $U^{\sigma_i^{\prime} \sigma_{i+1}^{\prime}\sigma_{i} \sigma_{i+1}}$ is a Trotter gate.
By carrying out a singular-value decomposition of $\Phi^{\sigma_i^{\prime} \sigma_{i+1}^{\prime} }_{\alpha_{i-1} \alpha_{i+1}}$ the updated tensors are obtained and one can continue with the following sites.

Since the time evolution, and many other DMRG based algorithms, often scale polynomially with the local Hilbert-space dimension,  they can become expensive when these are too large.  This is also the case in the Holstein model where the local dimension is $2(M+1)$. Therefore, a number of methods have been introduced for a more efficient treatment of such large local dimensions.\cite{Zhang1998,Jeckelmann1998,Kohler21,Stolppkoehler2020} In this work, we use local basis optimization,\cite{Zhang1998} which has already been combined with different numerical methods, including matrix-product states, e.g.,  in Refs.~\onlinecite{Friedman2000,Wong08,Guo2012,Brockt15,Brockt_17,Stolpp2020,Jansen2020,Jansen_Jooss_2021}. The idea of the local basis optimization can be intuitively understood by looking at the Holstein model in the strong-coupling limit.  For a site with a localized electron,  the phonon distribution is that of a coherent state. Thus, the site can be described by two states,  either the zero phonon or a coherent phonon state. This encourages the search for a basis where the state can be represented with a negligible error but with significantly fewer modes in generic parameter regimes. This basis can be found by diagonalizing the local, single-site reduced-density matrix $\rho=U^{\dagger}W U$, where $\rho^{\sigma_{i} \sigma_{i}^{\prime}}= \sum\limits_{\sigma_{i+1},\alpha_{i-1}, \alpha_{i+1}} \Phi^{ \sigma_i \sigma_{i+1} }_{\alpha_{i-1} \alpha_{i+1}} \Phi^{\dagger}{}^{\sigma_{i+1} \sigma_i^{\prime}}_{\alpha_{i-1} \alpha_{i+1}}$.  Here,  $W $ is a diagonal matrix with elements $w_{\alpha}$. The truncation can be done based on the $w_{\alpha}$'s which are the eigenvalues of $\rho$. The transformation matrices $U$ transform the site from the physical state $\ket{{\sigma}_i}$ to the optimal basis state $\ket{\tilde{\sigma_{i}}}=\sum\limits_{\sigma_{i}} U^{  \tilde{\sigma_{i}}}{}^{\sigma_{i}} \ket{{\sigma}_i}$.  When combining the local basis optimization with DMRG, which we refer to as DMRG-LBO, the transformation matrices can be applied to $\Phi^{\sigma_i^{\prime} \sigma_{i+1}^{\prime} }_{\alpha_{i-1} \alpha_{i+1}}$ from Eq.~\eqref{eq:aoogate} before the subsequent singular-value decomposition.  This singular-value decomposition is done before moving to the next set of sites and applying the next Trotter gate. Due to the extra cost of the transformations,  the procedure is only beneficial if the local basis can be truncated such that the dimension of the optimal basis is significantly  smaller than the one  of the physical basis (see Ref.~\onlinecite{Brockt15} for technical details). 

\begin{figure}
	\centering
	\includegraphics{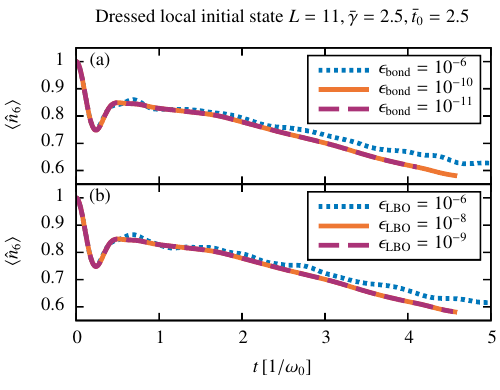}
	\caption{Convergence of the DMRG data with $\epsilon_{\rm bond}$ and $\epsilon_{\rm LBO}$. We show the electron occupation on the central site in the Holstein chain with $L=11,\bar \gamma=\bar t_0=2.5$, starting from the dressed local initial state. (a) Fixed $\epsilon_{\rm LBO}=10^{-8}$ and different $\epsilon_{\rm bond}$.  (b) Fixed $\epsilon_{\rm bond}=10^{-10}$ and different $\epsilon_{\rm LBO}$. We further use $M=40$ and $\Delta t=0.004/\omega_0$.\label{fig:DMRG_conv}}
\end{figure}

When diagonalizing the reduced density matrix to obtain the optimal basis,  the smallest eigenvalues $w_{\alpha}$ are discarded such that the truncation error is below a threshold $\epsilon_{\rm LBO}$ so that: 
\begin{equation}
\label{eq:trunc_svd}
\sum\limits_{\textrm{discarded} \: \alpha  } w_{\alpha}/(\sum\limits_{\textrm{all} \: \beta} w_{\beta})<\epsilon_{\rm LBO}.
\end{equation} 
For the standard DMRG truncation done in the time-evolution scheme, we discard all singular values with a threshold $\epsilon_{\rm bond}$ such that: 
\begin{equation}
\label{eq:trunc_lbo}
\sum\limits_{\textrm{discarded}   \: \alpha} s_{\alpha}^2/(\sum\limits_{\textrm{all} \:  \beta } s_{\beta}^2)<\epsilon_{\rm bond}.
\end{equation}  
Note that the MPS representation of the state in Eq.~\eqref{MPS_def} can indeed be exact  for a fixed phonon cutoff $M$. This would, however, require exponentially large matrices and is unfeasible.  By introducing a finite truncation, the representation is no longer exact, but is an approximation of the state. Since this error can be made arbitrarily small in principle and often also in practical applications, we refer to DMRG as an exact method and convergence of the calculations must always  be controlled with respect to $\epsilon_{\rm bond}$ and $\epsilon_{\rm LBO}$. How to do this is demonstrated for one example in Fig.~\ref{fig:DMRG_conv}(a) for $\epsilon_{\rm bond}$ and Fig.~\ref{fig:DMRG_conv}(b) for $\epsilon_{\rm LBO}$. 
Figure~\ref{fig:DMRG_conv}(a) shows the electron density for a fixed $\epsilon_{\rm LBO}$ with different $\epsilon_{\rm bond}$.  For  $\epsilon_{\rm bond}$ chosen too large (here,  when $\epsilon_{\rm bond}=10^{-6}$), too many states are truncated away and the time-dependent expectation value of the observable is inaccurately captured by the MPS.  However, if $\epsilon_{\rm bond}$ is set to be sufficiently small, the data becomes independent of the truncation and the approximated state can be used to correctly calculate the expectation values.  A similar check must be done for $\epsilon_{\rm LBO}$. This is illustrated in Fig.~\ref{fig:DMRG_conv}(b) with a fixed $\epsilon_{\rm bond}$.  Now, a poorly chosen $\epsilon_{\rm LBO}$ leads to an erroneous representation of the local state and the expectation value of the observable deviates from its correct form. As is the case for $\epsilon_{\rm bond}$,  below a sufficiently small  $\epsilon_{\rm LBO}$, the expectation values become independent thereof and the data are converged up to a certain accuracy. 
This convergence must be checked for each observable individually. We find that in particular the reduced mean-squared displacement (RMSD) and the total phonon number need smaller cutoffs to reach convergence.

Note that a convergence check with respect to the local phonon truncation $M$ is also needed. This must be carried out by varying $M$ as illustrated for exact diagonalization in Fig.~\ref{fig:trimer_conv_M} in Sec.~\ref{Sec:ED}.
The DMRG-LBO calculations in this work are done using the ITensor Software Library.\cite{itensor}

	\section{Independent trajectory methods\label{Sec:Trajectory}}

In this section, we first briefly review the phase-space approach to quantum mechanics and the ansatz to describe the nuclear dynamics by a swarm of independent trajectories. We then describe the two independent-trajectory methods used in this work: multitrajectory Ehrenfest (MTE)\cite{Ehrenfest1927,Tully1998,Kirrander2020} and fewest-switches surface hopping (FSSH).\cite{Tully1990} For the latter we also review some possible improvements to the algorithm, both for its general description and specific ones for its application to large systems. We conclude this section with a short discussion of our implementation of both methods and their internal convergence. 

\subsection{Phase space representation of quantum mechanics\label{Sec:QMPhaseSpace}}

The independent-trajectory methods combine a (classical) phase-space description of the nuclear degrees of freedom with a quantum mechanical wave-function formalism of the electronic subsystem. The partial phase-space description by itself is not an approximation and can be understood in the framework of the partial Wigner transform.\cite{Kapral1999,Grunwald2009,Ando2002}
The Wigner transform (sometimes called Weyl transform) maps an operator $\hat O$ to a function of phase-space $O_W(R,P)$.\cite{Wigner1932,Weyl1927,Moyal1949,Hillery1984,Case2008} 
For a composite electron-nuclear system one can define the partial Wigner transform acting on the nuclear degrees of freedom by:\cite{Ando2002,Horenko2001,Ando2003}
\begin{align}
O_{a,b:W}(R,P)&=\int \textup{d}Z \Bigg[e^{i P \cdot Z/\hbar} \label{Eq:PhaseSpace:Rep_then_WignerTransform}\\
&\times \left\langle R - \frac{Z}{2}, \phi_a\left(R-\frac{Z}{2}\right) \middle| \hat{O} \middle| R +\frac{Z}{2},\phi_b\left(R+\frac{Z}{2}\right)\right\rangle\Bigg],\nonumber
\end{align}
where the selected electronic basis $|\phi_a(R)\rangle$ can refer to both
diabatic or adiabatic states and the integral is taken over the whole nuclear position space $Z$. Of special importance is the partial Wigner transform of the density matrix of the full system $\hat \rho=\ket{\Psi}\bra{\Psi}$, which allows us to calculate expectation values of any operator via a phase-space average:
\begin{align}
\braket{\hat O}=\sum_{a,b}\int \textup{d}R \int \textup{d}P\ O_{a,b:W}(R,P) \frac{\rho_{b,a:W}(R,P)}{(2\pi\hbar)^{n}}.\label{Eq:PhaseSpace:ExpectationValue}
\end{align}
The term $W_{b,a}(R,P)\coloneqq\frac{\rho_{b,a:W}(R,P)}{(2\pi\hbar)^{n}}$ looks like a probability distribution in phase space, but can contain negative values. When we speak of the partially Wigner-transformed density matrix we always refer to this object $W_{b,a}(R,P)$.
For a pure state expanded in the same basis as used in Eq.~\eqref{Eq:PhaseSpace:Rep_then_WignerTransform} it can be written as:
\begin{align}
W_{a,b}(R,P)=\int \textup{d}Z \frac{e^{i P \cdot Z/\hbar}}{(2\pi\hbar)^{n}} \Psi_a(R-\frac{Z}{2})\Psi_b^*(R+\frac{Z}{2}).\label{Eq:PhaseSpace:Wignerfunction}
\end{align}
This form of the partial Wigner transform is basis dependent and allows us to describe both phonon-independent (diabatic) and adiabatic states easily, since only the wave function in the respective basis $\Psi_a^{(d)}$ or $\Psi_a^{(a)}$, see Eqs.~\eqref{Eq:BH:WFexpansion:adiabatic} and \eqref{Eq:BH:WFexpansion:diabatic}, is needed. 
However, one also needs to represent the observable in the same basis (see Eq.~\eqref{Eq:PhaseSpace:Rep_then_WignerTransform}), which can be non-trivial in an adiabatic basis. 

An alternative definition of the partial Wigner transform exists,\cite{Kapral1999,Subotnik2013,Grunwald2009} in which the electronic states in Eq.~\eqref{Eq:PhaseSpace:Rep_then_WignerTransform} do not depend on $Z$ and are evaluated at position $R$. For a phonon-independent electronic basis both definitions are equivalent, while the $R$-dependence of adiabatic states is treated differently. One can think of this alternative, the so called ``Wigner-then-adiabatic''\cite{Ryabinkin2014}-definition, as if one first performs the partial Wigner transform in a phonon-independent basis, and afterward represents it in another, e.g., adiabatic, basis. It is frequently used when relating the quantum-classical Liouville approach,\cite{Kapral1999} which is an approximate solution to the time evolution of the Wigner function, to trajectory-based descriptions of the phase-space evolution, such as MTE and FSSH.\cite{Subotnik2013,Grunwald2009}

Adiabatic initial states can be much easier represented in the variant of Eq.~\eqref{Eq:PhaseSpace:Wignerfunction} and this definition will be used later when sampling the independent-trajectory methods, see Secs.~\ref{Sec:Independent-trajectory:General} and \ref{Sec:Indep-Traj:Convergence}. 

\subsection{Independent-trajectory methods\label{Sec:Independent-trajectory:General}}

Independent-trajectory methods approximate the partially Wigner-transformed density matrix (\ref{Eq:PhaseSpace:Wignerfunction}) by an ensemble average over a number $N_{traj}$ of classical nuclear trajectories (indexed by $r$) described in nuclear phase space by $R^r(t),P^r(t)$, each with an attached electronic density matrix $\hat \rho^r_{el}$, and potentially a weight-factor $w^r$:
\begin{align}
\hat{W} (R,P,t) \approx \frac{1}{{N_{traj}}} \sum_r^{N_{traj}} w^r \delta(R-R^r(t))\delta(P-P^r(t)) \cdot \hat{\rho}_{el}^{r}(t).\label{Eq:Independent_trajectory_average}
\end{align}
Expectation values of observables can then be calculated via Eq.~\eqref{Eq:PhaseSpace:ExpectationValue}.
The two most common approaches to obtain initial conditions in the trajectory-simulations are to sample them either from the partial Wigner transform of a given density matrix, or from a classical molecular dynamics simulation, where often the former approach is preferred.\cite{Brown1981,Persico2014,Barbatti2016,Suchan2018,Mai2018}
The implementation of the Wigner sampling used in this work is described in more detail in Sec.~\ref{Sec:Indep-Traj:Convergence}.

The nuclear trajectory $r$ is propagated via classical Newtonian equations of motion under some method-specific classical Hamiltonian $H^{nuc,r}$, which is independently obtained for each trajectory:
\begin{equation}
\frac{\partial R^{r}_{k}}{\partial t} =  \frac{\partial H^{nuc,r}}{\partial P^{r}_{k}},\quad \frac{\partial P^{r}_{k}}{\partial t} = - \frac{\partial H^{nuc, r}}{\partial R^{r}_{k}}.\label{Eq:Semiclassics:Classical_EOM}
\end{equation}
Here, $R^{r}_{k}$ and $P^{r}_{k}$ are the $k$th-component of the nuclear position and momentum of the trajectory $r$. Without the attached electronic density matrix and when using a Wigner sampling of the initial nuclear quantum state, this approach is equivalent to the truncated Wigner approximation in a phase-space representation.\cite{Polkovnikov2010}
The electronic density matrix is time evolved under the influence of the classical nuclear positions $R^r(t)$ and momenta $P^r(t)$, the explicit form is given in the method sections, Secs.~\ref{Sec:MTE} and \ref{Sec:FSSH}.

As the name implies, different trajectories with their attached electronic density matrices are not allowed to interact during the time evolution. This approximation is at the core of these methods and allows easy, distributed parallel calculations. 
In this paper, we will use two well-established independent-trajectory methods with different underlying approximations and equations of motion: multitrajectory Ehrenfest\cite{Ehrenfest1927,Tully1998,Kirrander2020} and fewest-switches surface hopping.\cite{Tully1990}
More details on the initial sampling used in this work and a convergence analysis of both methods is given at the end of this section (see Sec.~\ref{Sec:Indep-Traj:Convergence}).

\subsection{Multitrajectory Ehrenfest dynamics\label{Sec:MTE}}

In the first method, the multitrajectory Ehrenfest (MTE) approach, the coupling of each nuclear trajectory with its attached electronic density matrix is described on a mean-field level, called Ehrenfest dynamics,\cite{Ehrenfest1927,Tully1998,Kirrander2020} or sometimes mixed quantum/classical time-dependent self-consistent-field.\cite{Garciavela1992,Topaler1998} Here, the nuclear trajectory is propagated via the Newtonian equations of motion (\ref{Eq:Semiclassics:Classical_EOM}) according to the mean-field nuclear Hamiltonian:
\begin{align}
H^{nuc,r}=T^{nuc}(P^r) + \textup{Tr} \left[\hat H^{BO}(R^r)\hat \rho^r_{el} \right]_e,\label{Eq:MTE:nuc}
\end{align}
where $\textup{Tr}[\cdot]_e$ is a trace in the electronic subsystem only and $T^{nuc}(P)$ denotes the classical nuclear kinetic energy.
The electronic density matrix is then propagated by the time-dependent electronic Hamiltonian $\hat H^{BO}(R^r(t))$:
\begin{align}
\frac{d}{dt} \hat{\rho}^{r}_{el}(t) = -\frac{i}{\hbar}\Big[ \hat H^{BO}(R^r(t)), \hat{\rho}^{r}_{el}(t)\Big].\label{Eq:MTE:el}
\end{align}
Ehrenfest dynamics is independent of the choice of the electronic basis, as can be seen from Eqs.~(\ref{Eq:MTE:nuc}) and (\ref{Eq:MTE:el}), where we did not specify any basis.
Ehrenfest dynamics can be derived from the exact quantum dynamics by first applying a mean-field approximation, the time-dependent self-consistent field method, and afterwards a classical approximation for the nuclear coordinates, see also Refs.~\onlinecite{Marx2009,Tully1998} and references therein. 
Alternatively, one can start from the (approximate) quantum-classical Liouville equation and neglect electron-phonon correlations in the total density matrix.\cite{Grunwald2009,Gerasimenko1982} 
The time evolution is completely deterministic and the multitrajectory approach only serves as an accurate sampling of the initial state.

It is easier to see the limitations of the mean-field approximation and the ability of Ehrenfest dynamics to describe non-adiabatic effects if we write the electronic time-evolution in the (adiabatic) Born-Oppenheimer basis $\ket{\phi^{BO}_a(R(t))}$, which in the independent-trajectory approach is now explicitly time dependent. For a pure state described by $\ket{\Psi_{el}^r(t)} = \sum_a \Psi^{r,(a)}_{el,a}(t) \ket{\phi^{BO}_a(R^r(t))}$, the Schr\"odinger equation for the electronic wave function becomes:
\begin{align}
\begin{split}
i \hbar \partial_t \Psi^{r,(a)}_{el,a}(t) = \sum_b \Bigg(&\delta_{a,b} E^{BO}_b(R^r(t)) \\
&+ \sum_k \dot R^r_k(t) A_{a,b,(k)}(R^r(t))\Bigg) \Psi^{r,(a)}_{el,b}(t), \label{Eq:MTE:adiaBasis}
\end{split}
\end{align}
where $\dot R_k$ is the time-derivative of the $k$-th component of the nuclear coordinate and $A_{a,b,(k)}$ are the derivative couplings defined in Eq.~\eqref{Eq:BO:DerivCoup}.\cite{Tully1990}

Similar to the Schr\"odinger equation in the Born-Huang approach, Eq.~(\ref{Eq:BO:NucSchrEq}), the derivative couplings are responsible for non-adiabatic transitions between different Born-Oppenheimer surfaces. However, these transitions are now described by a purely electronic Schr\"odinger equation (\ref{Eq:MTE:adiaBasis}), where the nuclear momenta enter only via the velocity term $\dot R_k$ in front of the derivative couplings. Hence, non-adiabatic transitions between Born-Oppenheimer surfaces are possible in Ehrenfest dynamics and are again peaked where the derivative couplings become large.

One problem of the mean-field description in Ehrenfest dynamics becomes apparent when considering an electron-nuclear wave packet initially localized to a certain Born-Oppenheimer state, which then passes through a region with strong non-adiabatic coupling. 
Beyond that region, the wave packet will have contributions on more than one surface, and, as shown in the exact Schr\"odinger equation (\ref{Eq:BO:NucSchrEq}), these contributions should evolve independently from each other, when the derivative couplings become negligible again. This is not reproduced in Ehrenfest dynamics, where each nuclear trajectory will evolve in one effective potential built from the electronic contributions in all occupied surfaces, leading to possibly unphysical paths when the force-contributions of the involved surfaces differ strongly.\cite{Tully1990,Yonehara2012,Jasper2006}
For more information on the implementation of MTE in this work and a convergence analysis, see Sec.~\ref{Sec:Indep-Traj:Convergence}.

\subsection{Fewest-switches surface hopping\label{Sec:FSSH}}

Surface-hopping methods\cite{Bjerre1967,Tully1971} try to circumvent the mean-field averaging problem by calculating the forces acting on the nuclei from a single energy surface $E_{\lambda^r}(R^r)$ in each time step, with the possibility of stochastic hops between surfaces.
While it is not a mean-field method, it still uses an independent-trajectory approach, which combined with the classical path assumption for the nuclei leads to the same electronic time-evolution equations (\ref{Eq:MTE:el}) and (\ref{Eq:MTE:adiaBasis}) as MTE.\cite{Tully1998,Tully1990}

Surface-hopping methods interpret the resulting electronic populations $|\Psi^r_{el,a}(t)|^2$ as the probability of the trajectory $r$ to be on the electronic surface $E^r_a(R)=\braket{\phi_a|\hat H^{BO}(R^r)|\phi_a}$. Since the forces on the nuclei are calculated from a single diagonal entry of the electronic density matrix, they are strongly basis dependent. The typical choice is the Born-Oppenheimer basis (see, for example, the discussion in Ref.~\onlinecite{Tully1998}), which leads to the nuclear Hamiltonian:
\begin{align}
H^{nuc,r}=T^{nuc}(P^r)+E^{BO}_{\lambda^r}(R^r),\label{Eq:FSSH:NucHam}
\end{align}
where $T^{nuc}$ is the kinetic energy term of the classical nuclear Hamiltonian and $\lambda^r$ is the currently active Born-Oppenheimer surface of trajectory $r$.
The hopping algorithm is designed such that the distribution of trajectories on the surfaces approximately reproduces the electronic populations given by $|\Psi^r_{el,a}(t)|^2$. The most common form for the hopping algorithm is the fewest-switches surface hopping (FSSH),\cite{Tully1990} which is also used in this work.

The time derivative of the electronic populations $|\Psi^r_{el,a}|^2=\rho^r_{a,a}$ in the Born-Oppenheimer basis can be expressed by using Eq.~(\ref{Eq:MTE:adiaBasis}) as $\dot \rho^r_{a,a}=\sum_{b\ne a} b^r_{a,b}$ with $b^r_{a,b}=-2 \Re \left[\frac{i}{\hbar}(\Psi^r_{el,a})^* \Psi^r_{el,b} \sum_k \dot{R}^r_k \cdot A_{a,b,(k)}(R^r)\right]$, where $\Re$ refers to the real part.
From this expression, FSSH estimates the change of electronic population from surface $a$ to surface $b$ within a time step $\Delta t$ as $-b^r_{a,b} \Delta t$. 
The FSSH algorithm, as the name fewest-switches implies, tries to use the minimum number of hops between surfaces to satisfy this relation. For that reason, a hop from an active surface $\lambda$ to another surface $\lambda'$ within a time step $\Delta t$ is allowed with a probability equal to $p^r_{\lambda \rightarrow \lambda'}=\max\left\{\frac{-\Delta t b^r_{\lambda,\lambda'}}{\rho^r_{\lambda,\lambda}},0\right\}$, i.e., hops are only allowed in one direction between two surfaces.

In order to conserve the total energy of the system, a surface hop in the fewest-switches algorithm\cite{Tully1990} is accompanied by a velocity adjustment of the nuclear degrees of freedom. The velocity adjustment happens in the direction of the derivative coupling $\vec A_{\lambda,\lambda'}$ between the surfaces. If the corresponding momentum is not sufficient to compensate for the energy increment of a hop, the hop is frustrated and ignored. When dealing with relatively low nuclear kinetic energies, this can lead to significant deviations of the distribution of trajectories from the propagated electronic populations $|\Psi^r_{el,a}(t)|^2$.
The occurrence of frustrated hops can render FSSH inferior to an Ehrenfest approach in some cases.\cite{Muller1997}
Even without frustrated hops one cannot guarantee the internal consistency between the electronic populations and the trajectory distributions due to missing decoherence effects,\cite{Granucci2007} which might be remedied by introducing a decoherence correction, see Sec.~\ref{Sec:FSSH:DC}.

It seems that each nuclear trajectory in FSSH has two quantities describing the electronic state: the electronic amplitudes $\Psi_{el}^r$, which determine the switching probabilities, and the currently active surface $\lambda^r$, which determines the nuclear Hamiltonian in Eq.~(\ref{Eq:FSSH:NucHam}).\cite{Landry2013}
Typically, the active-surface distributions are used to calculate populations of the Born-Oppenheimer states: $\braket{\hat n^{BO}_a}=\frac{1}{N_{traj}} \sum_r^{N_{traj}} \delta_{a,\lambda^r}$, as they approximately obey detailed balance.\cite{Parandekar2005,Schmidt2008}
In contrast, the calculation of electronic properties in other basis sets requires a proper definition of the full electronic density matrix $\hat \rho_{el}^r$ used in Eq.~(\ref{Eq:Independent_trajectory_average}) to approximate the partially Wigner-transformed density matrix.
In particular, a consistent approach to calculate diabatic populations $n_a=\textup{Tr}\left[\hat n_a\right]=\textup{Tr}\left[\ket{\chi_a}\bra{\chi_a}\right]$, is not easily found in FSSH, see, for example, the discussions in Refs.~\onlinecite{Subotnik2013,Landry2013,Subotnik2016}. 
Two simple possible definitions for the matrix elements of the density matrix in the Born-Oppenheimer basis are:\cite{Subotnik2013,Landry2013,Subotnik2016,Carof2019}
\begin{align}
&\rho_{el,a,b}^{r,\textup{(AS)}}=\delta_{\lambda^r(t),a}\delta_{a,b}\label{Eq:FSSH:AS_Dens} \\
&\rho_{el,a,b}^{r,\textup{(WF)}}=\Psi_{el,a}^r \Psi_{el,b}^{r,*}.\label{Eq:FSSH:WF_Dens}
\end{align}

The first definition $\hat \rho^{\textup{(AS)}}_{el}$ corresponds to using only the distribution of active surfaces for the calculation of the electronic density matrix. With this ansatz, however, only a diagonal density matrix in the basis of Born-Oppenheimer states can be described for every nuclear configuration $R$. Thus, for an initial state that is a coherent superposition of several Born-Oppenheimer states, such as is the case for a typical diabatic state, this ansatz cannot capture the correct initial diabatic populations. The second ansatz $\hat \rho^{\textup{(WF)}}_{el}$ relies on the electronic wave function amplitudes. Similar to MTE, electronic populations in any basis and for any initial state can be described with this ansatz. On the downside, the electronic amplitudes are not guaranteed to obey detailed balance and are unreliable for longer times.\cite{Parandekar2005,Schmidt2008,Landry2013,Subotnik2016}

Based on an approximate derivation of surface hopping from the quantum-classical Liouville equation,\cite{Subotnik2013} Landry \textit{et al.} proposed the use of a mixed diabatic electronic density,\cite{Landry2013} which corresponds to using
\begin{align}
\rho_{el,a,b}^{r,\textup{(mixed)}} =
\begin{cases}
\delta_{\lambda^r(t),a},\ &\textup{for } a=b\\
\Psi_{el,a}^r \Psi_{el,b}^{r,*},\ &\textup{for } a\ne b,
\end{cases}\label{Eq:FSSH:Mixed_Dens}
\end{align}
for the electronic density matrix.
Here, the active surface distribution is used for the diagonal elements of the electronic density matrix, while the off-diagonal elements are constructed from the electronic amplitudes. 
This expression combines the strengths of the long-time detailed balance given for the active-surface distributions and the short-time coherences of the electronic amplitudes.\cite{Landry2013}

We note that the forces on the nuclei are still calculated from the active surfaces only, ignoring any coherences between adiabatic states, i.e., off-diagonal elements of the electronic density matrix in the adiabatic basis. When the coherences have a significant influence on the wave function dynamics, this will lead to deviations of the nuclear motion from the exact dynamics. This is, for example, the case for the (site-)local initial states studied in this work, which have non-zero coherences from the beginning. This deficiency is one consequence of the basis-dependence of surface hopping, which works best only when starting from an adiabatic (Born-Oppenheimer) initial state. 

The mixed electronic density matrix defined by Eq.~\eqref{Eq:FSSH:Mixed_Dens} provides a consistent way to calculate electronic populations in any basis. Unless stated otherwise in this work, we will use this definition for the calculation of diabatic populations. There is, however, one caveat to using the mixed electronic density matrix: the diabatic populations are not guaranteed to be positive.\cite{Landry2013}
We find this to be of relevance only in large systems starting from a local initial state, at sites far away from the initially occupied site with very low populations. Observables that put special focus on these small diabatic populations, such as the later investigated reduced mean-squared displacement (see Eq.~\eqref{Eq:Extended_Sys:MSD}), cannot be calculated reliably by using the mixed electronic density matrix of Eq.~\eqref{Eq:FSSH:Mixed_Dens}, see Appendix~\ref{Sec:FSSH_Large_Sys}. In these cases, one should consider to resort to the aforementioned definitions $\hat \rho^{\textup{(AS)}}_{el}$ (Eq.~\eqref{Eq:FSSH:AS_Dens}) or $\hat \rho^{\textup{(WF)}}_{el}$ (Eq.~\eqref{Eq:FSSH:WF_Dens}).
More information on our implementation of FSSH and a convergence analysis is given in Sec.~\ref{Sec:Indep-Traj:Convergence}.

\subsection{Improvements to fewest-switches surface hopping\label{Sec:Surfhop_improvements}}

Many improvements for surface-hopping algorithms have been proposed in the last years, as portrayed in the reviews Refs.~\onlinecite{Wang2016,Smith_2019}. We consider here only the very common decoherence correction, and two further corrections specifically proposed for large systems, see also Ref.~\onlinecite{Wang2020}. In the following the trajectory index $r$ is omitted for clarity.

\subsubsection{Decoherence correction\label{Sec:FSSH:DC}}

The assumption of independent-trajectories within the surface-hopping approach does not only discard the phase relation between different trajectories, but it also leads to overcoherence within the individual trajectories: the electronic amplitudes are evolved via Eq.~(\ref{Eq:MTE:adiaBasis}) and without interactions with other trajectories. Therefore, they will keep the phase relation between different surfaces, which can lead to self-interference effects at later times. Already in the first proposal of the fewest-switches algorithm,\cite{Tully1990} Tully mentioned the possibility of adding coherence damping terms to the time evolution of the electronic amplitudes. Since then, a large variety of decoherence corrections have been proposed, see, for example, Refs.~\onlinecite{Crespo-Otero2018,Wang2016,Subotnik2016,Wang2020}.
We use a force-based decoherence rate, proposed in Refs.~\onlinecite{Schwartz1996,Bittner1995}, on the basis of a frozen Gaussian method,\cite{Heller1975,Neria1993} where the electronic amplitudes of all non-active states $a\ne\lambda$ decay exponentially in each time step via $\Psi_a'=\Psi_a\cdot\exp(-\Delta t/\tau_a)$, with the decoherence rate
\begin{align}
\frac{1}{\tau_a} = \sqrt{\sum_k \left(F_k^\lambda-F_k^a\right)^2/(4 a_k \hbar^2)}.\label{Eq:Decoherence_rate}
\end{align}
Here, $F_k^a$ is the $k$-th component of the force acting on the $a$-th potential energy surface and $a_k$ is the width of the Gaussians used in the frozen Gaussian ansatz, which can be calculated in the Holstein model via $a_k=\frac{m\omega_0}{\hbar}$ (see also Ref.~\onlinecite{Wang2020}).

Subotnik \textit{et al.} have proposed an advanced expression for the decoherence rates of on- and off-diagonal elements of the electronic density matrix by deriving the surface-hopping approach from the quantum-classical Liouville equation.\cite{Subotnik2013} 
We will not use their expression in this work, as it requires the propagation of additional variables.

Using a decoherence correction, in the following denoted by FSSH+D, simplifies the ambiguity of the electronic density matrix in FSSH, mentioned above in Sec.~\ref{Sec:FSSH}. Since the electronic amplitudes are dampened towards the active surface distributions, all definitions of the density matrix become the same for long times. In order to recover the correct coherences of local initial states, one should thus resort to the mixed $\hat \rho^{\textup{(mixed)}}_{el}$ (Eq.~\eqref{Eq:FSSH:Mixed_Dens}) or the electronic amplitude (wave function) $\hat \rho^{\textup{(WF)}}_{el}$ (Eq.~\eqref{Eq:FSSH:WF_Dens}) definition. 

\subsubsection{Avoiding derivative couplings\label{Sec:FSSH:avoid_deriv_coup}}

Special care needs to be taken when dealing with large systems in FSSH, see Ref.~\onlinecite{Wang2020}.
Here, different energy surfaces can come very close or even cross when different adiabatic states are localized to far separated regions of the system and are only very weakly coupled. 
The derivative couplings (see Eq.~(\ref{Eq:BH:DerivCoup_Energy_expression}) or, for the dimer, Eq.~\eqref{Eq:BO:DimerDerivCoup}) then become very localized, and cannot be sampled reliably unless very small time steps are used.\cite{Granucci2001,Fernandez-Alberti2012,Wang2014,Mai2020} 
In the FSSH algorithm, the derivative couplings are used in three steps: the electronic propagation according to Eq.~\eqref{Eq:MTE:adiaBasis}, the calculation of the hopping probabilities, and the direction of the velocity adjustment after a successful surface hop. 

Since the electronic propagation is basis independent (see Eq.~\eqref{Eq:MTE:el}), it can be carried out in a diabatic basis $\ket{\chi_a}$, to avoid the derivative couplings altogether. In each time step the adiabatic amplitudes are then calculated via $\Psi_{el,a}^{(a)}(t)=\sum_b \braket{\phi^{BO}_a(R(t))|\chi_b} \Psi_{el,b}^{(d)}(t)$ (see Eqs.~\eqref{Eq:BH:WFexpansion:adiabatic},\eqref{Eq:BH:WFexpansion:diabatic}).\cite{Granucci2001}

For calculating numerically stable hopping probabilities, also close to surface crossings, different schemes have been proposed, see Ref.~\onlinecite{Wang2020} and references therein. 
In this work, we use an alternative expression for the hopping probability $p_{\lambda \rightarrow a}=\max\{g_a^{(\lambda)},0\}$ in systems with $L\ge3$, which relies only on the general unitary adiabatic time propagator $\Psi_a(t+\Delta t)=\sum_b P_{a,b}(t,t+\Delta t) \Psi_b(t)$:
\begin{align}
g_a^{(\lambda)}
=
&\frac{|\Psi_\lambda( t)|^2-|\Psi_\lambda( t+\Delta  t)|^2}{|\Psi_\lambda( t)|^2} \nonumber\\
& \times \frac{\Re \left[P_{a,\lambda}^*(t,t+\Delta t) \Psi_a(t+\Delta t) \Psi^*_\lambda(t)\right]}{|\Psi_\lambda( t)|^2 - \Re \left[P_{\lambda,\lambda}^*(t,t+\Delta t) \Psi_\lambda(t+\Delta t) \Psi^*_\lambda(t)\right]},	\label{Eq:FSSH:locally_diab_hopp_prob}
\end{align}
and which was proposed in Ref.~\onlinecite{Mai2015} as a numerically more stable variant of a similar expression proposed in Ref.~\onlinecite{Granucci2001}. 

The hopping probabilities defined by Eq.~\eqref{Eq:FSSH:locally_diab_hopp_prob} fulfill the sum rule $\frac{|\Psi_\lambda(t)|^2-|\Psi_\lambda(t+\Delta t)|^2}{|\Psi_\lambda(t)|^2}
=  \sum_{a\ne\lambda} g_a^{(\lambda)}$ in each time step exactly. This is necessary to obtain a self-consistent description between electronic amplitudes and active-surface distributions, as in the original FSSH algorithm.\cite{Tully1990} Eq.~\eqref{Eq:FSSH:locally_diab_hopp_prob} can be evaluated without resorting to derivative couplings once the electronic wave-function propagation is obtained in the diabatic basis.

We compared our results also to the ``crossing-corrected'' algorithm proposed in Ref.~\onlinecite{Qiu2018} as an extension of a self-consistent correction suggested in Ref.~\onlinecite{Wang2014}, which also relies on exactly enforcing the sum-rule decomposition mentioned earlier. 
This method relies on identifying surface intersections, or near intersections, which can be insufficient for asymptotically approaching Born-Oppenheimer surfaces, as observed in the Holstein trimer (see Fig.~\ref{Fig:BH_Trimer:Trimer_Surf}). Then again, the crossing-corrected algorithm alleviates the difficult calculation of the velocity-adjustment while passing a surface intersection (by neglecting the velocity-adjustment).
We could not observe a significant difference between using the crossing-corrected scheme and using Eq.~\eqref{Eq:FSSH:locally_diab_hopp_prob} and employed the latter in this work. Hence, in our implementation, we still rely on the derivative couplings in the calculation of the velocity-adjustment.

\subsubsection{Decoherence enhanced spurious charge transfer\label{Sec:FSSH:Spurious_Charge_Transfer}}

Another problem appears when applying the decoherence correction in large systems: the ``decoherence correction enhanced trivial crossing problem''\cite{Bai2018,Wang2020} or also called ``spurious charge transfer''.\cite{Giannini2018,Carof2019} In large systems, different Born-Oppenheimer states might be localized to completely different parts of the system. A surface hop between these states, although rare, will correspond to an instantaneous jump in space. This becomes even more severe when surface crossings are not properly accounted for\cite{Wang2020} (see Sec.~\ref{Sec:FSSH:avoid_deriv_coup}). The electronic propagation via Eq.~\eqref{Eq:MTE:el} is only indirectly, through the changing nuclear trajectory, affected by the jump, and subsequent hopping events can often still be well described. Hence, in normal FSSH, the erroneous jumps have only little effect, especially when calculating diabatic populations from the electronic amplitudes only ($\rho^{\textup{(WF)}}$, see Eq.~\eqref{Eq:FSSH:WF_Dens}). 
The decoherence correction, however, collapses the electronic amplitudes to the active surface, resulting in the spurious charge transfer described in Refs.~\onlinecite{Bai2018,Wang2020,Giannini2018,Carof2019}. 

We note that this problem is even more severe when analyzing the real-time evolution of a local initial state in a large system, as done in this work. An example is the bare local state at the central site, with all phonon degrees of freedom in their ground state. This initial state has an almost equal weight on all Born-Oppenheimer surfaces. Thus, already the initial state has a large mismatch between local densities calculated according to the active surfaces (where the initial state is far spread), and the densities calculated according to the electronic amplitudes (where the local initial state is properly recovered). The decoherence correction then quickly removes the correct short-time coherences captured by the electronic amplitudes, resulting in an unphysical super-fast spreading of the wave function. This is shown in Appendix~\ref{Sec:FSSH_Large_Sys}.

Two ways to restrict the decoherence correction to avoid spurious charge transfer have been suggested: (i) allowing decoherence only when the currently active surface has a wave-function population above a threshold,\cite{Bai2018} or (ii) restricting the decoherence correction to a certain ``active space'' in the diabatic basis.\cite{Giannini2018} When describing the short-time evolution of local initial states, the first approach will have nearly no effect, since, as mentioned earlier, the adiabatic populations of most Born-Oppenheimer states will be above any reasonable threshold. We are thus left with the active-space ansatz of Ref.~\onlinecite{Giannini2018}, which was already applied to calculating mobilities in organic semiconductors.\cite{Giannini2018,Carof2019}

This approach\cite{Giannini2018} introduces an additional step in the time evolution of every independent trajectory: After propagating the nuclei and electronic wave functions, but before the decoherence correction is applied, an ``active region'' in the diabatic states is defined, which should contain at least a fraction of $R=0.999$ of the electronic charge density. 
In our implementation, we construct this region by subsequently adding diabatic basis states with decreasing electronic population to an active region subset, until the total electronic population of the subset exceeds the threshold $R$. Afterwards, the decoherence correction is carried out, but all changes of the diabatic electronic amplitudes outside the active region are ignored, while the amplitudes inside the active region are rescaled to conserve the norm. Since $R$ is close to 1, only diabatic states with very low electronic population should be affected. As also pointed out in the original paper,\cite{Giannini2018} this active region only influences the decoherence correction step, while the propagation of the electronic wave functions is carried out in the usual way. This approach is well suited for our local initial states, where initially only a single diabatic state is occupied. We found quantitative changes in the time evolution depending on the exact value of the threshold $R$ used, see Appendix~\ref{Sec:FSSH_Large_Sys}, and in general, a value of $R=0.99$ seems to improve the results over the originally suggested $R=0.999$ for the local initial states in the large systems.

We can go one step further by completely turning off the decoherence correction in the short-time regime of a local initial state, when we know that the coherences between Born-Oppenheimer states are still important. In this work, we take a simple approach of delaying the use of a decoherence correction in large systems ($L\ge 11$) until a time of $\pi/(2\omega_0)$, i.e., a quarter phonon oscillation period, has passed.
The effect of this delayed decoherence is shown in Appendix~\ref{Sec:FSSH_Large_Sys}. 
FSSH with this form of the restricted decoherence, i.e., an active region restriction with $R=0.99$ and the delayed decoherence, will be denoted as FSSH+RD.

From the previous discussion it is clear that in order to correctly describe the initial electronic state, all coherences between the adiabatic states need to be included. In the example of our local states in large systems, the initial adiabatic states are delocalized over many sites and only the coherences in the electronic wave function recover an electron density that is initially localized to a single site. This is also illustrated in Appendix~\ref{Sec:FSSH_Large_Sys}.
Studies focusing on extracting mobilities and steady-state properties often discard these initial coherences or directly start from a relaxed adiabatic state.\cite{Bai2018,Carof2019} Using this ansatz has the benefit that all definitions of the electronic density matrix are equivalent in the beginning and both types of restricted decoherence correction mentioned earlier can be used. 
Furthermore, it has been shown that when using a crossing-corrected algorithm (see Sec.~\ref{Sec:FSSH:avoid_deriv_coup}) and by including only a subset of all adiabatic states in the algorithm, system-size independent results could be obtained with surface hopping in very large systems, also without limitations to the decoherence correction.\cite{Qiu2019} Similar to the first restricted decoherence ansatz mentioned earlier, this approach is not suitable for our bare local initial states, which have an almost equal contribution on all adiabatic states. 
An alternative extension to ensure inherently system-size independent dynamics, which also works for our initial states, is a subsystem surface hopping ansatz,\cite{Wang2013} which combines a surface hopping description for a dynamical subset of the lattice sites with molecular dynamics on the other sites. This can even be combined with an additional subsystem described on a purely statistical level.\cite{Qiu2022}
In this work, where we focus on the initial real-time non-adiabatic dynamics, we stay with a pure surface hopping description, include all coherences for the local initial states, and use the restricted decoherence approach outlined earlier.

\subsection{Computational details of the independent trajectory algorithms and convergence\label{Sec:Indep-Traj:Convergence}}

The calculations presented in this work always start with the nuclear oscillators on every site in a coherent state and the electron either in a local or adiabatic state.
The partially Wigner-transformed density matrix (Eq.~\eqref{Eq:PhaseSpace:Wignerfunction}) has non-zero entries only in one diagonal matrix element of the diabatic or adiabatic electron basis, respectively: $W(R,P)\coloneqq W_{a,a}(R,P)\ne 0$, corresponding to a normal Wigner transform of this component of the nuclear wave function ($\Psi_a^{(d)}(R)$ or $\Psi_a^{(a)}(R)$). 
In both independent trajectory methods, we sample the initial nuclear positions and momenta from that Wigner transform, which for the coherent states leads to uncorrelated Gaussians in the positions and momenta, centered around their quantum mechanical averages $\braket{\hat x_i}$ and $\braket{\hat p_i}$. In the dimensionless variables this can be expressed as:
\begin{align}
W(\bar x,\bar p)=\frac{1}{(\pi)^L} \prod_{i=1}^{L} 
\exp\left(-(\bar x_i - \braket{\hat{\bar x}_i})^2\right) \exp\left(-(\bar p_i - \braket{\hat{\bar p}_i})^2\right).
\end{align}
Here, $L$ is the number of sites. In particular, for all states studied here, we have $\braket{\hat{\bar p}_i}=0,\,\forall i$. For the bare states (see Figs.~\ref{Fig:Introduction:States_Sketch}(b),(d) and Fig.~\ref{Fig:Introduction:States_Sketch2}(a)), all oscillators are centered around zero $\braket{\hat{\bar x}_i}=0,\,\forall i$, while the dressed and adiabatic states have $\braket{\hat{\bar x}_k}=\sqrt{2}\bar \gamma$ for one site $k$ (polaron states, Figs.~\ref{Fig:Introduction:States_Sketch}(a),(c),(e)) or every second site (dressed CDW state, Fig.~\ref{Fig:Introduction:States_Sketch2}(b)).
Taking $W(\bar x,\bar p)$ as the probability distribution for starting at a certain phase-space point, there is no need to keep track of any weighting factor in Eq.~\eqref{Eq:Independent_trajectory_average} for the different trajectories and all are weighted equally for calculating observables ($w^r=1,\,\forall r$).
All randomly drawn positions and momenta are moved and scaled to reproduce the correct mean and variance of the nuclear positions and momenta on each site. 
Only after the nuclear sampling, the electronic amplitudes are set to the predefined initial state $\Psi_{el}^{initial}(R)$, which, for an adiabatic state, depends on the nuclear positions $R$. In FSSH, the active surfaces are set to the initial adiabatic surface (for adiabatic initial states), or randomly sampled from the overlaps between the initial state with all adiabatic states at that nuclear position $R$ (for local initial states).

After the initialization, in each time step, we first perform the integration of the nuclear equation of motion, and afterwards that of the electronic wave function. In MTE (in a phonon-independent (diabatic) basis), we can integrate the nuclear positions on each site exactly for a constant electron density $n$ on that site via: $\bar x(t+\Delta t)=\bar x(t)\cos(\omega_0 \Delta t) + \bar p(t)\sin(\omega_0 \Delta t) + \sqrt{2}\bar \gamma n \left[1-\cos(\omega_0 \Delta t)\right]$, and correspondingly for the momenta. 
Afterwards the vector of all new nuclear positions $\vec x \coloneqq \vec x(t+\Delta t)$ is used to represent the electronic time-propagator in Born-Oppenheimer states to obtain: $\ket{\Psi_{el}(t+\Delta t)} = \sum_a \ket{\phi^{BO}_a(\vec x)} \exp(-i/\hbar \Delta t E^{BO}_a(\vec x)) \braket{\phi^{BO}_a (\vec x) | \Psi_{el}(t)}$, i.e., the electronic wave function is propagated for fixed nuclear positions.

In FSSH (in an adiabatic basis), the electron densities determining the nuclear forces change with the nuclear position $\vec x$ and we resort to a velocity-Verlet integration for the nuclear time step. To avoid using derivative couplings in the electronic integration (see Sec.~\ref{Sec:FSSH:avoid_deriv_coup}), we propagate the electronic wave function in the diabatic basis and can use the same electronic propagation as for MTE. The overlap of Born-Oppenheimer eigenstates at different time steps (and thus different $\vec x$) is obtained in the diabatic basis. In addition, we make sure that the basis transformation between the diabatic and the Born-Oppenheimer basis, which is obtained by a numerical diagonalization of the Born-Oppenheimer Hamiltonian, does not change sign between subsequent time steps. 

After the propagation, in FSSH, we allow for the surface hops, where the hopping probabilities in systems with $L\ge 3$ are calculated according to Eq.~\eqref{Eq:FSSH:locally_diab_hopp_prob}, with the adiabatic time propagator $P_{a,b}$ determined from the electron propagation step. Lastly, if used, the decoherence correction is carried out in FSSH, with the restrictions mentioned in Sec.~\ref{Sec:FSSH:Spurious_Charge_Transfer} for large systems ($L\ge 11$).

\begin{figure}
	\centering
	\includegraphics{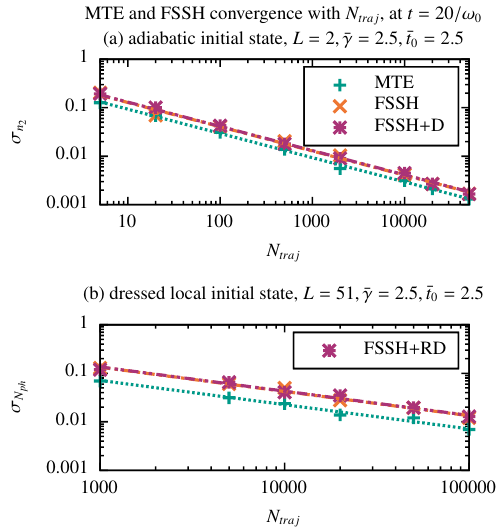}
	\caption{Convergence of the absolute statistical error in MTE, FSSH and FSSH+D/FSSH+RD with the number of used trajectories $N_{traj}$ for $\bar \gamma=\bar t_0=2.5$ and $\Delta t=0.001/\omega_0$ at $t=20/\omega_0$ for two example systems and example observables. Dashed lines indicate linear fits of the log-log data with slope $-0.5$, corresponding to the expected inverse square-root decay of the statistical error with the number of used trajectories. The statistical error is calculated as the standard deviation from $N_r=50$ simulation runs for data points with $N_{traj}\le 20000$ and from $N_r=25$ runs for a higher number of used trajectories. 
		(a) Adiabatic initial state in the Holstein dimer, with the observable being the population on the second dimer site $n_2$. From the fitted intercepts of the dashed lines we estimate  that FSSH needs $\approx 1.88$ times more trajectories than MTE for a similar convergence, and FSSH+D $\approx 1.96$ times more trajectories than MTE. (b) Dressed local initial state in the Holstein chain with $L=51$, where we analyze the total phonon number $N_{ph}$. The fitted intercepts indicate that FSSH needs $\approx 3.5$ times more trajectories than MTE for a similar convergence, and FSSH+D $\approx 3.6$ times more trajectories than MTE. For $L=51$, the restricted version of the decoherence correction is used, see Sec.~\ref{Sec:FSSH:Spurious_Charge_Transfer}. Note that MTE, FSSH and FSSH+D/RD converge to different values, especially for the second case, see Fig.~\ref{Fig:Semiclassics:convergence_timestep}.
	}
	\label{Fig:Semiclassics:convergence_stat_error}
\end{figure}
One needs to ensure internal convergence of the independent trajectory methods, both with the number of used trajectories and the used time step. However, even then, the calculated observables will not necessarily converge to the exact values. Since FSSH needs to sample both the initial state and the random surface hops, we can expect it to converge slower with the number of used trajectories than MTE, which only needs to sample the initial state.
We can analyze the quality of the trajectory-ensemble average of the methods by repeating a simulation $N_r$ times and investigating the standard deviation $\sigma_O$ of an observable of interest $O$ across the different runs, also called the statistical error in the observable. 
This is illustrated for two typical cases in Fig.~\ref{Fig:Semiclassics:convergence_stat_error}. 
The obtained data points can be well described by an inverse square-root dependence of the statistical error on the number of used trajectories. In general, we found that MTE needs fewer trajectories to obtain the same absolute statistical error as the FSSH methods, the factor depending on the system parameters, initial state, and investigated observables. Typical values obtained from linear fits of the log-log data are included in Fig.~\ref{Fig:Semiclassics:convergence_stat_error}. Relative statistical errors might deviate from this observation, as the different methods converge the observables to different values. 
For the comparison to the ED, DMRG-LBO and MCE results presented later, we use 20000 trajectories for MTE and FSSH for all systems, with the exceptions of the dressed initial state in large systems $L\ge 11$ and for the data presented in Appendix~\ref{Sec:One-Electron_reduced_el_hopp}, where we could still observe a slight visible improvement by increasing to 50000 trajectories.

\begin{figure}
	\centering
	\includegraphics{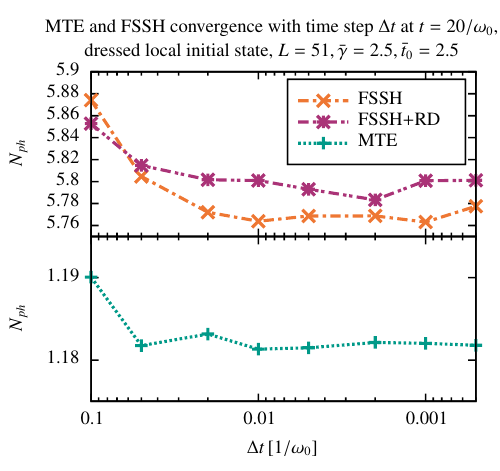}
	\caption{Convergence of the total phonon number with the used integration time step $\Delta t$ for MTE, FSSH and FSSH+RD for $L=51,\bar\gamma=\bar t_0=2.5$ with $N_{traj}=100000$ at $t=20/\omega_0$. The lines are guides for the eye. 
	}
	\label{Fig:Semiclassics:convergence_timestep}
\end{figure}

In addition, a suitable integration time step needs to be used. One of the more difficult systems to converge with respect to the time step is shown in Fig.~\ref{Fig:Semiclassics:convergence_timestep}, with $L=51,\bar\gamma=\bar t_0=2.5$ and $N_{traj}=100000$. 
We can still observe variations in the total number of phonons for time steps smaller than $\Delta t=0.01/\omega_0$, which are, however, in the order of magnitude of the statistical error for the 100000 trajectories used (see Fig.~\ref{Fig:Semiclassics:convergence_stat_error}). For results for the Holstein dimer $L=2$ and for the many-electron calculations in Sec.~\ref{Sec:Many-Electron_Extended}, we use $\Delta t=0.01/\omega_0$, while for all one-electron results in systems with $L>2$, we resort to a smaller time step of $\Delta t=0.001/\omega_0$.

	\section{Multiconfigurational Ehrenfest method\label{Sec:Multi}}

In this section, we explain the method called ``Multiconfigurational Ehrenfest'' (MCE).\cite{Shalashilin2009,Shalashilin2010} The method
is based on: (i) a wave-function ansatz constructed from multiple configurations and (ii) a Gaussian-state basis for oscillators which
is guided by Ehrenfest dynamics.  

The MCE method is not a fully variational method, as opposed to, e.g., Gaussian-based Multi-Configuration time-dependent
Hartree (G-MCTDH),\cite{Burghardt99,Burghardt03,Burghardt08,Gonzalez20} variational Multiconfigurational Gaussian (vMCG) \cite{Worth04,Worth03,Richings15,Gonzalez20} and the Davydov D2 ansatz.\cite{Davydov_1982,Cruzeiro-Hansson94,Zhao2012,Zhou15,Chen2019}
The non-variational character, on the one hand, leads to a less complex equation of motion, while on the other hand, the resulting 
dynamics conserve energy only approximately.\cite{Ma18} MCE, in principle, converges to exact dynamics if the configurations 
form a complete, or sufficiently large, basis set.\cite{Ma18} 
A characteristic feature of MCE, which needs to be kept in mind,
is that the convergence has different rates for different observables. 
Observables linear in position or momenta converge much 
faster than observables with higher powers of position or momenta, e.g., the energy of the oscillators. As we confirm below, convergence of these more difficult observables, as well as total energy conservation, can easier be reached for short time intervals or small systems in practical applications. Even when the total energy still drifts,  the average nuclear position and the electronic densities close to the initially occupied sites are captured quite well even for later times.

The MCE algorithm comes in two flavors called MCEv1,\cite{Shalashilin2009} where all 
electronic coefficients are coupled across the configurations, and MCEv2,\cite{Shalashilin2010} where each configuration has one additional global coefficient and only these coefficients are coupled across different configurations.
The MCEv1 is recommended for model Hamiltonians\cite{Symonds2018} such as  the Holstein model to which it was already 
applied.\cite{Chen2019} In this work, only MCEv1 results are included.
However, some of the implementation strategies devised for the MCEv2, e.g., the initialization of the total state, are adapted to MCEv1.

\subsection{Ansatz for the total state}
The time-dependent basis used in both versions of MCE is constructed from coherent states and an electronic wave function.\cite{Shalashilin2009}
For a single configuration $r$, the coherent state product of the oscillators can be written as
\begin{equation}
|\mathbf{z}^r\rangle \coloneqq \exp\left(\sum_{i=1}^{L} \left(z^r_{i}\hat{b}_{i}^\dagger-{z^r_i}^*\hat{b}_{i} -i \Re(z^r_{i})\Im(z^r_{i})\right)\right) |0\rangle_{ph},
\end{equation}
and is fully characterized by the set of complex numbers $\{z^r_{1}, \dots, z^r_{L}\}$ defined as the eigenvalues
of the annihilation operators $\hat{b}_{i}|\mathbf{z}^r \rangle = z^r_{i}|\mathbf{z}^r\rangle$.
The additional global phase factor $\Re(z^r_{i})\Im(z^r_{i})$, with $\Re$ and $\Im$ referring to the real and imaginary part, is introduced to simplify the phononic overlap matrix.
A full configuration $r$ is constructed by attaching an electronic state $|\phi^r\rangle$ to the coherent state product $|\mathbf{z}^r\rangle$, 
resulting in the total state $ |\psi^r\rangle=|\phi^r\rangle\otimes|\mathbf{z}^r\rangle $ for the $r$-th configuration.

In MCE, the ansatz can be written as $|\phi^r\rangle=\sum_{i=1}^L a^r_{i}(t)\hat{c}_{i}^\dagger|0\rangle_{el}$.\cite{Bramley2019} 
We note that in this paper, we consider only single-particle electronic states.  Representing many-body electronic states is possible, however, 
they require the use of a many-body basis instead of the single-particle basis.
Putting all together, the ansatz for the total state, constructed from $N_c$ configurations $ |\psi^r\rangle $, can be written as \cite{Chen2019}
\begin{equation}\label{Eq:MCE:wave_function}
|\Psi(t)\rangle = \sum_{r=1}^{N_c} |\psi^r\rangle =  \sum_{r=1}^{N_c}\left(\sum_{i=1}^{L} a^r_{i}(t)\hat{c}_{i}^\dagger|0\rangle_{el}\right)\otimes|\mathbf{z}^r(t)\rangle.
\end{equation}
The individual configurations in MCEv1 are not normalized $ \langle\psi^r |\psi^r\rangle \ne 1$, 
but one can normalize the full state $ |\Psi\rangle $.

\subsection{Initialization\label{Sec:MCE:Initialization}}
Initialization of the configurations in Eq.~\eqref{Eq:MCE:wave_function} is important to accurately represent the initial state $ |\Psi_{initial}\rangle $ with the state ansatz of Eq.~\eqref{Eq:MCE:wave_function}. Choosing a good subset $ \{\ket{\mathbf{z}^r}\}_{r=1,\ldots, N_c} $ is crucial for the correctness and efficiency of MCE.\cite{Symonds2018} 
We have observed that MCEv1 is somewhat less sensitive to the initial sampling than MCEv2 due to the coupling between the configurations, see also Ref.$~$\onlinecite{Bramley2019}.

For the initialization, first a set of normalized, but in general nonorthogonal, initial configurations $\ket{ \psi^r}$ need to be found, onto which the initial state can be projected:\cite{Symonds2018}
$|\Psi\rangle = \sum_{r=1}^{N_c}| \psi^r\rangle \sum_{s=1}^{N_c} \left(S^{-1}\right)^{rs}\langle \psi^s|\Psi_{initial}\rangle = \sum_{r=1}^{N_c}| \psi^r\rangle A^r$, with $S^{rs}=\langle\psi^r|\psi^s\rangle$. If the configurations form a complete basis, we have $\ket{\Psi}=\ket{\Psi_{initial}}$.
The coefficients $A^r$ are then absorbed into the electronic coefficients $a^r_{i}$ of the configurations to obtain the non-normalized initial configurations used in the state ansatz of MCEv1, Eq.~\eqref{Eq:MCE:wave_function}. The norm of the resulting state deviates from unity $ \langle\Psi|\Psi\rangle \ne 1 $ in general, but the deviation from 1 can be used as a measure of how well $ |\Psi_{initial}\rangle $ is represented.
To find the set of initial configurations $ \{\ket{\psi^r}\}_{r=1,\ldots, N_c} $, both the coherent state products $\mathbf{z}^r$ and the electronic coefficients ${a}^r_{i}$ need to be specified. 

We first start with the specification of  the coherent state products $\mathbf{z}^r$. 
As mentioned in the computational details for the independent-trajectory methods, Sec.~\ref{Sec:Indep-Traj:Convergence}, the nuclear wave functions of all initial states studied in this work correspond to a coherent state on every site. 
In the coherent phonon basis of MCE, this can easily be described by coherent phonon states on every site $i$ with $z_i^{ini}$. 
According to the ``compressed coherent state swarms''-method proposed in Ref.~\onlinecite{Shalashilin2008}, we sample the values of the coherent state products $\{z^r_{1}, \dots, z^r_{L}\}$ from a Gaussian distribution around these initial values, i.e., for the $r$-th configuration:
\begin{align}
P(\mathbf{z}^r) \propto \prod_{i}\exp\left(-2 \alpha |z^r_{i} - z_i^{ini}|^2\right).\label{Eq:MCE:sampling}
\end{align}
Here, $\alpha$ describes a compression parameter of the sampling width of the Gaussian, with $\alpha=1$ corresponding to a sampling width equal to the width of the coherent states itself.
As investigated in Ref.~\onlinecite{Shalashilin2008}, increasing the compression parameter $\alpha$ can improve the sampling of the initial state and reduce the number of configurations needed to obtain a norm of the projected initial state close to 1. This can lead to more accurate dynamics at later times,\cite{Ma18} but also the opposite might be the case.\cite{Shalashilin2008} 

We adjust the $\alpha$-parameter in an iterative scheme, as in Ref.~\onlinecite{Symonds2018}, here done by iteratively multiplying $\alpha$ by a constant factor $\gamma>1$ until the total norm of the state reaches $ 1 - \langle\Psi|\Psi\rangle < \num{2e-5} $. In addition, all sampled coherent state products are shifted and scaled to reproduce the correct mean and standard deviation of Eq.~\eqref{Eq:MCE:sampling}. For small systems with a large number of configurations, the sampled coherent states can be very dense and cause numerical problems in the inversion of the overlap matrix. To overcome this problem, the sampled configurations in systems with $L\le3$ are all moved very slightly according to a repulsive force, exponentially decaying with the distance between the coherent states. 
Afterwards, if the overlap matrix is still ill-conditioned due to the dense configurations, which might result in erroneous total norms larger than 1, we allow for an expansion of the sampling region by iteratively reducing the compression parameter $\alpha$, possibly even below 1. This never occurred for system sizes $L>3$.

In each iteration of the compression scheme, after the $\mathbf{z}^r$ are obtained, the electronic coefficients are chosen and normalized within each configuration $r$, for which several approaches are possible. 
One option is to choose the electronic coefficients as they are given by the initial wave function at the center point of the sampled coherent state.
We observed a better convergence when using a random sampling of the electron state and employed the so-called ``quantum superposition sampling'' 
(QSS).\cite{Bramley2019} 
In this approach, the electronic coefficients are chosen randomly and also initially unoccupied basis states are sampled. We use a slight variation in not choosing the electronic coefficients completely at random, but Gauss-distributed around its values at the coherent-state center point, with a subsequent normalization. For the electronic Gaussian function we choose a standard deviation of $\sigma_{el}=0.5/\sqrt{\alpha}$, which changes iteratively with the compression scheme of the nuclear sampling process.
A value of $\sigma_{el}=0$ corresponds to the deterministic electronic coefficients obtained from $\ket{\Psi_{initial}}$ at the center point of the selected phonon coherent state, while a value of $\sigma_{el}=\infty$ corresponds to completely random electronic coefficients. While we do not perform a systematic study, we find the choice $\sigma_{el}=0.5/\sqrt{\alpha}$ to be a reasonable choice for most of our investigated initial states. The exception is the tunneling process (see Fig.~\ref{Fig:Results:Holstein_Dimer:Gammabar2_tbar0.5:dressed}), where a value of $\sigma_{el}=1/\sqrt{\alpha}$ produces better results, as more basis states on the initially unoccupied site are helpful in recovering the tunneling transition. 

In general, treating adiabatic initial states and related observables, such as the occupation of the Born-Oppenheimer states $\braket{\hat n^{BO}_a}$, is computationally demanding in our implementation of MCE. A single configuration in the state ansatz used here, Eq.~\eqref{Eq:MCE:wave_function}, contains one diabatic electronic state and a coherent phonon-state, which has a Gaussian shape in a phonon coordinate representation $R$. Thus, Gaussian integrals in the phonon coordinate space are needed to evaluate projections onto the Born-Oppenheimer states, which become demanding for large systems and are only done for small systems $L\le3$ in this work. 
Alternatively, one could directly implement MCE with an adiabatic electronic basis,\cite{Makhov2017} which is, however, not pursued in this work.

Finally, for large systems, we find it beneficial to use a site-dependent compression value $\alpha_i$, similar to the pancake sampling.\cite{Shalashilin2008} In the pancake-sampling, the nuclear initial conditions on sites that are regarded as less important for the dynamics (e.g., bath sites in comparison to system sites) are sampled with a higher compression.
We use a pancake-like sampling for systems with $L>3$, where the compression parameters $\alpha_i$ on each site $i$ are scaled with an exponential of the distance from the initially occupied (central) site $c$, similar to the formula used in Ref.~\onlinecite{Ronto2013}: $\alpha_i=\alpha\cdot\exp\left(|i-c|/\tau\right)^2$, with the characteristic decay-scale $\tau$, which is set to 3 in this work.  
When using this pancake-like sampling for our large system studies ($L>3$), we find that the iterative compression of the coherent state sampling stopped very close to a value of $\alpha=1$, i.e., the total norm of the state is already accurately described by the pancake compression, without any additional global compression.

\subsection{Time propagation\label{Sec:MCE:Propagation}}
The equation of motion for the electronic coefficients $a^r_{i} $ can be obtained from the Dirac-Frenkel time-dependent variational principle\cite{Itzykson2012} (using $\hbar=1$ in this subsection)
with the Lagrangian $\mathcal{L} = \langle\Psi|i\frac{{\partial}}{\partial t}-\hat{H}|\Psi\rangle $. Performing the variations 
$\frac{\partial\mathcal{L}}{\partial {a^r_i}^*} - \frac{\mathrm{d}}{\mathrm{d}t}\frac{\partial\mathcal{L}}{\partial {\dot{a}^{r*}_i}} = 0$, the equation of 
motion for the electronic coefficients $a^s_{i} $ reads\cite{Chen2019}
\begin{equation}\label{Eq:MCE:eom_electron_variational}
\begin{split}
i\sum_{s}R^{rs} \dot{a}^s_{i} =
& \frac{\partial}{\partial {\dot{a}^{r*}_i}} \langle\Psi|\hat{H}|\Psi\rangle\\ &-i\sum_{s}R^{rs}a^s_{i}\sum_{j}\left(z^{r*}_{j}\dot{z}^s_{j} - \frac{\dot{z}^{s*}_{j} z^s_{j}+z^{s*}_{j}\dot{z}^s_{j}}{2}\right),
\end{split}
\end{equation}
with $R^{rs}=\braket{\mathbf{z}^r|\mathbf{z}^s}$.

We note that, at this point, we could apply the variational principle also to the ${z^r_{i}}^*$ coordinates to derive the equation of motion for $z^r_{i}$. 
This would lead to a fully variational method, the so-called multiple Davydov $ D_2 $ ansatz.\cite{Davydov_1982,Cruzeiro-Hansson94,Zhao2012,Zhou15,Chen2019}
Instead, in the MCE method, a different choice for the equation of motion for $z^r_{i}$ is used.
Concretely, the Ehrenfest forces, $\dot{x}^r_{i} =\partial H^{nuc,r}/\partial p^r_{i}$ and  $\dot{p}^r_{i} = -\partial H^{nuc,r}/\partial x^r_{i}$, from each configuration-averaged Hamiltonian $H^{nuc,r} = \langle \psi^r|\hat{H}|\psi^r\rangle/\langle \psi^r|\psi^r\rangle$ are used to evolve the coefficients $z^r_{i}$.
For the Holstein Hamiltonian (Eq. \eqref{Eq:Holstein_model:Hamiltonian:2ndQ}), this can be compactly written as
\begin{equation}\label{Eq:MCE:eom_phonon}
i\dot{z}^r_{i} = \frac{\partial H^{nuc,r}}{\partial\dot{z}^{r*}_{i}} = \omega_0 z^r_{i} -\gamma \underbrace{\frac{|a^r_{i}|^2}{\sum_{j}|a^r_{j}|^2}}_{n^r_{i}},
\end{equation}
where the second term is the force proportional to the electronic density $n^r_{i}$ on site $ i $ in the configuration $r$.
Inserting the equation of motion \eqref{Eq:MCE:eom_phonon} into Eq.~\eqref{Eq:MCE:eom_electron_variational} and applying this to our Holstein model, we get a linear coupled system of equations of motion:
\begin{equation}\label{Eq:MCE:eom_electron}
\begin{split}
i\sum_{s}R^{rs} \dot{a}^s_{i} =
&-t_0\sum_{s} R^{rs} (a^s_{n-1} + a^s_{n+1}) \\
&-\gamma\left(z^{r*}_{i}\sum_{s}R^{rs}a^s_{i}+\sum_{s}R^{rs}a^s_{i}z^s_{i}\right)\\
&+\gamma\sum_{s} \left(R^{rs}\sum_{j}z^{r*}_{j}n^s_{j}\right)a^s_{i}\\
&+i\gamma\sum_{s}R^{rs}a^s_{i}\sum_{j} n^s_{j}\Im(z^s_{j}),
\end{split}
\end{equation}
resulting in a norm-conserving dynamics.\cite{Ma18} In this work, the equations of motion in Eq.~\eqref{Eq:MCE:eom_phonon} and 
Eq.~\eqref{Eq:MCE:eom_electron} are integrated with an adaptive time-step Runge-Kutta-Fehlberg4(5) integrator. 
The error tolerance is chosen sufficiently small to conserve the norm up to \num{1e-6}.

\subsection{Convergence properties and benchmark of MCE\label{Sec:MCE:Convergence}}

Investigating the internal convergence in the MCE algorithm is important, since the calculated observables should approach the exact values, when the basis at a certain time step becomes sufficiently large.
Previous convergence tests of MCE and benchmarks to other methods exist, see, e.g., for the spin-boson model in Ref.~\onlinecite{Shalashilin2009} and for a donor-acceptor charge transfer system in Ref.~\onlinecite{Ma18}. A comparative study in an extended periodic dispersive Holstein chain was carried out recently in Ref.~\onlinecite{Chen2019} for up to 16 lattice sites. 
For the results presented in our work, we analyze the convergence of MCE for Holstein chains with up to $L=51$ sites with a benchmark obtained from the numerically exact DMRG-LBO (denoted as DMRG) simulations.

We find that the convergence works well for small systems and also in large systems for certain observables, such as the average phonon position and electronic densities close to the initial position of the electron. 
However, as mentioned before, a good convergence is difficult to reach for long times in large systems for other observables,\cite{Ma18} e.g., subsystem energies and local quantities far away from the initially occupied sites.
Since the aforementioned violation of total energy conservation is induced by the incomplete basis,\cite{Ma18} one can attempt to use the total-energy drift as an internal convergence criterion. 
For the convergence analysis, it is again helpful to repeat each simulation $N_r$ times, see Sec.~\ref{Sec:Indep-Traj:Convergence}. We then obtain both the run-averaged observable $\braket{O}^r$ and its standard deviation $\sigma_O$. 

\begin{figure}
	\centering
	\includegraphics{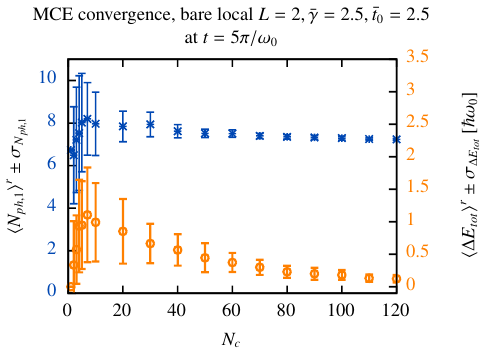}
	\caption{MCE convergence of the phonon number on the initially occupied site $N_{ph,1}$ (blue stars) and total energy drift $\Delta E_{tot}$ (orange circles) versus the number of used configurations $N_c$ for the bare local initial state in the Holstein dimer with $\bar\gamma=2.5,\bar t_0=2.5$. We show the results at time $t=5\pi/\omega_0$. The observables are run-averaged over $N_r=50$ runs and the resulting standard deviation is displayed as error bars.}
	\label{Fig:MCE:convergence_example}
\end{figure}

As a typical scenario obtained for small systems, Fig.~\ref{Fig:MCE:convergence_example} shows the run-averaged number of phonons with its standard deviation as error bars over the number of used configurations at a time $t=5\pi/\omega_0$, starting from the bare local state (see Fig.~\ref{Fig:Introduction:States_Sketch}(b)) in the Holstein dimer with $\bar \gamma=\bar t_0=2.5$. This is compared to the run-averaged total energy drift, with its standard deviation as error bars.

\begin{figure}
	\centering
	\includegraphics{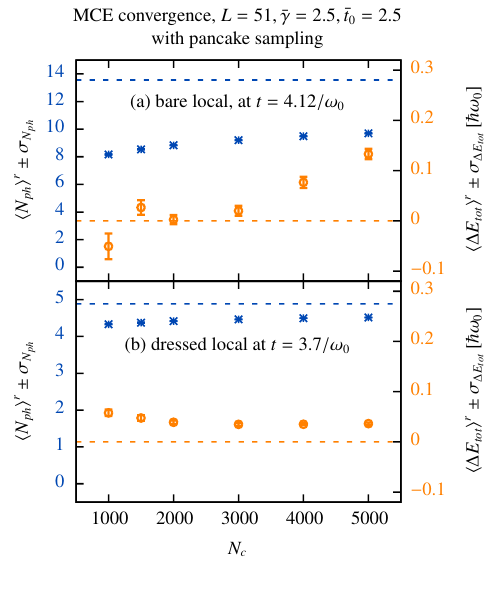}
	\caption{MCE convergence of the total phonon number $N_{ph}=\braket{\sum_i \hat b_i^\dagger \hat b_i}$ (blue stars) and total energy drift $\Delta E_{tot}$ (orange circles) over the number of used configurations $N_c$ for the (a) bare and (b) dressed local initial state in the Holstein chain with $L=51,\bar\gamma=2.5,\bar t_0=2.5$. The MCE initialization is done with the pancake-like sampling, which improves convergence.
		The observables are run-averaged over $N_r=10$ runs and the resulting standard deviation is included as error bars. The desired converged results are indicated by dashed lines: zero energy drift, and the total phonon number obtained with DMRG at the selected time.
		We show the results at time (a) $t=4.12/\omega_0$ and (b) $t=3.7/\omega_0$, corresponding to the first major maximum (bare) or minimum (dressed) in the total phonon number obtained with DMRG, see Figs.~\ref{Fig:Results_Extended:bare_local_Nph} and \ref{Fig:Results_Extended:dressed_local_Nph}.}
	\label{Fig:MCE:convergence_example_L51}
\end{figure}

For a large number of configurations $N_c$, the total phonon number converges to a constant value and the energy drift tends to zero, as expected. In addition, the standard deviation across different runs becomes negligible and the dynamics thus independent of the random initial sampling. Interestingly, the behavior is different for small $N_c$, where, at first, the energy drift and its standard deviation increase with the number of configurations. 
We can understand this from the fact that for far-separated configurations in the phonon Hilbert space, their dynamics become almost independent and similar to the MTE approach introduced in Sec.~\ref{Sec:MTE}. Only when the configurations come close to each other, but are not yet dense enough to form a nearly complete basis, an energy drift is introduced. 
We note that we also enforce the correct mean and variance of the initial phonon sampling, up to the compression scheme mentioned in Sec.~\ref{Sec:MCE:Initialization}, which influences the low-$N_c$ regime. 
The total phonon number and its standard deviation converge only in the large-$N_c$ limit, when the energy drift approaches zero again. This is relevant for large systems, where this limit is difficult to reach (see Fig.~\ref{Fig:MCE:convergence_example_L51}): only looking at the energy drift, without its trend with increasing the number of configurations, could give a wrong impression of the internal convergence of MCE. 

For small systems, $L{=}2$ and $L{=}3$, we could always reach the large-$N_c$ limit. Here, we identify the energy drift as a useful tool to assess internal convergence. It grows with the simulation time $t$, so that an increasing number of configurations $N_c$ is needed to obtain correct results for later times. In our results for $L=2$ and $L=3$ we indicate the time $t^{MCE}_{0.2\hbar\omega_0}$, where the run-averaged energy drift exceeds $0.2\,\hbar\omega$.
This is an arbitrary value, useful only for comparing the convergence between different system and method parameters. 

For larger systems, we could only reach the large-$N_c$ limit for very short times, while for most of the dynamics, the simulations are just beyond the energy-drift peak (for the dressed initial states), or even in the rising energy-drift regime (for the bare initial states). The MCE convergence of the total phonon number $N_{ph}=\braket{\sum_i \hat b_i^\dagger \hat b_i}$ and the total energy drift in the Holstein chain with $L=51,\bar \gamma=\bar t_0=2.5$ is compared to the value obtained with DMRG (see Sec.~\ref{Sec:DMRG}) in Fig.~\ref{Fig:MCE:convergence_example_L51}. 
We analyze the convergence at the time where the total phonon number obtained from DMRG reaches the first major maximum (bare), or minimum (dressed), see our results in Figs.~\ref{Fig:Results_Extended:bare_local_Nph} and \ref{Fig:Results_Extended:dressed_local_Nph} for the obtained time evolution. 

For both the bare and dressed initial state in Fig.~\ref{Fig:MCE:convergence_example_L51}, we could not converge the total phonon number to the DMRG value for the number of configurations available. For the bare local initial state, the total energy drift is negative for small $N_c$, which in our simulations always indicates a poor sampling. For the dressed initial state, the total energy drift is already in the falling branch, and only shows a small absolute total energy drift $\braket{\Delta E_{tot}^r(N_c=5000)}\approx0.036\,\hbar\omega_0$. While for small systems, such a value is completely sufficient, this is not the case  for the Holstein chain ($L \gg 3$) with only one localized electron: the total initial energy stays constant with the number of sites (ignoring the zero-point energy of the phonon harmonic oscillators) and thus the average energy per site becomes very small for large systems. A global quantity like the total phonon number is very difficult to converge in this case. 

We will see later that some observables are still very well recovered in these regimes, for example, local observables around the initially occupied site. 
Thus, we consider the MCEv1 method used in this work as a promising technique, especially for small systems and some local observables with a larger energy scale, while global observables and questions of energy transfer between subsystems seem to be much more difficult to obtain, or would require more extensive computational resources.

	\section{Results for the Holstein dimer\label{Sec:Dimer}}

Before analyzing the  Holstein chain, we first review the non-adiabatic dynamics in the Holstein dimer. 
As a prototypical system for an avoided crossing in a confining potential, this and similar models are already well studied in the literature,\cite{Stock2005} see also Sec.~\ref{Sec:Intro}. Here, we concentrate on a few example cases to illustrate the influence of non-adiabatic effects on electron-nuclear dynamics and the ability of the trajectory-based methods to capture these effects. We will use our insights later in our interpretation of our results for the Holstein trimer and the  Holstein chain, as many observations carry over to these systems. 
For the Holstein dimer, we compare the independent-trajectory methods multitrajectory Ehrenfest (MTE) and fewest-switches surface hopping (FSSH) to exact diagonalization (ED) and the multiconfigurational Ehrenfest approach (MCE). For the ED calculation, we include the results from the grid-based calculation in the Born-Oppenheimer basis (see Sec.~\ref{Sec:NucWFEigenstates}), which is, in all tested systems, identical to ED in second quantization (Sec.~\ref{Sec:ED}).

We analyze the three different initial conditions shown in Figs.~\ref{Fig:Introduction:States_Sketch}(a)-(c).
(i) The wave function has contributions only on the upper Born-Oppenheimer surface, which we call an adiabatic initial state, and then enters a region of significant derivative coupling, thus allowing for transitions between the surfaces (Fig.~\ref{Fig:Introduction:States_Sketch}(a)). 
(ii) The electron is initially localized to a single dimer site (Fig.~\ref{Fig:Introduction:States_Sketch}(b),(c)). 
Here, the initial electron state does thus not depend on the nuclear position and populates a single (trivial) diabatic state. It is in a coherent superposition of both adiabatic states, which provides a challenge for the surface-hopping approach (see Sec.~\ref{Sec:FSSH}).
One example of a comparative study in a similar system with initial conditions not restricted to a single adiabatic state is Ref.~\onlinecite{Landry2013}, where also the mixed calculation of diabatic populations in FSSH was suggested (see Sec.~\ref{Sec:FSSH}).
In both the adiabatic and local initial states, the nuclear wave function in the respective adiabatic or diabatic basis (see Eqs.~\eqref{Eq:BH:WFexpansion:adiabatic} and \eqref{Eq:BH:WFexpansion:diabatic}), has only one non-zero component $\Psi_a(q)=\delta_{a,b}\Psi_{nuc}(q)$. For this component of the nuclear wave function we always choose a coherent state centered around some $\bar q_0\coloneqq \braket{\hat {\bar q}(t{=}0)}$ and with $\bar p_{q,0}=0$.

For the local initial conditions (ii), we analyze two types of such initial states: First, the bare local initial state, where the system starts with all phonon sites in the ground state $\bar q_0=0$, corresponding to a bare charge carrier injected locally into a Holstein chain, where no electron-phonon coupling or electron spreading could yet develop, see Fig.~\ref{Fig:Introduction:States_Sketch}(b). 
Second, the localized electron can start with local phonon excitations already developed on the initial site $\bar q_0=\bar \gamma$, which is called the dressed local initial state, see Fig.~\ref{Fig:Introduction:States_Sketch}(c). For the Holstein dimer, we consider this state mainly to showcase the effect of a tunneling transition, but dressed local states will be analyzed in more detail for the larger systems. The dressed local state can, for example, be realized by a sudden parameter quench, where the electronic hopping matrix element is turned on for times $t>0$.

\subsection{Adiabatic initial state: Transition through the avoided crossing\label{Sec:Results_dimer:Adiabatic}}

In the adiabatic initial state, the electron initially occupies the upper Born-Oppenheimer state, while we choose the phonon distribution as a coherent state around $\bar q_0=\bar \gamma$. In the spirit of the language of chemical bonds, this initial configuration corresponds to an antibonding initial state.
As parameters, we choose an intermediate regime with $\bar \gamma=\bar t_0=2.5$, which allows for non-adiabatic transitions between the Born-Oppenheimer surfaces when the wave packet approaches the avoided crossing point at $\bar q=0$.

\begin{figure*}
	\includegraphics{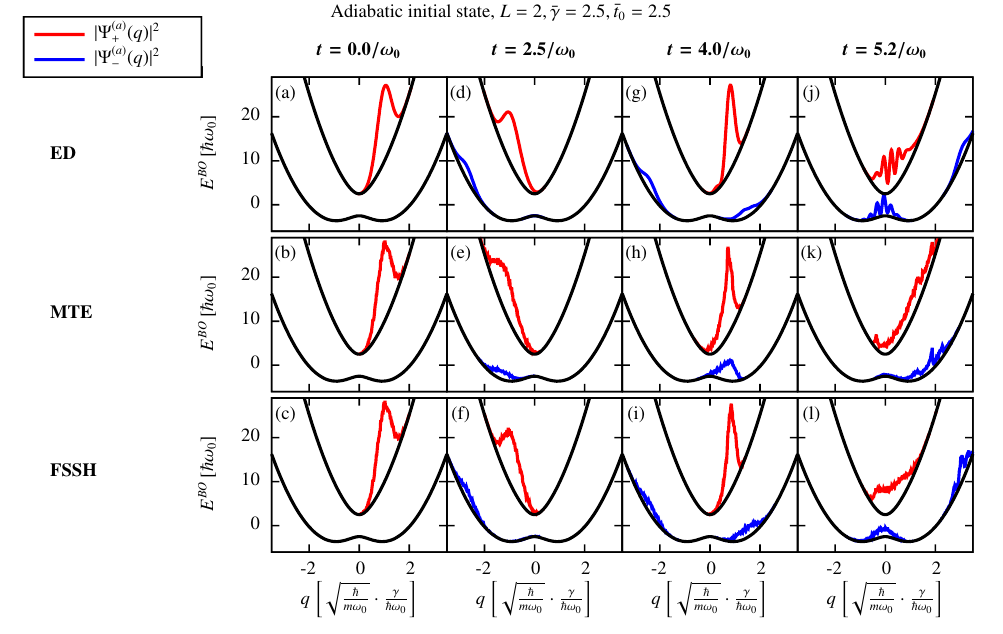}
	\caption{Snapshots of the Born-Oppenheimer (adiabatic) probability densities $|\Psi^{(a)}_\pm(q,t)|^2$, see Eq.~\eqref{Eq:BH:WFexpansion:adiabatic}, for three different methods starting from the adiabatic initial state in the Holstein dimer with $\bar \gamma=\bar t_0=2.5$. In the first row, we show the numerically exact results obtained from exact diagonalization using the approach outlined in Sec.~\ref{Sec:NucWFEigenstates}. This is compared to the two independent-trajectory methods multitrajectory Ehrenfest (MTE) and fewest-switches surface hopping (FSSH), here without the decoherence correction, in the two rows below. For each method, the Born-Oppenheimer probability densities are depicted for four different time snapshots, (a)-(c) $t=0/\omega_0$, (d)-(f) $t=2.5/\omega_0$, (g)-(i) $t=4/\omega_0$ and (j)-(l) $t=5.2/\omega_0$. In each snapshot, the density on the upper and lower Born-Oppenheimer surface are represented as a red and blue line. Both are drawn on top of the respective Born-Oppenheimer energy surfaces with an arbitrary but constant scaling to improve visibility. MTE and FSSH use 20000 trajectories and $\Delta t=0.01/\omega_0$, and the ED method 250 eigenstates (see Secs.~\ref{Sec:NucWFEigenstates} and \ref{Sec:Trajectory}).}\label{Fig:Dimer_NucWF_TimeEvolution}
\end{figure*}

To get an impression of the time-evolution predicted by the exact and independent-trajectory methods, we show the probability densities on the two Born-Oppenheimer surfaces in nuclear coordinate space $|\Psi^{(a)}_\pm(q,t)|^2$ (see Eq.~(\ref{Eq:BH:WFexpansion:adiabatic})) in Fig.~\ref{Fig:Dimer_NucWF_TimeEvolution}. For the independent-trajectory methods MTE and FSSH (here without decoherence correction), they are obtained from $|\Psi^{(a)}_\pm(q,t)|^2=\int \textup{d}p_q \braket{\phi^{BO}_\pm(q)|\hat W(q,p_q,t)|\phi^{BO}_\pm(q)}$ using the approximate partially Wigner-transformed density matrix from Eq.~(\ref{Eq:Independent_trajectory_average}). 
For each method, the Born-Oppenheimer probability densities are shown for four different time snapshots.

\begin{figure}[]
	\includegraphics{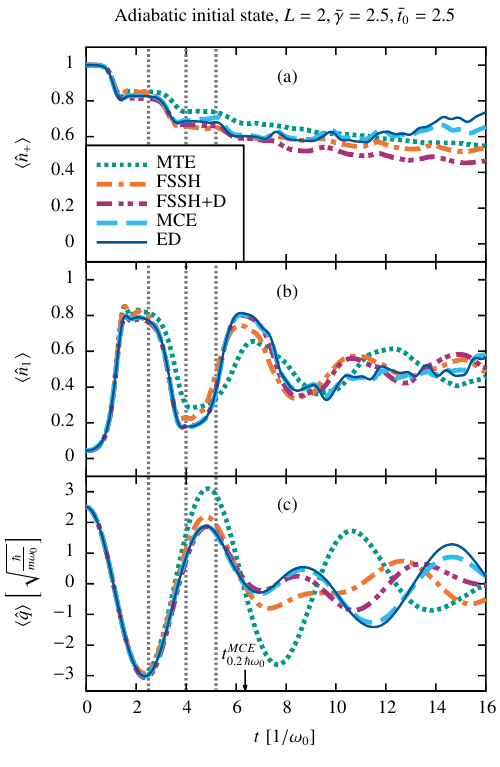}
	\caption{Time evolution of three observables starting from the adiabatic initial state in the Holstein dimer with $\bar \gamma=2.5$ and $\bar t_0=2.5$.
		We show the (a) electronic occupation of the upper Born-Oppenheimer surface $\braket{\hat n_+}$, (b) electronic occupation of the first dimer site $\braket{\hat n_1}$, and (c) average phonon coordinate $\braket{\hat q}$ for five different methods: Exact diagonalization (ED) as a dark blue solid line, multiconfigurational Ehrenfest (MCE) as a dashed light blue line, fewest-switches surface hopping without decoherence correction (FSSH) and with decoherence correction (FSSH+D) as dashed-dotted magenta and orange lines, and multitrajectory Ehrenfest (MTE) as a dotted green line. MTE, FSSH and FSSH+D use 20000 trajectories and $\Delta t=0.01/\omega_0$, MCE uses 300 configurations, and ED 250 eigenstates.
		For the descriptions of the methods, see Secs.~\ref{Sec:NucWFEigenstates},\ref{Sec:Trajectory}, and \ref{Sec:Multi}. The energy drift of MCE reaches $0.2\,\hbar\omega_0$ at $t^{MCE}_{0.2\,\hbar \omega_0}=6.37/\omega_0$. The times of the snapshots shown in Fig.~\ref{Fig:Dimer_NucWF_TimeEvolution} are indicated with vertical gray dashed lines.}\label{Fig:Results:Holstein_Dimer:Antibond_Gammabar2.5_tbar2.5:combined}
\end{figure}

The first snapshot at $t=0/\omega_0$ illustrates the starting configuration of each method, Figs.~\ref{Fig:Dimer_NucWF_TimeEvolution}(a)-(c), where the initial Gaussian shape is roughly recovered by the random sampling of the 20000 trajectories of the independent-trajectory methods. At $t=2.5/\omega_0$ (Figs.~\ref{Fig:Dimer_NucWF_TimeEvolution}(d)-(f)), the initial wave packet has passed through the avoided crossing for the first time and now shows contributions on both Born-Oppenheimer surfaces. The FSSH method reproduces the separation of the upper and lower component of the nuclear wave function, while they are centered around the same $q$ in the MTE method. 
Thus, MTE already fails to correctly describe the very first transition through the avoided crossing. The same holds for the second splitting at $t=4/\omega_0$ (Figs.~\ref{Fig:Dimer_NucWF_TimeEvolution}(g)-(i)). At this time, the mean nuclear position $\braket{\hat q}$ of MTE also starts to deviate significantly from the exact results, as illustrated later in Fig.~\ref{Fig:Results:Holstein_Dimer:Antibond_Gammabar2.5_tbar2.5:combined}.
Finally, at $t=5.2/\omega_0$ (Figs.~\ref{Fig:Dimer_NucWF_TimeEvolution}(j)-(l)), the remaining upper part of the nuclear wave function interferes with its recurring lower component. The following interference pattern is not accurately captured by the surface-hopping method and observables will start to deviate from the exact results at later times.

We can investigate the transition probabilities back into the lower Born-Oppenheimer state by looking at the occupation of the upper Born-Oppenheimer surface $\braket{\hat n_+}(t)$. This is depicted in Fig.~\ref{Fig:Results:Holstein_Dimer:Antibond_Gammabar2.5_tbar2.5:combined}(a), which also includes the results from the fewest-switches surface hopping with the force-based decoherence rate of Eq.~(\ref{Eq:Decoherence_rate}), and the multiconfigurational Ehrenfest method (see Sec.~\ref{Sec:Multi}).
We can see the first three transitions through the avoided crossing, where every time a portion of the upper nuclear wave function passes to the lower surface, while $\braket{\hat n_+}(t)$ stays plateau-like between the transitions. Already after the first transition, the MTE method overestimates the occupation of the upper Born-Oppenheimer surface, i.e., it underestimates the relaxation probability. FSSH exhibits deviations after the first transition as well, but is capable of still giving qualitatively accurate results for the second, third, and fourth transition. The decoherence correction (FSSH+D) improves the quantitative descriptions of the first three plateaus, but overestimates the relaxation for longer times more than the conventional FSSH. It was reported before that decoherence corrections can induce problems when describing recoherences, where previously separated parts of the wave function meet again.\cite{Subotnik2016}
All of the independent-trajectory methods fall short of correctly describing the long-time behavior for times $t>10/\omega_0$, when the wave function is spread across most of the nuclear coordinate space. 

In addition, we can look at the electronic densities on the two dimer sites, corresponding for the one-electron case to the populations in our diabatic basis, which is displayed for the first site $\braket{\hat n_1}$ in Fig.~\ref{Fig:Results:Holstein_Dimer:Antibond_Gammabar2.5_tbar2.5:combined}(b).
We observe that in the first few transitions through the avoided crossing, the electron, for the most part, switches the occupied site, while for later times, it is mostly delocalized between both dimer sites. Again, MTE reproduces the exact results to the least extent. The surface-hopping methods improve on that, with the decoherence correction again leading to a better quantitative description for short times. For times $t>8/\omega_0$, both FSSH with and without decoherence deviate significantly from the exact results. The time evolution of both the diabatic and adiabatic populations is very close to the results obtained in Ref.~\onlinecite{Stock2005} for a similar initial state.

Finally, we also compare a nuclear observable, for which we choose the average nuclear position $\braket{\hat q}$ (see Fig.~\ref{Fig:Results:Holstein_Dimer:Antibond_Gammabar2.5_tbar2.5:combined}(c)). This observable is slightly more robust than the electronic occupations, with even MTE showing larger deviations only for $t>4/\omega_0$. FSSH+D reproduces the nuclear position to the largest extent.

MCE coincides with the ED results for most of the time evolution and only slight deviations can be seen at the end of the shown time interval for all observables in Fig.~\ref{Fig:Results:Holstein_Dimer:Antibond_Gammabar2.5_tbar2.5:combined}. It is thus a significant improvement over the MTE method.
The energy drift of MCE reaches $0.2\,\hbar\omega_0$ at $t^{MCE}_{0.2\,\hbar \omega_0}=6.37/\omega_0$.

\subsection{Bare local initial state with large electron hopping\label{Sec:Results_Dimer:bare_large_hopping}}

The bare local initial states start with the electron localized to one of the two dimer sites, while the phonons are in a coherent state around $\bar q_0=0$. For these small values of $\bar q$, the Born-Oppenheimer wave functions of the dimer, Eq.~(\ref{Eq:BH-Dimer:WF}), have almost an equal weight in both Born-Oppenheimer states. Thus, this initial state has, already from the beginning, coherent contributions in both Born-Oppenheimer surfaces.
To disentangle the effects of the initial coherent superposition of the state and later non-adiabatic transitions, we choose the  parameters $\bar \gamma=4, \bar t_0=10$. In this large electron hopping parameter regime, non-adiabatic transitions are mostly excluded, see also the Landau-Zener formula (\ref{Eq:BO:LZ}). 

\begin{figure}[]
	\includegraphics{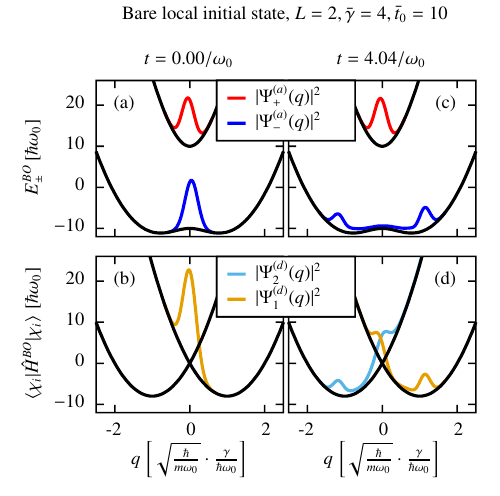}
	\caption{Probability densities obtained from exact diagonalization (see Sec.~\ref{Sec:NucWFEigenstates}) in the Holstein dimer, starting from the bare local initial state with $\bar\gamma=4,\bar t_0=10$. We show snapshots for the initial state $t=0/\omega_0$, and at time $t=4.04/\omega_0$. 
		We display the Born-Oppenheimer (adiabatic) probability densities $|\Psi^{(a)}_\pm(q)|^2$ ((a) and (c)) and local (diabatic) probability densities $|\Psi^{(d)}_i(q)|^2$ ((b) and (d)) (see Eqs.~(\ref{Eq:BH:WFexpansion:adiabatic}) and (\ref{Eq:BH:WFexpansion:diabatic})). 
		The probability densities are drawn on top of the diagonal elements of the Born-Oppenheimer Hamiltonian in the respective electronic bases $E^{BO}_\pm(q)=\braket{\phi^{BO}_\pm(q)|\hat H^{BO}(q)|\phi^{BO}_\pm(q)}$ and $\braket{\chi_i|\hat H^{BO}(q)|\chi_i}$ with an arbitrary but constant scaling to improve visibility. \label{Fig:Results:Holstein_Dimer:Bare_Localized_Gammabar4_tbar10:exact_WF}}
\end{figure}

The probability densities of the initial state in both the adiabatic and diabatic basis (see Eqs.~(\ref{Eq:BH:WFexpansion:adiabatic}) and (\ref{Eq:BH:WFexpansion:diabatic})) are shown in  Figs.~\ref{Fig:Results:Holstein_Dimer:Bare_Localized_Gammabar4_tbar10:exact_WF}(a) and \ref{Fig:Results:Holstein_Dimer:Bare_Localized_Gammabar4_tbar10:exact_WF}(b), obtained from exact diagonalization.
Fig.~\ref{Fig:Results:Holstein_Dimer:Bare_Localized_Gammabar4_tbar10:exact_WF} visualizes the problem of the basis-dependence of the FSSH method: since the forces on the atoms are calculated from the adiabatic electron densities only (Fig.~\ref{Fig:Results:Holstein_Dimer:Bare_Localized_Gammabar4_tbar10:exact_WF}(a)), ignoring the coherent superposition of the two states, we cannot expect FSSH to reproduce the atomic movement exactly, even for short times. 

However, for later times the nuclear wave function components on the two Born-Oppenheimer surfaces will evolve nearly independently from each other, due to the large electron hopping, as can be seen from Fig.~\ref{Fig:Results:Holstein_Dimer:Bare_Localized_Gammabar4_tbar10:exact_WF}(c). The antibonding contribution stays localized around $q=0$, while the bonding (lower Born-Oppenheimer state) contribution separates into both potential energy minima. 
The electron in the bonding contribution is mostly localized on the different Holstein dimer sites, while in the antibonding contribution, the electron rapidly oscillates between the sites. The snapshot of the diabatic probability densities in Fig.~\ref{Fig:Results:Holstein_Dimer:Bare_Localized_Gammabar4_tbar10:exact_WF}(d) captures a moment in which the electron in the antibonding contribution is completely delocalized.

\begin{figure}
	\includegraphics{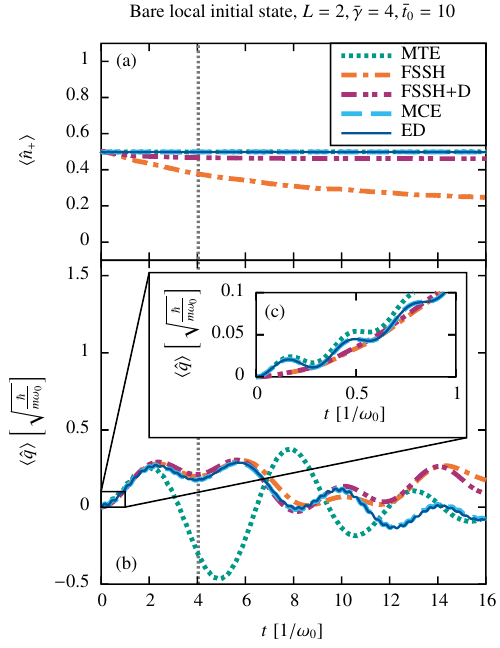}
	\caption{Time evolution of the (a) electronic occupation of the upper Born-Oppenheimer surface $\braket{\hat n_+}$ and (b) average phonon coordinate $\braket{\hat q}$, starting from the bare local initial state in the Holstein dimer with $\bar \gamma=4$ and $\bar t_0=10$, for the methods MTE, FSSH, FSSH+D, ED, and MCE (see Secs.~\ref{Sec:NucWFEigenstates},\ref{Sec:Trajectory}, and \ref{Sec:Multi}). The inset (c) depicts the magnified short-time evolution of the average phonon coordinate. MTE, FSSH and FSSH+D use 20000 trajectories and $\Delta t=0.01/\omega_0$, MCE uses 60 configurations, and ED 250 eigenstates.
		The energy drift of MCE does not reach $0.2\,\hbar\omega_0$ in the time window shown. The vertical gray line indicates the time of the snapshot shown in Figs.~\ref{Fig:Results:Holstein_Dimer:Bare_Localized_Gammabar4_tbar10:exact_WF}(c),(d).}
	\label{Fig:Results:Holstein_Dimer:Bare_Localized_Gammabar4_tbar10:combined}
\end{figure}

This specific local state poses an additional problem for the FSSH method: the occurrence of frustrated hops. Since the nuclear wave function is initialized around $q=0$, almost all classical trajectories from the lower surface will never have enough kinetic energy to perform a hop to the upper surface. Only trajectories starting on the upper surface can hop to the lower surface and, since they keep their high kinetic energy, later hop back to the upper surface. 
This asymmetry leads to incorrect electronic occupations of the Born-Oppenheimer surfaces, as seen from Fig.~\ref{Fig:Results:Holstein_Dimer:Bare_Localized_Gammabar4_tbar10:combined}(a). The decoherence correction dampens out this effect to keep the electronic occupations of the Born-Oppenheimer surfaces close to $0.5$, as ED, MCE, and also the MTE method obtain. 

In addition to the long-time effect of the frustrated hops, the surface-hopping methods cannot reproduce the short-time evolution of the average phonon position $\braket{\hat q}$ (see the inset Fig.~\ref{Fig:Results:Holstein_Dimer:Bare_Localized_Gammabar4_tbar10:combined}(c)) and fail to describe the initial small-scale oscillations of the average nuclear position. This deviation is caused by the aforementioned shortcoming of the adiabatic FSSH method to correctly calculate nuclear forces for an initial state with coherences between the adiabatic states. Intriguingly, the surface-hopping methods reproduce the average phonon position for longer times much better than MTE, see Fig.~\ref{Fig:Results:Holstein_Dimer:Bare_Localized_Gammabar4_tbar10:combined}(b). 
The failure of the MTE method can be understood from its difficulties in reproducing the separation of the upper and lower part of the nuclear wave function visible in Fig.~\ref{Fig:Results:Holstein_Dimer:Bare_Localized_Gammabar4_tbar10:exact_WF}, as already seen for the adiabatic initial state in Sec.~\ref{Sec:Results_dimer:Adiabatic}. 

None of the independent-trajectory methods is able to reproduce the small-scale oscillations of $\braket{\hat q}$ of the exact results for longer times. They correspond to the fast oscillation of the anti-bonding contribution of the nuclear wave function and thus to a fast oscillation of the electron between the Holstein sites. Thus, the long-term oscillations of the electron densities on the Holstein sites are also not reproduced (not shown here). The FSSH+D method loses this fast oscillation the quickest, while it is able to describe the average long-term behavior better than MTE and FSSH without decoherence.

Nonetheless, this initial configuration demonstrates that adiabatic surface-hopping methods have significant difficulties in describing initial states that already start with strong coherences between different Born-Oppenheimer states, i.e., off-diagonal elements in the adiabatic representation of the electronic density matrix. In addition, we see the effect of frustrated hops, which is the most severe when the nuclear trajectories have little momentum compared to the potential energy splitting in the non-adiabatic region.

MCE does not reach an energy drift of $0.2\,\hbar\omega_0$ in the time interval shown and is in excellent agreement with the exact results. Thus, it appears to be the method of choice.

\subsection{Bare local initial state with small electron hopping\label{Sec:Results_Dimer:bare_small_hopping}}

Next, we stay with the bare local initial state, but study a system with reduced electron hopping $\bar t_0=0.5$, also called the resonant regime (see, e.g., Ref.~\onlinecite{Sato18}). This is far away from an adiabatic parameter regime and surface transitions around $\bar q=0$ are very likely. We stay in the strong-coupling regime with $\bar \gamma=2$.
Starting from the bare charge-density wave, published results for larger systems\cite{Stolpp2020} suggest the formation of plateaus in the electronic occupations for these parameters. They can occur when most of the nuclear wave packet evolves away from $\bar q=0$ to the local minima of the lower energy surface $\bar q\approx\sqrt{\bar{\gamma}^2-\bar{t}_0^2/\bar{\gamma}^2}$. Therefore, the plateaus correspond to a transient local electron trapping.
However, we can only expect them to occur for short times, before the nuclear wave packet spreads in the whole nuclear configuration space. 

\begin{figure}[t]
	\includegraphics{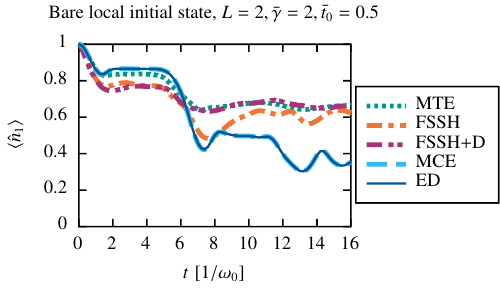}
	\caption{Electronic occupation of the first, initially occupied, dimer site $\braket{\hat n_1}$ for the bare local initial state in the Holstein dimer with $\bar \gamma=2$ and $\bar t_0=0.5$. We compare the results of the methods MTE, FSSH, FSSH+D, ED, and MCE (see Secs.~\ref{Sec:NucWFEigenstates},\ref{Sec:Trajectory}, and \ref{Sec:Multi}). MTE, FSSH and FSSH+D use 20000 trajectories and $\Delta t=0.01/\omega_0$, MCE uses 150 configurations and ED 250 eigenstates.
		The energy drift of MCE does not reach $0.2\,\hbar\omega_0$ in the time window shown.}\label{Fig:Results:Holstein_Dimer:Gammabar2_tbar0.5:bare}
\end{figure}

The comparison of ED, MCE, and the independent-trajectory methods is shown in Fig.~\ref{Fig:Results:Holstein_Dimer:Gammabar2_tbar0.5:bare}. 
We observe that among the independent-trajectory methods, only MTE reproduces the first electronic population plateau at roughly the correct height, while the surface-hopping methods both obtain a too low value. After the first plateau, all of the independent-trajectory methods start to largely deviate from the numerically exact results. In particular, the decoherence correction of FSSH has a detrimental effect on the accuracy of describing the electron density $\braket{\hat n_1}$: Since the wave function is relaxed to the adiabatic states, the coherences between these states are lost even faster than in the other independent-trajectory methods. The failure of the FSSH methods to describe the short-time evolution illustrates the difficulties of these methods to deal with these coherences, especially in such a fast-phonon parameter regime. We note that this initial state is a difficult scenario and that local initial states that start away from $\bar q=0$ are already easier to describe with FSSH. Similar to the large electron-hopping case studied before (Sec.~\ref{Sec:Results_Dimer:bare_large_hopping}), the energy drift of MCE does not reach $0.2\,\hbar\omega_0$ in the time interval shown, and we observe a very good agreement between MCE and exact diagonalization. 

\subsection{Tunneling transition of a dressed local initial state\label{Sec:Results_Dimer:tunneling}}

\begin{figure}
	\includegraphics{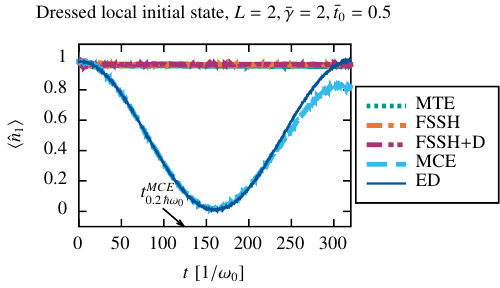}
	\caption{Electronic occupation of the first, initially occupied, dimer site $\braket{\hat n_1}$ for the dressed local initial state in the Holstein dimer with $\bar \gamma=2$ and $\bar t_0=0.5$. We show the long-time results for the methods MTE, FSSH, FSSH+D, ED, and MCE (see Secs.~\ref{Sec:NucWFEigenstates},\ref{Sec:Trajectory}, and \ref{Sec:Multi}).
		MTE, FSSH and FSSH+D use 20000 trajectories and $\Delta t=0.01/\omega_0$, MCE uses 200 configurations and a larger standard-deviation in the initial quantum superposition sampling of the electronic coefficients of $\sigma_{el}=1/\sqrt{\alpha}$, with the compression parameter $\alpha$, see Sec.~\ref{Sec:MCE:Initialization} for details. ED uses 250 eigenstates.
		The energy drift of MCE reaches $0.2\,\hbar\omega_0$ at $t^{MCE}_{0.2\,\hbar \omega_0}=123.7/\omega_0$.}\label{Fig:Results:Holstein_Dimer:Gammabar2_tbar0.5:dressed}
\end{figure}

As a last example  for the dimer, we analyze the even more difficult question of the slow tunneling of an electron between the two dimer sites for the dressed local initial state. The long-time results are displayed in Fig.~\ref{Fig:Results:Holstein_Dimer:Gammabar2_tbar0.5:dressed}, again for $\bar \gamma=2$ and $\bar t_0=0.5$.
We observe slow tunneling with a period of $T\approx 100\,\pi/\omega_0$ between the two dimer sites. This fits the calculated eigenenergies obtained from our exact-diagonalization method, which predicts an energy splitting of the two lowest eigenstates of $\Delta E\approx0.02\,\hbar\omega_0$. The dressed local initial state can be described well as a coherent superposition of these two eigenstates, which are a symmetric and an anti-symmetric combination of Gaussians localized to the two potential energy minima of the Holstein dimer, with only very little contribution on the upper Born-Oppenheimer surface.
Unsurprisingly, all of the independent-trajectory methods stay in the initial potential minimum and do not reproduce the tunneling to the other minimum at all. The classical nuclear kinetic energy of the bonding part of the nuclear wave function is too low to overcome the tunneling barrier. In contrast, the MCE method is able to partially capture the tunneling effect to the other dimer site. This is only possible because some configurations move to the other dimer site, which can be further facilitated in this method by using a larger initial electronic spread for the quantum superposition sampling (see Sec.~\ref{Sec:MCE:Initialization}). We note that using the MCEv2 method (see Sec.~\ref{Sec:Multi}) with additional initial configurations mirrored from the initially occupied to the unoccupied site, we could observe a complete recurrence of the density $\braket{\hat n_1}\rightarrow1$ after one tunneling period (results not shown here). This and other extensions of MCE that include tunneling, see, for example, Refs.~\onlinecite{Makhov2014,Makhov2016}, are not discussed further in this work. Extensions of the independent-trajectory methods to include tunneling effects\cite{Zheng2014,Zheng2014b} are beyond the scope of this work as well, where we will focus on the short-time dynamics in the following examples. However, one should keep the inherent lack of tunneling effects in the independent-trajectory methods in mind. 

\subsection{Summary for the Holstein dimer}

In this section, we compared MTE, MCE and FSSH to numerically exact results in the Holstein dimer. The presented examples illustrate some of the typical non-adiabatic effects trajectory-based methods need to look out for. This includes non-adiabatic transitions of an excited initial state (Sec.~\ref{Sec:Results_dimer:Adiabatic}), built-in coherences between different adiabatic basis states due to the choice of the initial state (Sec.~\ref{Sec:Results_Dimer:bare_large_hopping}), and a combination of both, here in a fast-phonon regime (Sec.~\ref{Sec:Results_Dimer:bare_small_hopping}). MTE, which is basis-independent, is able to describe the short-time dynamics in all cases, but fails to capture independent dynamics on different adiabatic surfaces. Thus, as soon as this becomes relevant in the real-time dynamics, e.g., after the first surface transition in the adiabatic initial state (Fig.~\ref{Fig:Results:Holstein_Dimer:Antibond_Gammabar2.5_tbar2.5:combined}), after the initial build-up of phonons in the bare local initial state (Fig.~\ref{Fig:Results:Holstein_Dimer:Bare_Localized_Gammabar4_tbar10:combined}), or after the end of the first local-trapping plateau in the fast-phonon regime (Fig.~\ref{Fig:Results:Holstein_Dimer:Gammabar2_tbar0.5:bare}), the MTE results deviate significantly from the exact results.

The FSSH methods, in contrast, are well suited to describe independent components of the nuclear wave function on different energy surfaces and can capture non-adiabatic transitions between surfaces well. They are less accurate, however, when the coherences between the adiabatic states become important, either because previously separated portions of the nuclear wave function meet again (Fig.~\ref{Fig:Dimer_NucWF_TimeEvolution}) or because the initial state is already in a highly coherent superposition of different adiabatic states, as for the bare local electron (Figs.~\ref{Fig:Results:Holstein_Dimer:Bare_Localized_Gammabar4_tbar10:combined} and \ref{Fig:Results:Holstein_Dimer:Gammabar2_tbar0.5:bare}). 
Intriguingly, the problematic coherences of the local initial states have a tendency to decay for later times, most pronounced in an approximately adiabatic parameter regime (Sec.~\ref{Sec:Results_Dimer:bare_large_hopping}). Here, the qualities of FSSH, approximately obeying detailed balance and being able to describe a wave-function splitting, can lead to a much better long-time agreement with numerically exact results. The fast-phonon (slow electron) example (Fig.~\ref{Fig:Results:Holstein_Dimer:Gammabar2_tbar0.5:bare}) shows that this is not true for all cases and one should analyze both system parameters and the initial state to make an assessment of the accuracy of the FSSH methods. The decoherence correction improves the average phonon quantities in the long-time regime and is able to remedy some of the frustrated-hop problem of FSSH for the very low nuclear kinetic energies studied in this work (Figs.~\ref{Fig:Results:Holstein_Dimer:Antibond_Gammabar2.5_tbar2.5:combined} and \ref{Fig:Results:Holstein_Dimer:Bare_Localized_Gammabar4_tbar10:combined}). However, the artificial removal of coherences between the adiabatic states can lead to detrimental results, when exactly these coherences are important for describing the initial state (Fig.~\ref{Fig:Results:Holstein_Dimer:Gammabar2_tbar0.5:bare}).
Both MTE and the FSSH methods are not able to describe a slow tunneling transition in the classical energetically forbidden regime (Fig.~\ref{Fig:Results:Holstein_Dimer:Gammabar2_tbar0.5:dressed}). Combined with the problem of frustrated hops and, in general, the quantum nature of phonons, the independent-trajectory methods seem more promising if the phonons are initially prepared in a thermal state, which is, however, not studied in this work.

The MCE method produces remarkable results for all initial states and a very good agreement with the ED data. In addition, it is, for the system and method parameters used, in most cases computationally even cheaper than the independent-trajectory methods (which, however, are easier to parallelize). In this small system, MCE, as an example of a coupled-trajectory method, seems to be the method of choice. It is even able to capture some of the tunneling transition, but special care must be taken in the choice of the initial basis and a perfect transition is difficult to achieve, at least in MCEv1 (Fig.~\ref{Fig:Results:Holstein_Dimer:Gammabar2_tbar0.5:dressed}).

	\section{Results for the Holstein trimer\label{Sec:Results:Trimer}}

For the extended systems ($L>2$) with one electron, we concentrate on the parameters $\bar\gamma=\bar t_0=2.5$, which have already been used for the adiabatic initial state in the Holstein dimer (see Sec.~\ref{Sec:Results_dimer:Adiabatic}) and for the calculation of the Born-Oppenheimer surfaces of the Holstein trimer in Sec.~\ref{Sec:BH-Trimer}. With these parameters we stay in the strong-coupling regime and an intermediate electron hopping, which is neither in the near-adiabatic regime displayed in Sec.~\ref{Sec:Results_Dimer:bare_large_hopping} nor in the rather unphysical fast-phonon regime of Sec.~\ref{Sec:Results_Dimer:bare_small_hopping}. These parameters put all studied methods to a test, as both non-adiabatic transitions and high phonon numbers need to be captured correctly.

For the Holstein trimer, the Born-Oppenheimer surfaces are still easy to visualize, as illustrated in Fig.~\ref{Fig:BH_Trimer:BO-contour} and Fig.~\ref{Fig:BH_Trimer:Trimer_Surf}, which is useful to understand the qualitatively different behavior of the independent-trajectory methods MTE and FSSH, also for the Holstein chain (Sec.~\ref{Sec:One-Electron_Extended}). These two methods are again compared to MCE and ED, where for the exact diagonalization, we now resort to the description in second quantization from Sec.~\ref{Sec:ED}. The occupation of the Born-Oppenheimer states $\braket{\hat n^{BO}_a}$, as well as adiabatic initial states, are not calculated in our implementation of this method. We nonetheless start with one adiabatic initial condition, as one would typically study in a quantum-chemistry problem and then turn  to two different local initial states.
These local states are typical initial conditions of one-electron problems studied in recent quantum-many body investigations, for example, in Refs.~\onlinecite{Stolpp2020,Kloss19} for the Holstein chain, and will also be investigated for the Holstein chain in Sec.~\ref{Sec:One-Electron_Extended}.

\subsection{Adiabatic initial state\label{Sec:Results:Trimer_adiabatic}}

\begin{figure}[t]
	\includegraphics{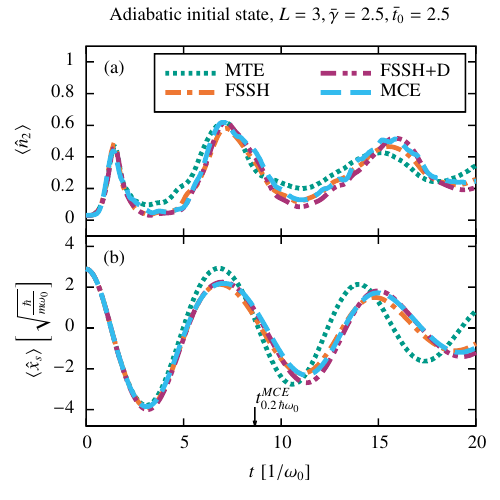}
	\caption{Time evolution of (a) the electronic population on the central site $\braket{\hat n_2}$ and (b) the symmetric phonon mode $\braket{\hat x_s}$, starting from the non-bonding initial state ($\braket{\hat n^{BO}_1(\bar t{=}0)}{=}1$) around $\bar x_a{=}0, \bar x_s{=}2 \bar \gamma/\sqrt{3}$ in the Holstein trimer with $\bar \gamma=2.5$ and $\bar t_0=2.5$. 
		We show results obtained from the methods MTE, FSSH, FSSH+D, and MCE (see Secs.~\ref{Sec:Trajectory} and \ref{Sec:Multi}).
		The independent trajectory methods use 20000 trajectories and $\Delta t=0.001/\omega_0$, and MCE uses 1200 configurations. The energy drift of MCE reaches $0.2\,\hbar\omega_0$ at $t^{MCE}_{0.2\,\hbar \omega_0}=8.66/\omega_0$.}\label{Fig:Results:Holstein_Trimer:nonbonding_Gammabar2.5_tbar2.5}
\end{figure}

We start with an example where the wave function is initially restricted to one of the three Born-Oppenheimer energy surfaces of Fig.~\ref{Fig:BH_Trimer:Trimer_Surf}, which we again call an adiabatic initial state.
We choose the middle, ``non-bonding'', Born-Oppenheimer surface $E^{BO}_1$. For the phonons, we choose a coherent state around $\bar x_a{=}0, \bar x_s{=}2 \bar \gamma/\sqrt{3}\approx2.887$, corresponding to a phonon cloud present on the central Holstein site, while the edge phonon modes are in their ground state. This phonon state is equivalent to the dressed local state investigated later (Sec.~\ref{Sec:Results:Trimer_diabatic}). 
We can understand this state as a dressed local state, where the electron has suddenly been removed and placed into the non-bonding adiabatic state, with its electron density concentrated on the edge sites.
We depict the electronic density on the central site and the average symmetric phonon distortion in Fig.~\ref{Fig:Results:Holstein_Trimer:nonbonding_Gammabar2.5_tbar2.5}.

The figure illustrates that about half of the electronic density quickly moves to the central site, as the symmetric phonon mode goes through zero, but oscillates back as the phonon mode continues to larger negative values. From the shape of the Born-Oppenheimer surfaces (Fig.~\ref{Fig:BH_Trimer:Trimer_Surf}), we know that the non-bonding and the bonding electronic state for these negative values of $\braket{\hat x_s}$ (and with $\braket{\hat x_a}\approx0$) are almost equivalent and have only very little electronic contribution on the central site. After this initial transition, the electronic population on the central site and the symmetric phonon mode oscillate synchronously since now most of the electronic population has relaxed to the ground state. 

We conclude that FSSH, especially with the decoherence correction, is able to describe this time evolution predicted from MCE very well. In addition, MTE exhibits not too large deviations from the MCE results for short times. However, for longer times, it is not able to predict the same oscillation frequency of the symmetric phonon mode $\hat x_s$ as obtained from FSSH and MCE.
For this adiabatic initial state, FSSH appears to be  a very suitable independent-trajectory method, with even higher accuracy for later times than in the Holstein dimer (Fig.~\ref{Fig:Results:Holstein_Dimer:Antibond_Gammabar2.5_tbar2.5:combined}). We attribute this to the larger system size and the beneficial shape of the Born-Oppenheimer surfaces: The region where the non-bonding and the bonding surface come close to each other is also the region where they result in mostly the same forces. This might also lead to the rather good performance of MTE: even if the surface splitting is not perfectly described, the effect on the nuclear dynamics is only minor for short times. For later times, of course, this is not true anymore.
We note that MCE shows an energy drift (with $t^{MCE}_{0.2\,\hbar \omega_0}=8.66/\omega_0$) and data at later times should be interpreted with some caution. 

\subsection{Local initial states\label{Sec:Results:Trimer_diabatic}}

Next, we analyze two local initial states, the bare and dressed local state, see Figs.~\ref{Fig:Introduction:States_Sketch}(d) and (e).
We start with the bare local initial state, where all phonon harmonic oscillators are in the ground state and the electron is localized to one trimer site. We choose the electron to be localized initially to the central site: $\braket{\hat n_2(t{=}0)}=1$. The results presented here differ only quantitatively when we choose an edge site instead.

The ground-state configuration of the phonons corresponds to a Gaussian distribution on each site centered around $\bar x_i=0$. In the reduced trimer coordinates defined in Eq.~\eqref{Eq:Model_Dimer_Trimer:Trimer_phonon_transform}, it is centered around $\bar x_s{=}\bar x_a{=}0$. As described in Sec.~\ref{Sec:BH-Trimer}, the second, non-bonding Born-Oppenheimer state has no electronic contribution on the central site for zero anti-symmetric distortion $\bar x_a=0$ (the red lines in Fig.~\ref{Fig:BH_Trimer:Trimer_Surf}). Here, the localized electron occupies only the lowest, bonding and highest, anti-bonding Born-Oppenheimer state, in fact with an equal weight. Due to the finite width of the Gaussian distribution, the electron will nonetheless have a (smaller) contribution in the non-bonding Born-Oppenheimer state. Similar to the bare local state in the dimer (Secs.~\ref{Sec:Results_Dimer:bare_large_hopping} and \ref{Sec:Results_Dimer:bare_small_hopping}), this initial state is rather distinct from a pure adiabatic state and coherences between the adiabatic basis states are important.

\begin{figure}[t]
	\includegraphics{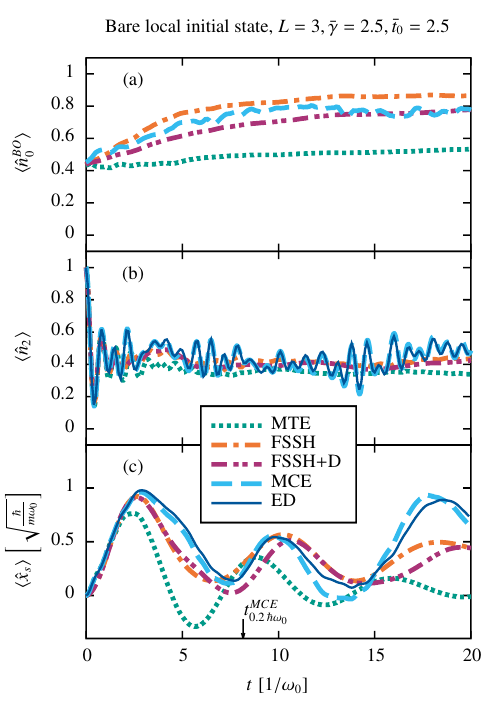}
	\caption{Time evolution of three observables starting from the bare local initial state at the central site in the Holstein trimer with $\bar \gamma=2.5$ and $\bar t_0=2.5$. 
		We show (a) the occupation of the lowest Born-Oppenheimer state $\braket{\hat n^{BO}_0}$, (b) the electronic population on the central site $\braket{\hat n_2}$, and (c) the symmetric phonon mode $\braket{\hat  x_s}$, for the methods MTE, FSSH, FSSH+D, ED, and MCE (see Secs.~\ref{Sec:ED},\ref{Sec:Trajectory}, and \ref{Sec:Multi}). The independent trajectory methods use 20000 trajectories and $\Delta t=0.001/\omega_0$, ED uses 26 local phonon states, and MCE 1000 configurations. The energy drift of MCE reaches $0.2\,\hbar\omega_0$ at $t^{MCE}_{0.2\,\hbar \omega_0}=8.16/\omega_0$.}\label{Fig:Results:Holstein_Trimer:bare_localized_center_Gammabar2.5_tbar2.5}
\end{figure}

The occupation of the lowest Born-Oppenheimer state is depicted in Fig.~\ref{Fig:Results:Holstein_Trimer:bare_localized_center_Gammabar2.5_tbar2.5}(a). 
MTE builds up a much smaller occupation of the lowest Born-Oppenheimer state compared to the surface-hopping methods or MCE. MCE predicts occupations of the lowest Born-Oppenheimer state between FSSH with and FSSH without decoherence for intermediate times.
Note that the energy drift of MCE reaches $0.2\,\hbar \omega_0$ at $t^{MCE}_{0.2\,\hbar \omega_0}=8.16/\omega_0$ (see Sec.~\ref{Sec:MCE:Convergence}). FSSH without decoherence shows the largest relaxation to the lowest Born-Oppenheimer state, similar to the case of large electron hopping, investigated for the Holstein dimer in Sec.~\ref{Sec:Results_Dimer:bare_large_hopping}. 

For the dynamics of the electronic population on the central Holstein site, the independent-trajectory methods are again not able to reproduce the persisting fast oscillations between the sites at late times (see Fig.~\ref{Fig:Results:Holstein_Trimer:bare_localized_center_Gammabar2.5_tbar2.5}(b)), however, the FSSH methods seem to describe the average occupation slightly better than MTE. 

Lastly, we observe large deviations in the average symmetric phonon distortion $\braket{\hat  x_s}$ in Fig.~\ref{Fig:Results:Holstein_Trimer:bare_localized_center_Gammabar2.5_tbar2.5}(c). MTE deviates strongly from ED already after $t\approx2.5/\omega_0$ and also the surface-hopping methods do not quantitatively reproduce the later time evolution. However, they recover the qualitative oscillations in the observable until $t\approx15/\omega_0$. We also see noticeable deviations between ED and MCE when the energy drift becomes too large at times $t>t^{MCE}_{0.2\,\hbar \omega_0}$.

Next, we consider the dressed local initial state, where a coherent phonon state on the initially occupied central site dresses the electron. This corresponds to a Gaussian distribution around zero for the edge sites and around $\bar x_2{=}\sqrt{2}\bar \gamma$ for the central site, or equivalently in the trimer coordinates, $\bar x_a{=}0$ and $\bar x_s{=}2 \bar \gamma/\sqrt{3}\approx2.887$, as for the adiabatic state studied before (Sec.~\ref{Sec:Results:Trimer_adiabatic}).

\begin{figure}[t]
	\includegraphics{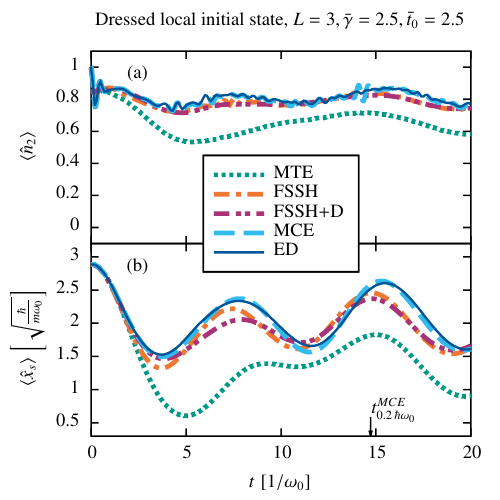}
	\caption{Time evolution of (a) the electronic population on the central site $\braket{\hat n_2}$ and (b) the symmetric phonon mode $\braket{\hat x_s}$ starting from the dressed local initial state at the central site in the Holstein trimer with $\bar \gamma=2.5$ and $\bar t_0=2.5$, for MTE, FSSH, FSSH+D, ED, and MCE (see Secs.~\ref{Sec:ED},\ref{Sec:Trajectory}, and \ref{Sec:Multi}). The independent trajectory methods use 20000 trajectories and $\Delta t=0.001/\omega_0$, ED uses 21 local phonon states, and MCE 1000 configurations. The energy drift of MCE reaches $0.2\,\hbar\omega_0$ at $t^{MCE}_{0.2\,\hbar \omega_0}=14.72/\omega_0$.}\label{Fig:Results:Holstein_Trimer:dressed_localized_center_Gammabar2.5_tbar2.5}
\end{figure}

The occupation of the initially occupied central Holstein site is displayed in Fig.~\ref{Fig:Results:Holstein_Trimer:dressed_localized_center_Gammabar2.5_tbar2.5}(a). 
As always for the local initial states, the independent-trajectory methods cannot reproduce the small-scale oscillations of the electronic density for long times. MTE is also not able to reproduce the average electron density, i.e., too much electronic weight is shifted to the edge sites. 
Moreover, the average symmetric phonon distortion $\braket{\hat x_s}$ in Fig.~\ref{Fig:Results:Holstein_Trimer:dressed_localized_center_Gammabar2.5_tbar2.5}(b) is not correctly described in MTE, while the FSSH methods show qualitatively the correct large-scale oscillations. 
There are no large deviations between ED and MCE, consistent with a small energy drift that reaches $0.2\,\hbar\omega_0$ only at $t^{MCE}_{0.2\,\hbar \omega_0}\approx14.72$.

For both local initial states, MTE has problems to even qualitatively describe the correct long-time behavior.
We attribute this failure to the inability of MTE to correctly describe a wave-function splitting (see Fig.~\ref{Fig:Dimer_NucWF_TimeEvolution}) because of the mean-field description of the electron-phonon coupling.
Describing the wave-function splitting correctly is even important for these local initial states, as illustrated in the Holstein dimer (see Fig.~\ref{Fig:Results:Holstein_Dimer:Bare_Localized_Gammabar4_tbar10:exact_WF}). 
For the bare local initial state (Fig.~\ref{Fig:Results:Holstein_Trimer:bare_localized_center_Gammabar2.5_tbar2.5}(a)), we observe nearly no relaxation towards the lowest Born-Oppenheimer state in the MTE simulation. The dynamics are influenced too much by the large initial occupation of the highest Born-Oppenheimer state, which has its potential energy minimum at $\bar x_a=\bar x_s=0$. In the dressed local initial state, too much electronic weight in the MTE simulation is able to escape the local trapping on the central site (Fig.~\ref{Fig:Results:Holstein_Trimer:dressed_localized_center_Gammabar2.5_tbar2.5}). While the state has a large initial contribution in the lowest Born-Oppenheimer state $\braket{\hat n^{BO}_0(t{=}0)}\approx0.92225$, the remaining weight in the higher states shifts part of the electron from the potential energy minimum corresponding to the central site ($\bar x_s=\bar \gamma$) to the other two minima (see the Born-Oppenheimer surfaces, Figs.~\ref{Fig:BH_Trimer:BO-contour} and \ref{Fig:BH_Trimer:Trimer_Surf}).
In summary, the dynamics in MTE are influenced too much by the higher Born-Oppenheimer states, as also observed in other comparative studies, e.g., Ref.~\onlinecite{Jasper2001}, and the method has difficulties to describe the formation of stable localized charge carriers. 

From the Holstein dimer we have seen that FSSH faces challenges as well, in particular for local initial states, which resurface in the Holstein trimer. The low kinetic energy of the phonons leads to frustrated hops, most pronounced without the decoherence correction, and thus to an overestimation of the relaxation to the lower Born-Oppenheimer surfaces (see Fig.~\ref{Fig:Results:Holstein_Trimer:bare_localized_center_Gammabar2.5_tbar2.5}(a)). The effect is less severe here, due to the moderate strength of the electron-hopping parameter and the different structure of the Born-Oppenheimer surfaces of the trimer (see Fig.~\ref{Fig:BH_Trimer:Trimer_Surf}). Second, FSSH has difficulties describing the correct short-time behavior of both electron densities and nuclear oscillations, due to the missing coherences between the Born-Oppenheimer states for the calculation of the nuclear forces. The decoherence correction worsens this deficiency for the electron density, as it actively dampens out these coherences. 
This effect seems to be much less relevant for the two cases investigated here, compared to the Holstein dimer. The dressed local initial state is less affected by both problems, as it is much closer to an adiabatic initial state. 

For the parameters chosen here, and in general for sufficiently fast electrons, the independent-trajectory methods seem to perform better if the wave function has initially contributions only in a single adiabatic state.
Despite the difficulties of the surface-hopping methods in describing a local initial state, we observe a good qualitative agreement of the long-time behavior of the investigated observables, in contrast to the MTE method. The coupled-trajectory method MCE outperforms both independent-trajectory methods. As long as the energy drift is sufficiently small, it reveals a very good agreement with the ED results, at a still very low computational cost.

	\section{Results for the Holstein chain - one electron\label{Sec:One-Electron_Extended}}

We now turn to the spatial spreading of a single localized electron in Holstein chains of lengths $L=11$ and 51. All observables and methods were also compared for $L=101$, which, however, provides no new insight over the $L=51$ results and is not shown here. 
We stay in the intermediate regime of $\bar\gamma=\bar t_0=2.5$, analyzed before for the Holstein trimer. 

In our study, we want to investigate the detailed short-time evolution of a localized charge carrier, when the phonon subsystem is not yet adjusted to the electron (bare initial state) or is only locally equilibrated (dressed initial state). These two cases correspond precisely to the ``Franck-Condon excitation'' and ``relaxed excitation'' analyzed in Ref.~\onlinecite{Kloss19} with a matrix-product state method for $\bar t_0=1$. 
A comparison to the data of Ref.~\onlinecite{Kloss19} is included in Appendix~\ref{Sec:One-Electron_reduced_el_hopp}. Our conclusions for the trajectory-based methods studied here also hold for that parameter set.

The evolution of such localized states has previously  been investigated in a periodic dispersive Holstein model with the variational Davydov D2 ansatz for 32 sites,\cite{Zhao2012} which also considered off-diagonal electron-phonon coupling, 
and with the hierarchical equations of motion method for 10 sites.\cite{Chen15} The latter was compared to MCE,\cite{Chen2019} with a good agreement of both methods for the system parameters investigated in that study. 
Furthermore, the spreading, and relatedly the mobility, of a local initial state in a large system has  been investigated using the surface-hopping methods, see, e.g., Refs.~\onlinecite{Bai2018,Carof2019}, and references therein. Typically, in these studies, the phonons are initially prepared in a thermal state and the long-time behavior is analyzed.
For an extensive discussion of such initial states and the classification of transport in strongly correlated 1D systems from the perspective of condensed matter theory and quantum many-body methods, see Ref.~\onlinecite{Bertini2021}.

In all following sections we compare the trajectory-based methods to the numerically exact DMRG-LBO results, which we will denote simply as DMRG. 
In FSSH without decoherence, we use the mixed definition of the electronic density matrix $\hat \rho^{\textup{(mixed)}}_{el}$ (see Eq.~\eqref{Eq:FSSH:Mixed_Dens}), while with the restricted decoherence correction, we use the wave-function definition of the density matrix $\hat \rho^{\textup{(WF)}}_{el}$ (see Eq.~\eqref{Eq:FSSH:WF_Dens}). 
In contrast to the small systems studied before, we reach convergence of the MCE simulation for most observables only for very short times (see Sec.~\ref{Sec:MCE:Convergence}). Identifying a useful total energy-drift convergence criterion as for the dimer and trimer is not easily done here, as the relevant energy scales of many of the studied observables become very small. The MCE simulations in this section are done with $N_c=5000$ configurations. As can be seen from the convergence study in Fig.~\ref{Fig:MCE:convergence_example_L51}, this is not enough to consider MCE as an exact method for all observables studied here, and further convergence for even higher $N_c$ is very slow. Similar to MTE and FSSH, we will thus judge the accuracy of the MCE results by comparing them to the DMRG simulation.

\subsection{Bare local initial state\label{Sec:Results:Extended_one_el:Bare}}

\begin{figure*}
	\includegraphics{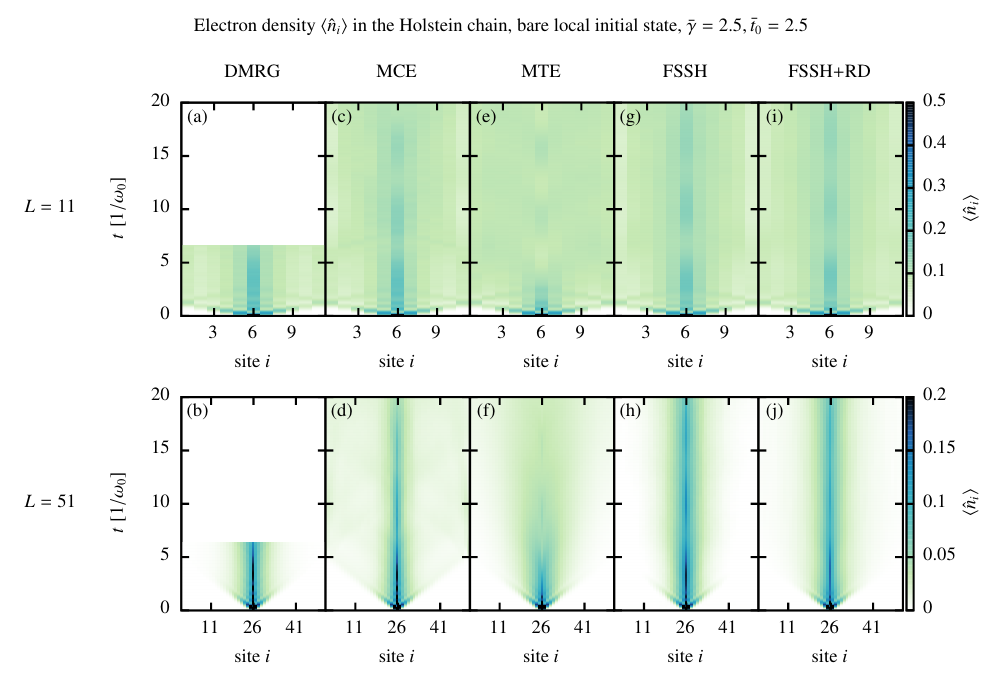}
	\caption{Electron densities $\braket{\hat n_i}$ in the Holstein chain, starting from the bare local electron on the central site with $\bar\gamma=\bar t_0=2.5$. We show the results for $L=11$ and 51 lattice sites in the two rows, obtained from the methods DMRG ((a)-(b)), MCE ((c)-(d)), MTE ((e)-(f)), and FSSH without ((g)-(h)) and with restricted decoherence ((i)-(j)) in the five columns (see Secs.~\ref{Sec:DMRG},\ref{Sec:Trajectory}, and \ref{Sec:Multi}). For better visibility, the maximum color range is set to $\braket{\hat n_i}=0.5$ for $L=11$ and to $\braket{\hat n_i}=0.2$ for $L=51$. In DMRG, we use $\epsilon_{\rm LBO}=10^{-8},\epsilon_{\rm bond}=10^{-8}, \Delta t=0.004/\omega_0$ and $M=40$. For MCE, we use 5000 configurations initialized with the pancake-like sampling, and for the independent trajectory methods, we use 20000 trajectories and $\Delta t=0.001/\omega_0$.}\label{Fig:Results_Extended:bare_local_all_n}
\end{figure*}
We start with the bare local initial state, with the electron placed on the central site of the Holstein chain (see Fig.~\ref{Fig:Introduction:States_Sketch}(d)).
We show the time evolution of the electron densities on the different sites in Fig.~\ref{Fig:Results_Extended:bare_local_all_n} for the two system sizes obtained from the methods DMRG, MCE, MTE, and FSSH without and with restricted decoherence. 

The DMRG results are only available for short times for all system sizes. During that time window a part of the electron density first shows a ballistic spreading until $t\approx2/\omega_0$, while it stays mostly localized after that. For $L=11$ (Fig.~\ref{Fig:Results_Extended:bare_local_all_n}(a)), the electron density already reaches the boundaries of the system by that time and is reflected there, while for $L=51$ (Fig.~\ref{Fig:Results_Extended:bare_local_all_n}(b)), it stays localized in the center for later times and does not reach the chain boundary in the simulated time interval. 

All other methods reproduce the initial ballistic spreading at short times, but deviate from DMRG for later times. 
The electron densities obtained from MCE agree very well with the DMRG results for $L=11$ (Fig.~\ref{Fig:Results_Extended:bare_local_all_n}(c)), where almost no quantitative difference is visible. For $L=51$ (Fig.~\ref{Fig:Results_Extended:bare_local_all_n}(d)), we find a similar excellent agreement for the electron densities on sites close to the initially occupied central site. In contrast, MCE, at least with the number of configurations used here, is not able to contain the electron density in a localized region for later times in this larger system. Instead, a small portion of the electron density continues to spread ballistically through the system and reaches the chain boundaries. This ballistic escape comes along with a weak periodic modulation with the phonon oscillation period. 
The situation becomes much worse for MTE (see Figs.~\ref{Fig:Results_Extended:bare_local_all_n}(e),(f)), which displays a broad delocalization of the electron density for times $t>2/\omega_0$ for both system sizes, not observed in any of the other methods.
Since the nuclear trajectories of MCE follow the Ehrenfest equations of motion, this might explain the poor convergence of MCE in this setup, as illustrated in Sec.~\ref{Sec:MCE:Convergence}.

FSSH without (Figs.~\ref{Fig:Results_Extended:bare_local_all_n}(g),(h)) and with decoherence (Figs.~\ref{Fig:Results_Extended:bare_local_all_n}(i),(j)) predict a very similar electronic density for $L=11$ as MCE and DMRG. For $L=51$, both FSSH methods reproduce the localization after the initial ballistic expansion, also observed in the DMRG results.
The width of this localized electron density does, however, not match the DMRG results exactly, as analyzed later with the reduced mean-squared displacement (Fig.~\ref{Fig:Results_Extended:bare_local_MSD}). FSSH without decoherence recovers the dynamics of the electron density slightly better than with decoherence, which was already observed in the dimer (see Sec.~\ref{Sec:Results_Dimer:bare_small_hopping}). 
The electron density obtained from FSSH without decoherence has small fluctuations on the sites far away from the center for $L=51$, which are not visible in Fig.~\ref{Fig:Results_Extended:bare_local_all_n}(h) because of their very small amplitude. They are caused by the mixed calculation of the electron density matrix via Eq.~\eqref{Eq:FSSH:Mixed_Dens}. While they seem unproblematic in Fig.~\ref{Fig:Results_Extended:bare_local_all_n}(h), they prevent an accurate calculation of the reduced mean-squared displacement (see Appendix~\ref{Sec:FSSH_Large_Sys}). 
We note that the close agreement of FSSH with FSSH+RD in Fig. 21 is only achieved by using this mixed definition for calculating the electronic density matrix in FSSH without decoherence. The electron density calculated from the wave function spreads similar as in the MTE simulation (see Appendix \ref{Sec:FSSH_Large_Sys}), but the distribution of active surfaces is very similar to the FSSH+RD simulation, mostly due to the occurrence of frustrated hops. As pointed out in Ref.~\onlinecite{Subotnik2016}, frustrated hops help to approximately recover detailed balance. For initial states starting from a thermal phonon distribution, frustrated hops might occur less often and FSSH without restricted decoherence could deviate strongly from the FSSH+RD results.

\begin{figure}
	\includegraphics{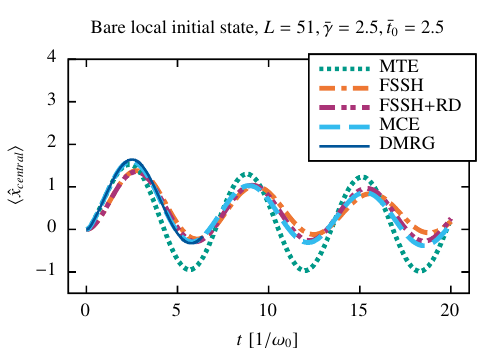}
	\caption{Phonon position $\braket{\hat x_{central}}$ on the central, initially occupied site in the Holstein chain, starting from the bare local electron on the central site with $\bar\gamma=\bar t_0=2.5$ for $L=51$, obtained from the methods MTE, FSSH, FSSH+RD, DMRG, and MCE (see Secs.~\ref{Sec:DMRG},\ref{Sec:Trajectory}, and \ref{Sec:Multi}). We use the same method parameters as in Fig.~\ref{Fig:Results_Extended:bare_local_all_n}.}\label{Fig:Results_Extended:bare_local_central_x_L51}
\end{figure}

\begin{figure}
	\includegraphics{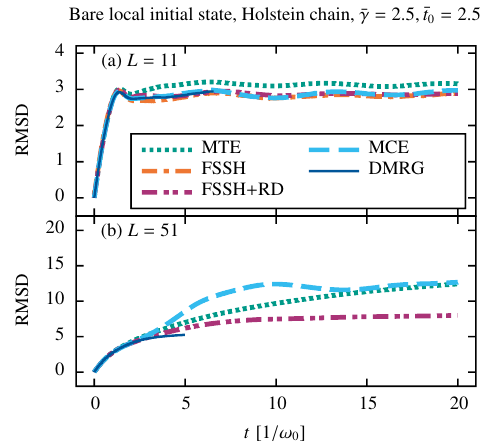}
	\caption{Reduced mean-squared displacement $\textup{RMSD}$ in the Holstein chain given via Eq.~(\ref{Eq:Extended_Sys:MSD}), starting with the bare local electron on the central site with $\bar\gamma=\bar t_0=2.5$ for (a) $L=11$ and (b) $L=51$, obtained from the methods MTE, FSSH+RD, DMRG, and MCE (see Secs.~\ref{Sec:DMRG},\ref{Sec:Trajectory}, and \ref{Sec:Multi}). For FSSH without decoherence, only the results for $L=11$ are included, see Appendix~\ref{Sec:FSSH_Large_Sys}. We use the same method parameters as in Fig.~\ref{Fig:Results_Extended:bare_local_all_n}, with the exception of the DMRG cutoffs, which are decreased to $\epsilon_{\rm LBO}=10^{-9},\epsilon_{\rm bond}=10^{-9}$ for $L=51$.}\label{Fig:Results_Extended:bare_local_MSD}
\end{figure}

The agreement of MCE and the FSSH methods with the DMRG results appears to be more reliable close to the initially occupied central site in Fig.~\ref{Fig:Results_Extended:bare_local_all_n}.
We analyze this further for the time evolution of the phonon position on that central site $x_{central}$ in Fig.~\ref{Fig:Results_Extended:bare_local_central_x_L51} for $L=51$ sites. We do not include the $L=11$ results here, which differ only slightly from the $L=51$ results for this observable.
In this observable, there is no significant difference between MCE and DMRG for the times available.
Similar to most local initial states observed in this work, MTE describes the evolution at very short times the best among the independent-trajectory methods. For later times, however, the phonon position obtained with MTE oscillates almost around zero, indicating  that the electron does not stay sufficiently localized, with most of the electron lost in the broad spreading observed before (Fig.~\ref{Fig:Results_Extended:bare_local_all_n}(f)).
In contrast, FSSH+RD reveals a remarkable agreement with MCE for the phonon position for later times, albeit not capturing the first maximum at the correct height. FSSH without decoherence correction deviates stronger for later times, but still captures the qualitative oscillations of the MCE calculation much better than MTE. Similar to our study in the dimer (Fig.~\ref{Fig:Results:Holstein_Dimer:Bare_Localized_Gammabar4_tbar10:combined}), both FSSH methods can reproduce the long-time dynamics surprisingly well, even if the initial state includes coherences between different adiabatic states, thus preventing a completely accurate description of the short-time dynamics.

A good quantitative description of the spread of the electron density is given by the reduced mean-squared displacement (RMSD):
\begin{align}
\textup{RMSD}(t)=\sqrt{\sum_{i=1}^L \braket{\hat n_i(t)} \left(a i - x^{el}_0\right)^2},\label{Eq:Extended_Sys:MSD}
\end{align} 
where $a\cdot i$ denotes the position of the $i$-th site in the Holstein chain, with $a$ the distance between Holstein sites. $x^{el}_0=\left[\sum_{i=1}^L \braket{\hat n_i(t=0)} a i\right]$ is the average initial electron position, i.e., here the center of the chain.
In the following, we set $a=1$. 
The reduced mean-squared displacement is depicted in Fig.~\ref{Fig:Results_Extended:bare_local_MSD}. Note that the reduced mean-squared displacement weights the electronic density on the edge sites much higher than on the central site (with a factor of 625 for the $L=51$ system). Thus, deviations in the electron density away from the initially occupied site are magnified. This also makes the use of FSSH without decoherence difficult, see Appendix~\ref{Sec:FSSH_Large_Sys}.

For $L=11$ (Fig.~\ref{Fig:Results_Extended:bare_local_MSD}(a)), the FSSH, MCE, and DMRG methods show a very good agreement and seem to converge to the same value. MTE, in contrast, obtains a too high long-time value, consistent with the loss of the local electron on the central site, also visible in Fig.~\ref{Fig:Results_Extended:bare_local_all_n}(e).
For $L=51$ (Fig.~\ref{Fig:Results_Extended:bare_local_MSD}(b)), we observe that the RMSD obtained from MTE follows a mostly square-root behavior for later times, indicating that the spreading of the density is of a diffusive type. The FSSH+RD data follow the same curve for some time, while the RMSD stays almost constant and thus localized for later times. 
The long-time value is slightly higher than what is obtained with DMRG for the times available.
Note that we sometimes use different cutoffs for the DMRG simulations, since some observables, e.g., the RMSD, converge slower than others.
The MCE data exhibit a strong increase of the RMSD for times greater than $t\approx 5/\omega_0$ in this larger system, representative of the ballistic escape of the electron density at those times. We note that this ballistic behavior is observed only for a small portion of the electron density, which, however, becomes the dominant term in the RMSD due to the aforementioned quadratic scaling with the distance from the center of the chain. 

\begin{figure}
	\includegraphics{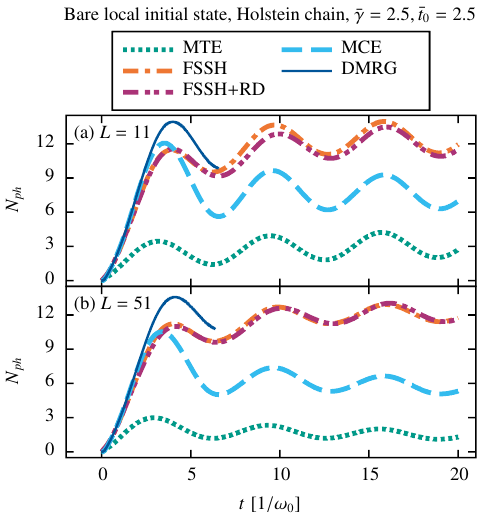}
	\caption{Total phonon number $N_{ph}=\braket{\sum_i \hat b^\dagger_i \hat b_i}$ in the Holstein chain, starting with the bare local electron on the central site with $\bar\gamma=\bar t_0=2.5$ for (a) $L=11$ and (b) $L=51$, obtained from the methods MTE, FSSH, FSSH+RD, DMRG, and MCE (see Secs.~\ref{Sec:DMRG},\ref{Sec:Trajectory}, and \ref{Sec:Multi}). We use the same method parameters as in Fig.~\ref{Fig:Results_Extended:bare_local_all_n}.}\label{Fig:Results_Extended:bare_local_Nph}
\end{figure}

Finally, in Fig.~\ref{Fig:Results_Extended:bare_local_Nph}, we analyze the total phonon number in the system, which is proportional to the energy stored in the phonon subsystem alone (discounting the zero-point oscillator energies). 
First,  the total phonon number in the system is mostly independent of the system size, as expected, since the total initial energy, without the zero-point phonon energy, is the same for all system sizes. No method is able to reproduce the total phonon number obtained in the DMRG simulations, not even for the $L=11$ system. Even MCE has difficulties in computing this quantity, as it is quadratic in the phonon positions and momenta and is directly influenced by the energy drift (see Sec.~\ref{Sec:MCE:Convergence}). The most significant deviation is observed in MTE, where the total phonon number stays very small. 
In contrast, the FSSH methods obtain a similar phonon number as MCE for the first maximum, but then stay at these high values, corresponding to the long-time localization observed before. Unfortunately, we cannot unambiguously determine which long-time behavior is correct, as the DMRG data are not available until late times, and all other methods deviate from DMRG before that. From the figure, we observe that FSSH methods are the closest to the DMRG data
and we may therefore speculate that the FSSH methods also approximate the long-time behavior the closest.
However, numerically exact simulations for longer times are needed for a further analysis.

Overall, in the global quantities of the RMSD and the total phonon number, no consistent picture arises from the trajectory-based methods. 
It remains a promising question for future studies to further investigate the validity of FSSH in the long-time limit, building on existing research on steady-state properties, such as mobilities at finite temperatures.\cite{Bai2018,Carof2019}
As illustrated in various examples in the Holstein dimer and trimer (Secs.~\ref{Sec:Dimer} and \ref{Sec:Results:Trimer}), the difficulties of the FSSH method in describing the short-time behavior stem from the local initial state. Here, the bare local initial state is a near-worst-case scenario, as it has an almost equal electronic weight in all Born-Oppenheimer states.
Since MTE fails in describing any of the analyzed global quantities, a pure mean-field solution seems to be of limited use, for the parameters studied here. A mixture of MTE with a form of decoherence, e.g., the coherent switching with decay-of-mixing method,\cite{Zhu2004b} might be a promising compromise. Finally, even a coupled-trajectory method such as MCE can fail in such large systems, where a good sampling of the relevant Hilbert space is non-trivial. Using an alternative algorithm, which is not based on MTE-guided trajectories, might be beneficial here (see Sec.~\ref{Sec:Multi}). However, for $L=11$, MCE still recovers the electron densities, including the RMSD, very well and the phonon position at the central site shows excellent agreement with DMRG for both system sizes. Our assessment of the trajectory-based methods also holds for the parameter regime $\bar t_0=1$, shown in Appendix~\ref{Sec:One-Electron_reduced_el_hopp} in a comparison with the data from Ref.~\onlinecite{Kloss19}.

\subsection{Dressed local initial state\label{Sec:Results:Extended_one_el:Dressed}} 

\begin{figure*}
	\includegraphics{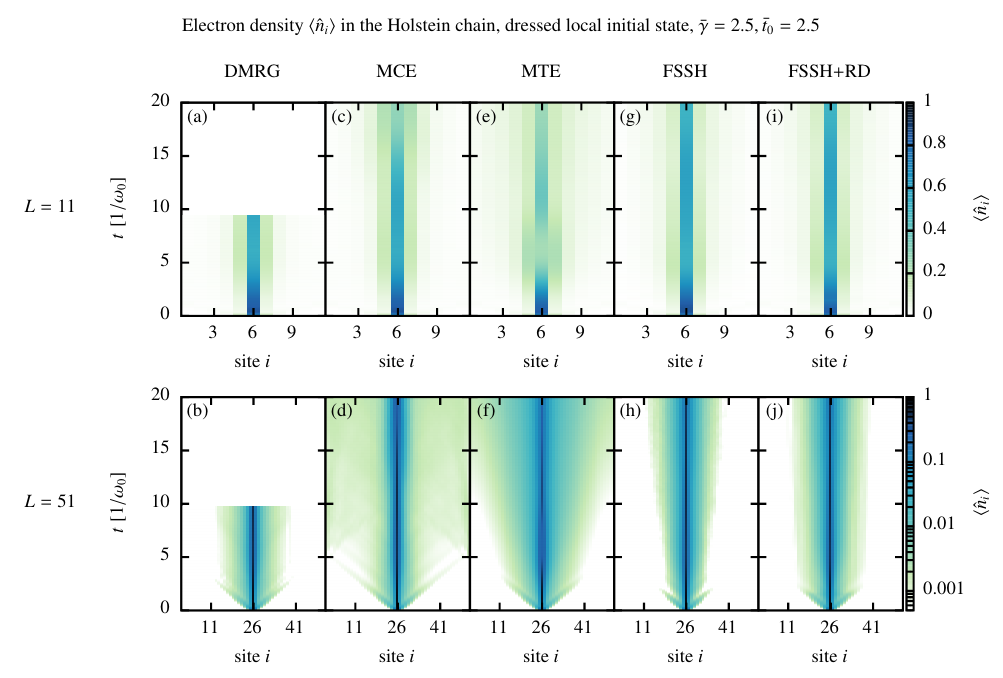}
	\caption{Electron densities $\braket{\hat n_i}$ in the Holstein chain, starting from the dressed local electron on the central site with $\bar\gamma=\bar t_0=2.5$. We show the results for $L=11$ and 51 lattice sites in the two rows, obtained from the methods DMRG ((a)-(b)), MCE ((c)-(d)), MTE ((e)-(f)), FSSH ((g)-(h)), and FSSH+RD ((i)-(j)) (see Secs.~\ref{Sec:DMRG},\ref{Sec:Trajectory}, and \ref{Sec:Multi}) in the five columns. In DMRG, we use $\epsilon_{\rm LBO}=10^{-8},\epsilon_{\rm bond}=10^{-8}, \Delta t=0.004/\omega_0$ and $M=40$. MCE uses 5000 configurations initialized with the pancake-like sampling and the independent trajectory methods use 50000 trajectories and $\Delta t=0.001/\omega_0$.}\label{Fig:Results_Extended:dressed_local_all_n}
\end{figure*}
We now turn to the dressed initial state, again with the electron initially localized to the central site and $\bar\gamma=\bar t_0=2.5$ for $L=11$ and $L=51$. We start with the evolution of the electron density, displayed in Fig.~\ref{Fig:Results_Extended:dressed_local_all_n}.
In contrast to the bare local initial state, we obtain a much stronger localization in all methods and therefore, we switch to a logarithmic scale for $L=51$. 
Most of the observations from the bare local initial state carry over to the dressed initial state.

For $L=11$, the MCE, FSSH, and FSSH+RD results agree well with the DMRG data for the times available. 
MTE shows a broad spreading for late times, which deviates from all other methods starting after the first half phonon period in both system sizes. This is similar to the dressed initial state in the trimer (Fig.~\ref{Fig:Results:Holstein_Trimer:dressed_localized_center_Gammabar2.5_tbar2.5}), where the initial localization of the electron was partially lost in the time evolution.
For $L=51$, MCE reproduces the electron density around the center well, while it features a small ballistic escape of the electron density to the chain boundaries, not predicted from the DMRG calculation.  The width of the electronic density distribution at longer times for $L=51$ appears to be better recovered with FSSH+RD here (Fig.~\ref{Fig:Results_Extended:dressed_local_MSD}(h)) than for the bare local initial state studied before. FSSH without decoherence, however, apparently produces a more localized electron density compared to the DMRG data after the initial ballistic expansion, and features a slow spreading of the electronic density for later times. Note that due to the logarithmic scale very small electronic contributions (below 0.0005) are cut off. This hides very small electronic populations developing outside of the localized region in the FSSH method without decoherence (similar to the bare local initial state), only relevant for observables such as the RMSD.

\begin{figure}
	\includegraphics{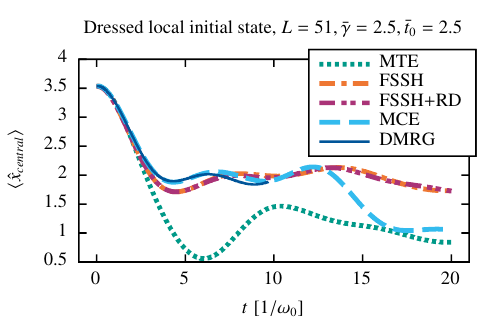}
	\caption{Phonon position $\braket{\hat x_{central}}$ on the central, initially occupied site in the Holstein chain, starting from the dressed local electron on the central site with $\bar\gamma=\bar t_0=2.5$ for $L=51$, obtained from the methods MTE, FSSH, FSSH+RD, DMRG, and MCE (see Secs.~\ref{Sec:DMRG},\ref{Sec:Trajectory}, and \ref{Sec:Multi}). We use the same method parameters as in Fig.~\ref{Fig:Results_Extended:dressed_local_all_n}.}\label{Fig:Results_Extended:dressed_local_central_x_L51}
\end{figure}

\begin{figure}
	\includegraphics{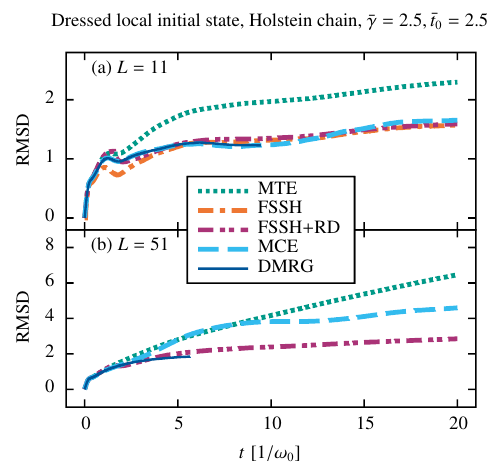}
	\caption{Reduced mean-squared displacement $\textup{RMSD}$ in the Holstein chain given via Eq.~(\ref{Eq:Extended_Sys:MSD}), starting with the dressed local electron on the central site with $\bar\gamma=\bar t_0=2.5$ for (a) $L=11$ and (b) $L=51$. We show the results obtained from DMRG, MCE, MTE and FSSH+RD (see Secs.~\ref{Sec:DMRG},\ref{Sec:Trajectory}, and \ref{Sec:Multi}). For FSSH without decoherence, only the results for $L=11$ are included, see Appendix~\ref{Sec:FSSH_Large_Sys}. We use the same method parameters as in Fig.~\ref{Fig:Results_Extended:dressed_local_all_n}, with the exception of the DMRG cutoffs, which are decreased to $\epsilon_{\rm LBO}=10^{-9},\epsilon_{\rm bond}=10^{-9}$ for $L=51$.
	}\label{Fig:Results_Extended:dressed_local_MSD}
\end{figure}

Next, we analyze the nuclear position on the central site, for $L=51$ in Fig.~\ref{Fig:Results_Extended:dressed_local_central_x_L51}.
Again, the MCE results for this observable are very close to the DMRG simulation for the times available, while MTE predicts much too low values for the phonon position, similar to the bare local case.
We observe that the FSSH methods follow the dynamics of DMRG closely, and even outperform MTE in the short-time regime. This is in contrast to the bare local state studied before, where MTE provides a better description for the first phonon oscillation (see Fig.~\ref{Fig:Results_Extended:bare_local_central_x_L51}), reflecting that the dressed initial state is easier to describe for the FSSH methods than the bare initial state. 
For long times, MCE, at least for a short duration, appears to come closer to the MTE simulation. We cannot judge whether this is captured correctly, since MCE is less reliable for these times, as mentioned in the convergence analysis from Sec.~\ref{Sec:MCE:Convergence}. 

For the global evolution, we again look at the reduced mean-squared displacement in Fig.~\ref{Fig:Results_Extended:dressed_local_MSD}.
For $L=11$, we observe a good agreement of FSSH+RD, MCE, and DMRG, similar to the bare local case. FSSH without decoherence correction displays slight deviations and MTE predicts much too high values. 
For $L=51$, we can see a continuous expansion of the RMSD in the MTE data, which increases slightly faster than a pure diffusion curve (a linear fit of the log-log curve reveals a slope of $\approx0.6$ for times $t>1/\omega_0$).
FSSH+RD is the closest to the DMRG results from the trajectory-based methods and shows only a very small spreading for later times. In the MCE calculation, we have a second increment of the RMSD for intermediate times, similar to the bare local case, but then a transition to a mostly flat value soon after the ballistic part of the density reaches the boundaries of the system. 

\begin{figure}
	\includegraphics{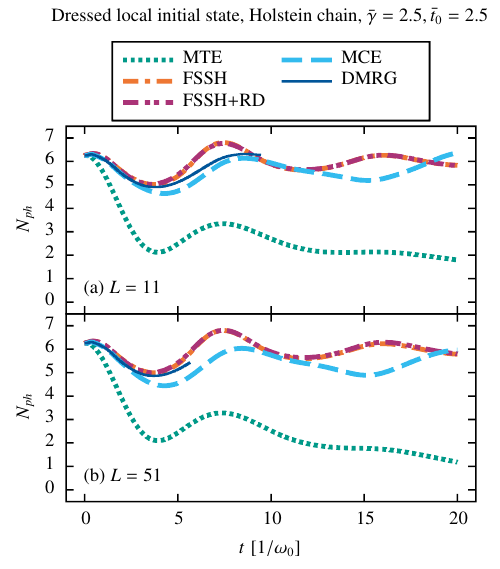}
	\caption{Total phonon number $N_{ph}=\braket{\sum_i \hat b^\dagger_i \hat b_i}$ in the Holstein chain, starting with the dressed local electron on the central site with $\bar\gamma=\bar t_0=2.5$ for (a) $L=11$ and (b) $L=51$, obtained from the methods MTE, FSSH, FSSH+RD, DMRG, and MCE (see Secs.~\ref{Sec:DMRG},\ref{Sec:Trajectory}, and \ref{Sec:Multi}). We use the same method parameters as in Fig.~\ref{Fig:Results_Extended:dressed_local_all_n}, with the exception of the DMRG cutoffs, which are decreased to $\epsilon_{\rm LBO}=10^{-9},\epsilon_{\rm bond}=10^{-9}$ for $L=51$.}\label{Fig:Results_Extended:dressed_local_Nph}
\end{figure}

As the second global quantity, we analyze the total phonon number in Fig.~\ref{Fig:Results_Extended:dressed_local_Nph}. Here, DMRG, MCE, FSSH, and FSSH+RD stay close to the initial value, while MTE completely loses the energy initially stored in the phonon sector, consistent with the broad spreading of the electronic density observed before. 
Similar to the case of the bare local state, the evolution of the total phonon number is almost independent of the system size. For the dressed local state studied here, however, MCE, FSSH, and FSSH+RD are able to recover the short-time dynamics of DMRG much better.

In summary, for both local initial states, MTE does not form a stable localized phonon-dressed electron distribution. In the Born-Huang notation, we interpret this as an insufficient relaxation to the lowest Born-Oppenheimer state, as already observed in the Holstein trimer (see Sec.~\ref{Sec:Results:Trimer}). 
This also poses a challenge for coupled-trajectory methods based on MTE, such as the analyzed MCEv1, where convergence is very difficult to achieve in some observables.
While the results of the FSSH methods are still more in a qualitative rather than a quantitative agreement with the DMRG data, at least for $L=51$, we see a strong improvement of the short-time dynamics of the dressed local state, compared to the bare local case analyzed before (Sec.~\ref{Sec:Results:Extended_one_el:Bare}). 
We note that a good description of FSSH relies on the improvements mentioned in Sec.~\ref{Sec:Surfhop_improvements}, especially the decoherence correction, which allows a consistent calculation of the diabatic electronic populations and of the RMSD, and the restriction of exactly that decoherence to avoid an unphysical super-fast spreading of the wave function (see Appendix \ref{Sec:FSSH_Large_Sys}).

From the results presented here, we conclude that MTE is not well suited to describe the real-time dynamics of an initially localized electron, neither with nor without a phonon dressed state, at least for the system parameters studied here. FSSH is, for many observables, the closest to DMRG and we may thus expect it
to provide a reasonable account of the long-time dynamics. Moreover, FSSH works better for the dressed than the bare initial state. The promising coupled-trajectory technique MCE is in excellent agreement with the DMRG results where it can be converged. While the short-time dynamics are well represented for all cases, the Ehrenfest-guided dynamics of the underlying basis set seem insufficient to accurately reproduce the correct long-times values for some observables. Still, for $L=11$ it is the best trajectory-based method for all studied observables, except for the total phonon number, and for $L=51$ it shows the most accurate agreement with the DMRG results for the electron density and phonon position on the central site. 

The DMRG results give access only to a limited time scale, in which the results clearly do not show a diffusive behavior. This is different from some previous studies using, for example, surface hopping for an initial state with the phonons prepared in a thermal state.\cite{Bai2018,Carof2019} The general transport behavior of 1D correlated models is non-trivial to predict, see Ref.~\onlinecite{Bertini2021}. It is possible that our quenched initial states are still in a pre-thermal regime for the time-scales investigated, but with the methods and results available we cannot yet determine the correct physical long-time behavior of, e.g., the RMSD. Since for $L=11$ the electron density quickly reaches the boundaries of the system (see Figs.~\ref{Fig:Results_Extended:bare_local_all_n} and \ref{Fig:Results_Extended:dressed_local_all_n}), our results differ qualitatively between both studied system sizes.
Out of all studied examples studied in our work, the results for the time-evolution of local electron states in the Holstein chain (Sec.~\ref{Sec:One-Electron_Extended}) are the least satisfactory, and further studies are needed in the future. 

	\section{Results for the Holstein chain - charge-density waves\label{Sec:Many-Electron_Extended}}

\begin{figure}
	\centering
	\includegraphics{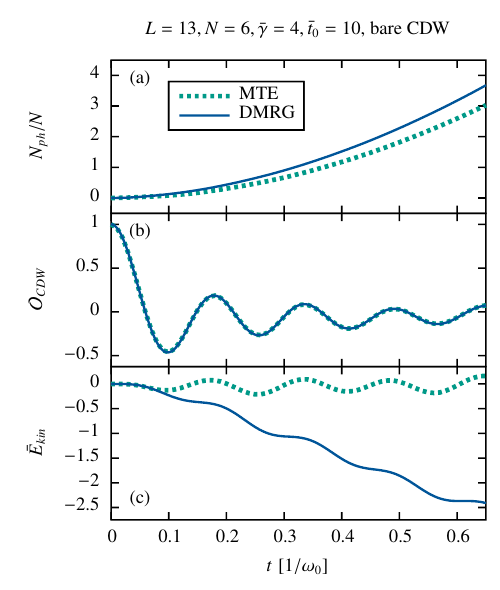}
	\caption{Observables of the Holstein chain with $L=13$,  $N=6$ electrons, $\bar{\gamma}=4$, and $\bar{t}_0=10$.  We show (a) the average phonon number, (b) the order parameter, and (c) the kinetic energy of the electrons. The initial state is the bare charge-density wave, see Sec.~\ref{Sec:Intro} for details. For the MTE method, we use 4000 trajectories and $\Delta t=0.01/\omega_0$, and for the DMRG data, we use $\epsilon_{\rm LBO}=10^{-8},\epsilon_{\rm bond}=10^{-8},\Delta t=0.001/\omega_0$ and $M=35$.}
	\label{fig:ExtCDW:bare_t10_gamma4}
\end{figure}

In this section, we compare the MTE method to DMRG data for the extended Holstein chain close to half filling.
This system is known to have a phase transition from a Tomonaga-Luttinger-liquid state to a charge-density wave.\cite{bursill_98,creffield_05}
The non-equilibrium breakdown of the charge-density wave has already been the subject of several studies, e.g.,  Refs.~\onlinecite{hashimoto_ishihara_17,Stolpp2020,Jansen_Jooss_2021}.
To test the MTE method, we compare the results to DMRG data in setups strongly motivated by the results of Ref.~\onlinecite{Stolpp2020}. Our main goal is to find out whether the MTE method captures the decay of the order parameter and the energy transfer from electronic to vibrational degrees of freedom.
The two initial states  are similar to those considered in previous sections.
Now, however,  both the bare and the dressed states have electrons on every other site, see Fig.~\ref{Fig:Introduction:States_Sketch2}. In Figs.~\ref{fig:ExtCDW:bare_t10_gamma4}(a)-(c), we show: (a) the average phonon number $N_{ph}/N$, with $N$ the number of electrons, 
(b) the CDW order parameter $O_{\textrm{CDW}}$, defined as
\begin{equation}\label{eq:OCDW_def}
O_{\rm CDW} = \frac{1}{N} \sum_{i=1}^{L} (-1)^{i} \langle \hat n_i\rangle,
\end{equation}
and (c) the electron kinetic energy $E_{kin}$, calculated with DMRG and MTE.
The initial state is the bare CDW and we use $\bar{\gamma}=4$ and $\bar{t}_0=10$.
In Fig.~\ref{fig:ExtCDW:bare_t10_gamma4}(b), we see that the order parameter is well described by the MTE method.
The rapid decay due to the large hopping amplitude and the following oscillations are all correctly captured.
In fact, the DMRG and MTE data are indistinguishable on the scale of the figure.
For the phonon number (Fig.~\ref{fig:ExtCDW:bare_t10_gamma4}(a)) and the kinetic energy (Fig.~\ref{fig:ExtCDW:bare_t10_gamma4}(c)), we observe a quite different picture.
The MTE results conserve the total energy, but this results from underestimating the phonon energy and at the same time overestimating the kinetic energy of the electron.
In fact, the physically very interesting relaxation process of the electrons to lower quasi-momentum states does not seem to be captured by MTE at all for the parameters chosen here. 

\begin{figure}
	\centering
	\includegraphics{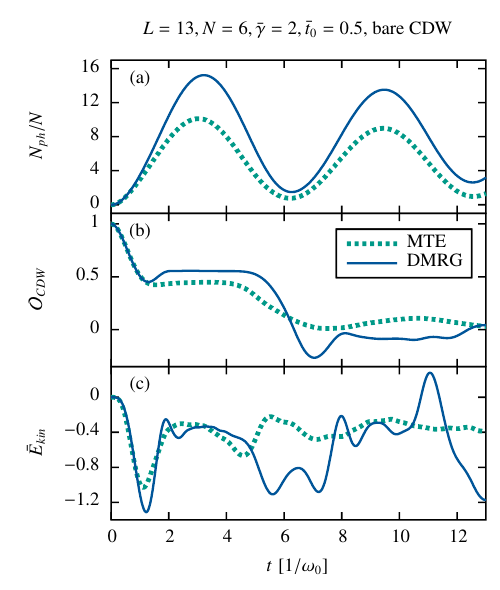}
	\caption{Observables of the Holstein chain with $L=13$,  $N=6$ electrons, $\bar{\gamma}=2$, and $\bar{t}_0=0.5$.  We show (a) the average phonon number, (b) the order parameter, and (c) the kinetic energy of the electrons. The initial state is the bare charge-density wave,  see Sec.~\ref{Sec:Intro} for details. For the MTE method, we use 4000 trajectories and $\Delta t=0.01/\omega_0$, and for the DMRG data, we use $\epsilon_{\rm LBO}=10^{-7},\epsilon_{\rm bond}=10^{-7},\Delta t=0.02/\omega_0$ and $M=35$.}
	\label{fig:ExtCDW:bare_t05_gamma2}
\end{figure}

In Figs.~\ref{fig:ExtCDW:bare_t05_gamma2}(a)-(c), we show the same observables for the same initial state but for $\bar{\gamma}=2$ and $\bar{t}_0=0.5$.
Here, we also observe significant differences between the MTE and the DMRG data.
The CDW order parameter is not simulated correctly by MTE, even though the physically interesting plateau is captured.  A similar picture as before emerges for the phonon number even though no clear trend can be seen for the kinetic energy.
Beyond small time scales, there is a significant deviation between the DMRG and MTE data. We note that for these parameters, much longer times are reached in the DMRG simulation.

\begin{figure}
	\centering
	\includegraphics{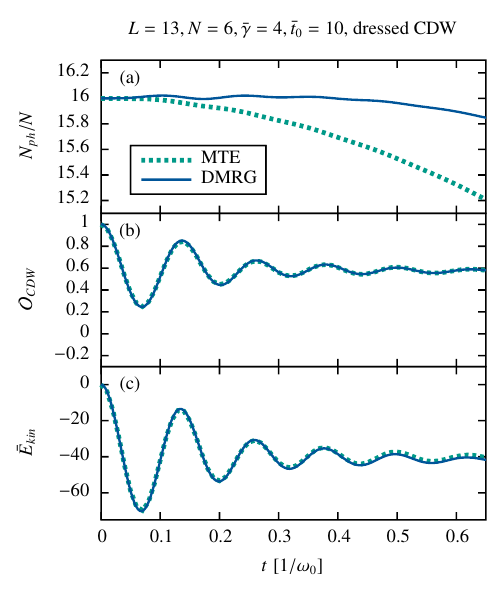}
	\caption{Observables of the Holstein chain with $L=13$,  $N=6$ electrons, $\bar{\gamma}=4$, and $\bar{t}_0=10$.  We show (a) the average phonon number, (b) the order parameter, and (c) the kinetic energy of the electrons. The initial state is the dressed charge-density wave,  see Sec.~\ref{Sec:Intro} for details. For the MTE method, we use 4000 trajectories and $\Delta t=0.01/\omega_0$, and for the DMRG data, we use $\epsilon_{\rm LBO}=10^{-7},\epsilon_{\rm bond}=10^{-7},\Delta t=0.001/\omega_0$ and $M=40$.}
	\label{fig:ExtCDW:dressed_t10_gamma4}
\end{figure}

Lastly, we look at the $\bar{\gamma}=4$ and $\bar{t}_0=10$ data but starting from a dressed CDW.
The three observables are displayed in Figs.~\ref{fig:ExtCDW:dressed_t10_gamma4}(a)-(c). 
Apparently, MTE captures the correct physical behavior for these parameters in the time interval where DMRG results are available.
This includes the initial decay followed by a saturation for the CDW order parameter in Fig.~\ref{fig:ExtCDW:dressed_t10_gamma4}(b)  and the decrease of both the phonon energy in Fig.~\ref{fig:ExtCDW:dressed_t10_gamma4}(a) and the electron kinetic energy in Fig.~\ref{fig:ExtCDW:dressed_t10_gamma4}(c). 
Note that the relative error is much smaller in this setup, compared to the bare CDW shown in Fig.~\ref{fig:ExtCDW:bare_t10_gamma4}. Our data indicate that MTE captures the correct physics quantitatively for the dressed CDW with a large hopping amplitude, but fails to describe the correct physics of the electron relaxation of the bare CDW state, seen in the kinetic energy.
One reason might be that the dressed CDW is closer to an adiabatic initial state, which for the large electron-hopping parameters used in Fig.~\ref{fig:ExtCDW:dressed_t10_gamma4}
is easier to describe for MTE.\label{sec:multel_results}

	\section{Conclusions\label{Sec:Conclusions}}

In this work, we benchmarked several trajectory-based quantum-chemistry methods, MTE, FSSH, and MCE, against the numerically exact algorithms ED and DMRG-LBO. Our focus was on the real-time dynamics in the Holstein model, a prototypical electron-phonon system often studied in condensed-matter physics. We analyzed the methods in the framework of the Born-Huang formalism to better understand their qualitatively different behavior and the influence of non-adiabatic effects on their dynamics. 

Compared to previous benchmark studies of trajectory-based methods, see, e.g., Refs.~\onlinecite{Hoffmann19,Chen2016,Stock2005,Freixas2021,Krotz2022,Chen2019}, a special focus was put on the influence of the initial state and of coherences on the non-adiabatic dynamics and the trajectory-based methods.
We provide a new systematic comparison to exact DMRG-LBO data for quenched initial states in the extended Holstein model.
While the MTE simulation always recovered the ultrashort time dynamics of the exact data, we found in general a better long-time description using a surface hopping approach, even for initial states with significant coherences between many adiabatic states. The coupled-trajectory method MCE provides excellent results for small systems, but converging this method can become difficult for large systems. We presented a detailed convergence analysis of MCE based on the criterion of energy conservation and by comparing to the exact DMRG-LBO results. We will now conclude by discussing each method in detail.

In contrast to classical-trajectory methods that employ the Born-Oppenheimer approximation (see Sec.~\ref{Sec:BH:recap}), both MTE and FSSH are in principle able to capture some non-adiabatic effects, such as transitions between adiabatic energy surfaces (see Figs.~\ref{Fig:Dimer_NucWF_TimeEvolution} and \ref{Fig:Results:Holstein_Dimer:Antibond_Gammabar2.5_tbar2.5:combined}). 

MTE works independently of the initial state and can recover the ultrashort dynamics for all investigated states.
The method fails, however, in describing independent dynamics on different adiabatic energy surfaces, due to the inherent mean-field approximation (which is a known
result\cite{Tully1990,Kirrander2020}). This is true both, when the time-evolution leads to an avoided crossing and  a wave-function splitting (see Figs.~\ref{Fig:Dimer_NucWF_TimeEvolution}, \ref{Fig:Results:Holstein_Dimer:Antibond_Gammabar2.5_tbar2.5:combined}, and \ref{Fig:Results:Holstein_Trimer:nonbonding_Gammabar2.5_tbar2.5}), as well as when different adiabatic states are populated from the beginning, as for the local states studied in this work. Even in an adiabatic parameter regime, where the contributions on the different surfaces should influence each other only very little (see Sec.~\ref{Sec:BH:recap}), such initial states can lead to qualitatively wrong long-time dynamics (see, e.g., Fig.~\ref{Fig:Results:Holstein_Dimer:Bare_Localized_Gammabar4_tbar10:combined}). Our interpretation is that the dynamics are often too much influenced by the high Born-Oppenheimer states, as discussed for the Holstein trimer (Sec.~\ref{Sec:Results:Trimer}) and cannot correctly describe the relaxation 
back to low-energy surfaces.  This is also observed in the extended systems with one electron, where the bare and dressed local states both show a broad spreading of the electron density, not seen by the other methods.
Our results confirm the known fact\cite{Horsfield_2006} that MTE is a questionable choice for analyzing energy transfer between the electronic and phononic subsystems, as can be seen from the total phonon number, which is described poorly (see Figs.~\ref{Fig:Results_Extended:bare_local_Nph} and \ref{Fig:Results_Extended:dressed_local_all_n}, and Sec.~\ref{Sec:Many-Electron_Extended}).  
Then again, for the charge-density wave states in a half-filled system, the electronic order parameter is captured reasonably well (Sec.~\ref{Sec:Many-Electron_Extended}), especially in the adiabatic parameter regime. Here, for the dressed CDW, even the electron kinetic energy is accurately predicted, which we attribute to the fact that the state is close to an adiabatic state in an adiabatic parameter regime.
In addition, MTE is the computationally cheapest of all studied methods in the large systems and can easily be extended to many-electron systems. It is thus a straightforward first method to implement also in condensed-matter problems. However, if one wants to accurately study systems with relevant non-adiabatic effects, either due to the parameter choice, or due to initial states, one should consider using an improved method.
It might be beneficial to use a mixture of MTE and the adiabatic basis of surface hopping, by introducing some form of decoherence in an MTE approach,\cite{Zhu2004,Zhu2004b} which, however, reintroduces a basis-dependence and was not studied in this work. 

The FSSH method is strongly basis dependent and works best for adiabatic initial states, as shown in the Holstein dimer and trimer (see Figs.~\ref{Fig:Results:Holstein_Dimer:Antibond_Gammabar2.5_tbar2.5:combined} and \ref{Fig:Results:Holstein_Trimer:nonbonding_Gammabar2.5_tbar2.5}). There, it can correctly describe wave-function splitting and independent dynamics on the energy surfaces. One can question its use for initial states with non-zero off-diagonal elements of the adiabatic electronic density matrix, so-called ``coherences'', as for the local initial states studied in this work. FSSH cannot completely capture the correct short-time dynamics for these initial states, since the method ignores the coherences in its calculation of the nuclear forces, illustrated in Figs.~\ref{Fig:Results:Holstein_Dimer:Bare_Localized_Gammabar4_tbar10:exact_WF} and \ref{Fig:Results:Holstein_Dimer:Bare_Localized_Gammabar4_tbar10:combined}. It has been suggested that for such coherent initial states, e.g., created by an attosecond light-pulse, a mean-field description, such as MTE, might be a better choice.\cite{Kirrander2020} 
Nonetheless, we observed, in almost all systems, a better long-time description with FSSH  than with MTE. For not too slow phonons, the initial coherences dampen out fast enough to recover reasonable long-time dynamics. 
The decoherence corrections improve the internal consistency of the different definitions of the electronic density matrix available and in general seem to reproduce the nuclear trajectories better. It also reduces the problem of frustrated hops in setups with very low nuclear kinetic energy around the avoided crossing, as seen in Fig.~\ref{Fig:Results:Holstein_Dimer:Bare_Localized_Gammabar4_tbar10:combined}.
When using a restricted decoherence correction and with a proper treatment of the derivative couplings (see Sec.~\ref{Sec:Surfhop_improvements}), any unphysical spurious charge transfer in large systems can be avoided.
In these large systems, out of the tested trajectory-based methods, the FSSH+RD results for the RMSD were the closest to the DMRG results (see Figs.~\ref{Fig:Results_Extended:bare_local_MSD} and \ref{Fig:Results_Extended:dressed_local_MSD}) and the method reproduced the local trapping of the electron. 
Further comparisons to numerically exact methods, in particular in the
long-time limit and for large systems, are needed in the future.

Here, we studied the most common form of a surface-hopping method and included only the corrections necessary to obtain reasonable results for the extended systems (see Sec.~\ref{Sec:Surfhop_improvements}). The field of surface-hopping methods is still rapidly evolving, and many variations and improvements have been suggested in the recent years, for which we refer to the reviews Refs.~\onlinecite{Wang2016,Smith_2019}. 
In that regard, our study serves as a starting point for a possible future comparison of these methods against the simple FSSH and DMRG-LBO.

The theoretical foundation of FSSH works with general many-particle states\cite{Tully1990} and can readily be applied to many-electron systems. Due to the exponential growth of the electronic Hilbert space, this is limited in practice. 
The variant ``independent-electron surface hopping'' was suggested,\cite{Shenvi2009} for systems where the electrons are not subject to direct electronic correlations, as in the Holstein model. Here, a single Slater determinant is taken as the ansatz for the many-particle state and derivative couplings and hopping probabilities are expressed in terms of the single-particle orbitals, which drastically reduces the computational cost. While this was not studied in this work, it is a natural next step in the analysis of the charge-density wave states of Sec.~\ref{Sec:Many-Electron_Extended}.

The coupled-trajectory method MCE produces very accurate results whenever it can be converged and is even computationally cheap for small systems. It can also, at least partially, capture a tunneling transition in a classical energetically forbidden regime, which is not the case for our implementations of MTE and FSSH.
Such a coupled-trajectory method seems to be the quantum-chemistry method of choice for the small systems studied here.
For large systems, it is more difficult to converge for physically reasonable parameters and cannot be regarded as an exact method for the $N_c\sim 5\cdot10^3$ configurations used in this work. The computational cost scales with the third power of the number of configurations. While including more configurations is definitely feasible using longer computation times, the very slow convergence of MCE for $L=51$ analyzed in Fig.~\ref{Fig:MCE:convergence_example_L51} suggests that the implementation of alternative coupled-trajectory methods might also be worth trying.
Nevertheless, even though we could not reach convergence according to our criteria, local observables around the initially occupied sites were still recovered very well (see Figs.~\ref{Fig:Results_Extended:bare_local_central_x_L51} and \ref{Fig:Results_Extended:dressed_local_central_x_L51}). 
In the future, it would be interesting to apply MCE also to the charge-density wave states of Sec.~\ref{Sec:Many-Electron_Extended}.
Another direction could be the implementation of MCE in the adiabatic basis, see, e.g., Ref.~\onlinecite{Makhov2017}. 
As alternative coupled-trajectory methods, one might consider 
Gaussian-based Multi-Configuration time-dependent
Hartree (G-MCTDH),\cite{Burghardt99,Burghardt03,Burghardt08,Gonzalez20} variational Multiconfigurational Gaussian (vMCG) \cite{Worth04,Worth03,Richings15,Gonzalez20} the Davydov D2 ansatz,\cite{Davydov_1982,Cruzeiro-Hansson94,Zhao2012,Zhou15,Chen2019} MCEv2 with trajectory cloning,\cite{Makhov2014,Makhov2015} and more adiabatic-surface guided methods, such as full multiple spawning,\cite{Martinez1996} or ab-initio multiple spawning,\cite{Ben-nun2000} to name a few examples.
Furthermore, multilayer multiconfiguration time-dependent Hartree\cite{Wang09,Wang2015} is a promising coupled-trajectory method for benchmarks.

Out of the tested methods, DMRG-LBO, although costly and due to entanglement growth limited to short times,\cite{Schollwock2011} is the exact method of choice for the larger 1D systems. By construction, DMRG is a many-body technique and can therefore include electronic correlations. In contrast to MCE, the DMRG-LBO method used in our work could be converged in all cases, although the simulation time is limited. DMRG is, regardless of the specific algorithm, designed for many-electron systems, for which we showed examples in our work.
The recently introduced projected-purification method \cite{Kohler21} might be another avenue towards reaching longer times, but time-dependent simulations
of this method have not been systematically explored yet (see Ref.~\onlinecite{Mardazad21} for an application).
Extensions of DMRG or generalizations of matrix-product states to two-dimensional systems exist,\cite{Schollwock2005,Schollwock2011,Stoudenmire2012,Orus2014,Verstraete2004-2D,Zheng2017,Bruognolo2021} 
yet are more expensive algorithms and combinations with efficient treatments of phonons have not been attempted or systematically tested.
In the future, it would be interesting to apply the DMRG-LBO algorithm to systems with time-dependent external fields, e.g., for optical excitations, or interacting electrons.
Another direction would be to extend the initial conditions to phonon distributions at finite temperature.\cite{Jansen2020}

To conclude, in this work, we provide unbiased data with numerically exact DMRG-LBO for 1D Holstein chains of large system sizes that serve as a new benchmark for approximate quantum-chemistry or condensed-matter methods. We provide comparisons of several trajectory-based quantum-chemistry methods, MTE, FSSH, and MCE to the DMRG-LBO data.
This allows us to explicitly quantify their strengths and weaknesses beyond relying on internal consistency checks.
The initial conditions studied in this work start with many (or all) phonon oscillators in their ground state. In combination with the intermediate electron-phonon coupling and electron-hopping parameter choices, this provides a challenging testbed for all trajectory-based methods and we carefully studied  the influence of the initial conditions. 
The MCE method can be converged to exact data for many systems and observables and
is the most promising among the approximate methods tested here. More efficient implementations or alternative coupled-trajectory algorithms are needed for larger systems.
In the future, comparative large-system studies with the phonons prepared in a thermal state, possibly combined with initial conditions created from an explicit optical excitation, would yield additional insight into the qualities of the trajectory-based methods.
	
	\begin{acknowledgments}
		The authors are grateful to S. Kehrein, A. Osterkorn and E. Paprotzki for fruitful discussions. We thank Kloss \textit{et al.} for sending us their data.
		This paper is funded by the Deutsche Forschungsgemeinschaft (DFG, German Research Foundation) - SFB 1073 - 217133147 (projects B03, B09). 
	\end{acknowledgments}

	\section*{Data Availability}
	The data that support the findings of this study are openly available at \url{https://doi.org/10.25625/YDU1XT}, G\"ottingen Research Online / Data.
	
	\appendix
	
	\section{Surface hopping for local initial states in large systems\label{Sec:FSSH_Large_Sys}}

\begin{figure}
	\includegraphics{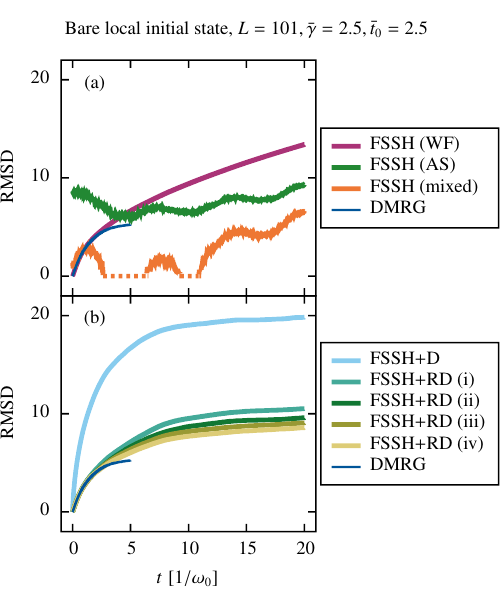}
	\caption{(a) Reduced mean-squared displacement (see Eq.~\eqref{Eq:Extended_Sys:MSD}) calculated with the three different possible definitions of the electronic density matrix (see Sec.~\ref{Sec:FSSH}) for FSSH without decoherence and compared to the DMRG results for the bare local initial state with $L=101,\bar\gamma=\bar t_0=2.5$. The dashed line in the mixed definition indicated negative values in the square-root. (b) The RMSD for FSSH with a decoherence correction without restriction (FSSH+D) and with four different variations of restricted decoherence (see Sec.~\ref{Sec:FSSH:Spurious_Charge_Transfer}) (FSSH+RD): (i): $R=0.999$ without delay, (ii): $R=0.999$ with delay, (iii) $R=0.99$ without delay, and (iv): $R=0.99$ with delay. We use the same parameters as in Fig.~\ref{Fig:Results_Extended:bare_local_all_n}, with the exception of the DMRG cutoffs, which are decreased to $\epsilon_{\rm LBO}=10^{-9}$ and $\epsilon_{\rm bond}=10^{-9}$.}
	\label{Fig:Results_Extended:FSSH_large_systems}
\end{figure}

For almost all observables and systems investigated in this work, the mixed definition of the electronic density matrix for FSSH according to Eq.~\eqref{Eq:FSSH:Mixed_Dens} produces good results and could recover the diabatic populations at least approximately even for local initial states. The exceptions are very large systems with a local initial state, where small negative and positive electronic populations occur after some time throughout the whole system. While most of the density is still well recovered, see Fig.~\ref{Fig:Results_Extended:bare_local_all_n} and Fig.~\ref{Fig:Results_Extended:dressed_local_all_n}, this poses a problem for quantities that emphasize the small occupations on the borders of the chain such as the reduced mean-squared displacement (Eq.~\eqref{Eq:Extended_Sys:MSD}). This is shown in Fig.~\ref{Fig:Results_Extended:FSSH_large_systems}(a) for a very large example system with $L=101$ sites. 

The mixed definition partially even leads to negative values in the square-root for the RMSD and is also otherwise unreliable. One could resort to the active-surface (AS, Eq.~\eqref{Eq:FSSH:AS_Dens}) definition, which seems to provide better long-time results, but does not reproduce the correct initial value. In contrast, the wave-function (WF, Eq.~\eqref{Eq:FSSH:WF_Dens}) definition does not capture the localization for longer times. 
We note that if one discards the coherences in the initial state, all definitions of the electronic density matrix give the same result. In Fig.~\ref{Fig:Results_Extended:FSSH_large_systems}(a) this corresponds to the RMSD of the FSSH (AS) curve at $t=0/\omega_0$. With the electron density delocalized significantly already from the beginning, this simplification is not suitable for describing the short-time non-adiabatic dynamics in quenched large systems and was not used in this work.

Using a decoherence correction allows us to use the wave-function definition and still recover the long-time localization, as described in Sec.~\ref{Sec:FSSH}. However, since the adiabatic states in the bare local initial state are strongly delocalized, this decoherence correction leads to the enhanced effect of the spurious charge transfer mentioned in Sec.~\ref{Sec:FSSH:Spurious_Charge_Transfer}, which results in an unphysical fast spreading of the wave function, see Fig.~\ref{Fig:Results_Extended:FSSH_large_systems}(b), (FSSH+D). This can be remedied by using a restricted decoherence correction (FSSH+RD), as described in Sec.~\ref{Sec:FSSH:Spurious_Charge_Transfer}. We show results for four types of restrictions: (i) using the active-space threshold value of $R=0.999$, suggested by Ref.~\onlinecite{Giannini2018} without delay  and (ii) with delay, and  using a reduced threshold of $R=0.99$ (which corresponds to an even stronger restriction) without (iii) and with delay (iv). In this work, the restriction (iv) is used for system sizes of $L\ge 11$.

	\section{Reduced electron hopping parameter regime\texorpdfstring{, \boldmath{$\bar t_0=1$}}{}\label{Sec:One-Electron_reduced_el_hopp}}

\begin{figure}[t]
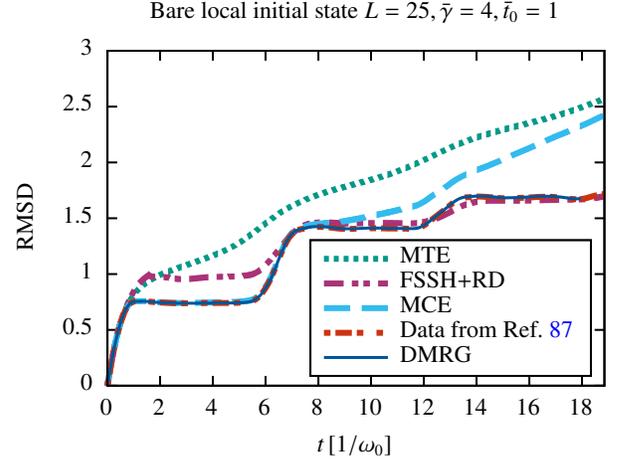

	\include{Figures/Extended_System_Reduced_el_hopp/gamma_4}
	\caption{Reduced mean-squared displacement (see Eq.~\eqref{Eq:Extended_Sys:MSD}) obtained from the bare local initial state in the Holstein chain with $L=25,\bar t_0=1, \bar \gamma=4$, obtained with MTE, FSSH+RD, MCE, and DMRG (see Secs.~\ref{Sec:DMRG},\ref{Sec:Trajectory}, and \ref{Sec:Multi}). This is compared to the data presented in Ref.~\protect\onlinecite{Kloss19}. 
		In the DMRG runs, we use $\epsilon_{\rm LBO}=10^{-8},\epsilon_{\rm bond}=10^{-8},\Delta t=0.01/\omega_0$, and $M=90$. MCE uses 4500 configurations initialized with the pancake-like sampling, and the independent trajectory methods use 50000 trajectories. In the FSSH+RD runs, the RMSD is calculated from the wave-function definition of the density matrix (see Eq.~\eqref{Eq:FSSH:WF_Dens}).
	}
	\label{Fig:Results_Ext_Sys_Red_Elhopp:gamma4}
\end{figure}

\begin{figure}[t]
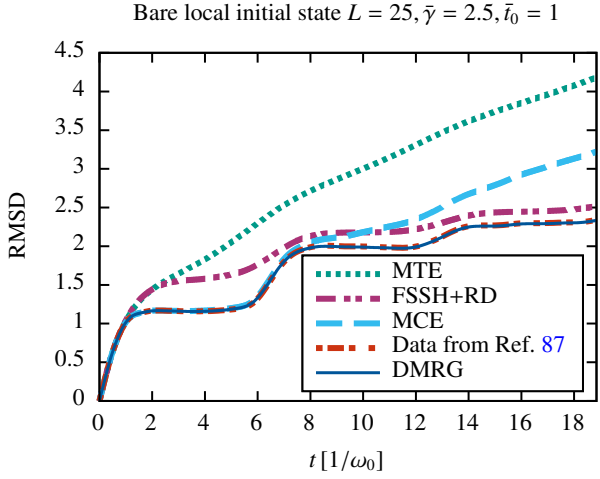

	\include{Figures/Extended_System_Reduced_el_hopp/gamma_2_5}
	\caption{Reduced mean-squared displacement (see Eq.~\eqref{Eq:Extended_Sys:MSD}) obtained from the bare local initial state in the Holstein chain with $L=25,\bar t_0=1, \bar \gamma=2.5$, obtained with MTE, FSSH+RD, MCE, and DMRG (see Secs.~\ref{Sec:DMRG},\ref{Sec:Trajectory}, and \ref{Sec:Multi}). This is compared to the data presented in Ref.~\protect\onlinecite{Kloss19}. 
		In our DMRG simulations, we use $\epsilon_{\rm LBO}=10^{-8},\epsilon_{\rm bond}=10^{-8},\Delta t=0.01/\omega_0$, and $M=35$. MCE uses 4500 configurations initialized with the pancake-like sampling, and the independent trajectory methods use 50000 trajectories. In the FSSH+RD runs, the RMSD is calculated from the wave-function definition of the density matrix (see Eq.~\eqref{Eq:FSSH:WF_Dens}).
	}
	\label{Fig:Results_Ext_Sys_Red_Elhopp:gamma2_5}
\end{figure}

Here, we compare MTE, FSSH+RD, MCE, and DMRG for $\bar t_0=1$, which was already studied in Ref.~\onlinecite{Kloss19} with another matrix-product state method designed for electron-phonon problems. We compute the reduced mean-squared displacement (see Eq.~\eqref{Eq:Extended_Sys:MSD}) in a Holstein chain with $L=25$ for both $\bar \gamma=4$ and $\bar \gamma=2.5$ for the bare local initial state. This initial state corresponds to the ``Franck-Condon'' excitation studied in Ref.~\onlinecite{Kloss19} for the same parameters and we compare our results to their data. The results are shown in Figs.~\ref{Fig:Results_Ext_Sys_Red_Elhopp:gamma4} and \ref{Fig:Results_Ext_Sys_Red_Elhopp:gamma2_5}.

We observe a slightly different time evolution than for the parameters investigated in Sec.~\ref{Sec:One-Electron_Extended}: The DMRG time evolution shows a step-like increment of the RMSD, similar to the intermediate plateau formation that we observe in the slow electron regime (see Figs.~\ref{Fig:Results:Holstein_Dimer:Gammabar2_tbar0.5:bare} and \ref{fig:ExtCDW:bare_t05_gamma2}), and which we attributed to a transient local trapping. The transient trapping resurfaces here with several steps in the RMSD at multiples of the phonon oscillation period, as already observed in Ref.~\onlinecite{Kloss19}. For the times available, DMRG and the data of Ref.~\onlinecite{Kloss19} agree very well. 

In both cases ($\bar \gamma=4$ and $\bar \gamma=2.5$), MCE displays the best short-time description of all trajectory-based methods. The method is again difficult to converge and cannot completely capture the formation of the second plateau for the used number of configurations, and the RMSD is drastically overestimated for later times. MTE cannot reproduce the formation of plateaus at all, which are only visible as wiggles in a growing RMSD. FSSH with the restricted decoherence correction (see Sec.~\ref{Sec:FSSH:Spurious_Charge_Transfer}) reproduces the formation of several plateaus. For short times, the method overestimates the height of the first RMSD plateau for both values of $\bar \gamma$, while the RMSD is better described for later times, especially for $\bar \gamma=4$, but also for $\bar \gamma=2.5$. 

Thus, all trajectory-based methods show a similar capability to describe the exact time evolution as already for the parameters investigated in Sec.~\ref{Sec:One-Electron_Extended}: MTE cannot capture the local (transient) trapping of the electron, MCE works the best for short times, but is very difficult to converge for longer times, when also MTE fails, and FSSH cannot correctly describe the short-time evolution due to the build-in coherences of the local initial state, but is able to predict a reasonable long-time behavior of the spreading of the electron. 

	\cleardoublepage
	\bibliography{ref}

\begin{thebibliography}{277}%
\makeatletter
\providecommand \@ifxundefined [1]{%
 \@ifx{#1\undefined}
}%
\providecommand \@ifnum [1]{%
 \ifnum #1\expandafter \@firstoftwo
 \else \expandafter \@secondoftwo
 \fi
}%
\providecommand \@ifx [1]{%
 \ifx #1\expandafter \@firstoftwo
 \else \expandafter \@secondoftwo
 \fi
}%
\providecommand \natexlab [1]{#1}%
\providecommand \enquote  [1]{``#1''}%
\providecommand \bibnamefont  [1]{#1}%
\providecommand \bibfnamefont [1]{#1}%
\providecommand \citenamefont [1]{#1}%
\providecommand \href@noop [0]{\@secondoftwo}%
\providecommand \href [0]{\begingroup \@sanitize@url \@href}%
\providecommand \@href[1]{\@@startlink{#1}\@@href}%
\providecommand \@@href[1]{\endgroup#1\@@endlink}%
\providecommand \@sanitize@url [0]{\catcode `\\12\catcode `\$12\catcode
  `\&12\catcode `\#12\catcode `\^12\catcode `\_12\catcode `\%12\relax}%
\providecommand \@@startlink[1]{}%
\providecommand \@@endlink[0]{}%
\providecommand \url  [0]{\begingroup\@sanitize@url \@url }%
\providecommand \@url [1]{\endgroup\@href {#1}{\urlprefix }}%
\providecommand \urlprefix  [0]{URL }%
\providecommand \Eprint [0]{\href }%
\providecommand \doibase [0]{http://dx.doi.org/}%
\providecommand \selectlanguage [0]{\@gobble}%
\providecommand \bibinfo  [0]{\@secondoftwo}%
\providecommand \bibfield  [0]{\@secondoftwo}%
\providecommand \translation [1]{[#1]}%
\providecommand \BibitemOpen [0]{}%
\providecommand \bibitemStop [0]{}%
\providecommand \bibitemNoStop [0]{.\EOS\space}%
\providecommand \EOS [0]{\spacefactor3000\relax}%
\providecommand \BibitemShut  [1]{\csname bibitem#1\endcsname}%
\let\auto@bib@innerbib\@empty
\bibitem [{\citenamefont {Gonz\'alez}\ and\ \citenamefont
  {Lindh}(2020)}]{Gonzalez20}%
  \BibitemOpen
  \bibfield  {author} {\bibinfo {author} {\bibfnamefont {L.}~\bibnamefont
  {Gonz\'alez}}\ and\ \bibinfo {author} {\bibfnamefont {R.}~\bibnamefont
  {Lindh}},\ }\href@noop {} {\emph {\bibinfo {title} {Quantum Chemistry and
  Dynamics of Excited States: Methods and Applications}}}\ (\bibinfo
  {publisher} {John Wiley \& Sons, Ltd},\ \bibinfo {year} {2020})\BibitemShut
  {NoStop}%
\bibitem [{\citenamefont {Curchod}\ and\ \citenamefont
  {Mart\'inez}(2018)}]{Curchod18}%
  \BibitemOpen
  \bibfield  {author} {\bibinfo {author} {\bibfnamefont {B.~F.~E.}\
  \bibnamefont {Curchod}}\ and\ \bibinfo {author} {\bibfnamefont {T.~J.}\
  \bibnamefont {Mart\'inez}},\ }\href {\doibase 10.1021/acs.chemrev.7b00423}
  {\bibfield  {journal} {\bibinfo  {journal} {Chem. Rev.}\ }\textbf {\bibinfo
  {volume} {118}},\ \bibinfo {pages} {3305} (\bibinfo {year}
  {2018})}\BibitemShut {NoStop}%
\bibitem [{\citenamefont {Agostini}\ and\ \citenamefont
  {Curchod}(2019)}]{Agostini19}%
  \BibitemOpen
  \bibfield  {author} {\bibinfo {author} {\bibfnamefont {F.}~\bibnamefont
  {Agostini}}\ and\ \bibinfo {author} {\bibfnamefont {B.~F.~E.}\ \bibnamefont
  {Curchod}},\ }\href {\doibase 10.1002/wcms.1417} {\bibfield  {journal}
  {\bibinfo  {journal} {Wiley Interdiscip. Rev.: Comput. Mol. Sci.}\ }\textbf
  {\bibinfo {volume} {9}},\ \bibinfo {pages} {e1417} (\bibinfo {year}
  {2019})}\BibitemShut {NoStop}%
\bibitem [{\citenamefont {Domcke}, \citenamefont {Yarkony},\ and\ \citenamefont
  {K{\"o}ppel}(2011)}]{Domcke2011}%
  \BibitemOpen
  \bibfield  {author} {\bibinfo {author} {\bibfnamefont {W.}~\bibnamefont
  {Domcke}}, \bibinfo {author} {\bibfnamefont {D.~R.}\ \bibnamefont {Yarkony}},
  \ and\ \bibinfo {author} {\bibfnamefont {H.}~\bibnamefont {K{\"o}ppel}},\
  }\href {\doibase 10.1142/7803} {\emph {\bibinfo {title} {Conical
  {Intersections}: {Theory}, {Computation} and {Experiment}}}},\ \bibinfo
  {series} {Advanced {Series} in {Physical} {Chemistry}}, Vol.~\bibinfo
  {volume} {17}\ (\bibinfo  {publisher} {World Scientific},\ \bibinfo {year}
  {2011})\BibitemShut {NoStop}%
\bibitem [{\citenamefont {Yarkony}(1996)}]{Yarkony1996}%
  \BibitemOpen
  \bibfield  {author} {\bibinfo {author} {\bibfnamefont {D.~R.}\ \bibnamefont
  {Yarkony}},\ }\href {\doibase 10.1103/RevModPhys.68.985} {\bibfield
  {journal} {\bibinfo  {journal} {Rev. Mod. Phys.}\ }\textbf {\bibinfo {volume}
  {68}},\ \bibinfo {pages} {985} (\bibinfo {year} {1996})}\BibitemShut
  {NoStop}%
\bibitem [{\citenamefont {Nelson}\ \emph {et~al.}(2011)\citenamefont {Nelson},
  \citenamefont {Fernandez-Alberti}, \citenamefont {Chernyak}, \citenamefont
  {Roitberg},\ and\ \citenamefont {Tretiak}}]{Nelson2011}%
  \BibitemOpen
  \bibfield  {author} {\bibinfo {author} {\bibfnamefont {T.}~\bibnamefont
  {Nelson}}, \bibinfo {author} {\bibfnamefont {S.}~\bibnamefont
  {Fernandez-Alberti}}, \bibinfo {author} {\bibfnamefont {V.}~\bibnamefont
  {Chernyak}}, \bibinfo {author} {\bibfnamefont {A.~E.}\ \bibnamefont
  {Roitberg}}, \ and\ \bibinfo {author} {\bibfnamefont {S.}~\bibnamefont
  {Tretiak}},\ }\href {\doibase 10.1021/jp109522g} {\bibfield  {journal}
  {\bibinfo  {journal} {J. Phys. Chem. B}\ }\textbf {\bibinfo {volume} {115}},\
  \bibinfo {pages} {5402} (\bibinfo {year} {2011})}\BibitemShut {NoStop}%
\bibitem [{\citenamefont {Coles}\ \emph {et~al.}(2014)\citenamefont {Coles},
  \citenamefont {Yang}, \citenamefont {Wang}, \citenamefont {Grant},
  \citenamefont {Taylor}, \citenamefont {Saikin}, \citenamefont {Aspuru-Guzik},
  \citenamefont {Lidzey}, \citenamefont {Tang},\ and\ \citenamefont
  {Smith}}]{Coles2014}%
  \BibitemOpen
  \bibfield  {author} {\bibinfo {author} {\bibfnamefont {D.~M.}\ \bibnamefont
  {Coles}}, \bibinfo {author} {\bibfnamefont {Y.}~\bibnamefont {Yang}},
  \bibinfo {author} {\bibfnamefont {Y.}~\bibnamefont {Wang}}, \bibinfo {author}
  {\bibfnamefont {R.~T.}\ \bibnamefont {Grant}}, \bibinfo {author}
  {\bibfnamefont {R.~A.}\ \bibnamefont {Taylor}}, \bibinfo {author}
  {\bibfnamefont {S.~K.}\ \bibnamefont {Saikin}}, \bibinfo {author}
  {\bibfnamefont {A.}~\bibnamefont {Aspuru-Guzik}}, \bibinfo {author}
  {\bibfnamefont {D.~G.}\ \bibnamefont {Lidzey}}, \bibinfo {author}
  {\bibfnamefont {J.~K.-H.}\ \bibnamefont {Tang}}, \ and\ \bibinfo {author}
  {\bibfnamefont {J.~M.}\ \bibnamefont {Smith}},\ }\href {\doibase
  10.1038/ncomms6561} {\bibfield  {journal} {\bibinfo  {journal} {Nat.
  Commun.}\ }\textbf {\bibinfo {volume} {5}},\ \bibinfo {pages} {5561}
  (\bibinfo {year} {2014})}\BibitemShut {NoStop}%
\bibitem [{\citenamefont {Hutchison}\ \emph {et~al.}(2012)\citenamefont
  {Hutchison}, \citenamefont {Schwartz}, \citenamefont {Genet}, \citenamefont
  {Devaux},\ and\ \citenamefont {Ebbesen}}]{Hutchison2012}%
  \BibitemOpen
  \bibfield  {author} {\bibinfo {author} {\bibfnamefont {J.~A.}\ \bibnamefont
  {Hutchison}}, \bibinfo {author} {\bibfnamefont {T.}~\bibnamefont {Schwartz}},
  \bibinfo {author} {\bibfnamefont {C.}~\bibnamefont {Genet}}, \bibinfo
  {author} {\bibfnamefont {E.}~\bibnamefont {Devaux}}, \ and\ \bibinfo {author}
  {\bibfnamefont {T.~W.}\ \bibnamefont {Ebbesen}},\ }\href {\doibase
  10.1002/anie.201107033} {\bibfield  {journal} {\bibinfo  {journal} {Angew.
  Chem., Int. Ed.}\ }\textbf {\bibinfo {volume} {51}},\ \bibinfo {pages} {1592}
  (\bibinfo {year} {2012})}\BibitemShut {NoStop}%
\bibitem [{\citenamefont {Orgiu}\ \emph {et~al.}(2015)\citenamefont {Orgiu},
  \citenamefont {George}, \citenamefont {Hutchison}, \citenamefont {Devaux},
  \citenamefont {Dayen}, \citenamefont {Doudin}, \citenamefont {Stellacci},
  \citenamefont {Genet}, \citenamefont {Schachenmayer}, \citenamefont {Genes},
  \citenamefont {Pupillo}, \citenamefont {Samor{\`\i}},\ and\ \citenamefont
  {Ebbesen}}]{Orgiu2015}%
  \BibitemOpen
  \bibfield  {author} {\bibinfo {author} {\bibfnamefont {E.}~\bibnamefont
  {Orgiu}}, \bibinfo {author} {\bibfnamefont {J.}~\bibnamefont {George}},
  \bibinfo {author} {\bibfnamefont {J.~A.}\ \bibnamefont {Hutchison}}, \bibinfo
  {author} {\bibfnamefont {E.}~\bibnamefont {Devaux}}, \bibinfo {author}
  {\bibfnamefont {J.~F.}\ \bibnamefont {Dayen}}, \bibinfo {author}
  {\bibfnamefont {B.}~\bibnamefont {Doudin}}, \bibinfo {author} {\bibfnamefont
  {F.}~\bibnamefont {Stellacci}}, \bibinfo {author} {\bibfnamefont
  {C.}~\bibnamefont {Genet}}, \bibinfo {author} {\bibfnamefont
  {J.}~\bibnamefont {Schachenmayer}}, \bibinfo {author} {\bibfnamefont
  {C.}~\bibnamefont {Genes}}, \bibinfo {author} {\bibfnamefont
  {G.}~\bibnamefont {Pupillo}}, \bibinfo {author} {\bibfnamefont
  {P.}~\bibnamefont {Samor{\`\i}}}, \ and\ \bibinfo {author} {\bibfnamefont
  {T.~W.}\ \bibnamefont {Ebbesen}},\ }\href {\doibase 10.1038/nmat4392}
  {\bibfield  {journal} {\bibinfo  {journal} {Nat. Mater.}\ }\textbf {\bibinfo
  {volume} {14}},\ \bibinfo {pages} {1123} (\bibinfo {year}
  {2015})}\BibitemShut {NoStop}%
\bibitem [{\citenamefont {Thomas}\ \emph {et~al.}(2016)\citenamefont {Thomas},
  \citenamefont {George}, \citenamefont {Shalabney}, \citenamefont {Dryzhakov},
  \citenamefont {Varma}, \citenamefont {Moran}, \citenamefont {Chervy},
  \citenamefont {Zhong}, \citenamefont {Devaux}, \citenamefont {Genet},
  \citenamefont {Hutchison},\ and\ \citenamefont {Ebbesen}}]{Thomas2016}%
  \BibitemOpen
  \bibfield  {author} {\bibinfo {author} {\bibfnamefont {A.}~\bibnamefont
  {Thomas}}, \bibinfo {author} {\bibfnamefont {J.}~\bibnamefont {George}},
  \bibinfo {author} {\bibfnamefont {A.}~\bibnamefont {Shalabney}}, \bibinfo
  {author} {\bibfnamefont {M.}~\bibnamefont {Dryzhakov}}, \bibinfo {author}
  {\bibfnamefont {S.~J.}\ \bibnamefont {Varma}}, \bibinfo {author}
  {\bibfnamefont {J.}~\bibnamefont {Moran}}, \bibinfo {author} {\bibfnamefont
  {T.}~\bibnamefont {Chervy}}, \bibinfo {author} {\bibfnamefont
  {X.}~\bibnamefont {Zhong}}, \bibinfo {author} {\bibfnamefont
  {E.}~\bibnamefont {Devaux}}, \bibinfo {author} {\bibfnamefont
  {C.}~\bibnamefont {Genet}}, \bibinfo {author} {\bibfnamefont {J.~A.}\
  \bibnamefont {Hutchison}}, \ and\ \bibinfo {author} {\bibfnamefont {T.~W.}\
  \bibnamefont {Ebbesen}},\ }\href {\doibase 10.1002/anie.201605504} {\bibfield
   {journal} {\bibinfo  {journal} {Angew. Chem., Int. Ed.}\ }\textbf {\bibinfo
  {volume} {55}},\ \bibinfo {pages} {11462} (\bibinfo {year}
  {2016})}\BibitemShut {NoStop}%
\bibitem [{\citenamefont {Bienfait}\ \emph {et~al.}(2016)\citenamefont
  {Bienfait}, \citenamefont {Pla}, \citenamefont {Kubo}, \citenamefont {Zhou},
  \citenamefont {Stern}, \citenamefont {Lo}, \citenamefont {Weis},
  \citenamefont {Schenkel}, \citenamefont {Vion}, \citenamefont {Esteve},
  \citenamefont {Morton},\ and\ \citenamefont {Bertet}}]{Bienfait2016}%
  \BibitemOpen
  \bibfield  {author} {\bibinfo {author} {\bibfnamefont {A.}~\bibnamefont
  {Bienfait}}, \bibinfo {author} {\bibfnamefont {J.~J.}\ \bibnamefont {Pla}},
  \bibinfo {author} {\bibfnamefont {Y.}~\bibnamefont {Kubo}}, \bibinfo {author}
  {\bibfnamefont {X.}~\bibnamefont {Zhou}}, \bibinfo {author} {\bibfnamefont
  {M.}~\bibnamefont {Stern}}, \bibinfo {author} {\bibfnamefont {C.~C.}\
  \bibnamefont {Lo}}, \bibinfo {author} {\bibfnamefont {C.~D.}\ \bibnamefont
  {Weis}}, \bibinfo {author} {\bibfnamefont {T.}~\bibnamefont {Schenkel}},
  \bibinfo {author} {\bibfnamefont {D.}~\bibnamefont {Vion}}, \bibinfo {author}
  {\bibfnamefont {D.}~\bibnamefont {Esteve}}, \bibinfo {author} {\bibfnamefont
  {J.~J.~L.}\ \bibnamefont {Morton}}, \ and\ \bibinfo {author} {\bibfnamefont
  {P.}~\bibnamefont {Bertet}},\ }\href {\doibase 10.1038/nature16944}
  {\bibfield  {journal} {\bibinfo  {journal} {Nature}\ }\textbf {\bibinfo
  {volume} {531}},\ \bibinfo {pages} {74} (\bibinfo {year} {2016})}\BibitemShut
  {NoStop}%
\bibitem [{\citenamefont {Flick}\ \emph {et~al.}(2017)\citenamefont {Flick},
  \citenamefont {Ruggenthaler}, \citenamefont {Appel},\ and\ \citenamefont
  {Rubio}}]{Flick2017}%
  \BibitemOpen
  \bibfield  {author} {\bibinfo {author} {\bibfnamefont {J.}~\bibnamefont
  {Flick}}, \bibinfo {author} {\bibfnamefont {M.}~\bibnamefont {Ruggenthaler}},
  \bibinfo {author} {\bibfnamefont {H.}~\bibnamefont {Appel}}, \ and\ \bibinfo
  {author} {\bibfnamefont {A.}~\bibnamefont {Rubio}},\ }\href {\doibase
  10.1073/pnas.1615509114} {\bibfield  {journal} {\bibinfo  {journal} {Proc.
  Natl. Acad. Sci.}\ }\textbf {\bibinfo {volume} {114}},\ \bibinfo {pages}
  {3026} (\bibinfo {year} {2017})}\BibitemShut {NoStop}%
\bibitem [{\citenamefont {Nabok}, \citenamefont {Bl{\"u}gel},\ and\
  \citenamefont {Friedrich}(2021)}]{Nabok2021}%
  \BibitemOpen
  \bibfield  {author} {\bibinfo {author} {\bibfnamefont {D.}~\bibnamefont
  {Nabok}}, \bibinfo {author} {\bibfnamefont {S.}~\bibnamefont {Bl{\"u}gel}}, \
  and\ \bibinfo {author} {\bibfnamefont {C.}~\bibnamefont {Friedrich}},\ }\href
  {\doibase 10.1038/s41524-021-00649-8} {\bibfield  {journal} {\bibinfo
  {journal} {npj Comput. Mater.}\ }\textbf {\bibinfo {volume} {7}},\ \bibinfo
  {pages} {178} (\bibinfo {year} {2021})}\BibitemShut {NoStop}%
\bibitem [{\citenamefont {Giustino}(2017)}]{Giustino17}%
  \BibitemOpen
  \bibfield  {author} {\bibinfo {author} {\bibfnamefont {F.}~\bibnamefont
  {Giustino}},\ }\href {\doibase 10.1103/RevModPhys.89.015003} {\bibfield
  {journal} {\bibinfo  {journal} {Rev. Mod. Phys.}\ }\textbf {\bibinfo {volume}
  {89}},\ \bibinfo {pages} {015003} (\bibinfo {year} {2017})}\BibitemShut
  {NoStop}%
\bibitem [{\citenamefont {Franchini}\ \emph {et~al.}(2021)\citenamefont
  {Franchini}, \citenamefont {Reticcioli}, \citenamefont {Setvin},\ and\
  \citenamefont {Diebold}}]{Franchini21}%
  \BibitemOpen
  \bibfield  {author} {\bibinfo {author} {\bibfnamefont {C.}~\bibnamefont
  {Franchini}}, \bibinfo {author} {\bibfnamefont {M.}~\bibnamefont
  {Reticcioli}}, \bibinfo {author} {\bibfnamefont {M.}~\bibnamefont {Setvin}},
  \ and\ \bibinfo {author} {\bibfnamefont {U.}~\bibnamefont {Diebold}},\ }\href
  {\doibase 10.1038/s41578-021-00289-w} {\bibfield  {journal} {\bibinfo
  {journal} {Nat. Rev. Mater.}\ }\textbf {\bibinfo {volume} {6}},\ \bibinfo
  {pages} {560} (\bibinfo {year} {2021})}\BibitemShut {NoStop}%
\bibitem [{\citenamefont {Kang}\ \emph {et~al.}(2021)\citenamefont {Kang},
  \citenamefont {Du}, \citenamefont {Zhou}, \citenamefont {Gu}, \citenamefont
  {Chen}, \citenamefont {Xu}, \citenamefont {Zhang}, \citenamefont {Sun},
  \citenamefont {Yin}, \citenamefont {Li}, \citenamefont {Pei}, \citenamefont
  {Zhang}, \citenamefont {Gu}, \citenamefont {Wang}, \citenamefont {Liu},
  \citenamefont {Xiong}, \citenamefont {Shi}, \citenamefont {Zhang},
  \citenamefont {Chen},\ and\ \citenamefont {Yang}}]{Kang21}%
  \BibitemOpen
  \bibfield  {author} {\bibinfo {author} {\bibfnamefont {L.}~\bibnamefont
  {Kang}}, \bibinfo {author} {\bibfnamefont {X.}~\bibnamefont {Du}}, \bibinfo
  {author} {\bibfnamefont {J.~S.}\ \bibnamefont {Zhou}}, \bibinfo {author}
  {\bibfnamefont {X.}~\bibnamefont {Gu}}, \bibinfo {author} {\bibfnamefont
  {Y.~J.}\ \bibnamefont {Chen}}, \bibinfo {author} {\bibfnamefont {R.~Z.}\
  \bibnamefont {Xu}}, \bibinfo {author} {\bibfnamefont {Q.~Q.}\ \bibnamefont
  {Zhang}}, \bibinfo {author} {\bibfnamefont {S.~C.}\ \bibnamefont {Sun}},
  \bibinfo {author} {\bibfnamefont {Z.~X.}\ \bibnamefont {Yin}}, \bibinfo
  {author} {\bibfnamefont {Y.~W.}\ \bibnamefont {Li}}, \bibinfo {author}
  {\bibfnamefont {D.}~\bibnamefont {Pei}}, \bibinfo {author} {\bibfnamefont
  {J.}~\bibnamefont {Zhang}}, \bibinfo {author} {\bibfnamefont {R.~K.}\
  \bibnamefont {Gu}}, \bibinfo {author} {\bibfnamefont {Z.~G.}\ \bibnamefont
  {Wang}}, \bibinfo {author} {\bibfnamefont {Z.~K.}\ \bibnamefont {Liu}},
  \bibinfo {author} {\bibfnamefont {R.}~\bibnamefont {Xiong}}, \bibinfo
  {author} {\bibfnamefont {J.}~\bibnamefont {Shi}}, \bibinfo {author}
  {\bibfnamefont {Y.}~\bibnamefont {Zhang}}, \bibinfo {author} {\bibfnamefont
  {Y.~L.}\ \bibnamefont {Chen}}, \ and\ \bibinfo {author} {\bibfnamefont
  {L.~X.}\ \bibnamefont {Yang}},\ }\href {\doibase 10.1038/s41467-021-26078-1}
  {\bibfield  {journal} {\bibinfo  {journal} {Nat. Commun.}\ }\textbf {\bibinfo
  {volume} {12}},\ \bibinfo {pages} {6183} (\bibinfo {year}
  {2021})}\BibitemShut {NoStop}%
\bibitem [{\citenamefont {Raja}\ \emph {et~al.}(2018)\citenamefont {Raja},
  \citenamefont {Selig}, \citenamefont {Bergh{\"a}user}, \citenamefont {Yu},
  \citenamefont {Hill}, \citenamefont {Rigosi}, \citenamefont {Brus},
  \citenamefont {Knorr}, \citenamefont {Heinz}, \citenamefont {Malic},\ and\
  \citenamefont {Chernikov}}]{Raja2018}%
  \BibitemOpen
  \bibfield  {author} {\bibinfo {author} {\bibfnamefont {A.}~\bibnamefont
  {Raja}}, \bibinfo {author} {\bibfnamefont {M.}~\bibnamefont {Selig}},
  \bibinfo {author} {\bibfnamefont {G.}~\bibnamefont {Bergh{\"a}user}},
  \bibinfo {author} {\bibfnamefont {J.}~\bibnamefont {Yu}}, \bibinfo {author}
  {\bibfnamefont {H.~M.}\ \bibnamefont {Hill}}, \bibinfo {author}
  {\bibfnamefont {A.~F.}\ \bibnamefont {Rigosi}}, \bibinfo {author}
  {\bibfnamefont {L.~E.}\ \bibnamefont {Brus}}, \bibinfo {author}
  {\bibfnamefont {A.}~\bibnamefont {Knorr}}, \bibinfo {author} {\bibfnamefont
  {T.~F.}\ \bibnamefont {Heinz}}, \bibinfo {author} {\bibfnamefont
  {E.}~\bibnamefont {Malic}}, \ and\ \bibinfo {author} {\bibfnamefont
  {A.}~\bibnamefont {Chernikov}},\ }\href {\doibase
  10.1021/acs.nanolett.8b01793} {\bibfield  {journal} {\bibinfo  {journal}
  {Nano Lett.}\ }\textbf {\bibinfo {volume} {18}},\ \bibinfo {pages} {6135}
  (\bibinfo {year} {2018})}\BibitemShut {NoStop}%
\bibitem [{\citenamefont {Shree}\ \emph {et~al.}(2018)\citenamefont {Shree},
  \citenamefont {Semina}, \citenamefont {Robert}, \citenamefont {Han},
  \citenamefont {Amand}, \citenamefont {Balocchi}, \citenamefont {Manca},
  \citenamefont {Courtade}, \citenamefont {Marie}, \citenamefont {Taniguchi},
  \citenamefont {Watanabe}, \citenamefont {Glazov},\ and\ \citenamefont
  {Urbaszek}}]{Shree2018}%
  \BibitemOpen
  \bibfield  {author} {\bibinfo {author} {\bibfnamefont {S.}~\bibnamefont
  {Shree}}, \bibinfo {author} {\bibfnamefont {M.}~\bibnamefont {Semina}},
  \bibinfo {author} {\bibfnamefont {C.}~\bibnamefont {Robert}}, \bibinfo
  {author} {\bibfnamefont {B.}~\bibnamefont {Han}}, \bibinfo {author}
  {\bibfnamefont {T.}~\bibnamefont {Amand}}, \bibinfo {author} {\bibfnamefont
  {A.}~\bibnamefont {Balocchi}}, \bibinfo {author} {\bibfnamefont
  {M.}~\bibnamefont {Manca}}, \bibinfo {author} {\bibfnamefont
  {E.}~\bibnamefont {Courtade}}, \bibinfo {author} {\bibfnamefont
  {X.}~\bibnamefont {Marie}}, \bibinfo {author} {\bibfnamefont
  {T.}~\bibnamefont {Taniguchi}}, \bibinfo {author} {\bibfnamefont
  {K.}~\bibnamefont {Watanabe}}, \bibinfo {author} {\bibfnamefont {M.~M.}\
  \bibnamefont {Glazov}}, \ and\ \bibinfo {author} {\bibfnamefont
  {B.}~\bibnamefont {Urbaszek}},\ }\href {\doibase 10.1103/PhysRevB.98.035302}
  {\bibfield  {journal} {\bibinfo  {journal} {Phys. Rev. B}\ }\textbf {\bibinfo
  {volume} {98}},\ \bibinfo {pages} {035302} (\bibinfo {year}
  {2018})}\BibitemShut {NoStop}%
\bibitem [{\citenamefont {Werdehausen}\ \emph {et~al.}(2018)\citenamefont
  {Werdehausen}, \citenamefont {Takayama}, \citenamefont {H\"oppner},
  \citenamefont {Albrecht}, \citenamefont {Rost}, \citenamefont {Lu},
  \citenamefont {Manske}, \citenamefont {Takagi},\ and\ \citenamefont
  {Kaiser}}]{Werdehausen2018}%
  \BibitemOpen
  \bibfield  {author} {\bibinfo {author} {\bibfnamefont {D.}~\bibnamefont
  {Werdehausen}}, \bibinfo {author} {\bibfnamefont {T.}~\bibnamefont
  {Takayama}}, \bibinfo {author} {\bibfnamefont {M.}~\bibnamefont {H\"oppner}},
  \bibinfo {author} {\bibfnamefont {G.}~\bibnamefont {Albrecht}}, \bibinfo
  {author} {\bibfnamefont {A.~W.}\ \bibnamefont {Rost}}, \bibinfo {author}
  {\bibfnamefont {Y.}~\bibnamefont {Lu}}, \bibinfo {author} {\bibfnamefont
  {D.}~\bibnamefont {Manske}}, \bibinfo {author} {\bibfnamefont
  {H.}~\bibnamefont {Takagi}}, \ and\ \bibinfo {author} {\bibfnamefont
  {S.}~\bibnamefont {Kaiser}},\ }\href {\doibase 10.1126/sciadv.aap8652}
  {\bibfield  {journal} {\bibinfo  {journal} {Sci. Adv.}\ }\textbf {\bibinfo
  {volume} {4}},\ \bibinfo {pages} {eaap8652} (\bibinfo {year}
  {2018})}\BibitemShut {NoStop}%
\bibitem [{\citenamefont {Chen}, \citenamefont {Sangalli},\ and\ \citenamefont
  {Bernardi}(2020)}]{Chen2020}%
  \BibitemOpen
  \bibfield  {author} {\bibinfo {author} {\bibfnamefont {H.-Y.}\ \bibnamefont
  {Chen}}, \bibinfo {author} {\bibfnamefont {D.}~\bibnamefont {Sangalli}}, \
  and\ \bibinfo {author} {\bibfnamefont {M.}~\bibnamefont {Bernardi}},\ }\href
  {\doibase 10.1103/PhysRevLett.125.107401} {\bibfield  {journal} {\bibinfo
  {journal} {Phys. Rev. Lett.}\ }\textbf {\bibinfo {volume} {125}},\ \bibinfo
  {pages} {107401} (\bibinfo {year} {2020})}\BibitemShut {NoStop}%
\bibitem [{\citenamefont {Mor}\ \emph {et~al.}(2021)\citenamefont {Mor},
  \citenamefont {Gosetti}, \citenamefont {Molina-S\'anchez}, \citenamefont
  {Sangalli}, \citenamefont {Achilli}, \citenamefont {Agekyan}, \citenamefont
  {Franceschini}, \citenamefont {Giannetti}, \citenamefont {Sangaletti},\ and\
  \citenamefont {Pagliara}}]{Mor2021}%
  \BibitemOpen
  \bibfield  {author} {\bibinfo {author} {\bibfnamefont {S.}~\bibnamefont
  {Mor}}, \bibinfo {author} {\bibfnamefont {V.}~\bibnamefont {Gosetti}},
  \bibinfo {author} {\bibfnamefont {A.}~\bibnamefont {Molina-S\'anchez}},
  \bibinfo {author} {\bibfnamefont {D.}~\bibnamefont {Sangalli}}, \bibinfo
  {author} {\bibfnamefont {S.}~\bibnamefont {Achilli}}, \bibinfo {author}
  {\bibfnamefont {V.~F.}\ \bibnamefont {Agekyan}}, \bibinfo {author}
  {\bibfnamefont {P.}~\bibnamefont {Franceschini}}, \bibinfo {author}
  {\bibfnamefont {C.}~\bibnamefont {Giannetti}}, \bibinfo {author}
  {\bibfnamefont {L.}~\bibnamefont {Sangaletti}}, \ and\ \bibinfo {author}
  {\bibfnamefont {S.}~\bibnamefont {Pagliara}},\ }\href {\doibase
  10.1103/PhysRevResearch.3.043175} {\bibfield  {journal} {\bibinfo  {journal}
  {Phys. Rev. Research}\ }\textbf {\bibinfo {volume} {3}},\ \bibinfo {pages}
  {043175} (\bibinfo {year} {2021})}\BibitemShut {NoStop}%
\bibitem [{\citenamefont {Li}\ \emph {et~al.}(2021)\citenamefont {Li},
  \citenamefont {Trovatello}, \citenamefont {Dal~Conte}, \citenamefont
  {Nu{\ss}}, \citenamefont {Soavi}, \citenamefont {Wang}, \citenamefont
  {Ferrari}, \citenamefont {Cerullo},\ and\ \citenamefont {Brixner}}]{Li2021}%
  \BibitemOpen
  \bibfield  {author} {\bibinfo {author} {\bibfnamefont {D.}~\bibnamefont
  {Li}}, \bibinfo {author} {\bibfnamefont {C.}~\bibnamefont {Trovatello}},
  \bibinfo {author} {\bibfnamefont {S.}~\bibnamefont {Dal~Conte}}, \bibinfo
  {author} {\bibfnamefont {M.}~\bibnamefont {Nu{\ss}}}, \bibinfo {author}
  {\bibfnamefont {G.}~\bibnamefont {Soavi}}, \bibinfo {author} {\bibfnamefont
  {G.}~\bibnamefont {Wang}}, \bibinfo {author} {\bibfnamefont {A.~C.}\
  \bibnamefont {Ferrari}}, \bibinfo {author} {\bibfnamefont {G.}~\bibnamefont
  {Cerullo}}, \ and\ \bibinfo {author} {\bibfnamefont {T.}~\bibnamefont
  {Brixner}},\ }\href {\doibase 10.1038/s41467-021-20895-0} {\bibfield
  {journal} {\bibinfo  {journal} {Nat. Commun.}\ }\textbf {\bibinfo {volume}
  {12}},\ \bibinfo {pages} {954} (\bibinfo {year} {2021})}\BibitemShut
  {NoStop}%
\bibitem [{\citenamefont {Frost}, \citenamefont {Whalley},\ and\ \citenamefont
  {Walsh}(2017)}]{Frost2017}%
  \BibitemOpen
  \bibfield  {author} {\bibinfo {author} {\bibfnamefont {J.~M.}\ \bibnamefont
  {Frost}}, \bibinfo {author} {\bibfnamefont {L.~D.}\ \bibnamefont {Whalley}},
  \ and\ \bibinfo {author} {\bibfnamefont {A.}~\bibnamefont {Walsh}},\ }\href
  {\doibase 10.1021/acsenergylett.7b00862} {\bibfield  {journal} {\bibinfo
  {journal} {ACS Energy Lett.}\ }\textbf {\bibinfo {volume} {2}},\ \bibinfo
  {pages} {2647} (\bibinfo {year} {2017})}\BibitemShut {NoStop}%
\bibitem [{\citenamefont {Raiser}\ \emph {et~al.}(2017)\citenamefont {Raiser},
  \citenamefont {Mildner}, \citenamefont {Ifland}, \citenamefont {Sotoudeh},
  \citenamefont {Bl{\"o}chl}, \citenamefont {Techert},\ and\ \citenamefont
  {Jooss}}]{Raiser2017}%
  \BibitemOpen
  \bibfield  {author} {\bibinfo {author} {\bibfnamefont {D.}~\bibnamefont
  {Raiser}}, \bibinfo {author} {\bibfnamefont {S.}~\bibnamefont {Mildner}},
  \bibinfo {author} {\bibfnamefont {B.}~\bibnamefont {Ifland}}, \bibinfo
  {author} {\bibfnamefont {M.}~\bibnamefont {Sotoudeh}}, \bibinfo {author}
  {\bibfnamefont {P.}~\bibnamefont {Bl{\"o}chl}}, \bibinfo {author}
  {\bibfnamefont {S.}~\bibnamefont {Techert}}, \ and\ \bibinfo {author}
  {\bibfnamefont {C.}~\bibnamefont {Jooss}},\ }\href {\doibase
  10.1002/aenm.201602174} {\bibfield  {journal} {\bibinfo  {journal} {Adv.
  Energy Mater.}\ }\textbf {\bibinfo {volume} {7}},\ \bibinfo {pages} {1602174}
  (\bibinfo {year} {2017})}\BibitemShut {NoStop}%
\bibitem [{\citenamefont {Park}\ \emph {et~al.}(2018)\citenamefont {Park},
  \citenamefont {Neukirch}, \citenamefont {Reyes-Lillo}, \citenamefont {Lai},
  \citenamefont {Ellis}, \citenamefont {Dietze}, \citenamefont {Neaton},
  \citenamefont {Yang}, \citenamefont {Tretiak},\ and\ \citenamefont
  {Mathies}}]{Park18}%
  \BibitemOpen
  \bibfield  {author} {\bibinfo {author} {\bibfnamefont {M.}~\bibnamefont
  {Park}}, \bibinfo {author} {\bibfnamefont {A.~J.}\ \bibnamefont {Neukirch}},
  \bibinfo {author} {\bibfnamefont {S.~E.}\ \bibnamefont {Reyes-Lillo}},
  \bibinfo {author} {\bibfnamefont {M.}~\bibnamefont {Lai}}, \bibinfo {author}
  {\bibfnamefont {S.~R.}\ \bibnamefont {Ellis}}, \bibinfo {author}
  {\bibfnamefont {D.}~\bibnamefont {Dietze}}, \bibinfo {author} {\bibfnamefont
  {J.~B.}\ \bibnamefont {Neaton}}, \bibinfo {author} {\bibfnamefont
  {P.}~\bibnamefont {Yang}}, \bibinfo {author} {\bibfnamefont {S.}~\bibnamefont
  {Tretiak}}, \ and\ \bibinfo {author} {\bibfnamefont {R.~A.}\ \bibnamefont
  {Mathies}},\ }\href {\doibase 10.1038/s41467-018-04946-7} {\bibfield
  {journal} {\bibinfo  {journal} {Nat. Commun.}\ }\textbf {\bibinfo {volume}
  {9}},\ \bibinfo {pages} {2525} (\bibinfo {year} {2018})}\BibitemShut
  {NoStop}%
\bibitem [{\citenamefont {Ghosh}\ \emph {et~al.}(2020)\citenamefont {Ghosh},
  \citenamefont {Welch}, \citenamefont {Neukirch}, \citenamefont {Zakhidov},\
  and\ \citenamefont {Tretiak}}]{Ghosh20}%
  \BibitemOpen
  \bibfield  {author} {\bibinfo {author} {\bibfnamefont {D.}~\bibnamefont
  {Ghosh}}, \bibinfo {author} {\bibfnamefont {E.}~\bibnamefont {Welch}},
  \bibinfo {author} {\bibfnamefont {A.~J.}\ \bibnamefont {Neukirch}}, \bibinfo
  {author} {\bibfnamefont {A.}~\bibnamefont {Zakhidov}}, \ and\ \bibinfo
  {author} {\bibfnamefont {S.}~\bibnamefont {Tretiak}},\ }\href {\doibase
  10.1021/acs.jpclett.0c00018} {\bibfield  {journal} {\bibinfo  {journal} {J.
  Phys. Chem. Lett.}\ }\textbf {\bibinfo {volume} {11}},\ \bibinfo {pages}
  {3271} (\bibinfo {year} {2020})}\BibitemShut {NoStop}%
\bibitem [{\citenamefont {Kressdorf}\ \emph {et~al.}(2020)\citenamefont
  {Kressdorf}, \citenamefont {Meyer}, \citenamefont {Belenchuk}, \citenamefont
  {Shapoval}, \citenamefont {ten Brink}, \citenamefont {Melles}, \citenamefont
  {Ross}, \citenamefont {Hoffmann}, \citenamefont {Moshnyaga}, \citenamefont
  {Seibt}, \citenamefont {Bl\"ochl},\ and\ \citenamefont
  {Jooss}}]{Kressdorf20}%
  \BibitemOpen
  \bibfield  {author} {\bibinfo {author} {\bibfnamefont {B.}~\bibnamefont
  {Kressdorf}}, \bibinfo {author} {\bibfnamefont {T.}~\bibnamefont {Meyer}},
  \bibinfo {author} {\bibfnamefont {A.}~\bibnamefont {Belenchuk}}, \bibinfo
  {author} {\bibfnamefont {O.}~\bibnamefont {Shapoval}}, \bibinfo {author}
  {\bibfnamefont {M.}~\bibnamefont {ten Brink}}, \bibinfo {author}
  {\bibfnamefont {S.}~\bibnamefont {Melles}}, \bibinfo {author} {\bibfnamefont
  {U.}~\bibnamefont {Ross}}, \bibinfo {author} {\bibfnamefont {J.}~\bibnamefont
  {Hoffmann}}, \bibinfo {author} {\bibfnamefont {V.}~\bibnamefont {Moshnyaga}},
  \bibinfo {author} {\bibfnamefont {M.}~\bibnamefont {Seibt}}, \bibinfo
  {author} {\bibfnamefont {P.}~\bibnamefont {Bl\"ochl}}, \ and\ \bibinfo
  {author} {\bibfnamefont {C.}~\bibnamefont {Jooss}},\ }\href {\doibase
  10.1103/PhysRevApplied.14.054006} {\bibfield  {journal} {\bibinfo  {journal}
  {Phys. Rev. Applied}\ }\textbf {\bibinfo {volume} {14}},\ \bibinfo {pages}
  {054006} (\bibinfo {year} {2020})}\BibitemShut {NoStop}%
\bibitem [{\citenamefont {Berk}\ \emph {et~al.}(2019)\citenamefont {Berk},
  \citenamefont {Jaris}, \citenamefont {Yang}, \citenamefont {Dhuey},
  \citenamefont {Cabrini},\ and\ \citenamefont {Schmidt}}]{Berk2019}%
  \BibitemOpen
  \bibfield  {author} {\bibinfo {author} {\bibfnamefont {C.}~\bibnamefont
  {Berk}}, \bibinfo {author} {\bibfnamefont {M.}~\bibnamefont {Jaris}},
  \bibinfo {author} {\bibfnamefont {W.}~\bibnamefont {Yang}}, \bibinfo {author}
  {\bibfnamefont {S.}~\bibnamefont {Dhuey}}, \bibinfo {author} {\bibfnamefont
  {S.}~\bibnamefont {Cabrini}}, \ and\ \bibinfo {author} {\bibfnamefont
  {H.}~\bibnamefont {Schmidt}},\ }\href {\doibase 10.1038/s41467-019-10545-x}
  {\bibfield  {journal} {\bibinfo  {journal} {Nat. Commun.}\ }\textbf {\bibinfo
  {volume} {10}},\ \bibinfo {pages} {2652} (\bibinfo {year}
  {2019})}\BibitemShut {NoStop}%
\bibitem [{\citenamefont {Godejohann}\ \emph {et~al.}(2020)\citenamefont
  {Godejohann}, \citenamefont {Scherbakov}, \citenamefont {Kukhtaruk},
  \citenamefont {Poddubny}, \citenamefont {Yaremkevich}, \citenamefont {Wang},
  \citenamefont {Nadzeyka}, \citenamefont {Yakovlev}, \citenamefont
  {Rushforth}, \citenamefont {Akimov},\ and\ \citenamefont
  {Bayer}}]{Akimov2020}%
  \BibitemOpen
  \bibfield  {author} {\bibinfo {author} {\bibfnamefont {F.}~\bibnamefont
  {Godejohann}}, \bibinfo {author} {\bibfnamefont {A.~V.}\ \bibnamefont
  {Scherbakov}}, \bibinfo {author} {\bibfnamefont {S.~M.}\ \bibnamefont
  {Kukhtaruk}}, \bibinfo {author} {\bibfnamefont {A.~N.}\ \bibnamefont
  {Poddubny}}, \bibinfo {author} {\bibfnamefont {D.~D.}\ \bibnamefont
  {Yaremkevich}}, \bibinfo {author} {\bibfnamefont {M.}~\bibnamefont {Wang}},
  \bibinfo {author} {\bibfnamefont {A.}~\bibnamefont {Nadzeyka}}, \bibinfo
  {author} {\bibfnamefont {D.~R.}\ \bibnamefont {Yakovlev}}, \bibinfo {author}
  {\bibfnamefont {A.~W.}\ \bibnamefont {Rushforth}}, \bibinfo {author}
  {\bibfnamefont {A.~V.}\ \bibnamefont {Akimov}}, \ and\ \bibinfo {author}
  {\bibfnamefont {M.}~\bibnamefont {Bayer}},\ }\href {\doibase
  10.1103/PhysRevB.102.144438} {\bibfield  {journal} {\bibinfo  {journal}
  {Phys. Rev. B}\ }\textbf {\bibinfo {volume} {102}},\ \bibinfo {pages}
  {144438} (\bibinfo {year} {2020})}\BibitemShut {NoStop}%
\bibitem [{\citenamefont {Frietsch}\ \emph {et~al.}(2020)\citenamefont
  {Frietsch}, \citenamefont {Donges}, \citenamefont {Carley}, \citenamefont
  {Teichmann}, \citenamefont {Bowlan}, \citenamefont {D{\"o}brich},
  \citenamefont {Carva}, \citenamefont {Legut}, \citenamefont {Oppeneer},
  \citenamefont {Nowak},\ and\ \citenamefont {Weinelt}}]{Frietsch2020}%
  \BibitemOpen
  \bibfield  {author} {\bibinfo {author} {\bibfnamefont {B.}~\bibnamefont
  {Frietsch}}, \bibinfo {author} {\bibfnamefont {A.}~\bibnamefont {Donges}},
  \bibinfo {author} {\bibfnamefont {R.}~\bibnamefont {Carley}}, \bibinfo
  {author} {\bibfnamefont {M.}~\bibnamefont {Teichmann}}, \bibinfo {author}
  {\bibfnamefont {J.}~\bibnamefont {Bowlan}}, \bibinfo {author} {\bibfnamefont
  {K.}~\bibnamefont {D{\"o}brich}}, \bibinfo {author} {\bibfnamefont
  {K.}~\bibnamefont {Carva}}, \bibinfo {author} {\bibfnamefont
  {D.}~\bibnamefont {Legut}}, \bibinfo {author} {\bibfnamefont {P.~M.}\
  \bibnamefont {Oppeneer}}, \bibinfo {author} {\bibfnamefont {U.}~\bibnamefont
  {Nowak}}, \ and\ \bibinfo {author} {\bibfnamefont {M.}~\bibnamefont
  {Weinelt}},\ }\href {\doibase 10.1126/sciadv.abb1601} {\bibfield  {journal}
  {\bibinfo  {journal} {Sci. Adv.}\ }\textbf {\bibinfo {volume} {6}},\ \bibinfo
  {pages} {eabb1601} (\bibinfo {year} {2020})}\BibitemShut {NoStop}%
\bibitem [{\citenamefont {Orenstein}(2012)}]{Orenstein2012}%
  \BibitemOpen
  \bibfield  {author} {\bibinfo {author} {\bibfnamefont {J.}~\bibnamefont
  {Orenstein}},\ }\href {\doibase 10.1063/PT.3.1717} {\bibfield  {journal}
  {\bibinfo  {journal} {Physics Today}\ }\textbf {\bibinfo {volume} {65}},\
  \bibinfo {pages} {44} (\bibinfo {year} {2012})}\BibitemShut {NoStop}%
\bibitem [{\citenamefont {Giannetti}\ \emph {et~al.}(2016)\citenamefont
  {Giannetti}, \citenamefont {Capone}, \citenamefont {Fausti}, \citenamefont
  {Fabrizio}, \citenamefont {Parmigiani},\ and\ \citenamefont
  {Mihailovic}}]{Giannetti2016}%
  \BibitemOpen
  \bibfield  {author} {\bibinfo {author} {\bibfnamefont {C.}~\bibnamefont
  {Giannetti}}, \bibinfo {author} {\bibfnamefont {M.}~\bibnamefont {Capone}},
  \bibinfo {author} {\bibfnamefont {D.}~\bibnamefont {Fausti}}, \bibinfo
  {author} {\bibfnamefont {M.}~\bibnamefont {Fabrizio}}, \bibinfo {author}
  {\bibfnamefont {F.}~\bibnamefont {Parmigiani}}, \ and\ \bibinfo {author}
  {\bibfnamefont {D.}~\bibnamefont {Mihailovic}},\ }\href {\doibase
  10.1080/00018732.2016.1194044} {\bibfield  {journal} {\bibinfo  {journal}
  {Advances in Physics}\ }\textbf {\bibinfo {volume} {65}},\ \bibinfo {pages}
  {58} (\bibinfo {year} {2016})}\BibitemShut {NoStop}%
\bibitem [{\citenamefont {Lloyd-Hughes}\ \emph {et~al.}(2021)\citenamefont
  {Lloyd-Hughes}, \citenamefont {Oppeneer}, \citenamefont {{Pereira dos
  Santos}}, \citenamefont {Schleife}, \citenamefont {Meng}, \citenamefont
  {Sentef}, \citenamefont {Ruggenthaler}, \citenamefont {Rubio}, \citenamefont
  {Radu}, \citenamefont {Murnane}, \citenamefont {Shi}, \citenamefont
  {Kapteyn}, \citenamefont {Stadtm{\"u}ller}, \citenamefont {Dani},
  \citenamefont {da~Jornada}, \citenamefont {Prinz}, \citenamefont
  {Aeschlimann}, \citenamefont {Milot}, \citenamefont {Burdanova},
  \citenamefont {Boland}, \citenamefont {Cocker},\ and\ \citenamefont
  {Hegmann}}]{Lloyd_Hughes_2021}%
  \BibitemOpen
  \bibfield  {author} {\bibinfo {author} {\bibfnamefont {J.}~\bibnamefont
  {Lloyd-Hughes}}, \bibinfo {author} {\bibfnamefont {P.~M.}\ \bibnamefont
  {Oppeneer}}, \bibinfo {author} {\bibfnamefont {T.}~\bibnamefont {{Pereira dos
  Santos}}}, \bibinfo {author} {\bibfnamefont {A.}~\bibnamefont {Schleife}},
  \bibinfo {author} {\bibfnamefont {S.}~\bibnamefont {Meng}}, \bibinfo {author}
  {\bibfnamefont {M.~A.}\ \bibnamefont {Sentef}}, \bibinfo {author}
  {\bibfnamefont {M.}~\bibnamefont {Ruggenthaler}}, \bibinfo {author}
  {\bibfnamefont {A.}~\bibnamefont {Rubio}}, \bibinfo {author} {\bibfnamefont
  {I.}~\bibnamefont {Radu}}, \bibinfo {author} {\bibfnamefont {M.}~\bibnamefont
  {Murnane}}, \bibinfo {author} {\bibfnamefont {X.}~\bibnamefont {Shi}},
  \bibinfo {author} {\bibfnamefont {H.}~\bibnamefont {Kapteyn}}, \bibinfo
  {author} {\bibfnamefont {B.}~\bibnamefont {Stadtm{\"u}ller}}, \bibinfo
  {author} {\bibfnamefont {K.~M.}\ \bibnamefont {Dani}}, \bibinfo {author}
  {\bibfnamefont {F.~H.}\ \bibnamefont {da~Jornada}}, \bibinfo {author}
  {\bibfnamefont {E.}~\bibnamefont {Prinz}}, \bibinfo {author} {\bibfnamefont
  {M.}~\bibnamefont {Aeschlimann}}, \bibinfo {author} {\bibfnamefont {R.~L.}\
  \bibnamefont {Milot}}, \bibinfo {author} {\bibfnamefont {M.}~\bibnamefont
  {Burdanova}}, \bibinfo {author} {\bibfnamefont {J.}~\bibnamefont {Boland}},
  \bibinfo {author} {\bibfnamefont {T.}~\bibnamefont {Cocker}}, \ and\ \bibinfo
  {author} {\bibfnamefont {F.}~\bibnamefont {Hegmann}},\ }\href {\doibase
  10.1088/1361-648x/abfe21} {\bibfield  {journal} {\bibinfo  {journal} {J.
  Phys. Condens. Matter}\ }\textbf {\bibinfo {volume} {33}},\ \bibinfo {pages}
  {353001} (\bibinfo {year} {2021})}\BibitemShut {NoStop}%
\bibitem [{\citenamefont {Mankowsky}\ \emph {et~al.}(2014)\citenamefont
  {Mankowsky}, \citenamefont {Subedi}, \citenamefont {F{\"o}rst}, \citenamefont
  {Mariager}, \citenamefont {Chollet}, \citenamefont {Lemke}, \citenamefont
  {Robinson}, \citenamefont {Glownia}, \citenamefont {Minitti}, \citenamefont
  {Frano}, \citenamefont {Fechner}, \citenamefont {Spaldin}, \citenamefont
  {Loew}, \citenamefont {Keimer}, \citenamefont {Georges},\ and\ \citenamefont
  {Cavalleri}}]{Mankowsky2014}%
  \BibitemOpen
  \bibfield  {author} {\bibinfo {author} {\bibfnamefont {R.}~\bibnamefont
  {Mankowsky}}, \bibinfo {author} {\bibfnamefont {A.}~\bibnamefont {Subedi}},
  \bibinfo {author} {\bibfnamefont {M.}~\bibnamefont {F{\"o}rst}}, \bibinfo
  {author} {\bibfnamefont {S.~O.}\ \bibnamefont {Mariager}}, \bibinfo {author}
  {\bibfnamefont {M.}~\bibnamefont {Chollet}}, \bibinfo {author} {\bibfnamefont
  {H.~T.}\ \bibnamefont {Lemke}}, \bibinfo {author} {\bibfnamefont {J.~S.}\
  \bibnamefont {Robinson}}, \bibinfo {author} {\bibfnamefont {J.~M.}\
  \bibnamefont {Glownia}}, \bibinfo {author} {\bibfnamefont {M.~P.}\
  \bibnamefont {Minitti}}, \bibinfo {author} {\bibfnamefont {A.}~\bibnamefont
  {Frano}}, \bibinfo {author} {\bibfnamefont {M.}~\bibnamefont {Fechner}},
  \bibinfo {author} {\bibfnamefont {N.~A.}\ \bibnamefont {Spaldin}}, \bibinfo
  {author} {\bibfnamefont {T.}~\bibnamefont {Loew}}, \bibinfo {author}
  {\bibfnamefont {B.}~\bibnamefont {Keimer}}, \bibinfo {author} {\bibfnamefont
  {A.}~\bibnamefont {Georges}}, \ and\ \bibinfo {author} {\bibfnamefont
  {A.}~\bibnamefont {Cavalleri}},\ }\href {\doibase 10.1038/nature13875}
  {\bibfield  {journal} {\bibinfo  {journal} {Nature}\ }\textbf {\bibinfo
  {volume} {516}},\ \bibinfo {pages} {71} (\bibinfo {year} {2014})}\BibitemShut
  {NoStop}%
\bibitem [{\citenamefont {Hu}\ \emph {et~al.}(2014)\citenamefont {Hu},
  \citenamefont {Kaiser}, \citenamefont {Nicoletti}, \citenamefont {Hunt},
  \citenamefont {Gierz}, \citenamefont {Hoffmann}, \citenamefont {Le~Tacon},
  \citenamefont {Loew}, \citenamefont {Keimer},\ and\ \citenamefont
  {Cavalleri}}]{Hu2014}%
  \BibitemOpen
  \bibfield  {author} {\bibinfo {author} {\bibfnamefont {W.}~\bibnamefont
  {Hu}}, \bibinfo {author} {\bibfnamefont {S.}~\bibnamefont {Kaiser}}, \bibinfo
  {author} {\bibfnamefont {D.}~\bibnamefont {Nicoletti}}, \bibinfo {author}
  {\bibfnamefont {C.~R.}\ \bibnamefont {Hunt}}, \bibinfo {author}
  {\bibfnamefont {I.}~\bibnamefont {Gierz}}, \bibinfo {author} {\bibfnamefont
  {M.~C.}\ \bibnamefont {Hoffmann}}, \bibinfo {author} {\bibfnamefont
  {M.}~\bibnamefont {Le~Tacon}}, \bibinfo {author} {\bibfnamefont
  {T.}~\bibnamefont {Loew}}, \bibinfo {author} {\bibfnamefont {B.}~\bibnamefont
  {Keimer}}, \ and\ \bibinfo {author} {\bibfnamefont {A.}~\bibnamefont
  {Cavalleri}},\ }\href {\doibase 10.1038/nmat3963} {\bibfield  {journal}
  {\bibinfo  {journal} {Nat. Mater.}\ }\textbf {\bibinfo {volume} {13}},\
  \bibinfo {pages} {705} (\bibinfo {year} {2014})}\BibitemShut {NoStop}%
\bibitem [{\citenamefont {Mitrano}\ \emph {et~al.}(2016)\citenamefont
  {Mitrano}, \citenamefont {Cantaluppi}, \citenamefont {Nicoletti},
  \citenamefont {Kaiser}, \citenamefont {Perucchi}, \citenamefont {Lupi},
  \citenamefont {Di~Pietro}, \citenamefont {Pontiroli}, \citenamefont
  {Ricc{\`o}}, \citenamefont {Clark}, \citenamefont {Jaksch},\ and\
  \citenamefont {Cavalleri}}]{Mitrano2016}%
  \BibitemOpen
  \bibfield  {author} {\bibinfo {author} {\bibfnamefont {M.}~\bibnamefont
  {Mitrano}}, \bibinfo {author} {\bibfnamefont {A.}~\bibnamefont {Cantaluppi}},
  \bibinfo {author} {\bibfnamefont {D.}~\bibnamefont {Nicoletti}}, \bibinfo
  {author} {\bibfnamefont {S.}~\bibnamefont {Kaiser}}, \bibinfo {author}
  {\bibfnamefont {A.}~\bibnamefont {Perucchi}}, \bibinfo {author}
  {\bibfnamefont {S.}~\bibnamefont {Lupi}}, \bibinfo {author} {\bibfnamefont
  {P.}~\bibnamefont {Di~Pietro}}, \bibinfo {author} {\bibfnamefont
  {D.}~\bibnamefont {Pontiroli}}, \bibinfo {author} {\bibfnamefont
  {M.}~\bibnamefont {Ricc{\`o}}}, \bibinfo {author} {\bibfnamefont {S.~R.}\
  \bibnamefont {Clark}}, \bibinfo {author} {\bibfnamefont {D.}~\bibnamefont
  {Jaksch}}, \ and\ \bibinfo {author} {\bibfnamefont {A.}~\bibnamefont
  {Cavalleri}},\ }\href {\doibase 10.1038/nature16522} {\bibfield  {journal}
  {\bibinfo  {journal} {Nature}\ }\textbf {\bibinfo {volume} {530}},\ \bibinfo
  {pages} {461} (\bibinfo {year} {2016})}\BibitemShut {NoStop}%
\bibitem [{\citenamefont {Sentef}\ \emph {et~al.}(2016)\citenamefont {Sentef},
  \citenamefont {Kemper}, \citenamefont {Georges},\ and\ \citenamefont
  {Kollath}}]{Sentef2016}%
  \BibitemOpen
  \bibfield  {author} {\bibinfo {author} {\bibfnamefont {M.~A.}\ \bibnamefont
  {Sentef}}, \bibinfo {author} {\bibfnamefont {A.~F.}\ \bibnamefont {Kemper}},
  \bibinfo {author} {\bibfnamefont {A.}~\bibnamefont {Georges}}, \ and\
  \bibinfo {author} {\bibfnamefont {C.}~\bibnamefont {Kollath}},\ }\href
  {\doibase 10.1103/PhysRevB.93.144506} {\bibfield  {journal} {\bibinfo
  {journal} {Phys. Rev. B}\ }\textbf {\bibinfo {volume} {93}},\ \bibinfo
  {pages} {144506} (\bibinfo {year} {2016})}\BibitemShut {NoStop}%
\bibitem [{\citenamefont {Babadi}\ \emph {et~al.}(2017)\citenamefont {Babadi},
  \citenamefont {Knap}, \citenamefont {Martin}, \citenamefont {Refael},\ and\
  \citenamefont {Demler}}]{Babadi2017}%
  \BibitemOpen
  \bibfield  {author} {\bibinfo {author} {\bibfnamefont {M.}~\bibnamefont
  {Babadi}}, \bibinfo {author} {\bibfnamefont {M.}~\bibnamefont {Knap}},
  \bibinfo {author} {\bibfnamefont {I.}~\bibnamefont {Martin}}, \bibinfo
  {author} {\bibfnamefont {G.}~\bibnamefont {Refael}}, \ and\ \bibinfo {author}
  {\bibfnamefont {E.}~\bibnamefont {Demler}},\ }\href {\doibase
  10.1103/PhysRevB.96.014512} {\bibfield  {journal} {\bibinfo  {journal} {Phys.
  Rev. B}\ }\textbf {\bibinfo {volume} {96}},\ \bibinfo {pages} {014512}
  (\bibinfo {year} {2017})}\BibitemShut {NoStop}%
\bibitem [{\citenamefont {Paeckel}\ \emph {et~al.}(2020)\citenamefont
  {Paeckel}, \citenamefont {Fauseweh}, \citenamefont {Osterkorn}, \citenamefont
  {K\"ohler}, \citenamefont {Manske},\ and\ \citenamefont
  {Manmana}}]{Paeckel2020}%
  \BibitemOpen
  \bibfield  {author} {\bibinfo {author} {\bibfnamefont {S.}~\bibnamefont
  {Paeckel}}, \bibinfo {author} {\bibfnamefont {B.}~\bibnamefont {Fauseweh}},
  \bibinfo {author} {\bibfnamefont {A.}~\bibnamefont {Osterkorn}}, \bibinfo
  {author} {\bibfnamefont {T.}~\bibnamefont {K\"ohler}}, \bibinfo {author}
  {\bibfnamefont {D.}~\bibnamefont {Manske}}, \ and\ \bibinfo {author}
  {\bibfnamefont {S.~R.}\ \bibnamefont {Manmana}},\ }\href {\doibase
  10.1103/PhysRevB.101.180507} {\bibfield  {journal} {\bibinfo  {journal}
  {Phys. Rev. B}\ }\textbf {\bibinfo {volume} {101}},\ \bibinfo {pages}
  {180507} (\bibinfo {year} {2020})}\BibitemShut {NoStop}%
\bibitem [{\citenamefont {Buzzi}\ \emph {et~al.}(2020)\citenamefont {Buzzi},
  \citenamefont {Nicoletti}, \citenamefont {Fechner}, \citenamefont
  {Tancogne-Dejean}, \citenamefont {Sentef}, \citenamefont {Georges},
  \citenamefont {Biesner}, \citenamefont {Uykur}, \citenamefont {Dressel},
  \citenamefont {Henderson}, \citenamefont {Siegrist}, \citenamefont
  {Schlueter}, \citenamefont {Miyagawa}, \citenamefont {Kanoda}, \citenamefont
  {Nam}, \citenamefont {Ardavan}, \citenamefont {Coulthard}, \citenamefont
  {Tindall}, \citenamefont {Schlawin}, \citenamefont {Jaksch},\ and\
  \citenamefont {Cavalleri}}]{Buzzi2020}%
  \BibitemOpen
  \bibfield  {author} {\bibinfo {author} {\bibfnamefont {M.}~\bibnamefont
  {Buzzi}}, \bibinfo {author} {\bibfnamefont {D.}~\bibnamefont {Nicoletti}},
  \bibinfo {author} {\bibfnamefont {M.}~\bibnamefont {Fechner}}, \bibinfo
  {author} {\bibfnamefont {N.}~\bibnamefont {Tancogne-Dejean}}, \bibinfo
  {author} {\bibfnamefont {M.~A.}\ \bibnamefont {Sentef}}, \bibinfo {author}
  {\bibfnamefont {A.}~\bibnamefont {Georges}}, \bibinfo {author} {\bibfnamefont
  {T.}~\bibnamefont {Biesner}}, \bibinfo {author} {\bibfnamefont
  {E.}~\bibnamefont {Uykur}}, \bibinfo {author} {\bibfnamefont
  {M.}~\bibnamefont {Dressel}}, \bibinfo {author} {\bibfnamefont
  {A.}~\bibnamefont {Henderson}}, \bibinfo {author} {\bibfnamefont
  {T.}~\bibnamefont {Siegrist}}, \bibinfo {author} {\bibfnamefont {J.~A.}\
  \bibnamefont {Schlueter}}, \bibinfo {author} {\bibfnamefont {K.}~\bibnamefont
  {Miyagawa}}, \bibinfo {author} {\bibfnamefont {K.}~\bibnamefont {Kanoda}},
  \bibinfo {author} {\bibfnamefont {M.-S.}\ \bibnamefont {Nam}}, \bibinfo
  {author} {\bibfnamefont {A.}~\bibnamefont {Ardavan}}, \bibinfo {author}
  {\bibfnamefont {J.}~\bibnamefont {Coulthard}}, \bibinfo {author}
  {\bibfnamefont {J.}~\bibnamefont {Tindall}}, \bibinfo {author} {\bibfnamefont
  {F.}~\bibnamefont {Schlawin}}, \bibinfo {author} {\bibfnamefont
  {D.}~\bibnamefont {Jaksch}}, \ and\ \bibinfo {author} {\bibfnamefont
  {A.}~\bibnamefont {Cavalleri}},\ }\href {\doibase 10.1103/PhysRevX.10.031028}
  {\bibfield  {journal} {\bibinfo  {journal} {Phys. Rev. X}\ }\textbf {\bibinfo
  {volume} {10}},\ \bibinfo {pages} {031028} (\bibinfo {year}
  {2020})}\BibitemShut {NoStop}%
\bibitem [{\citenamefont {Vogelgesang}\ \emph {et~al.}(2018)\citenamefont
  {Vogelgesang}, \citenamefont {Storeck}, \citenamefont {Horstmann},
  \citenamefont {Diekmann}, \citenamefont {Sivis}, \citenamefont {Schramm},
  \citenamefont {Rossnagel}, \citenamefont {Sch{\"a}fer},\ and\ \citenamefont
  {Ropers}}]{Vogelgesang18}%
  \BibitemOpen
  \bibfield  {author} {\bibinfo {author} {\bibfnamefont {S.}~\bibnamefont
  {Vogelgesang}}, \bibinfo {author} {\bibfnamefont {G.}~\bibnamefont
  {Storeck}}, \bibinfo {author} {\bibfnamefont {J.~G.}\ \bibnamefont
  {Horstmann}}, \bibinfo {author} {\bibfnamefont {T.}~\bibnamefont {Diekmann}},
  \bibinfo {author} {\bibfnamefont {M.}~\bibnamefont {Sivis}}, \bibinfo
  {author} {\bibfnamefont {S.}~\bibnamefont {Schramm}}, \bibinfo {author}
  {\bibfnamefont {K.}~\bibnamefont {Rossnagel}}, \bibinfo {author}
  {\bibfnamefont {S.}~\bibnamefont {Sch{\"a}fer}}, \ and\ \bibinfo {author}
  {\bibfnamefont {C.}~\bibnamefont {Ropers}},\ }\href {\doibase
  10.1038/nphys4309} {\bibfield  {journal} {\bibinfo  {journal} {Nat. Phys.}\
  }\textbf {\bibinfo {volume} {14}},\ \bibinfo {pages} {184} (\bibinfo {year}
  {2018})}\BibitemShut {NoStop}%
\bibitem [{\citenamefont {Storeck}\ \emph {et~al.}(2020)\citenamefont
  {Storeck}, \citenamefont {Horstmann}, \citenamefont {Diekmann}, \citenamefont
  {Vogelgesang}, \citenamefont {von Witte}, \citenamefont {Yalunin},
  \citenamefont {Rossnagel},\ and\ \citenamefont {Ropers}}]{Storeck20}%
  \BibitemOpen
  \bibfield  {author} {\bibinfo {author} {\bibfnamefont {G.}~\bibnamefont
  {Storeck}}, \bibinfo {author} {\bibfnamefont {J.~G.}\ \bibnamefont
  {Horstmann}}, \bibinfo {author} {\bibfnamefont {T.}~\bibnamefont {Diekmann}},
  \bibinfo {author} {\bibfnamefont {S.}~\bibnamefont {Vogelgesang}}, \bibinfo
  {author} {\bibfnamefont {G.}~\bibnamefont {von Witte}}, \bibinfo {author}
  {\bibfnamefont {S.~V.}\ \bibnamefont {Yalunin}}, \bibinfo {author}
  {\bibfnamefont {K.}~\bibnamefont {Rossnagel}}, \ and\ \bibinfo {author}
  {\bibfnamefont {C.}~\bibnamefont {Ropers}},\ }\href {\doibase
  10.1063/4.0000018} {\bibfield  {journal} {\bibinfo  {journal} {Struc. Dyn.}\
  }\textbf {\bibinfo {volume} {7}},\ \bibinfo {pages} {034304} (\bibinfo {year}
  {2020})}\BibitemShut {NoStop}%
\bibitem [{\citenamefont {Storeck}, \citenamefont {Rossnagel},\ and\
  \citenamefont {Ropers}(2021)}]{Storeck21}%
  \BibitemOpen
  \bibfield  {author} {\bibinfo {author} {\bibfnamefont {G.}~\bibnamefont
  {Storeck}}, \bibinfo {author} {\bibfnamefont {K.}~\bibnamefont {Rossnagel}},
  \ and\ \bibinfo {author} {\bibfnamefont {C.}~\bibnamefont {Ropers}},\ }\href
  {\doibase 10.1063/5.0052603} {\bibfield  {journal} {\bibinfo  {journal} {App.
  Phys. Lett.}\ }\textbf {\bibinfo {volume} {118}},\ \bibinfo {pages} {221603}
  (\bibinfo {year} {2021})}\BibitemShut {NoStop}%
\bibitem [{\citenamefont {Horstmann}\ \emph {et~al.}(2020)\citenamefont
  {Horstmann}, \citenamefont {B{\"o}ckmann}, \citenamefont {Wit}, \citenamefont
  {Kurtz}, \citenamefont {Storeck},\ and\ \citenamefont
  {Ropers}}]{Horstmann20}%
  \BibitemOpen
  \bibfield  {author} {\bibinfo {author} {\bibfnamefont {J.~G.}\ \bibnamefont
  {Horstmann}}, \bibinfo {author} {\bibfnamefont {H.}~\bibnamefont
  {B{\"o}ckmann}}, \bibinfo {author} {\bibfnamefont {B.}~\bibnamefont {Wit}},
  \bibinfo {author} {\bibfnamefont {F.}~\bibnamefont {Kurtz}}, \bibinfo
  {author} {\bibfnamefont {G.}~\bibnamefont {Storeck}}, \ and\ \bibinfo
  {author} {\bibfnamefont {C.}~\bibnamefont {Ropers}},\ }\href {\doibase
  10.1038/s41586-020-2440-4} {\bibfield  {journal} {\bibinfo  {journal}
  {Nature}\ }\textbf {\bibinfo {volume} {583}},\ \bibinfo {pages} {232}
  (\bibinfo {year} {2020})}\BibitemShut {NoStop}%
\bibitem [{\citenamefont {Leggett}\ \emph {et~al.}(1987)\citenamefont
  {Leggett}, \citenamefont {Chakravarty}, \citenamefont {Dorsey}, \citenamefont
  {Fisher}, \citenamefont {Garg},\ and\ \citenamefont {Zwerger}}]{Leggett87}%
  \BibitemOpen
  \bibfield  {author} {\bibinfo {author} {\bibfnamefont {A.~J.}\ \bibnamefont
  {Leggett}}, \bibinfo {author} {\bibfnamefont {S.}~\bibnamefont
  {Chakravarty}}, \bibinfo {author} {\bibfnamefont {A.~T.}\ \bibnamefont
  {Dorsey}}, \bibinfo {author} {\bibfnamefont {M.~P.~A.}\ \bibnamefont
  {Fisher}}, \bibinfo {author} {\bibfnamefont {A.}~\bibnamefont {Garg}}, \ and\
  \bibinfo {author} {\bibfnamefont {W.}~\bibnamefont {Zwerger}},\ }\href
  {\doibase 10.1103/RevModPhys.59.1} {\bibfield  {journal} {\bibinfo  {journal}
  {Rev. Mod. Phys.}\ }\textbf {\bibinfo {volume} {59}},\ \bibinfo {pages} {1}
  (\bibinfo {year} {1987})}\BibitemShut {NoStop}%
\bibitem [{\citenamefont {Makarov}\ and\ \citenamefont
  {Makri}(1994)}]{Makarov94}%
  \BibitemOpen
  \bibfield  {author} {\bibinfo {author} {\bibfnamefont {D.~E.}\ \bibnamefont
  {Makarov}}\ and\ \bibinfo {author} {\bibfnamefont {N.}~\bibnamefont
  {Makri}},\ }\href {\doibase 10.1016/0009-2614(94)00275-4} {\bibfield
  {journal} {\bibinfo  {journal} {Chem. Phys. Lett.}\ }\textbf {\bibinfo
  {volume} {221}},\ \bibinfo {pages} {482} (\bibinfo {year}
  {1994})}\BibitemShut {NoStop}%
\bibitem [{\citenamefont {Egger}\ and\ \citenamefont {Mak}(1994)}]{Egger94}%
  \BibitemOpen
  \bibfield  {author} {\bibinfo {author} {\bibfnamefont {R.}~\bibnamefont
  {Egger}}\ and\ \bibinfo {author} {\bibfnamefont {C.~H.}\ \bibnamefont
  {Mak}},\ }\href {\doibase 10.1103/PhysRevB.50.15210} {\bibfield  {journal}
  {\bibinfo  {journal} {Phys. Rev. B}\ }\textbf {\bibinfo {volume} {50}},\
  \bibinfo {pages} {15210} (\bibinfo {year} {1994})}\BibitemShut {NoStop}%
\bibitem [{\citenamefont {Kehrein}\ and\ \citenamefont
  {Mielke}(1996)}]{Kehrein96}%
  \BibitemOpen
  \bibfield  {author} {\bibinfo {author} {\bibfnamefont {S.~K.}\ \bibnamefont
  {Kehrein}}\ and\ \bibinfo {author} {\bibfnamefont {A.}~\bibnamefont
  {Mielke}},\ }\href {\doibase 10.1016/0375-9601(96)00475-6} {\bibfield
  {journal} {\bibinfo  {journal} {Phys. Lett. A}\ }\textbf {\bibinfo {volume}
  {219}},\ \bibinfo {pages} {313} (\bibinfo {year} {1996})}\BibitemShut
  {NoStop}%
\bibitem [{\citenamefont {Thompson}\ and\ \citenamefont
  {Makri}(1999)}]{Thompson99}%
  \BibitemOpen
  \bibfield  {author} {\bibinfo {author} {\bibfnamefont {K.}~\bibnamefont
  {Thompson}}\ and\ \bibinfo {author} {\bibfnamefont {N.}~\bibnamefont
  {Makri}},\ }\href {\doibase 10.1063/1.478011} {\bibfield  {journal} {\bibinfo
   {journal} {J. Chem. Phys.}\ }\textbf {\bibinfo {volume} {110}},\ \bibinfo
  {pages} {1343} (\bibinfo {year} {1999})}\BibitemShut {NoStop}%
\bibitem [{\citenamefont {Wang}(2000)}]{Wang00}%
  \BibitemOpen
  \bibfield  {author} {\bibinfo {author} {\bibfnamefont {H.}~\bibnamefont
  {Wang}},\ }\href {\doibase 10.1063/1.1323746} {\bibfield  {journal} {\bibinfo
   {journal} {J. Chem. Phys.}\ }\textbf {\bibinfo {volume} {113}},\ \bibinfo
  {pages} {9948} (\bibinfo {year} {2000})}\BibitemShut {NoStop}%
\bibitem [{\citenamefont {MacKernan}, \citenamefont {Kapral},\ and\
  \citenamefont {Ciccotti}(2002)}]{MacKernan02}%
  \BibitemOpen
  \bibfield  {author} {\bibinfo {author} {\bibfnamefont {D.}~\bibnamefont
  {MacKernan}}, \bibinfo {author} {\bibfnamefont {R.}~\bibnamefont {Kapral}}, \
  and\ \bibinfo {author} {\bibfnamefont {G.}~\bibnamefont {Ciccotti}},\ }\href
  {\doibase 10.1088/0953-8984/14/40/301} {\bibfield  {journal} {\bibinfo
  {journal} {J. Phys. Condens. Matter}\ }\textbf {\bibinfo {volume} {14}},\
  \bibinfo {pages} {9069} (\bibinfo {year} {2002})}\BibitemShut {NoStop}%
\bibitem [{\citenamefont {Wang}\ and\ \citenamefont {Thoss}(2003)}]{Wang2003}%
  \BibitemOpen
  \bibfield  {author} {\bibinfo {author} {\bibfnamefont {H.}~\bibnamefont
  {Wang}}\ and\ \bibinfo {author} {\bibfnamefont {M.}~\bibnamefont {Thoss}},\
  }\href {\doibase 10.1063/1.1580111} {\bibfield  {journal} {\bibinfo
  {journal} {J. Chem. Phys.}\ }\textbf {\bibinfo {volume} {119}},\ \bibinfo
  {pages} {1289} (\bibinfo {year} {2003})}\BibitemShut {NoStop}%
\bibitem [{\citenamefont {Wang}\ and\ \citenamefont {Thoss}(2008)}]{Wang08}%
  \BibitemOpen
  \bibfield  {author} {\bibinfo {author} {\bibfnamefont {H.}~\bibnamefont
  {Wang}}\ and\ \bibinfo {author} {\bibfnamefont {M.}~\bibnamefont {Thoss}},\
  }\href {\doibase 10.1088/1367-2630/10/11/115005} {\bibfield  {journal}
  {\bibinfo  {journal} {New J. Phys.}\ }\textbf {\bibinfo {volume} {10}},\
  \bibinfo {pages} {115005} (\bibinfo {year} {2008})}\BibitemShut {NoStop}%
\bibitem [{\citenamefont {Mac~Kernan}, \citenamefont {Ciccotti},\ and\
  \citenamefont {Kapral}(2008)}]{MacKernan08}%
  \BibitemOpen
  \bibfield  {author} {\bibinfo {author} {\bibfnamefont {D.}~\bibnamefont
  {Mac~Kernan}}, \bibinfo {author} {\bibfnamefont {G.}~\bibnamefont
  {Ciccotti}}, \ and\ \bibinfo {author} {\bibfnamefont {R.}~\bibnamefont
  {Kapral}},\ }\href {\doibase 10.1021/jp0761416} {\bibfield  {journal}
  {\bibinfo  {journal} {J. Phys. Chem. B}\ }\textbf {\bibinfo {volume} {112}},\
  \bibinfo {pages} {424} (\bibinfo {year} {2008})}\BibitemShut {NoStop}%
\bibitem [{\citenamefont {Wang}\ and\ \citenamefont {Thoss}(2009)}]{Wang09}%
  \BibitemOpen
  \bibfield  {author} {\bibinfo {author} {\bibfnamefont {H.}~\bibnamefont
  {Wang}}\ and\ \bibinfo {author} {\bibfnamefont {M.}~\bibnamefont {Thoss}},\
  }\enquote {\bibinfo {title} {Multilayer formulation of the multiconfiguration
  time-dependent {H}artree theory},}\ in\ \href {\doibase
  10.1002/9783527627400.ch14} {\emph {\bibinfo {booktitle} {Multidimensional
  Quantum Dynamics}}},\ \bibinfo {editor} {edited by\ \bibinfo {editor}
  {\bibfnamefont {H.-D.}\ \bibnamefont {Meyer}}, \bibinfo {editor}
  {\bibfnamefont {F.}~\bibnamefont {Gatti}}, \ and\ \bibinfo {editor}
  {\bibfnamefont {G.~A.}\ \bibnamefont {Worth}}}\ (\bibinfo  {publisher}
  {Wiley},\ \bibinfo {year} {2009})\ Chap.~\bibinfo {chapter} {14}, pp.\
  \bibinfo {pages} {131--147}\BibitemShut {NoStop}%
\bibitem [{\citenamefont {Kananenka}\ \emph {et~al.}(2016)\citenamefont
  {Kananenka}, \citenamefont {Hsieh}, \citenamefont {Cao},\ and\ \citenamefont
  {Geva}}]{Kananenka16}%
  \BibitemOpen
  \bibfield  {author} {\bibinfo {author} {\bibfnamefont {A.~A.}\ \bibnamefont
  {Kananenka}}, \bibinfo {author} {\bibfnamefont {C.-Y.}\ \bibnamefont
  {Hsieh}}, \bibinfo {author} {\bibfnamefont {J.}~\bibnamefont {Cao}}, \ and\
  \bibinfo {author} {\bibfnamefont {E.}~\bibnamefont {Geva}},\ }\href {\doibase
  10.1021/acs.jpclett.6b02389} {\bibfield  {journal} {\bibinfo  {journal} {J.
  Phys. Chem. Lett.}\ }\textbf {\bibinfo {volume} {7}},\ \bibinfo {pages}
  {4809} (\bibinfo {year} {2016})}\BibitemShut {NoStop}%
\bibitem [{\citenamefont {Chen}\ and\ \citenamefont
  {Reichman}(2016)}]{Chen2016}%
  \BibitemOpen
  \bibfield  {author} {\bibinfo {author} {\bibfnamefont {H.-T.}\ \bibnamefont
  {Chen}}\ and\ \bibinfo {author} {\bibfnamefont {D.~R.}\ \bibnamefont
  {Reichman}},\ }\href {\doibase 10.1063/1.4942867} {\bibfield  {journal}
  {\bibinfo  {journal} {J. Chem. Phys.}\ }\textbf {\bibinfo {volume} {144}},\
  \bibinfo {pages} {094104} (\bibinfo {year} {2016})}\BibitemShut {NoStop}%
\bibitem [{\citenamefont {Tully}(1990)}]{Tully1990}%
  \BibitemOpen
  \bibfield  {author} {\bibinfo {author} {\bibfnamefont {J.~C.}\ \bibnamefont
  {Tully}},\ }\href {\doibase 10.1063/1.459170} {\bibfield  {journal} {\bibinfo
   {journal} {J. Chem. Phys.}\ }\textbf {\bibinfo {volume} {93}},\ \bibinfo
  {pages} {1061} (\bibinfo {year} {1990})}\BibitemShut {NoStop}%
\bibitem [{\citenamefont {C.~Tully}(1998)}]{Tully1998}%
  \BibitemOpen
  \bibfield  {author} {\bibinfo {author} {\bibfnamefont {J.}~\bibnamefont
  {C.~Tully}},\ }\href {\doibase 10.1039/A801824C} {\bibfield  {journal}
  {\bibinfo  {journal} {Faraday Discuss.}\ }\textbf {\bibinfo {volume} {110}},\
  \bibinfo {pages} {407} (\bibinfo {year} {1998})}\BibitemShut {NoStop}%
\bibitem [{\citenamefont {Agostini}\ \emph {et~al.}(2016)\citenamefont
  {Agostini}, \citenamefont {Min}, \citenamefont {Abedi},\ and\ \citenamefont
  {Gross}}]{Agostini16}%
  \BibitemOpen
  \bibfield  {author} {\bibinfo {author} {\bibfnamefont {F.}~\bibnamefont
  {Agostini}}, \bibinfo {author} {\bibfnamefont {S.~K.}\ \bibnamefont {Min}},
  \bibinfo {author} {\bibfnamefont {A.}~\bibnamefont {Abedi}}, \ and\ \bibinfo
  {author} {\bibfnamefont {E.~K.~U.}\ \bibnamefont {Gross}},\ }\href {\doibase
  10.1021/acs.jctc.5b01180} {\bibfield  {journal} {\bibinfo  {journal} {J.
  Chem. Theory Comput.}\ }\textbf {\bibinfo {volume} {12}},\ \bibinfo {pages}
  {2127} (\bibinfo {year} {2016})}\BibitemShut {NoStop}%
\bibitem [{\citenamefont {Agostini}(2018)}]{Agostini18}%
  \BibitemOpen
  \bibfield  {author} {\bibinfo {author} {\bibfnamefont {F.}~\bibnamefont
  {Agostini}},\ }\href {\doibase 10.1140/epjb/e2018-90085-9} {\bibfield
  {journal} {\bibinfo  {journal} {Eur. Phys. J. B}\ }\textbf {\bibinfo {volume}
  {91}},\ \bibinfo {pages} {143} (\bibinfo {year} {2018})}\BibitemShut
  {NoStop}%
\bibitem [{\citenamefont {Ibele}\ and\ \citenamefont
  {Curchod}(2020)}]{Ibele2020}%
  \BibitemOpen
  \bibfield  {author} {\bibinfo {author} {\bibfnamefont {L.~M.}\ \bibnamefont
  {Ibele}}\ and\ \bibinfo {author} {\bibfnamefont {B.~F.~E.}\ \bibnamefont
  {Curchod}},\ }\href {\doibase 10.1039/D0CP01353F} {\bibfield  {journal}
  {\bibinfo  {journal} {Phys. Chem. Chem. Phys.}\ }\textbf {\bibinfo {volume}
  {22}},\ \bibinfo {pages} {15183} (\bibinfo {year} {2020})}\BibitemShut
  {NoStop}%
\bibitem [{\citenamefont {Shin}\ and\ \citenamefont {Metiu}(1995)}]{Shin95}%
  \BibitemOpen
  \bibfield  {author} {\bibinfo {author} {\bibfnamefont {S.}~\bibnamefont
  {Shin}}\ and\ \bibinfo {author} {\bibfnamefont {H.}~\bibnamefont {Metiu}},\
  }\href {\doibase 10.1063/1.468795} {\bibfield  {journal} {\bibinfo  {journal}
  {J. Chem. Phys.}\ }\textbf {\bibinfo {volume} {102}},\ \bibinfo {pages}
  {9285} (\bibinfo {year} {1995})}\BibitemShut {NoStop}%
\bibitem [{\citenamefont {Abedi}\ \emph {et~al.}(2013)\citenamefont {Abedi},
  \citenamefont {Agostini}, \citenamefont {Suzuki},\ and\ \citenamefont
  {Gross}}]{Abedi13}%
  \BibitemOpen
  \bibfield  {author} {\bibinfo {author} {\bibfnamefont {A.}~\bibnamefont
  {Abedi}}, \bibinfo {author} {\bibfnamefont {F.}~\bibnamefont {Agostini}},
  \bibinfo {author} {\bibfnamefont {Y.}~\bibnamefont {Suzuki}}, \ and\ \bibinfo
  {author} {\bibfnamefont {E.~K.~U.}\ \bibnamefont {Gross}},\ }\href {\doibase
  10.1103/PhysRevLett.110.263001} {\bibfield  {journal} {\bibinfo  {journal}
  {Phys. Rev. Lett.}\ }\textbf {\bibinfo {volume} {110}},\ \bibinfo {pages}
  {263001} (\bibinfo {year} {2013})}\BibitemShut {NoStop}%
\bibitem [{\citenamefont {Agostini}\ \emph {et~al.}(2015)\citenamefont
  {Agostini}, \citenamefont {Abedi}, \citenamefont {Suzuki}, \citenamefont
  {Min}, \citenamefont {Maitra},\ and\ \citenamefont {Gross}}]{Agostini15}%
  \BibitemOpen
  \bibfield  {author} {\bibinfo {author} {\bibfnamefont {F.}~\bibnamefont
  {Agostini}}, \bibinfo {author} {\bibfnamefont {A.}~\bibnamefont {Abedi}},
  \bibinfo {author} {\bibfnamefont {Y.}~\bibnamefont {Suzuki}}, \bibinfo
  {author} {\bibfnamefont {S.~K.}\ \bibnamefont {Min}}, \bibinfo {author}
  {\bibfnamefont {N.~T.}\ \bibnamefont {Maitra}}, \ and\ \bibinfo {author}
  {\bibfnamefont {E.~K.~U.}\ \bibnamefont {Gross}},\ }\href {\doibase
  10.1063/1.4908133} {\bibfield  {journal} {\bibinfo  {journal} {J. Chem.
  Phys.}\ }\textbf {\bibinfo {volume} {142}},\ \bibinfo {pages} {084303}
  (\bibinfo {year} {2015})}\BibitemShut {NoStop}%
\bibitem [{\citenamefont {Eich}\ and\ \citenamefont {Agostini}(2016)}]{Eich16}%
  \BibitemOpen
  \bibfield  {author} {\bibinfo {author} {\bibfnamefont {F.~G.}\ \bibnamefont
  {Eich}}\ and\ \bibinfo {author} {\bibfnamefont {F.}~\bibnamefont
  {Agostini}},\ }\href {\doibase 10.1063/1.4959962} {\bibfield  {journal}
  {\bibinfo  {journal} {J. Chem. Phys.}\ }\textbf {\bibinfo {volume} {145}},\
  \bibinfo {pages} {054110} (\bibinfo {year} {2016})}\BibitemShut {NoStop}%
\bibitem [{\citenamefont {Gossel}, \citenamefont {Lacombe},\ and\ \citenamefont
  {Maitra}(2019)}]{Gossel19}%
  \BibitemOpen
  \bibfield  {author} {\bibinfo {author} {\bibfnamefont {G.~H.}\ \bibnamefont
  {Gossel}}, \bibinfo {author} {\bibfnamefont {L.}~\bibnamefont {Lacombe}}, \
  and\ \bibinfo {author} {\bibfnamefont {N.~T.}\ \bibnamefont {Maitra}},\
  }\href {\doibase 10.1063/1.5090802} {\bibfield  {journal} {\bibinfo
  {journal} {J. Chem. Phys.}\ }\textbf {\bibinfo {volume} {150}},\ \bibinfo
  {pages} {154112} (\bibinfo {year} {2019})}\BibitemShut {NoStop}%
\bibitem [{\citenamefont {Martinez}\ \emph {et~al.}(2021)\citenamefont
  {Martinez}, \citenamefont {Rosenzweig}, \citenamefont {Hoffmann},
  \citenamefont {Lacombe},\ and\ \citenamefont {Maitra}}]{Martinez21}%
  \BibitemOpen
  \bibfield  {author} {\bibinfo {author} {\bibfnamefont {P.}~\bibnamefont
  {Martinez}}, \bibinfo {author} {\bibfnamefont {B.}~\bibnamefont
  {Rosenzweig}}, \bibinfo {author} {\bibfnamefont {N.~M.}\ \bibnamefont
  {Hoffmann}}, \bibinfo {author} {\bibfnamefont {L.}~\bibnamefont {Lacombe}}, \
  and\ \bibinfo {author} {\bibfnamefont {N.~T.}\ \bibnamefont {Maitra}},\
  }\href {\doibase 10.1063/5.0033386} {\bibfield  {journal} {\bibinfo
  {journal} {J. Chem. Phys.}\ }\textbf {\bibinfo {volume} {154}},\ \bibinfo
  {pages} {014102} (\bibinfo {year} {2021})}\BibitemShut {NoStop}%
\bibitem [{\citenamefont {Bostr\"om}, \citenamefont {Mikkelsen},\ and\
  \citenamefont {Verdozzi}(2016)}]{Bostrom2016a}%
  \BibitemOpen
  \bibfield  {author} {\bibinfo {author} {\bibfnamefont {E.}~\bibnamefont
  {Bostr\"om}}, \bibinfo {author} {\bibfnamefont {A.}~\bibnamefont
  {Mikkelsen}}, \ and\ \bibinfo {author} {\bibfnamefont {C.}~\bibnamefont
  {Verdozzi}},\ }\href {\doibase 10.1103/PhysRevB.93.195416} {\bibfield
  {journal} {\bibinfo  {journal} {Phys. Rev. B}\ }\textbf {\bibinfo {volume}
  {93}},\ \bibinfo {pages} {195416} (\bibinfo {year} {2016})}\BibitemShut
  {NoStop}%
\bibitem [{\citenamefont {Bostr\"om}\ \emph {et~al.}(2016)\citenamefont
  {Bostr\"om}, \citenamefont {Hopjan}, \citenamefont {Kartsev}, \citenamefont
  {Verdozzi},\ and\ \citenamefont {Almbladh}}]{Bostrom2016b}%
  \BibitemOpen
  \bibfield  {author} {\bibinfo {author} {\bibfnamefont {E.}~\bibnamefont
  {Bostr\"om}}, \bibinfo {author} {\bibfnamefont {M.}~\bibnamefont {Hopjan}},
  \bibinfo {author} {\bibfnamefont {A.}~\bibnamefont {Kartsev}}, \bibinfo
  {author} {\bibfnamefont {C.}~\bibnamefont {Verdozzi}}, \ and\ \bibinfo
  {author} {\bibfnamefont {C.-O.}\ \bibnamefont {Almbladh}},\ }\href {\doibase
  10.1088/1742-6596/696/1/012007} {\bibfield  {journal} {\bibinfo  {journal}
  {J. Phys.: Conf. Ser.}\ }\textbf {\bibinfo {volume} {696}},\ \bibinfo {pages}
  {012007} (\bibinfo {year} {2016})}\BibitemShut {NoStop}%
\bibitem [{\citenamefont {S\"akkinen}\ \emph
  {et~al.}(2015{\natexlab{a}})\citenamefont {S\"akkinen}, \citenamefont {Peng},
  \citenamefont {Appel},\ and\ \citenamefont {van Leeuwen}}]{Sakkinen15a}%
  \BibitemOpen
  \bibfield  {author} {\bibinfo {author} {\bibfnamefont {N.}~\bibnamefont
  {S\"akkinen}}, \bibinfo {author} {\bibfnamefont {Y.}~\bibnamefont {Peng}},
  \bibinfo {author} {\bibfnamefont {H.}~\bibnamefont {Appel}}, \ and\ \bibinfo
  {author} {\bibfnamefont {R.}~\bibnamefont {van Leeuwen}},\ }\href {\doibase
  10.1063/1.4936142} {\bibfield  {journal} {\bibinfo  {journal} {J. Chem.
  Phys.}\ }\textbf {\bibinfo {volume} {143}},\ \bibinfo {pages} {234101}
  (\bibinfo {year} {2015}{\natexlab{a}})}\BibitemShut {NoStop}%
\bibitem [{\citenamefont {S\"akkinen}\ \emph
  {et~al.}(2015{\natexlab{b}})\citenamefont {S\"akkinen}, \citenamefont {Peng},
  \citenamefont {Appel},\ and\ \citenamefont {van Leeuwen}}]{Sakkinen15b}%
  \BibitemOpen
  \bibfield  {author} {\bibinfo {author} {\bibfnamefont {N.}~\bibnamefont
  {S\"akkinen}}, \bibinfo {author} {\bibfnamefont {Y.}~\bibnamefont {Peng}},
  \bibinfo {author} {\bibfnamefont {H.}~\bibnamefont {Appel}}, \ and\ \bibinfo
  {author} {\bibfnamefont {R.}~\bibnamefont {van Leeuwen}},\ }\href {\doibase
  10.1063/1.4936143} {\bibfield  {journal} {\bibinfo  {journal} {J. Chem.
  Phys.}\ }\textbf {\bibinfo {volume} {143}},\ \bibinfo {pages} {234102}
  (\bibinfo {year} {2015}{\natexlab{b}})}\BibitemShut {NoStop}%
\bibitem [{\citenamefont {Karlsson}\ and\ \citenamefont {van
  Leeuwen}(2020)}]{Karlsson20}%
  \BibitemOpen
  \bibfield  {author} {\bibinfo {author} {\bibfnamefont {D.}~\bibnamefont
  {Karlsson}}\ and\ \bibinfo {author} {\bibfnamefont {R.}~\bibnamefont {van
  Leeuwen}},\ }\enquote {\bibinfo {title} {Non-equilibrium {G}reen's functions
  for coupled fermion-boson systems},}\ in\ \href {\doibase
  10.1007/978-3-319-44677-6_8} {\emph {\bibinfo {booktitle} {Handbook of
  Materials Modeling: Methods: Theory and Modeling}}},\ \bibinfo {editor}
  {edited by\ \bibinfo {editor} {\bibfnamefont {W.}~\bibnamefont {Andreoni}}\
  and\ \bibinfo {editor} {\bibfnamefont {S.}~\bibnamefont {Yip}}}\ (\bibinfo
  {publisher} {Springer International Publishing},\ \bibinfo {address} {Cham},\
  \bibinfo {year} {2020})\ pp.\ \bibinfo {pages} {367--395}\BibitemShut
  {NoStop}%
\bibitem [{\citenamefont {Sch\"uler}, \citenamefont {Berakdar},\ and\
  \citenamefont {Pavlyukh}(2016)}]{Schuler16}%
  \BibitemOpen
  \bibfield  {author} {\bibinfo {author} {\bibfnamefont {M.}~\bibnamefont
  {Sch\"uler}}, \bibinfo {author} {\bibfnamefont {J.}~\bibnamefont {Berakdar}},
  \ and\ \bibinfo {author} {\bibfnamefont {Y.}~\bibnamefont {Pavlyukh}},\
  }\href {\doibase 10.1103/PhysRevB.93.054303} {\bibfield  {journal} {\bibinfo
  {journal} {Phys. Rev. B}\ }\textbf {\bibinfo {volume} {93}},\ \bibinfo
  {pages} {054303} (\bibinfo {year} {2016})}\BibitemShut {NoStop}%
\bibitem [{\citenamefont {Karlsson}\ \emph {et~al.}(2021)\citenamefont
  {Karlsson}, \citenamefont {van Leeuwen}, \citenamefont {Pavlyukh},
  \citenamefont {Perfetto},\ and\ \citenamefont {Stefanucci}}]{Karlsson21}%
  \BibitemOpen
  \bibfield  {author} {\bibinfo {author} {\bibfnamefont {D.}~\bibnamefont
  {Karlsson}}, \bibinfo {author} {\bibfnamefont {R.}~\bibnamefont {van
  Leeuwen}}, \bibinfo {author} {\bibfnamefont {Y.}~\bibnamefont {Pavlyukh}},
  \bibinfo {author} {\bibfnamefont {E.}~\bibnamefont {Perfetto}}, \ and\
  \bibinfo {author} {\bibfnamefont {G.}~\bibnamefont {Stefanucci}},\ }\href
  {\doibase 10.1103/PhysRevLett.127.036402} {\bibfield  {journal} {\bibinfo
  {journal} {Phys. Rev. Lett.}\ }\textbf {\bibinfo {volume} {127}},\ \bibinfo
  {pages} {036402} (\bibinfo {year} {2021})}\BibitemShut {NoStop}%
\bibitem [{\citenamefont {Pavlyukh}\ \emph
  {et~al.}(2022{\natexlab{a}})\citenamefont {Pavlyukh}, \citenamefont
  {Perfetto}, \citenamefont {Karlsson}, \citenamefont {van Leeuwen},\ and\
  \citenamefont {Stefanucci}}]{Pavlyukh21a}%
  \BibitemOpen
  \bibfield  {author} {\bibinfo {author} {\bibfnamefont {Y.}~\bibnamefont
  {Pavlyukh}}, \bibinfo {author} {\bibfnamefont {E.}~\bibnamefont {Perfetto}},
  \bibinfo {author} {\bibfnamefont {D.}~\bibnamefont {Karlsson}}, \bibinfo
  {author} {\bibfnamefont {R.}~\bibnamefont {van Leeuwen}}, \ and\ \bibinfo
  {author} {\bibfnamefont {G.}~\bibnamefont {Stefanucci}},\ }\href {\doibase
  10.1103/PhysRevB.105.125135} {\bibfield  {journal} {\bibinfo  {journal}
  {Phys. Rev. B}\ }\textbf {\bibinfo {volume} {105}},\ \bibinfo {pages}
  {125135} (\bibinfo {year} {2022}{\natexlab{a}})}\BibitemShut {NoStop}%
\bibitem [{\citenamefont {Pavlyukh}\ \emph
  {et~al.}(2022{\natexlab{b}})\citenamefont {Pavlyukh}, \citenamefont
  {Perfetto}, \citenamefont {Karlsson}, \citenamefont {van Leeuwen},\ and\
  \citenamefont {Stefanucci}}]{Pavlyukh21b}%
  \BibitemOpen
  \bibfield  {author} {\bibinfo {author} {\bibfnamefont {Y.}~\bibnamefont
  {Pavlyukh}}, \bibinfo {author} {\bibfnamefont {E.}~\bibnamefont {Perfetto}},
  \bibinfo {author} {\bibfnamefont {D.}~\bibnamefont {Karlsson}}, \bibinfo
  {author} {\bibfnamefont {R.}~\bibnamefont {van Leeuwen}}, \ and\ \bibinfo
  {author} {\bibfnamefont {G.}~\bibnamefont {Stefanucci}},\ }\href {\doibase
  10.1103/PhysRevB.105.125134} {\bibfield  {journal} {\bibinfo  {journal}
  {Phys. Rev. B}\ }\textbf {\bibinfo {volume} {105}},\ \bibinfo {pages}
  {125134} (\bibinfo {year} {2022}{\natexlab{b}})}\BibitemShut {NoStop}%
\bibitem [{\citenamefont {Hoffmann}\ \emph {et~al.}(2019)\citenamefont
  {Hoffmann}, \citenamefont {Sch\"afer}, \citenamefont {S\"akkinen},
  \citenamefont {Rubio}, \citenamefont {Appel},\ and\ \citenamefont
  {Kelly}}]{Hoffmann19}%
  \BibitemOpen
  \bibfield  {author} {\bibinfo {author} {\bibfnamefont {N.~M.}\ \bibnamefont
  {Hoffmann}}, \bibinfo {author} {\bibfnamefont {C.}~\bibnamefont {Sch\"afer}},
  \bibinfo {author} {\bibfnamefont {N.}~\bibnamefont {S\"akkinen}}, \bibinfo
  {author} {\bibfnamefont {A.}~\bibnamefont {Rubio}}, \bibinfo {author}
  {\bibfnamefont {H.}~\bibnamefont {Appel}}, \ and\ \bibinfo {author}
  {\bibfnamefont {A.}~\bibnamefont {Kelly}},\ }\href {\doibase
  10.1063/1.5128076} {\bibfield  {journal} {\bibinfo  {journal} {J. Chem.
  Phys.}\ }\textbf {\bibinfo {volume} {151}},\ \bibinfo {pages} {244113}
  (\bibinfo {year} {2019})}\BibitemShut {NoStop}%
\bibitem [{\citenamefont {Bostr\"om}\ \emph {et~al.}(2019)\citenamefont
  {Bostr\"om}, \citenamefont {Helmer}, \citenamefont {Werner},\ and\
  \citenamefont {Verdozzi}}]{Bostrom19}%
  \BibitemOpen
  \bibfield  {author} {\bibinfo {author} {\bibfnamefont {E.}~\bibnamefont
  {Bostr\"om}}, \bibinfo {author} {\bibfnamefont {P.}~\bibnamefont {Helmer}},
  \bibinfo {author} {\bibfnamefont {P.}~\bibnamefont {Werner}}, \ and\ \bibinfo
  {author} {\bibfnamefont {C.}~\bibnamefont {Verdozzi}},\ }\href {\doibase
  10.1103/PhysRevResearch.1.013017} {\bibfield  {journal} {\bibinfo  {journal}
  {Phys. Rev. Research}\ }\textbf {\bibinfo {volume} {1}},\ \bibinfo {pages}
  {013017} (\bibinfo {year} {2019})}\BibitemShut {NoStop}%
\bibitem [{\citenamefont {Reinhard}\ \emph {et~al.}(2019)\citenamefont
  {Reinhard}, \citenamefont {Mordovina}, \citenamefont {Hubig}, \citenamefont
  {Kretchmer}, \citenamefont {Schollw\"ock}, \citenamefont {Appel},
  \citenamefont {Sentef},\ and\ \citenamefont {Rubio}}]{Reinhard19}%
  \BibitemOpen
  \bibfield  {author} {\bibinfo {author} {\bibfnamefont {T.~E.}\ \bibnamefont
  {Reinhard}}, \bibinfo {author} {\bibfnamefont {U.}~\bibnamefont {Mordovina}},
  \bibinfo {author} {\bibfnamefont {C.}~\bibnamefont {Hubig}}, \bibinfo
  {author} {\bibfnamefont {J.~S.}\ \bibnamefont {Kretchmer}}, \bibinfo {author}
  {\bibfnamefont {U.}~\bibnamefont {Schollw\"ock}}, \bibinfo {author}
  {\bibfnamefont {H.}~\bibnamefont {Appel}}, \bibinfo {author} {\bibfnamefont
  {M.~A.}\ \bibnamefont {Sentef}}, \ and\ \bibinfo {author} {\bibfnamefont
  {A.}~\bibnamefont {Rubio}},\ }\href {\doibase 10.1021/acs.jctc.8b01116}
  {\bibfield  {journal} {\bibinfo  {journal} {J. Chem. Theory Comput.}\
  }\textbf {\bibinfo {volume} {15}},\ \bibinfo {pages} {2221} (\bibinfo {year}
  {2019})}\BibitemShut {NoStop}%
\bibitem [{\citenamefont {Weber}\ and\ \citenamefont
  {Freericks}(2021)}]{weber2021field}%
  \BibitemOpen
  \bibfield  {author} {\bibinfo {author} {\bibfnamefont {M.}~\bibnamefont
  {Weber}}\ and\ \bibinfo {author} {\bibfnamefont {J.~K.}\ \bibnamefont
  {Freericks}},\ }\href {http://arxiv.org/abs/2107.04096} {\bibfield  {journal}
  {\bibinfo  {journal} {arXiv:2107.04096 [cond-mat, physics:quant-ph]}\ }
  (\bibinfo {year} {2021})}\BibitemShut {NoStop}%
\bibitem [{\citenamefont {Weber}\ and\ \citenamefont
  {Freericks}(2022)}]{weber2021realtime}%
  \BibitemOpen
  \bibfield  {author} {\bibinfo {author} {\bibfnamefont {M.}~\bibnamefont
  {Weber}}\ and\ \bibinfo {author} {\bibfnamefont {J.~K.}\ \bibnamefont
  {Freericks}},\ }\href {\doibase 10.1103/PhysRevE.105.025301} {\bibfield
  {journal} {\bibinfo  {journal} {Phys. Rev. E}\ }\textbf {\bibinfo {volume}
  {105}},\ \bibinfo {pages} {025301} (\bibinfo {year} {2022})}\BibitemShut
  {NoStop}%
\bibitem [{\citenamefont {Tavernelli}, \citenamefont {R{\"o}hrig},\ and\
  \citenamefont {Rothlisberger}(2005)}]{Tavernelli2005}%
  \BibitemOpen
  \bibfield  {author} {\bibinfo {author} {\bibfnamefont {I.}~\bibnamefont
  {Tavernelli}}, \bibinfo {author} {\bibfnamefont {U.~F.}\ \bibnamefont
  {R{\"o}hrig}}, \ and\ \bibinfo {author} {\bibfnamefont {U.}~\bibnamefont
  {Rothlisberger}},\ }\href {\doibase 10.1080/00268970512331339378} {\bibfield
  {journal} {\bibinfo  {journal} {Mol. Phys.}\ }\textbf {\bibinfo {volume}
  {103}},\ \bibinfo {pages} {963} (\bibinfo {year} {2005})}\BibitemShut
  {NoStop}%
\bibitem [{\citenamefont {Curchod}, \citenamefont {Rothlisberger},\ and\
  \citenamefont {Tavernelli}(2013)}]{Curchod2013}%
  \BibitemOpen
  \bibfield  {author} {\bibinfo {author} {\bibfnamefont {B.~F.~E.}\
  \bibnamefont {Curchod}}, \bibinfo {author} {\bibfnamefont {U.}~\bibnamefont
  {Rothlisberger}}, \ and\ \bibinfo {author} {\bibfnamefont {I.}~\bibnamefont
  {Tavernelli}},\ }\href {\doibase 10.1002/cphc.201200941} {\bibfield
  {journal} {\bibinfo  {journal} {ChemPhysChem}\ }\textbf {\bibinfo {volume}
  {14}},\ \bibinfo {pages} {1314} (\bibinfo {year} {2013})}\BibitemShut
  {NoStop}%
\bibitem [{\citenamefont {Pela}\ and\ \citenamefont {Draxl}(2022)}]{Pela2022}%
  \BibitemOpen
  \bibfield  {author} {\bibinfo {author} {\bibfnamefont {R.~R.}\ \bibnamefont
  {Pela}}\ and\ \bibinfo {author} {\bibfnamefont {C.}~\bibnamefont {Draxl}},\
  }\href {http://arxiv.org/abs/2201.12146} {\bibfield  {journal} {\bibinfo
  {journal} {arXiv:2201.12146 [cond-mat]}\ } (\bibinfo {year}
  {2022})}\BibitemShut {NoStop}%
\bibitem [{\citenamefont {Brockt}\ \emph {et~al.}(2015)\citenamefont {Brockt},
  \citenamefont {Dorfner}, \citenamefont {Vidmar}, \citenamefont
  {Heidrich-Meisner},\ and\ \citenamefont {Jeckelmann}}]{Brockt15}%
  \BibitemOpen
  \bibfield  {author} {\bibinfo {author} {\bibfnamefont {C.}~\bibnamefont
  {Brockt}}, \bibinfo {author} {\bibfnamefont {F.}~\bibnamefont {Dorfner}},
  \bibinfo {author} {\bibfnamefont {L.}~\bibnamefont {Vidmar}}, \bibinfo
  {author} {\bibfnamefont {F.}~\bibnamefont {Heidrich-Meisner}}, \ and\
  \bibinfo {author} {\bibfnamefont {E.}~\bibnamefont {Jeckelmann}},\ }\href
  {\doibase 10.1103/PhysRevB.92.241106} {\bibfield  {journal} {\bibinfo
  {journal} {Phys. Rev. B}\ }\textbf {\bibinfo {volume} {92}},\ \bibinfo
  {pages} {241106} (\bibinfo {year} {2015})}\BibitemShut {NoStop}%
\bibitem [{\citenamefont {Kloss}, \citenamefont {Reichman},\ and\ \citenamefont
  {Tempelaar}(2019)}]{Kloss19}%
  \BibitemOpen
  \bibfield  {author} {\bibinfo {author} {\bibfnamefont {B.}~\bibnamefont
  {Kloss}}, \bibinfo {author} {\bibfnamefont {D.~R.}\ \bibnamefont {Reichman}},
  \ and\ \bibinfo {author} {\bibfnamefont {R.}~\bibnamefont {Tempelaar}},\
  }\href {\doibase 10.1103/PhysRevLett.123.126601} {\bibfield  {journal}
  {\bibinfo  {journal} {Phys. Rev. Lett.}\ }\textbf {\bibinfo {volume} {123}},\
  \bibinfo {pages} {126601} (\bibinfo {year} {2019})}\BibitemShut {NoStop}%
\bibitem [{\citenamefont {Stolpp}\ \emph {et~al.}(2020)\citenamefont {Stolpp},
  \citenamefont {Herbrych}, \citenamefont {Dorfner}, \citenamefont {Dagotto},\
  and\ \citenamefont {Heidrich-Meisner}}]{Stolpp2020}%
  \BibitemOpen
  \bibfield  {author} {\bibinfo {author} {\bibfnamefont {J.}~\bibnamefont
  {Stolpp}}, \bibinfo {author} {\bibfnamefont {J.}~\bibnamefont {Herbrych}},
  \bibinfo {author} {\bibfnamefont {F.}~\bibnamefont {Dorfner}}, \bibinfo
  {author} {\bibfnamefont {E.}~\bibnamefont {Dagotto}}, \ and\ \bibinfo
  {author} {\bibfnamefont {F.}~\bibnamefont {Heidrich-Meisner}},\ }\href
  {\doibase 10.1103/PhysRevB.101.035134} {\bibfield  {journal} {\bibinfo
  {journal} {Phys. Rev. B}\ }\textbf {\bibinfo {volume} {101}},\ \bibinfo
  {pages} {035134} (\bibinfo {year} {2020})}\BibitemShut {NoStop}%
\bibitem [{\citenamefont {White}(1992)}]{White92}%
  \BibitemOpen
  \bibfield  {author} {\bibinfo {author} {\bibfnamefont {S.~R.}\ \bibnamefont
  {White}},\ }\href {\doibase 10.1103/PhysRevLett.69.2863} {\bibfield
  {journal} {\bibinfo  {journal} {Phys. Rev. Lett.}\ }\textbf {\bibinfo
  {volume} {69}},\ \bibinfo {pages} {2863} (\bibinfo {year}
  {1992})}\BibitemShut {NoStop}%
\bibitem [{\citenamefont {Schollw\"ock}(2005)}]{Schollwock2005}%
  \BibitemOpen
  \bibfield  {author} {\bibinfo {author} {\bibfnamefont {U.}~\bibnamefont
  {Schollw\"ock}},\ }\href {\doibase 10.1103/RevModPhys.77.259} {\bibfield
  {journal} {\bibinfo  {journal} {Rev. Mod. Phys.}\ }\textbf {\bibinfo {volume}
  {77}},\ \bibinfo {pages} {259} (\bibinfo {year} {2005})}\BibitemShut
  {NoStop}%
\bibitem [{\citenamefont {Schollw\"ock}(2011)}]{Schollwock2011}%
  \BibitemOpen
  \bibfield  {author} {\bibinfo {author} {\bibfnamefont {U.}~\bibnamefont
  {Schollw\"ock}},\ }\href {\doibase 10.1016/j.aop.2010.09.012} {\bibfield
  {journal} {\bibinfo  {journal} {Ann. Phys.}\ }\textbf {\bibinfo {volume}
  {326}},\ \bibinfo {pages} {96 } (\bibinfo {year} {2011})}\BibitemShut
  {NoStop}%
\bibitem [{\citenamefont {Wall}, \citenamefont {Safavi-Naini},\ and\
  \citenamefont {Rey}(2016)}]{wall_16}%
  \BibitemOpen
  \bibfield  {author} {\bibinfo {author} {\bibfnamefont {M.~L.}\ \bibnamefont
  {Wall}}, \bibinfo {author} {\bibfnamefont {A.}~\bibnamefont {Safavi-Naini}},
  \ and\ \bibinfo {author} {\bibfnamefont {A.~M.}\ \bibnamefont {Rey}},\ }\href
  {\doibase 10.1103/PhysRevA.94.053637} {\bibfield  {journal} {\bibinfo
  {journal} {Phys. Rev. A}\ }\textbf {\bibinfo {volume} {94}},\ \bibinfo
  {pages} {053637} (\bibinfo {year} {2016})}\BibitemShut {NoStop}%
\bibitem [{\citenamefont {K\"ohler}, \citenamefont {Stolpp},\ and\
  \citenamefont {Paeckel}(2021)}]{Kohler21}%
  \BibitemOpen
  \bibfield  {author} {\bibinfo {author} {\bibfnamefont {T.}~\bibnamefont
  {K\"ohler}}, \bibinfo {author} {\bibfnamefont {J.}~\bibnamefont {Stolpp}}, \
  and\ \bibinfo {author} {\bibfnamefont {S.}~\bibnamefont {Paeckel}},\ }\href
  {\doibase 10.21468/SciPostPhys.10.3.058} {\bibfield  {journal} {\bibinfo
  {journal} {SciPost Phys.}\ }\textbf {\bibinfo {volume} {10}},\ \bibinfo
  {pages} {058} (\bibinfo {year} {2021})}\BibitemShut {NoStop}%
\bibitem [{\citenamefont {Stolpp}\ \emph {et~al.}(2021)\citenamefont {Stolpp},
  \citenamefont {K\"ohler}, \citenamefont {Manmana}, \citenamefont
  {Jeckelmann}, \citenamefont {Heidrich-Meisner},\ and\ \citenamefont
  {Paeckel}}]{Stolppkoehler2020}%
  \BibitemOpen
  \bibfield  {author} {\bibinfo {author} {\bibfnamefont {J.}~\bibnamefont
  {Stolpp}}, \bibinfo {author} {\bibfnamefont {T.}~\bibnamefont {K\"ohler}},
  \bibinfo {author} {\bibfnamefont {S.~R.}\ \bibnamefont {Manmana}}, \bibinfo
  {author} {\bibfnamefont {E.}~\bibnamefont {Jeckelmann}}, \bibinfo {author}
  {\bibfnamefont {F.}~\bibnamefont {Heidrich-Meisner}}, \ and\ \bibinfo
  {author} {\bibfnamefont {S.}~\bibnamefont {Paeckel}},\ }\href {\doibase
  10.1016/j.cpc.2021.108106} {\bibfield  {journal} {\bibinfo  {journal}
  {Comput. Phys. Commun.}\ }\textbf {\bibinfo {volume} {269}},\ \bibinfo
  {pages} {108106} (\bibinfo {year} {2021})}\BibitemShut {NoStop}%
\bibitem [{\citenamefont {Jeckelmann}\ and\ \citenamefont
  {White}(1998)}]{Jeckelmann1998}%
  \BibitemOpen
  \bibfield  {author} {\bibinfo {author} {\bibfnamefont {E.}~\bibnamefont
  {Jeckelmann}}\ and\ \bibinfo {author} {\bibfnamefont {S.~R.}\ \bibnamefont
  {White}},\ }\href {\doibase 10.1103/PhysRevB.57.6376} {\bibfield  {journal}
  {\bibinfo  {journal} {Phys. Rev. B}\ }\textbf {\bibinfo {volume} {57}},\
  \bibinfo {pages} {6376} (\bibinfo {year} {1998})}\BibitemShut {NoStop}%
\bibitem [{\citenamefont {Zhang}, \citenamefont {Jeckelmann},\ and\
  \citenamefont {White}(1998)}]{Zhang1998}%
  \BibitemOpen
  \bibfield  {author} {\bibinfo {author} {\bibfnamefont {C.}~\bibnamefont
  {Zhang}}, \bibinfo {author} {\bibfnamefont {E.}~\bibnamefont {Jeckelmann}}, \
  and\ \bibinfo {author} {\bibfnamefont {S.~R.}\ \bibnamefont {White}},\ }\href
  {\doibase 10.1103/PhysRevLett.80.2661} {\bibfield  {journal} {\bibinfo
  {journal} {Phys. Rev. Lett.}\ }\textbf {\bibinfo {volume} {80}},\ \bibinfo
  {pages} {2661} (\bibinfo {year} {1998})}\BibitemShut {NoStop}%
\bibitem [{\citenamefont {Zhang}, \citenamefont {Jeckelmann},\ and\
  \citenamefont {White}(1999)}]{zhang99}%
  \BibitemOpen
  \bibfield  {author} {\bibinfo {author} {\bibfnamefont {C.}~\bibnamefont
  {Zhang}}, \bibinfo {author} {\bibfnamefont {E.}~\bibnamefont {Jeckelmann}}, \
  and\ \bibinfo {author} {\bibfnamefont {S.~R.}\ \bibnamefont {White}},\ }\href
  {\doibase 10.1103/PhysRevB.60.14092} {\bibfield  {journal} {\bibinfo
  {journal} {Phys. Rev. B}\ }\textbf {\bibinfo {volume} {60}},\ \bibinfo
  {pages} {14092} (\bibinfo {year} {1999})}\BibitemShut {NoStop}%
\bibitem [{\citenamefont {Guo}\ \emph {et~al.}(2012)\citenamefont {Guo},
  \citenamefont {Weichselbaum}, \citenamefont {von Delft},\ and\ \citenamefont
  {Vojta}}]{Guo2012}%
  \BibitemOpen
  \bibfield  {author} {\bibinfo {author} {\bibfnamefont {C.}~\bibnamefont
  {Guo}}, \bibinfo {author} {\bibfnamefont {A.}~\bibnamefont {Weichselbaum}},
  \bibinfo {author} {\bibfnamefont {J.}~\bibnamefont {von Delft}}, \ and\
  \bibinfo {author} {\bibfnamefont {M.}~\bibnamefont {Vojta}},\ }\href
  {\doibase 10.1103/PhysRevLett.108.160401} {\bibfield  {journal} {\bibinfo
  {journal} {Phys. Rev. Lett.}\ }\textbf {\bibinfo {volume} {108}},\ \bibinfo
  {pages} {160401} (\bibinfo {year} {2012})}\BibitemShut {NoStop}%
\bibitem [{\citenamefont {Friedman}(2000)}]{Friedman2000}%
  \BibitemOpen
  \bibfield  {author} {\bibinfo {author} {\bibfnamefont {B.}~\bibnamefont
  {Friedman}},\ }\href {\doibase 10.1103/PhysRevB.61.6701} {\bibfield
  {journal} {\bibinfo  {journal} {Phys. Rev. B}\ }\textbf {\bibinfo {volume}
  {61}},\ \bibinfo {pages} {6701} (\bibinfo {year} {2000})}\BibitemShut
  {NoStop}%
\bibitem [{\citenamefont {Wong}\ and\ \citenamefont {Chen}(2008)}]{Wong08}%
  \BibitemOpen
  \bibfield  {author} {\bibinfo {author} {\bibfnamefont {H.}~\bibnamefont
  {Wong}}\ and\ \bibinfo {author} {\bibfnamefont {Z.-D.}\ \bibnamefont
  {Chen}},\ }\href {\doibase 10.1103/PhysRevB.77.174305} {\bibfield  {journal}
  {\bibinfo  {journal} {Phys. Rev. B}\ }\textbf {\bibinfo {volume} {77}},\
  \bibinfo {pages} {174305} (\bibinfo {year} {2008})}\BibitemShut {NoStop}%
\bibitem [{\citenamefont {Jansen}, \citenamefont {Bon\v{c}a},\ and\
  \citenamefont {Heidrich-Meisner}(2020)}]{Jansen2020}%
  \BibitemOpen
  \bibfield  {author} {\bibinfo {author} {\bibfnamefont {D.}~\bibnamefont
  {Jansen}}, \bibinfo {author} {\bibfnamefont {J.}~\bibnamefont {Bon\v{c}a}}, \
  and\ \bibinfo {author} {\bibfnamefont {F.}~\bibnamefont {Heidrich-Meisner}},\
  }\href {\doibase 10.1103/PhysRevB.102.165155} {\bibfield  {journal} {\bibinfo
   {journal} {Phys. Rev. B}\ }\textbf {\bibinfo {volume} {102}},\ \bibinfo
  {pages} {165155} (\bibinfo {year} {2020})}\BibitemShut {NoStop}%
\bibitem [{\citenamefont {Jansen}, \citenamefont {Jooss},\ and\ \citenamefont
  {Heidrich-Meisner}(2021)}]{Jansen_Jooss_2021}%
  \BibitemOpen
  \bibfield  {author} {\bibinfo {author} {\bibfnamefont {D.}~\bibnamefont
  {Jansen}}, \bibinfo {author} {\bibfnamefont {C.}~\bibnamefont {Jooss}}, \
  and\ \bibinfo {author} {\bibfnamefont {F.}~\bibnamefont {Heidrich-Meisner}},\
  }\href {\doibase 10.1103/PhysRevB.104.195116} {\bibfield  {journal} {\bibinfo
   {journal} {Phys. Rev. B}\ }\textbf {\bibinfo {volume} {104}},\ \bibinfo
  {pages} {195116} (\bibinfo {year} {2021})}\BibitemShut {NoStop}%
\bibitem [{\citenamefont {Horsfield}\ \emph {et~al.}(2006)\citenamefont
  {Horsfield}, \citenamefont {Bowler}, \citenamefont {Ness}, \citenamefont
  {S{\'{a}}nchez}, \citenamefont {Todorov},\ and\ \citenamefont
  {Fisher}}]{Horsfield_2006}%
  \BibitemOpen
  \bibfield  {author} {\bibinfo {author} {\bibfnamefont {A.~P.}\ \bibnamefont
  {Horsfield}}, \bibinfo {author} {\bibfnamefont {D.~R.}\ \bibnamefont
  {Bowler}}, \bibinfo {author} {\bibfnamefont {H.}~\bibnamefont {Ness}},
  \bibinfo {author} {\bibfnamefont {C.~G.}\ \bibnamefont {S{\'{a}}nchez}},
  \bibinfo {author} {\bibfnamefont {T.~N.}\ \bibnamefont {Todorov}}, \ and\
  \bibinfo {author} {\bibfnamefont {A.~J.}\ \bibnamefont {Fisher}},\ }\href
  {\doibase 10.1088/0034-4885/69/4/r05} {\bibfield  {journal} {\bibinfo
  {journal} {Rep. Prog. Phys.}\ }\textbf {\bibinfo {volume} {69}},\ \bibinfo
  {pages} {1195} (\bibinfo {year} {2006})}\BibitemShut {NoStop}%
\bibitem [{\citenamefont {Wang}, \citenamefont {Akimov},\ and\ \citenamefont
  {Prezhdo}(2016)}]{Wang2016}%
  \BibitemOpen
  \bibfield  {author} {\bibinfo {author} {\bibfnamefont {L.}~\bibnamefont
  {Wang}}, \bibinfo {author} {\bibfnamefont {A.}~\bibnamefont {Akimov}}, \ and\
  \bibinfo {author} {\bibfnamefont {O.~V.}\ \bibnamefont {Prezhdo}},\ }\href
  {\doibase 10.1021/acs.jpclett.6b00710} {\bibfield  {journal} {\bibinfo
  {journal} {J. Phys. Chem. Lett.}\ }\textbf {\bibinfo {volume} {7}},\ \bibinfo
  {pages} {2100} (\bibinfo {year} {2016})}\BibitemShut {NoStop}%
\bibitem [{\citenamefont {Smith}\ and\ \citenamefont
  {Akimov}(2019)}]{Smith_2019}%
  \BibitemOpen
  \bibfield  {author} {\bibinfo {author} {\bibfnamefont {B.}~\bibnamefont
  {Smith}}\ and\ \bibinfo {author} {\bibfnamefont {A.~V.}\ \bibnamefont
  {Akimov}},\ }\href {\doibase 10.1088/1361-648x/ab5246} {\bibfield  {journal}
  {\bibinfo  {journal} {J. Phys.: Condens. Matter}\ }\textbf {\bibinfo {volume}
  {32}},\ \bibinfo {pages} {073001} (\bibinfo {year} {2019})}\BibitemShut
  {NoStop}%
\bibitem [{\citenamefont {Mardazad}\ \emph {et~al.}(2021)\citenamefont
  {Mardazad}, \citenamefont {Xu}, \citenamefont {Yang}, \citenamefont
  {Grundner}, \citenamefont {Schollw\"ock}, \citenamefont {Ma},\ and\
  \citenamefont {Paeckel}}]{Mardazad21}%
  \BibitemOpen
  \bibfield  {author} {\bibinfo {author} {\bibfnamefont {S.}~\bibnamefont
  {Mardazad}}, \bibinfo {author} {\bibfnamefont {Y.}~\bibnamefont {Xu}},
  \bibinfo {author} {\bibfnamefont {X.}~\bibnamefont {Yang}}, \bibinfo {author}
  {\bibfnamefont {M.}~\bibnamefont {Grundner}}, \bibinfo {author}
  {\bibfnamefont {U.}~\bibnamefont {Schollw\"ock}}, \bibinfo {author}
  {\bibfnamefont {H.}~\bibnamefont {Ma}}, \ and\ \bibinfo {author}
  {\bibfnamefont {S.}~\bibnamefont {Paeckel}},\ }\href {\doibase
  10.1063/5.0068292} {\bibfield  {journal} {\bibinfo  {journal} {J. Chem.
  Phys.}\ }\textbf {\bibinfo {volume} {155}},\ \bibinfo {pages} {194101}
  (\bibinfo {year} {2021})}\BibitemShut {NoStop}%
\bibitem [{\citenamefont {Stock}\ and\ \citenamefont
  {Thoss}(2005)}]{Stock2005}%
  \BibitemOpen
  \bibfield  {author} {\bibinfo {author} {\bibfnamefont {G.}~\bibnamefont
  {Stock}}\ and\ \bibinfo {author} {\bibfnamefont {M.}~\bibnamefont {Thoss}},\
  }\href {\doibase 10.1002/0471739464.ch5} {\bibfield  {journal} {\bibinfo
  {journal} {Adv. {Chem.} {Phys.}}\ }\textbf {\bibinfo {volume} {131}},\
  \bibinfo {pages} {243} (\bibinfo {year} {2005})}\BibitemShut {NoStop}%
\bibitem [{\citenamefont {Freixas}\ \emph {et~al.}(2021)\citenamefont
  {Freixas}, \citenamefont {White}, \citenamefont {Nelson}, \citenamefont
  {Song}, \citenamefont {Makhov}, \citenamefont {Shalashilin}, \citenamefont
  {Fernandez-Alberti},\ and\ \citenamefont {Tretiak}}]{Freixas2021}%
  \BibitemOpen
  \bibfield  {author} {\bibinfo {author} {\bibfnamefont {V.~M.}\ \bibnamefont
  {Freixas}}, \bibinfo {author} {\bibfnamefont {A.~J.}\ \bibnamefont {White}},
  \bibinfo {author} {\bibfnamefont {T.}~\bibnamefont {Nelson}}, \bibinfo
  {author} {\bibfnamefont {H.}~\bibnamefont {Song}}, \bibinfo {author}
  {\bibfnamefont {D.~V.}\ \bibnamefont {Makhov}}, \bibinfo {author}
  {\bibfnamefont {D.}~\bibnamefont {Shalashilin}}, \bibinfo {author}
  {\bibfnamefont {S.}~\bibnamefont {Fernandez-Alberti}}, \ and\ \bibinfo
  {author} {\bibfnamefont {S.}~\bibnamefont {Tretiak}},\ }\href {\doibase
  10.1021/acs.jpclett.1c00266} {\bibfield  {journal} {\bibinfo  {journal} {J.
  Phys. Chem. Lett.}\ }\textbf {\bibinfo {volume} {12}},\ \bibinfo {pages}
  {2970} (\bibinfo {year} {2021})}\BibitemShut {NoStop}%
\bibitem [{\citenamefont {Wang}\ \emph {et~al.}(2020)\citenamefont {Wang},
  \citenamefont {Qiu}, \citenamefont {Bai},\ and\ \citenamefont
  {Xu}}]{Wang2020}%
  \BibitemOpen
  \bibfield  {author} {\bibinfo {author} {\bibfnamefont {L.}~\bibnamefont
  {Wang}}, \bibinfo {author} {\bibfnamefont {J.}~\bibnamefont {Qiu}}, \bibinfo
  {author} {\bibfnamefont {X.}~\bibnamefont {Bai}}, \ and\ \bibinfo {author}
  {\bibfnamefont {J.}~\bibnamefont {Xu}},\ }\href {\doibase 10.1002/wcms.1435}
  {\bibfield  {journal} {\bibinfo  {journal} {Wiley Interdiscip. Rev.: Comput.
  Mol. Sci.}\ }\textbf {\bibinfo {volume} {10}},\ \bibinfo {pages} {e1435}
  (\bibinfo {year} {2020})}\BibitemShut {NoStop}%
\bibitem [{\citenamefont {Krotz}\ and\ \citenamefont
  {Tempelaar}(2022)}]{Krotz2022}%
  \BibitemOpen
  \bibfield  {author} {\bibinfo {author} {\bibfnamefont {A.}~\bibnamefont
  {Krotz}}\ and\ \bibinfo {author} {\bibfnamefont {R.}~\bibnamefont
  {Tempelaar}},\ }\href {\doibase 10.1063/5.0076070} {\bibfield  {journal}
  {\bibinfo  {journal} {J. Chem. Phys.}\ }\textbf {\bibinfo {volume} {156}},\
  \bibinfo {pages} {024105} (\bibinfo {year} {2022})}\BibitemShut {NoStop}%
\bibitem [{\citenamefont {Holstein}(1959)}]{Holstein1959a}%
  \BibitemOpen
  \bibfield  {author} {\bibinfo {author} {\bibfnamefont {T.}~\bibnamefont
  {Holstein}},\ }\href {\doibase 10.1016/0003-4916(59)90002-8} {\bibfield
  {journal} {\bibinfo  {journal} {Ann. Phys.}\ }\textbf {\bibinfo {volume}
  {8}},\ \bibinfo {pages} {325 } (\bibinfo {year} {1959})}\BibitemShut
  {NoStop}%
\bibitem [{\citenamefont {Shalashilin}(2009)}]{Shalashilin2009}%
  \BibitemOpen
  \bibfield  {author} {\bibinfo {author} {\bibfnamefont {D.~V.}\ \bibnamefont
  {Shalashilin}},\ }\href {\doibase 10.1063/1.3153302} {\bibfield  {journal}
  {\bibinfo  {journal} {J. Chem. Phys.}\ }\textbf {\bibinfo {volume} {130}},\
  \bibinfo {pages} {244101} (\bibinfo {year} {2009})}\BibitemShut {NoStop}%
\bibitem [{\citenamefont {Shalashilin}(2010)}]{Shalashilin2010}%
  \BibitemOpen
  \bibfield  {author} {\bibinfo {author} {\bibfnamefont {D.~V.}\ \bibnamefont
  {Shalashilin}},\ }\href {\doibase 10.1063/1.3442747} {\bibfield  {journal}
  {\bibinfo  {journal} {J. Chem. Phys.}\ }\textbf {\bibinfo {volume} {132}},\
  \bibinfo {pages} {244111} (\bibinfo {year} {2010})}\BibitemShut {NoStop}%
\bibitem [{\citenamefont {Tanimura}\ and\ \citenamefont
  {Kubo}(1989)}]{Tanimura1989}%
  \BibitemOpen
  \bibfield  {author} {\bibinfo {author} {\bibfnamefont {Y.}~\bibnamefont
  {Tanimura}}\ and\ \bibinfo {author} {\bibfnamefont {R.}~\bibnamefont
  {Kubo}},\ }\href {\doibase 10.1143/JPSJ.58.101} {\bibfield  {journal}
  {\bibinfo  {journal} {J. Phys. Soc. Jpn.}\ }\textbf {\bibinfo {volume}
  {58}},\ \bibinfo {pages} {101} (\bibinfo {year} {1989})}\BibitemShut
  {NoStop}%
\bibitem [{\citenamefont {Chen}, \citenamefont {Zhao},\ and\ \citenamefont
  {Tanimura}(2015)}]{Chen15}%
  \BibitemOpen
  \bibfield  {author} {\bibinfo {author} {\bibfnamefont {L.}~\bibnamefont
  {Chen}}, \bibinfo {author} {\bibfnamefont {Y.}~\bibnamefont {Zhao}}, \ and\
  \bibinfo {author} {\bibfnamefont {Y.}~\bibnamefont {Tanimura}},\ }\href
  {\doibase 10.1021/acs.jpclett.5b01368} {\bibfield  {journal} {\bibinfo
  {journal} {J. Phys. Chem. Lett.}\ }\textbf {\bibinfo {volume} {6}},\ \bibinfo
  {pages} {3110} (\bibinfo {year} {2015})}\BibitemShut {NoStop}%
\bibitem [{\citenamefont {Zhou}\ \emph {et~al.}(2015)\citenamefont {Zhou},
  \citenamefont {Huang}, \citenamefont {Zhu}, \citenamefont {Chernyak},\ and\
  \citenamefont {Zhao}}]{Zhou15}%
  \BibitemOpen
  \bibfield  {author} {\bibinfo {author} {\bibfnamefont {N.}~\bibnamefont
  {Zhou}}, \bibinfo {author} {\bibfnamefont {Z.}~\bibnamefont {Huang}},
  \bibinfo {author} {\bibfnamefont {J.}~\bibnamefont {Zhu}}, \bibinfo {author}
  {\bibfnamefont {V.}~\bibnamefont {Chernyak}}, \ and\ \bibinfo {author}
  {\bibfnamefont {Y.}~\bibnamefont {Zhao}},\ }\href {\doibase
  10.1063/1.4923009} {\bibfield  {journal} {\bibinfo  {journal} {J. Chem.
  Phys.}\ }\textbf {\bibinfo {volume} {143}},\ \bibinfo {pages} {014113}
  (\bibinfo {year} {2015})}\BibitemShut {NoStop}%
\bibitem [{\citenamefont {Chen}, \citenamefont {Gelin},\ and\ \citenamefont
  {Shalashilin}(2019)}]{Chen2019}%
  \BibitemOpen
  \bibfield  {author} {\bibinfo {author} {\bibfnamefont {L.}~\bibnamefont
  {Chen}}, \bibinfo {author} {\bibfnamefont {M.~F.}\ \bibnamefont {Gelin}}, \
  and\ \bibinfo {author} {\bibfnamefont {D.~V.}\ \bibnamefont {Shalashilin}},\
  }\href {\doibase 10.1063/1.5132341} {\bibfield  {journal} {\bibinfo
  {journal} {J. Chem. Phys.}\ }\textbf {\bibinfo {volume} {151}},\ \bibinfo
  {pages} {244116} (\bibinfo {year} {2019})}\BibitemShut {NoStop}%
\bibitem [{\citenamefont {Pekar}(1954)}]{Pekar1954}%
  \BibitemOpen
  \bibfield  {author} {\bibinfo {author} {\bibfnamefont {S.~I.}\ \bibnamefont
  {Pekar}},\ }\href@noop {} {\emph {\bibinfo {title} {Untersuchungen \"uber die
  {Elektronentheorie} der {Kristalle}.}}}\ (\bibinfo  {publisher}
  {Akademie-Verlag},\ \bibinfo {address} {Berlin},\ \bibinfo {year}
  {1954})\BibitemShut {NoStop}%
\bibitem [{\citenamefont {Feynman}(1955)}]{Feynman1955}%
  \BibitemOpen
  \bibfield  {author} {\bibinfo {author} {\bibfnamefont {R.~P.}\ \bibnamefont
  {Feynman}},\ }\href {\doibase 10.1103/PhysRev.97.660} {\bibfield  {journal}
  {\bibinfo  {journal} {Phys. Rev.}\ }\textbf {\bibinfo {volume} {97}},\
  \bibinfo {pages} {660} (\bibinfo {year} {1955})}\BibitemShut {NoStop}%
\bibitem [{\citenamefont {Peierls}(1955)}]{peirls_55}%
  \BibitemOpen
  \bibfield  {author} {\bibinfo {author} {\bibfnamefont {R.}~\bibnamefont
  {Peierls}},\ }\href@noop {} {\emph {\bibinfo {title} {Quantum Theory of
  Solids}}},\ \bibinfo {number} {pp. 108}\ (\bibinfo  {publisher} {Clarendon
  Press},\ \bibinfo {address} {Oxford},\ \bibinfo {year} {1955})\BibitemShut
  {NoStop}%
\bibitem [{\citenamefont {Hirsch}\ and\ \citenamefont
  {Fradkin}(1983)}]{Hirsch83}%
  \BibitemOpen
  \bibfield  {author} {\bibinfo {author} {\bibfnamefont {J.~E.}\ \bibnamefont
  {Hirsch}}\ and\ \bibinfo {author} {\bibfnamefont {E.}~\bibnamefont
  {Fradkin}},\ }\href {\doibase 10.1103/PhysRevB.27.4302} {\bibfield  {journal}
  {\bibinfo  {journal} {Phys. Rev. B}\ }\textbf {\bibinfo {volume} {27}},\
  \bibinfo {pages} {4302} (\bibinfo {year} {1983})}\BibitemShut {NoStop}%
\bibitem [{\citenamefont {Scalettar}, \citenamefont {Bickers},\ and\
  \citenamefont {Scalapino}(1989)}]{Scalletar89}%
  \BibitemOpen
  \bibfield  {author} {\bibinfo {author} {\bibfnamefont {R.~T.}\ \bibnamefont
  {Scalettar}}, \bibinfo {author} {\bibfnamefont {N.~E.}\ \bibnamefont
  {Bickers}}, \ and\ \bibinfo {author} {\bibfnamefont {D.~J.}\ \bibnamefont
  {Scalapino}},\ }\href {\doibase 10.1103/PhysRevB.40.197} {\bibfield
  {journal} {\bibinfo  {journal} {Phys. Rev. B}\ }\textbf {\bibinfo {volume}
  {40}},\ \bibinfo {pages} {197} (\bibinfo {year} {1989})}\BibitemShut
  {NoStop}%
\bibitem [{\citenamefont {Noack}, \citenamefont {Scalapino},\ and\
  \citenamefont {Scalettar}(1991)}]{Noack91}%
  \BibitemOpen
  \bibfield  {author} {\bibinfo {author} {\bibfnamefont {R.~M.}\ \bibnamefont
  {Noack}}, \bibinfo {author} {\bibfnamefont {D.~J.}\ \bibnamefont
  {Scalapino}}, \ and\ \bibinfo {author} {\bibfnamefont {R.~T.}\ \bibnamefont
  {Scalettar}},\ }\href {\doibase 10.1103/PhysRevLett.66.778} {\bibfield
  {journal} {\bibinfo  {journal} {Phys. Rev. Lett.}\ }\textbf {\bibinfo
  {volume} {66}},\ \bibinfo {pages} {778} (\bibinfo {year} {1991})}\BibitemShut
  {NoStop}%
\bibitem [{\citenamefont {Veki\ifmmode~\acute{c}\else \'{c}\fi{}},
  \citenamefont {Noack},\ and\ \citenamefont {White}(1992)}]{Noack92}%
  \BibitemOpen
  \bibfield  {author} {\bibinfo {author} {\bibfnamefont {M.}~\bibnamefont
  {Veki\ifmmode~\acute{c}\else \'{c}\fi{}}}, \bibinfo {author} {\bibfnamefont
  {R.~M.}\ \bibnamefont {Noack}}, \ and\ \bibinfo {author} {\bibfnamefont
  {S.~R.}\ \bibnamefont {White}},\ }\href {\doibase 10.1103/PhysRevB.46.271}
  {\bibfield  {journal} {\bibinfo  {journal} {Phys. Rev. B}\ }\textbf {\bibinfo
  {volume} {46}},\ \bibinfo {pages} {271} (\bibinfo {year} {1992})}\BibitemShut
  {NoStop}%
\bibitem [{\citenamefont {Bursill}, \citenamefont {McKenzie},\ and\
  \citenamefont {Hamer}(1998)}]{bursill_98}%
  \BibitemOpen
  \bibfield  {author} {\bibinfo {author} {\bibfnamefont {R.~J.}\ \bibnamefont
  {Bursill}}, \bibinfo {author} {\bibfnamefont {R.~H.}\ \bibnamefont
  {McKenzie}}, \ and\ \bibinfo {author} {\bibfnamefont {C.~J.}\ \bibnamefont
  {Hamer}},\ }\href {\doibase 10.1103/PhysRevLett.80.5607} {\bibfield
  {journal} {\bibinfo  {journal} {Phys. Rev. Lett.}\ }\textbf {\bibinfo
  {volume} {80}},\ \bibinfo {pages} {5607} (\bibinfo {year}
  {1998})}\BibitemShut {NoStop}%
\bibitem [{\citenamefont {Creffield}, \citenamefont {Sangiovanni},\ and\
  \citenamefont {Capone}(2005)}]{creffield_05}%
  \BibitemOpen
  \bibfield  {author} {\bibinfo {author} {\bibfnamefont {C.~E.}\ \bibnamefont
  {Creffield}}, \bibinfo {author} {\bibfnamefont {G.}~\bibnamefont
  {Sangiovanni}}, \ and\ \bibinfo {author} {\bibfnamefont {M.}~\bibnamefont
  {Capone}},\ }\href {\doibase 10.1140/epjb/e2005-00112-9} {\bibfield
  {journal} {\bibinfo  {journal} {Eur. Phys. J. B}\ }\textbf {\bibinfo {volume}
  {44}},\ \bibinfo {pages} {175} (\bibinfo {year} {2005})}\BibitemShut
  {NoStop}%
\bibitem [{\citenamefont {Bradley}, \citenamefont {Batrouni},\ and\
  \citenamefont {Scalettar}(2021)}]{Bradley21}%
  \BibitemOpen
  \bibfield  {author} {\bibinfo {author} {\bibfnamefont {O.}~\bibnamefont
  {Bradley}}, \bibinfo {author} {\bibfnamefont {G.~G.}\ \bibnamefont
  {Batrouni}}, \ and\ \bibinfo {author} {\bibfnamefont {R.~T.}\ \bibnamefont
  {Scalettar}},\ }\href {\doibase 10.1103/PhysRevB.103.235104} {\bibfield
  {journal} {\bibinfo  {journal} {Phys. Rev. B}\ }\textbf {\bibinfo {volume}
  {103}},\ \bibinfo {pages} {235104} (\bibinfo {year} {2021})}\BibitemShut
  {NoStop}%
\bibitem [{\citenamefont {Ara{\'u}jo}\ \emph {et~al.}(2022)\citenamefont
  {Ara{\'u}jo}, \citenamefont {de~Lima}, \citenamefont {Sorella},\ and\
  \citenamefont {Costa}}]{Araujo21}%
  \BibitemOpen
  \bibfield  {author} {\bibinfo {author} {\bibfnamefont {M.~V.}\ \bibnamefont
  {Ara{\'u}jo}}, \bibinfo {author} {\bibfnamefont {J.~P.}\ \bibnamefont
  {de~Lima}}, \bibinfo {author} {\bibfnamefont {S.}~\bibnamefont {Sorella}}, \
  and\ \bibinfo {author} {\bibfnamefont {N.~C.}\ \bibnamefont {Costa}},\ }\href
  {\doibase 10.1103/PhysRevB.105.165103} {\bibfield  {journal} {\bibinfo
  {journal} {Phys. Rev. B}\ }\textbf {\bibinfo {volume} {105}},\ \bibinfo
  {pages} {165103} (\bibinfo {year} {2022})}\BibitemShut {NoStop}%
\bibitem [{\citenamefont {Ranninger}\ and\ \citenamefont
  {Thibblin}(1992)}]{Ranninger92}%
  \BibitemOpen
  \bibfield  {author} {\bibinfo {author} {\bibfnamefont {J.}~\bibnamefont
  {Ranninger}}\ and\ \bibinfo {author} {\bibfnamefont {U.}~\bibnamefont
  {Thibblin}},\ }\href {\doibase 10.1103/PhysRevB.45.7730} {\bibfield
  {journal} {\bibinfo  {journal} {Phys. Rev. B}\ }\textbf {\bibinfo {volume}
  {45}},\ \bibinfo {pages} {7730} (\bibinfo {year} {1992})}\BibitemShut
  {NoStop}%
\bibitem [{\citenamefont {de~Mello}\ and\ \citenamefont
  {Ranninger}(1997)}]{deMello97}%
  \BibitemOpen
  \bibfield  {author} {\bibinfo {author} {\bibfnamefont {E.~V.~L.}\
  \bibnamefont {de~Mello}}\ and\ \bibinfo {author} {\bibfnamefont
  {J.}~\bibnamefont {Ranninger}},\ }\href {\doibase 10.1103/PhysRevB.55.14872}
  {\bibfield  {journal} {\bibinfo  {journal} {Phys. Rev. B}\ }\textbf {\bibinfo
  {volume} {55}},\ \bibinfo {pages} {14872} (\bibinfo {year}
  {1997})}\BibitemShut {NoStop}%
\bibitem [{\citenamefont {Firsov}\ and\ \citenamefont
  {Kudinov}(1997)}]{Firsov97}%
  \BibitemOpen
  \bibfield  {author} {\bibinfo {author} {\bibfnamefont {Y.~A.}\ \bibnamefont
  {Firsov}}\ and\ \bibinfo {author} {\bibfnamefont {E.~K.}\ \bibnamefont
  {Kudinov}},\ }\href {\doibase 10.1134/1.1130203} {\bibfield  {journal}
  {\bibinfo  {journal} {Phys. Solid State}\ }\textbf {\bibinfo {volume} {39}},\
  \bibinfo {pages} {1930} (\bibinfo {year} {1997})}\BibitemShut {NoStop}%
\bibitem [{\citenamefont {Chatterjee}\ and\ \citenamefont
  {Das}(2000)}]{Chatterjee00}%
  \BibitemOpen
  \bibfield  {author} {\bibinfo {author} {\bibfnamefont {J.}~\bibnamefont
  {Chatterjee}}\ and\ \bibinfo {author} {\bibfnamefont {A.~N.}\ \bibnamefont
  {Das}},\ }\href {\doibase 10.1103/PhysRevB.61.4592} {\bibfield  {journal}
  {\bibinfo  {journal} {Phys. Rev. B}\ }\textbf {\bibinfo {volume} {61}},\
  \bibinfo {pages} {4592} (\bibinfo {year} {2000})}\BibitemShut {NoStop}%
\bibitem [{\citenamefont {Hakio\v{g}lu}, \citenamefont {Ivanov},\ and\
  \citenamefont {Zhuravlev}(2000)}]{Hakioglu00}%
  \BibitemOpen
  \bibfield  {author} {\bibinfo {author} {\bibfnamefont {T.}~\bibnamefont
  {Hakio\v{g}lu}}, \bibinfo {author} {\bibfnamefont {V.~A.}\ \bibnamefont
  {Ivanov}}, \ and\ \bibinfo {author} {\bibfnamefont {M.~Y.}\ \bibnamefont
  {Zhuravlev}},\ }\href {\doibase 10.1016/S0378-4371(00)00111-4} {\bibfield
  {journal} {\bibinfo  {journal} {Physica A: Stat. Mech. Appl.}\ }\textbf
  {\bibinfo {volume} {284}},\ \bibinfo {pages} {172} (\bibinfo {year}
  {2000})}\BibitemShut {NoStop}%
\bibitem [{\citenamefont {Rongsheng}, \citenamefont {Zijing},\ and\
  \citenamefont {Kelin}(2002)}]{Rongsheng02}%
  \BibitemOpen
  \bibfield  {author} {\bibinfo {author} {\bibfnamefont {H.}~\bibnamefont
  {Rongsheng}}, \bibinfo {author} {\bibfnamefont {L.}~\bibnamefont {Zijing}}, \
  and\ \bibinfo {author} {\bibfnamefont {W.}~\bibnamefont {Kelin}},\ }\href
  {\doibase 10.1103/PhysRevB.65.174303} {\bibfield  {journal} {\bibinfo
  {journal} {Phys. Rev. B}\ }\textbf {\bibinfo {volume} {65}},\ \bibinfo
  {pages} {174303} (\bibinfo {year} {2002})}\BibitemShut {NoStop}%
\bibitem [{\citenamefont {Qing-Bao}\ and\ \citenamefont
  {Qing-Hu}(2005)}]{QingBao05}%
  \BibitemOpen
  \bibfield  {author} {\bibinfo {author} {\bibfnamefont {R.}~\bibnamefont
  {Qing-Bao}}\ and\ \bibinfo {author} {\bibfnamefont {C.}~\bibnamefont
  {Qing-Hu}},\ }\href {\doibase 10.1088/0253-6102/43/2/032} {\bibfield
  {journal} {\bibinfo  {journal} {Commun. Theor. Phys.}\ }\textbf {\bibinfo
  {volume} {43}},\ \bibinfo {pages} {357} (\bibinfo {year} {2005})}\BibitemShut
  {NoStop}%
\bibitem [{\citenamefont {Paganelli}\ and\ \citenamefont
  {Ciuchi}(2008{\natexlab{a}})}]{Paganelli08}%
  \BibitemOpen
  \bibfield  {author} {\bibinfo {author} {\bibfnamefont {S.}~\bibnamefont
  {Paganelli}}\ and\ \bibinfo {author} {\bibfnamefont {S.}~\bibnamefont
  {Ciuchi}},\ }\href {\doibase 10.1140/epjst/e2008-00737-4} {\bibfield
  {journal} {\bibinfo  {journal} {Eur. Phys. J. Spec. Top.}\ }\textbf {\bibinfo
  {volume} {160}},\ \bibinfo {pages} {343} (\bibinfo {year}
  {2008}{\natexlab{a}})}\BibitemShut {NoStop}%
\bibitem [{\citenamefont {Paganelli}\ and\ \citenamefont
  {Ciuchi}(2008{\natexlab{b}})}]{Paganelli08a}%
  \BibitemOpen
  \bibfield  {author} {\bibinfo {author} {\bibfnamefont {S.}~\bibnamefont
  {Paganelli}}\ and\ \bibinfo {author} {\bibfnamefont {S.}~\bibnamefont
  {Ciuchi}},\ }\href {\doibase 10.1088/0953-8984/20/23/235203} {\bibfield
  {journal} {\bibinfo  {journal} {J. Phys. Condens. Matter}\ }\textbf {\bibinfo
  {volume} {20}},\ \bibinfo {pages} {235203} (\bibinfo {year}
  {2008}{\natexlab{b}})}\BibitemShut {NoStop}%
\bibitem [{\citenamefont {Zhang}, \citenamefont {Wang},\ and\ \citenamefont
  {Chen}(2009)}]{Zhang09}%
  \BibitemOpen
  \bibfield  {author} {\bibinfo {author} {\bibfnamefont {Y.~Y.}\ \bibnamefont
  {Zhang}}, \bibinfo {author} {\bibfnamefont {X.~G.}\ \bibnamefont {Wang}}, \
  and\ \bibinfo {author} {\bibfnamefont {Q.~H.}\ \bibnamefont {Chen}},\ }\href
  {\doibase 10.1016/j.ssc.2009.08.009} {\bibfield  {journal} {\bibinfo
  {journal} {Solid State Commun.}\ }\textbf {\bibinfo {volume} {149}},\
  \bibinfo {pages} {2106} (\bibinfo {year} {2009})}\BibitemShut {NoStop}%
\bibitem [{\citenamefont {Sato}, \citenamefont {Kelly},\ and\ \citenamefont
  {Rubio}(2018)}]{Sato18}%
  \BibitemOpen
  \bibfield  {author} {\bibinfo {author} {\bibfnamefont {S.~A.}\ \bibnamefont
  {Sato}}, \bibinfo {author} {\bibfnamefont {A.}~\bibnamefont {Kelly}}, \ and\
  \bibinfo {author} {\bibfnamefont {A.}~\bibnamefont {Rubio}},\ }\href
  {\doibase 10.1103/PhysRevB.97.134308} {\bibfield  {journal} {\bibinfo
  {journal} {Phys. Rev. B}\ }\textbf {\bibinfo {volume} {97}},\ \bibinfo
  {pages} {134308} (\bibinfo {year} {2018})}\BibitemShut {NoStop}%
\bibitem [{\citenamefont {McKemmish}\ \emph {et~al.}(2015)\citenamefont
  {McKemmish}, \citenamefont {McKenzie}, \citenamefont {Hush},\ and\
  \citenamefont {Reimers}}]{McKemmish15}%
  \BibitemOpen
  \bibfield  {author} {\bibinfo {author} {\bibfnamefont {L.~K.}\ \bibnamefont
  {McKemmish}}, \bibinfo {author} {\bibfnamefont {R.~H.}\ \bibnamefont
  {McKenzie}}, \bibinfo {author} {\bibfnamefont {N.~S.}\ \bibnamefont {Hush}},
  \ and\ \bibinfo {author} {\bibfnamefont {J.~R.}\ \bibnamefont {Reimers}},\
  }\href {\doibase 10.1039/C5CP02239H} {\bibfield  {journal} {\bibinfo
  {journal} {Phys. Chem. Chem. Phys.}\ }\textbf {\bibinfo {volume} {17}},\
  \bibinfo {pages} {24666} (\bibinfo {year} {2015})}\BibitemShut {NoStop}%
\bibitem [{\citenamefont {Reimers}\ \emph {et~al.}(2015)\citenamefont
  {Reimers}, \citenamefont {McKemmish}, \citenamefont {McKenzie},\ and\
  \citenamefont {Hush}}]{Reimers15}%
  \BibitemOpen
  \bibfield  {author} {\bibinfo {author} {\bibfnamefont {J.~R.}\ \bibnamefont
  {Reimers}}, \bibinfo {author} {\bibfnamefont {L.~K.}\ \bibnamefont
  {McKemmish}}, \bibinfo {author} {\bibfnamefont {R.~H.}\ \bibnamefont
  {McKenzie}}, \ and\ \bibinfo {author} {\bibfnamefont {N.~S.}\ \bibnamefont
  {Hush}},\ }\href {\doibase 10.1039/C5CP02238J} {\bibfield  {journal}
  {\bibinfo  {journal} {Phys. Chem. Chem. Phys.}\ }\textbf {\bibinfo {volume}
  {17}},\ \bibinfo {pages} {24641} (\bibinfo {year} {2015})}\BibitemShut
  {NoStop}%
\bibitem [{\citenamefont {Subotnik}\ \emph {et~al.}(2016)\citenamefont
  {Subotnik}, \citenamefont {Jain}, \citenamefont {Landry}, \citenamefont
  {Petit}, \citenamefont {Ouyang},\ and\ \citenamefont
  {Bellonzi}}]{Subotnik2016}%
  \BibitemOpen
  \bibfield  {author} {\bibinfo {author} {\bibfnamefont {J.~E.}\ \bibnamefont
  {Subotnik}}, \bibinfo {author} {\bibfnamefont {A.}~\bibnamefont {Jain}},
  \bibinfo {author} {\bibfnamefont {B.}~\bibnamefont {Landry}}, \bibinfo
  {author} {\bibfnamefont {A.}~\bibnamefont {Petit}}, \bibinfo {author}
  {\bibfnamefont {W.}~\bibnamefont {Ouyang}}, \ and\ \bibinfo {author}
  {\bibfnamefont {N.}~\bibnamefont {Bellonzi}},\ }\href {\doibase
  10.1146/annurev-physchem-040215-112245} {\bibfield  {journal} {\bibinfo
  {journal} {Annu. Rev. Phys. Chem.}\ }\textbf {\bibinfo {volume} {67}},\
  \bibinfo {pages} {387} (\bibinfo {year} {2016})}\BibitemShut {NoStop}%
\bibitem [{\citenamefont {Capone}, \citenamefont {Stephan},\ and\ \citenamefont
  {Grilli}(1997)}]{capone_97}%
  \BibitemOpen
  \bibfield  {author} {\bibinfo {author} {\bibfnamefont {M.}~\bibnamefont
  {Capone}}, \bibinfo {author} {\bibfnamefont {W.}~\bibnamefont {Stephan}}, \
  and\ \bibinfo {author} {\bibfnamefont {M.}~\bibnamefont {Grilli}},\ }\href
  {\doibase 10.1103/PhysRevB.56.4484} {\bibfield  {journal} {\bibinfo
  {journal} {Phys. Rev. B}\ }\textbf {\bibinfo {volume} {56}},\ \bibinfo
  {pages} {4484} (\bibinfo {year} {1997})}\BibitemShut {NoStop}%
\bibitem [{\citenamefont {Ku}\ and\ \citenamefont {Trugman}(2007)}]{Ku07}%
  \BibitemOpen
  \bibfield  {author} {\bibinfo {author} {\bibfnamefont {L.-C.}\ \bibnamefont
  {Ku}}\ and\ \bibinfo {author} {\bibfnamefont {S.~A.}\ \bibnamefont
  {Trugman}},\ }\href {\doibase 10.1103/PhysRevB.75.014307} {\bibfield
  {journal} {\bibinfo  {journal} {Phys. Rev. B}\ }\textbf {\bibinfo {volume}
  {75}},\ \bibinfo {pages} {014307} (\bibinfo {year} {2007})}\BibitemShut
  {NoStop}%
\bibitem [{\citenamefont {Gole\ifmmode~\check{z}\else \v{z}\fi{}}\ \emph
  {et~al.}(2012)\citenamefont {Gole\ifmmode~\check{z}\else \v{z}\fi{}},
  \citenamefont {Bon\ifmmode~\check{c}\else \v{c}\fi{}a}, \citenamefont
  {Vidmar},\ and\ \citenamefont {Trugman}}]{Golez12}%
  \BibitemOpen
  \bibfield  {author} {\bibinfo {author} {\bibfnamefont {D.}~\bibnamefont
  {Gole\ifmmode~\check{z}\else \v{z}\fi{}}}, \bibinfo {author} {\bibfnamefont
  {J.}~\bibnamefont {Bon\ifmmode~\check{c}\else \v{c}\fi{}a}}, \bibinfo
  {author} {\bibfnamefont {L.}~\bibnamefont {Vidmar}}, \ and\ \bibinfo {author}
  {\bibfnamefont {S.~A.}\ \bibnamefont {Trugman}},\ }\href {\doibase
  10.1103/PhysRevLett.109.236402} {\bibfield  {journal} {\bibinfo  {journal}
  {Phys. Rev. Lett.}\ }\textbf {\bibinfo {volume} {109}},\ \bibinfo {pages}
  {236402} (\bibinfo {year} {2012})}\BibitemShut {NoStop}%
\bibitem [{\citenamefont {Jansen}\ \emph {et~al.}(2019)\citenamefont {Jansen},
  \citenamefont {Stolpp}, \citenamefont {Vidmar},\ and\ \citenamefont
  {Heidrich-Meisner}}]{jansen19}%
  \BibitemOpen
  \bibfield  {author} {\bibinfo {author} {\bibfnamefont {D.}~\bibnamefont
  {Jansen}}, \bibinfo {author} {\bibfnamefont {J.}~\bibnamefont {Stolpp}},
  \bibinfo {author} {\bibfnamefont {L.}~\bibnamefont {Vidmar}}, \ and\ \bibinfo
  {author} {\bibfnamefont {F.}~\bibnamefont {Heidrich-Meisner}},\ }\href
  {\doibase 10.1103/PhysRevB.99.155130} {\bibfield  {journal} {\bibinfo
  {journal} {Phys. Rev. B}\ }\textbf {\bibinfo {volume} {99}},\ \bibinfo
  {pages} {155130} (\bibinfo {year} {2019})}\BibitemShut {NoStop}%
\bibitem [{\citenamefont {Bon\v{c}a}, \citenamefont {Trugman},\ and\
  \citenamefont {Batisti\'{c}}(1999)}]{bonca99}%
  \BibitemOpen
  \bibfield  {author} {\bibinfo {author} {\bibfnamefont {J.}~\bibnamefont
  {Bon\v{c}a}}, \bibinfo {author} {\bibfnamefont {S.~A.}\ \bibnamefont
  {Trugman}}, \ and\ \bibinfo {author} {\bibfnamefont {I.}~\bibnamefont
  {Batisti\'{c}}},\ }\href {\doibase 10.1103/PhysRevB.60.1633} {\bibfield
  {journal} {\bibinfo  {journal} {Phys. Rev. B}\ }\textbf {\bibinfo {volume}
  {60}},\ \bibinfo {pages} {1633} (\bibinfo {year} {1999})}\BibitemShut
  {NoStop}%
\bibitem [{\citenamefont {Gole\v{z}}, \citenamefont {Bon\v{c}a},\ and\
  \citenamefont {Vidmar}(2012)}]{golez12a}%
  \BibitemOpen
  \bibfield  {author} {\bibinfo {author} {\bibfnamefont {D.}~\bibnamefont
  {Gole\v{z}}}, \bibinfo {author} {\bibfnamefont {J.}~\bibnamefont
  {Bon\v{c}a}}, \ and\ \bibinfo {author} {\bibfnamefont {L.}~\bibnamefont
  {Vidmar}},\ }\href {\doibase 10.1103/PhysRevB.85.144304} {\bibfield
  {journal} {\bibinfo  {journal} {Phys. Rev. B}\ }\textbf {\bibinfo {volume}
  {85}},\ \bibinfo {pages} {144304} (\bibinfo {year} {2012})}\BibitemShut
  {NoStop}%
\bibitem [{\citenamefont {Dorfner}\ \emph {et~al.}(2015)\citenamefont
  {Dorfner}, \citenamefont {Vidmar}, \citenamefont {Brockt}, \citenamefont
  {Jeckelmann},\ and\ \citenamefont {Heidrich-Meisner}}]{dorfner_vidmar_15}%
  \BibitemOpen
  \bibfield  {author} {\bibinfo {author} {\bibfnamefont {F.}~\bibnamefont
  {Dorfner}}, \bibinfo {author} {\bibfnamefont {L.}~\bibnamefont {Vidmar}},
  \bibinfo {author} {\bibfnamefont {C.}~\bibnamefont {Brockt}}, \bibinfo
  {author} {\bibfnamefont {E.}~\bibnamefont {Jeckelmann}}, \ and\ \bibinfo
  {author} {\bibfnamefont {F.}~\bibnamefont {Heidrich-Meisner}},\ }\href
  {\doibase 10.1103/PhysRevB.91.104302} {\bibfield  {journal} {\bibinfo
  {journal} {Phys. Rev. B}\ }\textbf {\bibinfo {volume} {91}},\ \bibinfo
  {pages} {104302} (\bibinfo {year} {2015})}\BibitemShut {NoStop}%
\bibitem [{\citenamefont {Kessing}\ \emph {et~al.}(2021)\citenamefont
  {Kessing}, \citenamefont {Yang}, \citenamefont {Manmana},\ and\ \citenamefont
  {Cao}}]{Kessing2021}%
  \BibitemOpen
  \bibfield  {author} {\bibinfo {author} {\bibfnamefont {R.~K.}\ \bibnamefont
  {Kessing}}, \bibinfo {author} {\bibfnamefont {P.-Y.}\ \bibnamefont {Yang}},
  \bibinfo {author} {\bibfnamefont {S.~R.}\ \bibnamefont {Manmana}}, \ and\
  \bibinfo {author} {\bibfnamefont {J.}~\bibnamefont {Cao}},\ }\href
  {https://arxiv.org/abs/2111.06137} {\bibfield  {journal} {\bibinfo  {journal}
  {arXiv:2111.06137 [cond-mat, physics:physics, physics:quant-ph]}\ } (\bibinfo
  {year} {2021})}\BibitemShut {NoStop}%
\bibitem [{\citenamefont {Wellein}, \citenamefont {R\"oder},\ and\
  \citenamefont {Fehske}(1996)}]{wellein96}%
  \BibitemOpen
  \bibfield  {author} {\bibinfo {author} {\bibfnamefont {G.}~\bibnamefont
  {Wellein}}, \bibinfo {author} {\bibfnamefont {H.}~\bibnamefont {R\"oder}}, \
  and\ \bibinfo {author} {\bibfnamefont {H.}~\bibnamefont {Fehske}},\ }\href
  {\doibase 10.1103/PhysRevB.53.9666} {\bibfield  {journal} {\bibinfo
  {journal} {Phys. Rev. B}\ }\textbf {\bibinfo {volume} {53}},\ \bibinfo
  {pages} {9666} (\bibinfo {year} {1996})}\BibitemShut {NoStop}%
\bibitem [{\citenamefont {Wellein}\ and\ \citenamefont
  {Fehske}(1998)}]{wellein98}%
  \BibitemOpen
  \bibfield  {author} {\bibinfo {author} {\bibfnamefont {G.}~\bibnamefont
  {Wellein}}\ and\ \bibinfo {author} {\bibfnamefont {H.}~\bibnamefont
  {Fehske}},\ }\href {\doibase 10.1103/PhysRevB.58.6208} {\bibfield  {journal}
  {\bibinfo  {journal} {Phys. Rev. B}\ }\textbf {\bibinfo {volume} {58}},\
  \bibinfo {pages} {6208} (\bibinfo {year} {1998})}\BibitemShut {NoStop}%
\bibitem [{\citenamefont {Vidmar}\ \emph {et~al.}(2011)\citenamefont {Vidmar},
  \citenamefont {Bon\ifmmode~\check{c}\else \v{c}\fi{}a}, \citenamefont
  {Tohyama},\ and\ \citenamefont {Maekawa}}]{Vidmar11}%
  \BibitemOpen
  \bibfield  {author} {\bibinfo {author} {\bibfnamefont {L.}~\bibnamefont
  {Vidmar}}, \bibinfo {author} {\bibfnamefont {J.}~\bibnamefont
  {Bon\ifmmode~\check{c}\else \v{c}\fi{}a}}, \bibinfo {author} {\bibfnamefont
  {T.}~\bibnamefont {Tohyama}}, \ and\ \bibinfo {author} {\bibfnamefont
  {S.}~\bibnamefont {Maekawa}},\ }\href {\doibase
  10.1103/PhysRevLett.107.246404} {\bibfield  {journal} {\bibinfo  {journal}
  {Phys. Rev. Lett.}\ }\textbf {\bibinfo {volume} {107}},\ \bibinfo {pages}
  {246404} (\bibinfo {year} {2011})}\BibitemShut {NoStop}%
\bibitem [{\citenamefont {Bon\ifmmode~\check{c}\else \v{c}\fi{}a},
  \citenamefont {Trugman},\ and\ \citenamefont {Berciu}(2019)}]{bonca2019}%
  \BibitemOpen
  \bibfield  {author} {\bibinfo {author} {\bibfnamefont {J.}~\bibnamefont
  {Bon\ifmmode~\check{c}\else \v{c}\fi{}a}}, \bibinfo {author} {\bibfnamefont
  {S.~A.}\ \bibnamefont {Trugman}}, \ and\ \bibinfo {author} {\bibfnamefont
  {M.}~\bibnamefont {Berciu}},\ }\href {\doibase 10.1103/PhysRevB.100.094307}
  {\bibfield  {journal} {\bibinfo  {journal} {Phys. Rev. B}\ }\textbf {\bibinfo
  {volume} {100}},\ \bibinfo {pages} {094307} (\bibinfo {year}
  {2019})}\BibitemShut {NoStop}%
\bibitem [{\citenamefont {Schr\"oder}\ and\ \citenamefont
  {Chin}(2016)}]{shroeder_16}%
  \BibitemOpen
  \bibfield  {author} {\bibinfo {author} {\bibfnamefont {F.~A. Y.~N.}\
  \bibnamefont {Schr\"oder}}\ and\ \bibinfo {author} {\bibfnamefont {A.~W.}\
  \bibnamefont {Chin}},\ }\href {\doibase 10.1103/PhysRevB.93.075105}
  {\bibfield  {journal} {\bibinfo  {journal} {Phys. Rev. B}\ }\textbf {\bibinfo
  {volume} {93}},\ \bibinfo {pages} {075105} (\bibinfo {year}
  {2016})}\BibitemShut {NoStop}%
\bibitem [{\citenamefont {Brockt}\ and\ \citenamefont
  {Jeckelmann}(2017)}]{Brockt_17}%
  \BibitemOpen
  \bibfield  {author} {\bibinfo {author} {\bibfnamefont {C.}~\bibnamefont
  {Brockt}}\ and\ \bibinfo {author} {\bibfnamefont {E.}~\bibnamefont
  {Jeckelmann}},\ }\href {\doibase 10.1103/PhysRevB.95.064309} {\bibfield
  {journal} {\bibinfo  {journal} {Phys. Rev. B}\ }\textbf {\bibinfo {volume}
  {95}},\ \bibinfo {pages} {064309} (\bibinfo {year} {2017})}\BibitemShut
  {NoStop}%
\bibitem [{\citenamefont {Jankovi{\'c}}\ and\ \citenamefont
  {Vukmirovi{\'c}}(2022)}]{Jankovic21}%
  \BibitemOpen
  \bibfield  {author} {\bibinfo {author} {\bibfnamefont {V.}~\bibnamefont
  {Jankovi{\'c}}}\ and\ \bibinfo {author} {\bibfnamefont {N.}~\bibnamefont
  {Vukmirovi{\'c}}},\ }\href {\doibase 10.1103/PhysRevB.105.054311} {\bibfield
  {journal} {\bibinfo  {journal} {Phys. Rev. B}\ }\textbf {\bibinfo {volume}
  {105}},\ \bibinfo {pages} {054311} (\bibinfo {year} {2022})}\BibitemShut
  {NoStop}%
\bibitem [{\citenamefont {Sun}, \citenamefont {Luo},\ and\ \citenamefont
  {Zhao}(2010)}]{Sun2010}%
  \BibitemOpen
  \bibfield  {author} {\bibinfo {author} {\bibfnamefont {J.}~\bibnamefont
  {Sun}}, \bibinfo {author} {\bibfnamefont {B.}~\bibnamefont {Luo}}, \ and\
  \bibinfo {author} {\bibfnamefont {Y.}~\bibnamefont {Zhao}},\ }\href {\doibase
  10.1103/PhysRevB.82.014305} {\bibfield  {journal} {\bibinfo  {journal} {Phys.
  Rev. B}\ }\textbf {\bibinfo {volume} {82}},\ \bibinfo {pages} {014305}
  (\bibinfo {year} {2010})}\BibitemShut {NoStop}%
\bibitem [{\citenamefont {Luo}, \citenamefont {Ye},\ and\ \citenamefont
  {Zhao}(2011)}]{Luo11}%
  \BibitemOpen
  \bibfield  {author} {\bibinfo {author} {\bibfnamefont {B.}~\bibnamefont
  {Luo}}, \bibinfo {author} {\bibfnamefont {J.}~\bibnamefont {Ye}}, \ and\
  \bibinfo {author} {\bibfnamefont {Y.}~\bibnamefont {Zhao}},\ }\href {\doibase
  10.1002/pssc.201000721} {\bibfield  {journal} {\bibinfo  {journal} {Phys.
  Status Solidi (c)}\ }\textbf {\bibinfo {volume} {8}},\ \bibinfo {pages} {70}
  (\bibinfo {year} {2011})}\BibitemShut {NoStop}%
\bibitem [{\citenamefont {Born}\ and\ \citenamefont
  {Huang}(1954)}]{BornHuang1954}%
  \BibitemOpen
  \bibfield  {author} {\bibinfo {author} {\bibfnamefont {M.}~\bibnamefont
  {Born}}\ and\ \bibinfo {author} {\bibfnamefont {K.}~\bibnamefont {Huang}},\
  }\href@noop {} {\emph {\bibinfo {title} {Dynamical Theory of Crystal
  Lattices}}}\ (\bibinfo  {publisher} {Clarendon Press},\ \bibinfo {year}
  {1954})\BibitemShut {NoStop}%
\bibitem [{\citenamefont {Born}\ and\ \citenamefont
  {Oppenheimer}(1927)}]{Born1927}%
  \BibitemOpen
  \bibfield  {author} {\bibinfo {author} {\bibfnamefont {M.}~\bibnamefont
  {Born}}\ and\ \bibinfo {author} {\bibfnamefont {R.}~\bibnamefont
  {Oppenheimer}},\ }\href {\doibase 10.1002/andp.19273892002} {\bibfield
  {journal} {\bibinfo  {journal} {Ann. Phys.}\ }\textbf {\bibinfo {volume}
  {389}},\ \bibinfo {pages} {457} (\bibinfo {year} {1927})}\BibitemShut
  {NoStop}%
\bibitem [{\citenamefont {Tully}(2000)}]{Tully2000}%
  \BibitemOpen
  \bibfield  {author} {\bibinfo {author} {\bibfnamefont {J.~C.}\ \bibnamefont
  {Tully}},\ }in\ \href {\doibase 10.1007/978-3-662-10421-7_3} {\emph {\bibinfo
  {booktitle} {Theoretical {Chemistry} {Accounts}: {New} {Century} {Issue}}}},\
  \bibinfo {editor} {edited by\ \bibinfo {editor} {\bibfnamefont {C.~J.}\
  \bibnamefont {Cramer}}\ and\ \bibinfo {editor} {\bibfnamefont {D.~G.}\
  \bibnamefont {Truhlar}}}\ (\bibinfo  {publisher} {Springer},\ \bibinfo
  {address} {Berlin, Heidelberg},\ \bibinfo {year} {2000})\ pp.\ \bibinfo
  {pages} {173--176}\BibitemShut {NoStop}%
\bibitem [{\citenamefont {Worth}\ and\ \citenamefont
  {Cederbaum}(2004)}]{Worth2004b}%
  \BibitemOpen
  \bibfield  {author} {\bibinfo {author} {\bibfnamefont {G.~A.}\ \bibnamefont
  {Worth}}\ and\ \bibinfo {author} {\bibfnamefont {L.~S.}\ \bibnamefont
  {Cederbaum}},\ }\href {\doibase 10.1146/annurev.physchem.55.091602.094335}
  {\bibfield  {journal} {\bibinfo  {journal} {Annu. Rev. Phys. Chem.}\ }\textbf
  {\bibinfo {volume} {55}},\ \bibinfo {pages} {127} (\bibinfo {year}
  {2004})}\BibitemShut {NoStop}%
\bibitem [{\citenamefont {Lichten}(1963)}]{Lichten1963}%
  \BibitemOpen
  \bibfield  {author} {\bibinfo {author} {\bibfnamefont {W.}~\bibnamefont
  {Lichten}},\ }\href {\doibase 10.1103/PhysRev.131.229} {\bibfield  {journal}
  {\bibinfo  {journal} {Phys. Rev.}\ }\textbf {\bibinfo {volume} {131}},\
  \bibinfo {pages} {229} (\bibinfo {year} {1963})}\BibitemShut {NoStop}%
\bibitem [{\citenamefont {Smith}(1969)}]{Smith1969}%
  \BibitemOpen
  \bibfield  {author} {\bibinfo {author} {\bibfnamefont {F.~T.}\ \bibnamefont
  {Smith}},\ }\href {\doibase 10.1103/PhysRev.179.111} {\bibfield  {journal}
  {\bibinfo  {journal} {Phys. Rev.}\ }\textbf {\bibinfo {volume} {179}},\
  \bibinfo {pages} {111} (\bibinfo {year} {1969})}\BibitemShut {NoStop}%
\bibitem [{\citenamefont {Baer}(1975)}]{Baer1975}%
  \BibitemOpen
  \bibfield  {author} {\bibinfo {author} {\bibfnamefont {M.}~\bibnamefont
  {Baer}},\ }\href {\doibase 10.1016/0009-2614(75)85599-0} {\bibfield
  {journal} {\bibinfo  {journal} {Chem. Phys. Lett.}\ }\textbf {\bibinfo
  {volume} {35}},\ \bibinfo {pages} {112} (\bibinfo {year} {1975})}\BibitemShut
  {NoStop}%
\bibitem [{\citenamefont {Mead}\ and\ \citenamefont
  {Truhlar}(1982)}]{Mead1982}%
  \BibitemOpen
  \bibfield  {author} {\bibinfo {author} {\bibfnamefont {C.~A.}\ \bibnamefont
  {Mead}}\ and\ \bibinfo {author} {\bibfnamefont {D.~G.}\ \bibnamefont
  {Truhlar}},\ }\href {\doibase 10.1063/1.443853} {\bibfield  {journal}
  {\bibinfo  {journal} {J. Chem. Phys.}\ }\textbf {\bibinfo {volume} {77}},\
  \bibinfo {pages} {6090} (\bibinfo {year} {1982})}\BibitemShut {NoStop}%
\bibitem [{\citenamefont {Van~Voorhis}\ \emph {et~al.}(2010)\citenamefont
  {Van~Voorhis}, \citenamefont {Kowalczyk}, \citenamefont {Kaduk},
  \citenamefont {Wang}, \citenamefont {Cheng},\ and\ \citenamefont
  {Wu}}]{Van_voorhis2010}%
  \BibitemOpen
  \bibfield  {author} {\bibinfo {author} {\bibfnamefont {T.}~\bibnamefont
  {Van~Voorhis}}, \bibinfo {author} {\bibfnamefont {T.}~\bibnamefont
  {Kowalczyk}}, \bibinfo {author} {\bibfnamefont {B.}~\bibnamefont {Kaduk}},
  \bibinfo {author} {\bibfnamefont {L.-P.}\ \bibnamefont {Wang}}, \bibinfo
  {author} {\bibfnamefont {C.-L.}\ \bibnamefont {Cheng}}, \ and\ \bibinfo
  {author} {\bibfnamefont {Q.}~\bibnamefont {Wu}},\ }\href {\doibase
  10.1146/annurev.physchem.012809.103324} {\bibfield  {journal} {\bibinfo
  {journal} {Annu. Rev. Phys. Chem.}\ }\textbf {\bibinfo {volume} {61}},\
  \bibinfo {pages} {149} (\bibinfo {year} {2010})}\BibitemShut {NoStop}%
\bibitem [{\citenamefont {K\"oppel}, \citenamefont {Domcke},\ and\
  \citenamefont {Cederbaum}(1984)}]{Koppel1984}%
  \BibitemOpen
  \bibfield  {author} {\bibinfo {author} {\bibfnamefont {H.}~\bibnamefont
  {K\"oppel}}, \bibinfo {author} {\bibfnamefont {W.}~\bibnamefont {Domcke}}, \
  and\ \bibinfo {author} {\bibfnamefont {L.~S.}\ \bibnamefont {Cederbaum}},\
  }\href {\doibase 10.1002/9780470142813.ch2} {\bibfield  {journal} {\bibinfo
  {journal} {Adv. Chem. Phys.}\ }\textbf {\bibinfo {volume} {57}},\ \bibinfo
  {pages} {59} (\bibinfo {year} {1984})}\BibitemShut {NoStop}%
\bibitem [{\citenamefont {Spencer}\ \emph {et~al.}(2016)\citenamefont
  {Spencer}, \citenamefont {Scalfi}, \citenamefont {Carof},\ and\ \citenamefont
  {Blumberger}}]{Spencer2016}%
  \BibitemOpen
  \bibfield  {author} {\bibinfo {author} {\bibfnamefont {J.}~\bibnamefont
  {Spencer}}, \bibinfo {author} {\bibfnamefont {L.}~\bibnamefont {Scalfi}},
  \bibinfo {author} {\bibfnamefont {A.}~\bibnamefont {Carof}}, \ and\ \bibinfo
  {author} {\bibfnamefont {J.}~\bibnamefont {Blumberger}},\ }\href {\doibase
  10.1039/C6FD00107F} {\bibfield  {journal} {\bibinfo  {journal} {Faraday
  Discuss.}\ }\textbf {\bibinfo {volume} {195}},\ \bibinfo {pages} {215}
  (\bibinfo {year} {2016})}\BibitemShut {NoStop}%
\bibitem [{\citenamefont {Giannini}\ and\ \citenamefont
  {Blumberger}(2022)}]{Giannini2022}%
  \BibitemOpen
  \bibfield  {author} {\bibinfo {author} {\bibfnamefont {S.}~\bibnamefont
  {Giannini}}\ and\ \bibinfo {author} {\bibfnamefont {J.}~\bibnamefont
  {Blumberger}},\ }\href {\doibase 10.1021/acs.accounts.1c00675} {\bibfield
  {journal} {\bibinfo  {journal} {Acc. Chem. Res.}\ }\textbf {\bibinfo {volume}
  {55}},\ \bibinfo {pages} {819} (\bibinfo {year} {2022})}\BibitemShut
  {NoStop}%
\bibitem [{\citenamefont {Landry}\ and\ \citenamefont
  {Subotnik}(2011)}]{Landry2011}%
  \BibitemOpen
  \bibfield  {author} {\bibinfo {author} {\bibfnamefont {B.~R.}\ \bibnamefont
  {Landry}}\ and\ \bibinfo {author} {\bibfnamefont {J.~E.}\ \bibnamefont
  {Subotnik}},\ }\href {\doibase 10.1063/1.3663870} {\bibfield  {journal}
  {\bibinfo  {journal} {J. Chem. Phys.}\ }\textbf {\bibinfo {volume} {135}},\
  \bibinfo {pages} {191101} (\bibinfo {year} {2011})}\BibitemShut {NoStop}%
\bibitem [{\citenamefont {Landry}\ and\ \citenamefont
  {Subotnik}(2012)}]{Landry2012}%
  \BibitemOpen
  \bibfield  {author} {\bibinfo {author} {\bibfnamefont {B.~R.}\ \bibnamefont
  {Landry}}\ and\ \bibinfo {author} {\bibfnamefont {J.~E.}\ \bibnamefont
  {Subotnik}},\ }\href {\doibase 10.1063/1.4733675} {\bibfield  {journal}
  {\bibinfo  {journal} {J. Chem. Phys.}\ }\textbf {\bibinfo {volume} {137}},\
  \bibinfo {pages} {22A513} (\bibinfo {year} {2012})}\BibitemShut {NoStop}%
\bibitem [{\citenamefont {Landau}(1932)}]{Landau1932}%
  \BibitemOpen
  \bibfield  {author} {\bibinfo {author} {\bibfnamefont {L.~D.}\ \bibnamefont
  {Landau}},\ }\href@noop {} {\bibfield  {journal} {\bibinfo  {journal} {Phys.
  Z. Sowjet.}\ }\textbf {\bibinfo {volume} {2}},\ \bibinfo {pages} {46}
  (\bibinfo {year} {1932})}\BibitemShut {NoStop}%
\bibitem [{\citenamefont {Zener}(1932)}]{Zener1932}%
  \BibitemOpen
  \bibfield  {author} {\bibinfo {author} {\bibfnamefont {C.}~\bibnamefont
  {Zener}},\ }\href {\doibase 10.1098/rspa.1932.0165} {\bibfield  {journal}
  {\bibinfo  {journal} {Proc. R. Soc. Lond. A}\ }\textbf {\bibinfo {volume}
  {137}},\ \bibinfo {pages} {696} (\bibinfo {year} {1932})}\BibitemShut
  {NoStop}%
\bibitem [{\citenamefont {Sandvik}(2010)}]{Sandvik2010}%
  \BibitemOpen
  \bibfield  {author} {\bibinfo {author} {\bibfnamefont {A.~W.}\ \bibnamefont
  {Sandvik}},\ }\href {\doibase 10.1063/1.3518900} {\bibfield  {journal}
  {\bibinfo  {journal} {AIP Conf. Proc.}\ }\textbf {\bibinfo {volume} {1297}},\
  \bibinfo {pages} {135} (\bibinfo {year} {2010})}\BibitemShut {NoStop}%
\bibitem [{\citenamefont {Car}\ and\ \citenamefont
  {Parrinello}(1985)}]{Car1985}%
  \BibitemOpen
  \bibfield  {author} {\bibinfo {author} {\bibfnamefont {R.}~\bibnamefont
  {Car}}\ and\ \bibinfo {author} {\bibfnamefont {M.}~\bibnamefont
  {Parrinello}},\ }\href {\doibase 10.1103/PhysRevLett.55.2471} {\bibfield
  {journal} {\bibinfo  {journal} {Phys. Rev. Lett.}\ }\textbf {\bibinfo
  {volume} {55}},\ \bibinfo {pages} {2471} (\bibinfo {year}
  {1985})}\BibitemShut {NoStop}%
\bibitem [{\citenamefont {Verlet}(1967)}]{Verlet1967}%
  \BibitemOpen
  \bibfield  {author} {\bibinfo {author} {\bibfnamefont {L.}~\bibnamefont
  {Verlet}},\ }\href {\doibase 10.1103/PhysRev.159.98} {\bibfield  {journal}
  {\bibinfo  {journal} {Phys. Rev.}\ }\textbf {\bibinfo {volume} {159}},\
  \bibinfo {pages} {98} (\bibinfo {year} {1967})}\BibitemShut {NoStop}%
\bibitem [{\citenamefont {Paeckel}\ \emph {et~al.}(2019)\citenamefont
  {Paeckel}, \citenamefont {K\"ohler}, \citenamefont {Swoboda}, \citenamefont
  {Manmana}, \citenamefont {Schollw\"ock},\ and\ \citenamefont
  {Hubig}}]{Paeckel2019}%
  \BibitemOpen
  \bibfield  {author} {\bibinfo {author} {\bibfnamefont {S.}~\bibnamefont
  {Paeckel}}, \bibinfo {author} {\bibfnamefont {T.}~\bibnamefont {K\"ohler}},
  \bibinfo {author} {\bibfnamefont {A.}~\bibnamefont {Swoboda}}, \bibinfo
  {author} {\bibfnamefont {S.~R.}\ \bibnamefont {Manmana}}, \bibinfo {author}
  {\bibfnamefont {U.}~\bibnamefont {Schollw\"ock}}, \ and\ \bibinfo {author}
  {\bibfnamefont {C.}~\bibnamefont {Hubig}},\ }\href {\doibase
  10.1016/j.aop.2019.167998} {\bibfield  {journal} {\bibinfo  {journal} {Ann.
  Phys.}\ }\textbf {\bibinfo {volume} {411}},\ \bibinfo {pages} {167998}
  (\bibinfo {year} {2019})}\BibitemShut {NoStop}%
\bibitem [{\citenamefont {Vidal}(2004)}]{Vidal2004}%
  \BibitemOpen
  \bibfield  {author} {\bibinfo {author} {\bibfnamefont {G.}~\bibnamefont
  {Vidal}},\ }\href {\doibase 10.1103/PhysRevLett.93.040502} {\bibfield
  {journal} {\bibinfo  {journal} {Phys. Rev. Lett.}\ }\textbf {\bibinfo
  {volume} {93}},\ \bibinfo {pages} {040502} (\bibinfo {year}
  {2004})}\BibitemShut {NoStop}%
\bibitem [{\citenamefont {Daley}\ \emph {et~al.}(2004)\citenamefont {Daley},
  \citenamefont {Kollath}, \citenamefont {Schollw{\"o}ck},\ and\ \citenamefont
  {Vidal}}]{Daley2004}%
  \BibitemOpen
  \bibfield  {author} {\bibinfo {author} {\bibfnamefont {A.~J.}\ \bibnamefont
  {Daley}}, \bibinfo {author} {\bibfnamefont {C.}~\bibnamefont {Kollath}},
  \bibinfo {author} {\bibfnamefont {U.}~\bibnamefont {Schollw{\"o}ck}}, \ and\
  \bibinfo {author} {\bibfnamefont {G.}~\bibnamefont {Vidal}},\ }\href
  {\doibase 10.1088/1742-5468/2004/04/P04005} {\bibfield  {journal} {\bibinfo
  {journal} {J. Stat. Mech.}\ }\textbf {\bibinfo {volume} {2004}},\ \bibinfo
  {pages} {P04005} (\bibinfo {year} {2004})}\BibitemShut {NoStop}%
\bibitem [{\citenamefont {White}\ and\ \citenamefont
  {Feiguin}(2004)}]{White04}%
  \BibitemOpen
  \bibfield  {author} {\bibinfo {author} {\bibfnamefont {S.~R.}\ \bibnamefont
  {White}}\ and\ \bibinfo {author} {\bibfnamefont {A.~E.}\ \bibnamefont
  {Feiguin}},\ }\href {\doibase 10.1103/PhysRevLett.93.076401} {\bibfield
  {journal} {\bibinfo  {journal} {Phys. Rev. Lett.}\ }\textbf {\bibinfo
  {volume} {93}},\ \bibinfo {pages} {076401} (\bibinfo {year}
  {2004})}\BibitemShut {NoStop}%
\bibitem [{\citenamefont {Haegeman}\ \emph {et~al.}(2011)\citenamefont
  {Haegeman}, \citenamefont {Cirac}, \citenamefont {Osborne}, \citenamefont
  {Pi\ifmmode~\check{z}\else \v{z}\fi{}orn}, \citenamefont {Verschelde},\ and\
  \citenamefont {Verstraete}}]{Haegeman2011}%
  \BibitemOpen
  \bibfield  {author} {\bibinfo {author} {\bibfnamefont {J.}~\bibnamefont
  {Haegeman}}, \bibinfo {author} {\bibfnamefont {J.~I.}\ \bibnamefont {Cirac}},
  \bibinfo {author} {\bibfnamefont {T.~J.}\ \bibnamefont {Osborne}}, \bibinfo
  {author} {\bibfnamefont {I.}~\bibnamefont {Pi\ifmmode~\check{z}\else
  \v{z}\fi{}orn}}, \bibinfo {author} {\bibfnamefont {H.}~\bibnamefont
  {Verschelde}}, \ and\ \bibinfo {author} {\bibfnamefont {F.}~\bibnamefont
  {Verstraete}},\ }\href {\doibase 10.1103/PhysRevLett.107.070601} {\bibfield
  {journal} {\bibinfo  {journal} {Phys. Rev. Lett.}\ }\textbf {\bibinfo
  {volume} {107}},\ \bibinfo {pages} {070601} (\bibinfo {year}
  {2011})}\BibitemShut {NoStop}%
\bibitem [{\citenamefont {Haegeman}\ \emph {et~al.}(2016)\citenamefont
  {Haegeman}, \citenamefont {Lubich}, \citenamefont {Oseledets}, \citenamefont
  {Vandereycken},\ and\ \citenamefont {Verstraete}}]{Haegeman2016}%
  \BibitemOpen
  \bibfield  {author} {\bibinfo {author} {\bibfnamefont {J.}~\bibnamefont
  {Haegeman}}, \bibinfo {author} {\bibfnamefont {C.}~\bibnamefont {Lubich}},
  \bibinfo {author} {\bibfnamefont {I.}~\bibnamefont {Oseledets}}, \bibinfo
  {author} {\bibfnamefont {B.}~\bibnamefont {Vandereycken}}, \ and\ \bibinfo
  {author} {\bibfnamefont {F.}~\bibnamefont {Verstraete}},\ }\href {\doibase
  10.1103/PhysRevB.94.165116} {\bibfield  {journal} {\bibinfo  {journal} {Phys.
  Rev. B}\ }\textbf {\bibinfo {volume} {94}},\ \bibinfo {pages} {165116}
  (\bibinfo {year} {2016})}\BibitemShut {NoStop}%
\bibitem [{\citenamefont {Verstraete}, \citenamefont {Garc\'{\i}a-Ripoll},\
  and\ \citenamefont {Cirac}(2004)}]{Verstrate2004}%
  \BibitemOpen
  \bibfield  {author} {\bibinfo {author} {\bibfnamefont {F.}~\bibnamefont
  {Verstraete}}, \bibinfo {author} {\bibfnamefont {J.~J.}\ \bibnamefont
  {Garc\'{\i}a-Ripoll}}, \ and\ \bibinfo {author} {\bibfnamefont {J.~I.}\
  \bibnamefont {Cirac}},\ }\href {\doibase 10.1103/PhysRevLett.93.207204}
  {\bibfield  {journal} {\bibinfo  {journal} {Phys. Rev. Lett.}\ }\textbf
  {\bibinfo {volume} {93}},\ \bibinfo {pages} {207204} (\bibinfo {year}
  {2004})}\BibitemShut {NoStop}%
\bibitem [{\citenamefont {Sirker}\ and\ \citenamefont
  {Kl\"umper}(2005)}]{Sirker2005}%
  \BibitemOpen
  \bibfield  {author} {\bibinfo {author} {\bibfnamefont {J.}~\bibnamefont
  {Sirker}}\ and\ \bibinfo {author} {\bibfnamefont {A.}~\bibnamefont
  {Kl\"umper}},\ }\href {\doibase 10.1103/PhysRevB.71.241101} {\bibfield
  {journal} {\bibinfo  {journal} {Phys. Rev. B}\ }\textbf {\bibinfo {volume}
  {71}},\ \bibinfo {pages} {241101} (\bibinfo {year} {2005})}\BibitemShut
  {NoStop}%
\bibitem [{\citenamefont {Feiguin}\ and\ \citenamefont
  {White}(2005)}]{Feiguin2005}%
  \BibitemOpen
  \bibfield  {author} {\bibinfo {author} {\bibfnamefont {A.~E.}\ \bibnamefont
  {Feiguin}}\ and\ \bibinfo {author} {\bibfnamefont {S.~R.}\ \bibnamefont
  {White}},\ }\href {\doibase 10.1103/PhysRevB.72.220401} {\bibfield  {journal}
  {\bibinfo  {journal} {Phys. Rev. B}\ }\textbf {\bibinfo {volume} {72}},\
  \bibinfo {pages} {220401} (\bibinfo {year} {2005})}\BibitemShut {NoStop}%
\bibitem [{\citenamefont {Stoudenmire}\ and\ \citenamefont
  {White}(2010)}]{Stoudenmire2010}%
  \BibitemOpen
  \bibfield  {author} {\bibinfo {author} {\bibfnamefont {E.~M.}\ \bibnamefont
  {Stoudenmire}}\ and\ \bibinfo {author} {\bibfnamefont {S.~R.}\ \bibnamefont
  {White}},\ }\href {\doibase 10.1088/1367-2630/12/5/055026} {\bibfield
  {journal} {\bibinfo  {journal} {New J. Phys.}\ }\textbf {\bibinfo {volume}
  {12}},\ \bibinfo {pages} {055026} (\bibinfo {year} {2010})}\BibitemShut
  {NoStop}%
\bibitem [{\citenamefont {Karrasch}, \citenamefont {Bardarson},\ and\
  \citenamefont {Moore}(2012)}]{Karrasch2012}%
  \BibitemOpen
  \bibfield  {author} {\bibinfo {author} {\bibfnamefont {C.}~\bibnamefont
  {Karrasch}}, \bibinfo {author} {\bibfnamefont {J.~H.}\ \bibnamefont
  {Bardarson}}, \ and\ \bibinfo {author} {\bibfnamefont {J.~E.}\ \bibnamefont
  {Moore}},\ }\href {\doibase 10.1103/PhysRevLett.108.227206} {\bibfield
  {journal} {\bibinfo  {journal} {Phys. Rev. Lett.}\ }\textbf {\bibinfo
  {volume} {108}},\ \bibinfo {pages} {227206} (\bibinfo {year}
  {2012})}\BibitemShut {NoStop}%
\bibitem [{\citenamefont {Karrasch}, \citenamefont {Bardarson},\ and\
  \citenamefont {Moore}(2013)}]{Karrasch_2013}%
  \BibitemOpen
  \bibfield  {author} {\bibinfo {author} {\bibfnamefont {C.}~\bibnamefont
  {Karrasch}}, \bibinfo {author} {\bibfnamefont {J.~H.}\ \bibnamefont
  {Bardarson}}, \ and\ \bibinfo {author} {\bibfnamefont {J.~E.}\ \bibnamefont
  {Moore}},\ }\href {\doibase 10.1088/1367-2630/15/8/083031} {\bibfield
  {journal} {\bibinfo  {journal} {New J. Phys.}\ }\textbf {\bibinfo {volume}
  {15}},\ \bibinfo {pages} {083031} (\bibinfo {year} {2013})}\BibitemShut
  {NoStop}%
\bibitem [{\citenamefont {Verstraete}\ and\ \citenamefont
  {Cirac}(2004)}]{Verstraete2004-2D}%
  \BibitemOpen
  \bibfield  {author} {\bibinfo {author} {\bibfnamefont {F.}~\bibnamefont
  {Verstraete}}\ and\ \bibinfo {author} {\bibfnamefont {J.~I.}\ \bibnamefont
  {Cirac}},\ }\href {http://arxiv.org/abs/cond-mat/0407066} {\bibfield
  {journal} {\bibinfo  {journal} {arXiv:cond-mat/0407066}\ } (\bibinfo {year}
  {2004})}\BibitemShut {NoStop}%
\bibitem [{\citenamefont {Stoudenmire}\ and\ \citenamefont
  {White}(2012)}]{Stoudenmire2012}%
  \BibitemOpen
  \bibfield  {author} {\bibinfo {author} {\bibfnamefont {E.~M.}\ \bibnamefont
  {Stoudenmire}}\ and\ \bibinfo {author} {\bibfnamefont {S.~R.}\ \bibnamefont
  {White}},\ }\href {\doibase 10.1146/annurev-conmatphys-020911-125018}
  {\bibfield  {journal} {\bibinfo  {journal} {Annu. Rev. Condens. Matter
  Phys.}\ }\textbf {\bibinfo {volume} {3}},\ \bibinfo {pages} {111} (\bibinfo
  {year} {2012})}\BibitemShut {NoStop}%
\bibitem [{\citenamefont {Or{\'u}s}(2014)}]{Orus2014}%
  \BibitemOpen
  \bibfield  {author} {\bibinfo {author} {\bibfnamefont {R.}~\bibnamefont
  {Or{\'u}s}},\ }\href {\doibase 10.1016/j.aop.2014.06.013} {\bibfield
  {journal} {\bibinfo  {journal} {Ann. Phys.}\ }\textbf {\bibinfo {volume}
  {349}},\ \bibinfo {pages} {117 } (\bibinfo {year} {2014})}\BibitemShut
  {NoStop}%
\bibitem [{\citenamefont {Zheng}\ \emph {et~al.}(2017)\citenamefont {Zheng},
  \citenamefont {Chung}, \citenamefont {Corboz}, \citenamefont {Ehlers},
  \citenamefont {Qin}, \citenamefont {Noack}, \citenamefont {Shi},
  \citenamefont {White}, \citenamefont {Zhang},\ and\ \citenamefont
  {Chan}}]{Zheng2017}%
  \BibitemOpen
  \bibfield  {author} {\bibinfo {author} {\bibfnamefont {B.-X.}\ \bibnamefont
  {Zheng}}, \bibinfo {author} {\bibfnamefont {C.-M.}\ \bibnamefont {Chung}},
  \bibinfo {author} {\bibfnamefont {P.}~\bibnamefont {Corboz}}, \bibinfo
  {author} {\bibfnamefont {G.}~\bibnamefont {Ehlers}}, \bibinfo {author}
  {\bibfnamefont {M.-P.}\ \bibnamefont {Qin}}, \bibinfo {author} {\bibfnamefont
  {R.~M.}\ \bibnamefont {Noack}}, \bibinfo {author} {\bibfnamefont
  {H.}~\bibnamefont {Shi}}, \bibinfo {author} {\bibfnamefont {S.~R.}\
  \bibnamefont {White}}, \bibinfo {author} {\bibfnamefont {S.}~\bibnamefont
  {Zhang}}, \ and\ \bibinfo {author} {\bibfnamefont {G.~K.-L.}\ \bibnamefont
  {Chan}},\ }\href {\doibase 10.1126/science.aam7127} {\bibfield  {journal}
  {\bibinfo  {journal} {Science}\ }\textbf {\bibinfo {volume} {358}},\ \bibinfo
  {pages} {1155} (\bibinfo {year} {2017})}\BibitemShut {NoStop}%
\bibitem [{\citenamefont {Bruognolo}\ \emph {et~al.}(2021)\citenamefont
  {Bruognolo}, \citenamefont {Li}, \citenamefont {von Delft},\ and\
  \citenamefont {Weichselbaum}}]{Bruognolo2021}%
  \BibitemOpen
  \bibfield  {author} {\bibinfo {author} {\bibfnamefont {B.}~\bibnamefont
  {Bruognolo}}, \bibinfo {author} {\bibfnamefont {J.-W.}\ \bibnamefont {Li}},
  \bibinfo {author} {\bibfnamefont {J.}~\bibnamefont {von Delft}}, \ and\
  \bibinfo {author} {\bibfnamefont {A.}~\bibnamefont {Weichselbaum}},\ }\href
  {\doibase 10.21468/SciPostPhysLectNotes.25} {\bibfield  {journal} {\bibinfo
  {journal} {SciPost Phys. Lect. Notes}\ ,\ \bibinfo {pages} {25}} (\bibinfo
  {year} {2021})}\BibitemShut {NoStop}%
\bibitem [{\citenamefont {Eisert}, \citenamefont {Cramer},\ and\ \citenamefont
  {Plenio}(2010)}]{Eisert2010}%
  \BibitemOpen
  \bibfield  {author} {\bibinfo {author} {\bibfnamefont {J.}~\bibnamefont
  {Eisert}}, \bibinfo {author} {\bibfnamefont {M.}~\bibnamefont {Cramer}}, \
  and\ \bibinfo {author} {\bibfnamefont {M.~B.}\ \bibnamefont {Plenio}},\
  }\href {\doibase 10.1103/RevModPhys.82.277} {\bibfield  {journal} {\bibinfo
  {journal} {Rev. Mod. Phys.}\ }\textbf {\bibinfo {volume} {82}},\ \bibinfo
  {pages} {277} (\bibinfo {year} {2010})}\BibitemShut {NoStop}%
\bibitem [{\citenamefont {Fishman}, \citenamefont {White},\ and\ \citenamefont
  {Stoudenmire}(2021)}]{itensor}%
  \BibitemOpen
  \bibfield  {author} {\bibinfo {author} {\bibfnamefont {M.}~\bibnamefont
  {Fishman}}, \bibinfo {author} {\bibfnamefont {S.~R.}\ \bibnamefont {White}},
  \ and\ \bibinfo {author} {\bibfnamefont {E.~M.}\ \bibnamefont
  {Stoudenmire}},\ }\href {http://arxiv.org/abs/2007.14822} {\bibfield
  {journal} {\bibinfo  {journal} {arXiv:2007.14822 [cond-mat,
  physics:physics]}\ } (\bibinfo {year} {2021})}\BibitemShut {NoStop}%
\bibitem [{\citenamefont {Ehrenfest}(1927)}]{Ehrenfest1927}%
  \BibitemOpen
  \bibfield  {author} {\bibinfo {author} {\bibfnamefont {P.}~\bibnamefont
  {Ehrenfest}},\ }\href {\doibase 10.1007/BF01329203} {\bibfield  {journal}
  {\bibinfo  {journal} {Z. Physik}\ }\textbf {\bibinfo {volume} {45}},\
  \bibinfo {pages} {455} (\bibinfo {year} {1927})}\BibitemShut {NoStop}%
\bibitem [{\citenamefont {Kirrander}\ and\ \citenamefont
  {Vacher}(2020)}]{Kirrander2020}%
  \BibitemOpen
  \bibfield  {author} {\bibinfo {author} {\bibfnamefont {A.}~\bibnamefont
  {Kirrander}}\ and\ \bibinfo {author} {\bibfnamefont {M.}~\bibnamefont
  {Vacher}},\ }\enquote {\bibinfo {title} {Ehrenfest {Methods} for {Electron}
  and {Nuclear} {Dynamics}},}\ in\ \href {\doibase 10.1002/9781119417774.ch15}
  {\emph {\bibinfo {booktitle} {Quantum {Chemistry} and {Dynamics} of {Excited}
  {States}}}},\ \bibinfo {editor} {edited by\ \bibinfo {editor} {\bibfnamefont
  {L.}~\bibnamefont {Gonz\'alez}}\ and\ \bibinfo {editor} {\bibfnamefont
  {R.}~\bibnamefont {Lindh}}}\ (\bibinfo  {publisher} {John Wiley \& Sons,
  Ltd},\ \bibinfo {year} {2020})\ pp.\ \bibinfo {pages} {469--497}\BibitemShut
  {NoStop}%
\bibitem [{\citenamefont {Kapral}\ and\ \citenamefont
  {Ciccotti}(1999)}]{Kapral1999}%
  \BibitemOpen
  \bibfield  {author} {\bibinfo {author} {\bibfnamefont {R.}~\bibnamefont
  {Kapral}}\ and\ \bibinfo {author} {\bibfnamefont {G.}~\bibnamefont
  {Ciccotti}},\ }\href {\doibase 10.1063/1.478811} {\bibfield  {journal}
  {\bibinfo  {journal} {J. Chem. Phys.}\ }\textbf {\bibinfo {volume} {110}},\
  \bibinfo {pages} {8919} (\bibinfo {year} {1999})}\BibitemShut {NoStop}%
\bibitem [{\citenamefont {Grunwald}, \citenamefont {Kelly},\ and\ \citenamefont
  {Kapral}(2009)}]{Grunwald2009}%
  \BibitemOpen
  \bibfield  {author} {\bibinfo {author} {\bibfnamefont {R.}~\bibnamefont
  {Grunwald}}, \bibinfo {author} {\bibfnamefont {A.}~\bibnamefont {Kelly}}, \
  and\ \bibinfo {author} {\bibfnamefont {R.}~\bibnamefont {Kapral}},\ }in\
  \href {\doibase 10.1007/978-3-642-02306-4_12} {\emph {\bibinfo {booktitle}
  {Energy {Transfer} {Dynamics} in {Biomaterial} {Systems}}}},\ \bibinfo
  {series and number} {Springer {Series} in {Chemical} {Physics}},\ \bibinfo
  {editor} {edited by\ \bibinfo {editor} {\bibfnamefont {I.}~\bibnamefont
  {Burghardt}}, \bibinfo {editor} {\bibfnamefont {V.}~\bibnamefont {May}},
  \bibinfo {editor} {\bibfnamefont {D.~A.}\ \bibnamefont {Micha}}, \ and\
  \bibinfo {editor} {\bibfnamefont {E.~R.}\ \bibnamefont {Bittner}}}\ (\bibinfo
   {publisher} {Springer},\ \bibinfo {address} {Berlin, Heidelberg},\ \bibinfo
  {year} {2009})\ pp.\ \bibinfo {pages} {383--413}\BibitemShut {NoStop}%
\bibitem [{\citenamefont {Ando}(2002)}]{Ando2002}%
  \BibitemOpen
  \bibfield  {author} {\bibinfo {author} {\bibfnamefont {K.}~\bibnamefont
  {Ando}},\ }\href {\doibase 10.1016/S0009-2614(02)00848-5} {\bibfield
  {journal} {\bibinfo  {journal} {Chem. Phys. Lett.}\ }\textbf {\bibinfo
  {volume} {360}},\ \bibinfo {pages} {240} (\bibinfo {year}
  {2002})}\BibitemShut {NoStop}%
\bibitem [{\citenamefont {Wigner}(1932)}]{Wigner1932}%
  \BibitemOpen
  \bibfield  {author} {\bibinfo {author} {\bibfnamefont {E.}~\bibnamefont
  {Wigner}},\ }\href {\doibase 10.1103/PhysRev.40.749} {\bibfield  {journal}
  {\bibinfo  {journal} {Phys. Rev.}\ }\textbf {\bibinfo {volume} {40}},\
  \bibinfo {pages} {749} (\bibinfo {year} {1932})}\BibitemShut {NoStop}%
\bibitem [{\citenamefont {Weyl}(1927)}]{Weyl1927}%
  \BibitemOpen
  \bibfield  {author} {\bibinfo {author} {\bibfnamefont {H.}~\bibnamefont
  {Weyl}},\ }\href {\doibase 10.1007/BF02055756} {\bibfield  {journal}
  {\bibinfo  {journal} {Z. Physik}\ }\textbf {\bibinfo {volume} {46}},\
  \bibinfo {pages} {1} (\bibinfo {year} {1927})}\BibitemShut {NoStop}%
\bibitem [{\citenamefont {Moyal}(1949)}]{Moyal1949}%
  \BibitemOpen
  \bibfield  {author} {\bibinfo {author} {\bibfnamefont {J.~E.}\ \bibnamefont
  {Moyal}},\ }\href {\doibase 10.1017/S0305004100000487} {\bibfield  {journal}
  {\bibinfo  {journal} {Math. Proc. Cambridge Philos. Soc.}\ }\textbf {\bibinfo
  {volume} {45}},\ \bibinfo {pages} {99} (\bibinfo {year} {1949})}\BibitemShut
  {NoStop}%
\bibitem [{\citenamefont {Hillery}\ \emph {et~al.}(1984)\citenamefont
  {Hillery}, \citenamefont {O'Connell}, \citenamefont {Scully},\ and\
  \citenamefont {Wigner}}]{Hillery1984}%
  \BibitemOpen
  \bibfield  {author} {\bibinfo {author} {\bibfnamefont {M.}~\bibnamefont
  {Hillery}}, \bibinfo {author} {\bibfnamefont {R.~F.}\ \bibnamefont
  {O'Connell}}, \bibinfo {author} {\bibfnamefont {M.~O.}\ \bibnamefont
  {Scully}}, \ and\ \bibinfo {author} {\bibfnamefont {E.~P.}\ \bibnamefont
  {Wigner}},\ }\href {\doibase 10.1016/0370-1573(84)90160-1} {\bibfield
  {journal} {\bibinfo  {journal} {Phys. Rep.}\ }\textbf {\bibinfo {volume}
  {106}},\ \bibinfo {pages} {121} (\bibinfo {year} {1984})}\BibitemShut
  {NoStop}%
\bibitem [{\citenamefont {Case}(2008)}]{Case2008}%
  \BibitemOpen
  \bibfield  {author} {\bibinfo {author} {\bibfnamefont {W.~B.}\ \bibnamefont
  {Case}},\ }\href {\doibase 10.1119/1.2957889} {\bibfield  {journal} {\bibinfo
   {journal} {Am. J. Phys.}\ }\textbf {\bibinfo {volume} {76}},\ \bibinfo
  {pages} {937} (\bibinfo {year} {2008})}\BibitemShut {NoStop}%
\bibitem [{\citenamefont {Horenko}, \citenamefont {Schmidt},\ and\
  \citenamefont {Sch{\"u}tte}(2001)}]{Horenko2001}%
  \BibitemOpen
  \bibfield  {author} {\bibinfo {author} {\bibfnamefont {I.}~\bibnamefont
  {Horenko}}, \bibinfo {author} {\bibfnamefont {B.}~\bibnamefont {Schmidt}}, \
  and\ \bibinfo {author} {\bibfnamefont {C.}~\bibnamefont {Sch{\"u}tte}},\
  }\href {\doibase 10.1063/1.1398577} {\bibfield  {journal} {\bibinfo
  {journal} {J. Chem. Phys.}\ }\textbf {\bibinfo {volume} {115}},\ \bibinfo
  {pages} {5733} (\bibinfo {year} {2001})}\BibitemShut {NoStop}%
\bibitem [{\citenamefont {Ando}\ and\ \citenamefont {Santer}(2003)}]{Ando2003}%
  \BibitemOpen
  \bibfield  {author} {\bibinfo {author} {\bibfnamefont {K.}~\bibnamefont
  {Ando}}\ and\ \bibinfo {author} {\bibfnamefont {M.}~\bibnamefont {Santer}},\
  }\href {\doibase 10.1063/1.1574015} {\bibfield  {journal} {\bibinfo
  {journal} {J. Chem. Phys.}\ }\textbf {\bibinfo {volume} {118}},\ \bibinfo
  {pages} {10399} (\bibinfo {year} {2003})}\BibitemShut {NoStop}%
\bibitem [{\citenamefont {Subotnik}, \citenamefont {Ouyang},\ and\
  \citenamefont {Landry}(2013)}]{Subotnik2013}%
  \BibitemOpen
  \bibfield  {author} {\bibinfo {author} {\bibfnamefont {J.~E.}\ \bibnamefont
  {Subotnik}}, \bibinfo {author} {\bibfnamefont {W.}~\bibnamefont {Ouyang}}, \
  and\ \bibinfo {author} {\bibfnamefont {B.~R.}\ \bibnamefont {Landry}},\
  }\href {\doibase 10.1063/1.4829856} {\bibfield  {journal} {\bibinfo
  {journal} {J. Chem. Phys.}\ }\textbf {\bibinfo {volume} {139}},\ \bibinfo
  {pages} {214107} (\bibinfo {year} {2013})}\BibitemShut {NoStop}%
\bibitem [{\citenamefont {Ryabinkin}\ \emph {et~al.}(2014)\citenamefont
  {Ryabinkin}, \citenamefont {Hsieh}, \citenamefont {Kapral},\ and\
  \citenamefont {Izmaylov}}]{Ryabinkin2014}%
  \BibitemOpen
  \bibfield  {author} {\bibinfo {author} {\bibfnamefont {I.~G.}\ \bibnamefont
  {Ryabinkin}}, \bibinfo {author} {\bibfnamefont {C.-Y.}\ \bibnamefont
  {Hsieh}}, \bibinfo {author} {\bibfnamefont {R.}~\bibnamefont {Kapral}}, \
  and\ \bibinfo {author} {\bibfnamefont {A.~F.}\ \bibnamefont {Izmaylov}},\
  }\href {\doibase 10.1063/1.4866366} {\bibfield  {journal} {\bibinfo
  {journal} {J. Chem. Phys.}\ }\textbf {\bibinfo {volume} {140}},\ \bibinfo
  {pages} {084104} (\bibinfo {year} {2014})}\BibitemShut {NoStop}%
\bibitem [{\citenamefont {Brown}\ and\ \citenamefont
  {Heller}(1981)}]{Brown1981}%
  \BibitemOpen
  \bibfield  {author} {\bibinfo {author} {\bibfnamefont {R.~C.}\ \bibnamefont
  {Brown}}\ and\ \bibinfo {author} {\bibfnamefont {E.~J.}\ \bibnamefont
  {Heller}},\ }\href {\doibase 10.1063/1.441822} {\bibfield  {journal}
  {\bibinfo  {journal} {J. Chem. Phys.}\ }\textbf {\bibinfo {volume} {75}},\
  \bibinfo {pages} {186} (\bibinfo {year} {1981})}\BibitemShut {NoStop}%
\bibitem [{\citenamefont {Persico}\ and\ \citenamefont
  {Granucci}(2014)}]{Persico2014}%
  \BibitemOpen
  \bibfield  {author} {\bibinfo {author} {\bibfnamefont {M.}~\bibnamefont
  {Persico}}\ and\ \bibinfo {author} {\bibfnamefont {G.}~\bibnamefont
  {Granucci}},\ }\href {\doibase 10.1007/s00214-014-1526-1} {\bibfield
  {journal} {\bibinfo  {journal} {Theor. Chem. Acc.}\ }\textbf {\bibinfo
  {volume} {133}},\ \bibinfo {pages} {1526} (\bibinfo {year}
  {2014})}\BibitemShut {NoStop}%
\bibitem [{\citenamefont {Barbatti}\ and\ \citenamefont
  {Sen}(2016)}]{Barbatti2016}%
  \BibitemOpen
  \bibfield  {author} {\bibinfo {author} {\bibfnamefont {M.}~\bibnamefont
  {Barbatti}}\ and\ \bibinfo {author} {\bibfnamefont {K.}~\bibnamefont {Sen}},\
  }\href {\doibase 10.1002/qua.25049} {\bibfield  {journal} {\bibinfo
  {journal} {Int. J. Quantum Chem.}\ }\textbf {\bibinfo {volume} {116}},\
  \bibinfo {pages} {762} (\bibinfo {year} {2016})}\BibitemShut {NoStop}%
\bibitem [{\citenamefont {Suchan}\ \emph {et~al.}(2018)\citenamefont {Suchan},
  \citenamefont {Hollas}, \citenamefont {Curchod},\ and\ \citenamefont
  {Slav{\'i}{\v c}ek}}]{Suchan2018}%
  \BibitemOpen
  \bibfield  {author} {\bibinfo {author} {\bibfnamefont {J.}~\bibnamefont
  {Suchan}}, \bibinfo {author} {\bibfnamefont {D.}~\bibnamefont {Hollas}},
  \bibinfo {author} {\bibfnamefont {B.~F.~E.}\ \bibnamefont {Curchod}}, \ and\
  \bibinfo {author} {\bibfnamefont {P.}~\bibnamefont {Slav{\'i}{\v c}ek}},\
  }\href {\doibase 10.1039/C8FD00088C} {\bibfield  {journal} {\bibinfo
  {journal} {Faraday Discuss.}\ }\textbf {\bibinfo {volume} {212}},\ \bibinfo
  {pages} {307} (\bibinfo {year} {2018})}\BibitemShut {NoStop}%
\bibitem [{\citenamefont {Mai}\ \emph {et~al.}(2018)\citenamefont {Mai},
  \citenamefont {Gattuso}, \citenamefont {Monari},\ and\ \citenamefont
  {Gonz{\'a}lez}}]{Mai2018}%
  \BibitemOpen
  \bibfield  {author} {\bibinfo {author} {\bibfnamefont {S.}~\bibnamefont
  {Mai}}, \bibinfo {author} {\bibfnamefont {H.}~\bibnamefont {Gattuso}},
  \bibinfo {author} {\bibfnamefont {A.}~\bibnamefont {Monari}}, \ and\ \bibinfo
  {author} {\bibfnamefont {L.}~\bibnamefont {Gonz{\'a}lez}},\ }\href
  {https://www.frontiersin.org/article/10.3389/fchem.2018.00495} {\bibfield
  {journal} {\bibinfo  {journal} {Front. Chem.}\ }\textbf {\bibinfo {volume}
  {6}},\ \bibinfo {pages} {495} (\bibinfo {year} {2018})}\BibitemShut {NoStop}%
\bibitem [{\citenamefont {Polkovnikov}(2010)}]{Polkovnikov2010}%
  \BibitemOpen
  \bibfield  {author} {\bibinfo {author} {\bibfnamefont {A.}~\bibnamefont
  {Polkovnikov}},\ }\href {\doibase 10.1016/j.aop.2010.02.006} {\bibfield
  {journal} {\bibinfo  {journal} {Ann. Phys.}\ }\textbf {\bibinfo {volume}
  {325}},\ \bibinfo {pages} {1790} (\bibinfo {year} {2010})}\BibitemShut
  {NoStop}%
\bibitem [{\citenamefont {Garc{\'i}a-Vela}, \citenamefont {Gerber},\ and\
  \citenamefont {Imre}(1992)}]{Garciavela1992}%
  \BibitemOpen
  \bibfield  {author} {\bibinfo {author} {\bibfnamefont {A.}~\bibnamefont
  {Garc{\'i}a-Vela}}, \bibinfo {author} {\bibfnamefont {R.~B.}\ \bibnamefont
  {Gerber}}, \ and\ \bibinfo {author} {\bibfnamefont {D.~G.}\ \bibnamefont
  {Imre}},\ }\href {\doibase 10.1063/1.463550} {\bibfield  {journal} {\bibinfo
  {journal} {J. Chem. Phys.}\ }\textbf {\bibinfo {volume} {97}},\ \bibinfo
  {pages} {7242} (\bibinfo {year} {1992})}\BibitemShut {NoStop}%
\bibitem [{\citenamefont {Topaler}\ \emph {et~al.}(1998)\citenamefont
  {Topaler}, \citenamefont {Allison}, \citenamefont {Schwenke},\ and\
  \citenamefont {Truhlar}}]{Topaler1998}%
  \BibitemOpen
  \bibfield  {author} {\bibinfo {author} {\bibfnamefont {M.~S.}\ \bibnamefont
  {Topaler}}, \bibinfo {author} {\bibfnamefont {T.~C.}\ \bibnamefont
  {Allison}}, \bibinfo {author} {\bibfnamefont {D.~W.}\ \bibnamefont
  {Schwenke}}, \ and\ \bibinfo {author} {\bibfnamefont {D.~G.}\ \bibnamefont
  {Truhlar}},\ }\href {\doibase 10.1063/1.477684} {\bibfield  {journal}
  {\bibinfo  {journal} {J. Chem. Phys.}\ }\textbf {\bibinfo {volume} {109}},\
  \bibinfo {pages} {3321} (\bibinfo {year} {1998})}\BibitemShut {NoStop}%
\bibitem [{\citenamefont {Marx}\ and\ \citenamefont {Hutter}(2009)}]{Marx2009}%
  \BibitemOpen
  \bibfield  {author} {\bibinfo {author} {\bibfnamefont {D.}~\bibnamefont
  {Marx}}\ and\ \bibinfo {author} {\bibfnamefont {J.}~\bibnamefont {Hutter}},\
  }\href {\doibase 10.1017/CBO9780511609633} {\emph {\bibinfo {title} {Ab
  {Initio} {Molecular} {Dynamics}: {Basic} {Theory} and {Advanced}
  {Methods}}}}\ (\bibinfo  {publisher} {Cambridge University Press},\ \bibinfo
  {address} {Cambridge},\ \bibinfo {year} {2009})\BibitemShut {NoStop}%
\bibitem [{\citenamefont {Gerasimenko}(1982)}]{Gerasimenko1982}%
  \BibitemOpen
  \bibfield  {author} {\bibinfo {author} {\bibfnamefont {V.~I.}\ \bibnamefont
  {Gerasimenko}},\ }\href {\doibase 10.1007/BF01027604} {\bibfield  {journal}
  {\bibinfo  {journal} {Theor. Math. Phys.}\ }\textbf {\bibinfo {volume}
  {50}},\ \bibinfo {pages} {49} (\bibinfo {year} {1982})}\BibitemShut {NoStop}%
\bibitem [{\citenamefont {Yonehara}, \citenamefont {Hanasaki},\ and\
  \citenamefont {Takatsuka}(2012)}]{Yonehara2012}%
  \BibitemOpen
  \bibfield  {author} {\bibinfo {author} {\bibfnamefont {T.}~\bibnamefont
  {Yonehara}}, \bibinfo {author} {\bibfnamefont {K.}~\bibnamefont {Hanasaki}},
  \ and\ \bibinfo {author} {\bibfnamefont {K.}~\bibnamefont {Takatsuka}},\
  }\href {\doibase 10.1021/cr200096s} {\bibfield  {journal} {\bibinfo
  {journal} {Chem. Rev.}\ }\textbf {\bibinfo {volume} {112}},\ \bibinfo {pages}
  {499} (\bibinfo {year} {2012})}\BibitemShut {NoStop}%
\bibitem [{\citenamefont {Jasper}\ \emph {et~al.}(2006)\citenamefont {Jasper},
  \citenamefont {Nangia}, \citenamefont {Zhu},\ and\ \citenamefont
  {Truhlar}}]{Jasper2006}%
  \BibitemOpen
  \bibfield  {author} {\bibinfo {author} {\bibfnamefont {A.~W.}\ \bibnamefont
  {Jasper}}, \bibinfo {author} {\bibfnamefont {S.}~\bibnamefont {Nangia}},
  \bibinfo {author} {\bibfnamefont {C.}~\bibnamefont {Zhu}}, \ and\ \bibinfo
  {author} {\bibfnamefont {D.~G.}\ \bibnamefont {Truhlar}},\ }\href {\doibase
  10.1021/ar040206v} {\bibfield  {journal} {\bibinfo  {journal} {Acc. Chem.
  Res.}\ }\textbf {\bibinfo {volume} {39}},\ \bibinfo {pages} {101} (\bibinfo
  {year} {2006})}\BibitemShut {NoStop}%
\bibitem [{\citenamefont {Bjerre}\ and\ \citenamefont
  {Nikitin}(1967)}]{Bjerre1967}%
  \BibitemOpen
  \bibfield  {author} {\bibinfo {author} {\bibfnamefont {A.}~\bibnamefont
  {Bjerre}}\ and\ \bibinfo {author} {\bibfnamefont {E.~E.}\ \bibnamefont
  {Nikitin}},\ }\href {\doibase 10.1016/0009-2614(67)85041-3} {\bibfield
  {journal} {\bibinfo  {journal} {Chem. Phys. Lett.}\ }\textbf {\bibinfo
  {volume} {1}},\ \bibinfo {pages} {179} (\bibinfo {year} {1967})}\BibitemShut
  {NoStop}%
\bibitem [{\citenamefont {Tully}\ and\ \citenamefont
  {Preston}(1971)}]{Tully1971}%
  \BibitemOpen
  \bibfield  {author} {\bibinfo {author} {\bibfnamefont {J.~C.}\ \bibnamefont
  {Tully}}\ and\ \bibinfo {author} {\bibfnamefont {R.~K.}\ \bibnamefont
  {Preston}},\ }\href {\doibase 10.1063/1.1675788} {\bibfield  {journal}
  {\bibinfo  {journal} {J. Chem. Phys.}\ }\textbf {\bibinfo {volume} {55}},\
  \bibinfo {pages} {562} (\bibinfo {year} {1971})}\BibitemShut {NoStop}%
\bibitem [{\citenamefont {M\"uller}\ and\ \citenamefont
  {Stock}(1997)}]{Muller1997}%
  \BibitemOpen
  \bibfield  {author} {\bibinfo {author} {\bibfnamefont {U.}~\bibnamefont
  {M\"uller}}\ and\ \bibinfo {author} {\bibfnamefont {G.}~\bibnamefont
  {Stock}},\ }\href {\doibase 10.1063/1.474288} {\bibfield  {journal} {\bibinfo
   {journal} {J. Chem. Phys.}\ }\textbf {\bibinfo {volume} {107}},\ \bibinfo
  {pages} {6230} (\bibinfo {year} {1997})}\BibitemShut {NoStop}%
\bibitem [{\citenamefont {Granucci}\ and\ \citenamefont
  {Persico}(2007)}]{Granucci2007}%
  \BibitemOpen
  \bibfield  {author} {\bibinfo {author} {\bibfnamefont {G.}~\bibnamefont
  {Granucci}}\ and\ \bibinfo {author} {\bibfnamefont {M.}~\bibnamefont
  {Persico}},\ }\href {\doibase 10.1063/1.2715585} {\bibfield  {journal}
  {\bibinfo  {journal} {J. Chem. Phys.}\ }\textbf {\bibinfo {volume} {126}},\
  \bibinfo {pages} {134114} (\bibinfo {year} {2007})}\BibitemShut {NoStop}%
\bibitem [{\citenamefont {Landry}, \citenamefont {Falk},\ and\ \citenamefont
  {Subotnik}(2013)}]{Landry2013}%
  \BibitemOpen
  \bibfield  {author} {\bibinfo {author} {\bibfnamefont {B.~R.}\ \bibnamefont
  {Landry}}, \bibinfo {author} {\bibfnamefont {M.~J.}\ \bibnamefont {Falk}}, \
  and\ \bibinfo {author} {\bibfnamefont {J.~E.}\ \bibnamefont {Subotnik}},\
  }\href {\doibase 10.1063/1.4837795} {\bibfield  {journal} {\bibinfo
  {journal} {J. Chem. Phys.}\ }\textbf {\bibinfo {volume} {139}},\ \bibinfo
  {pages} {211101} (\bibinfo {year} {2013})}\BibitemShut {NoStop}%
\bibitem [{\citenamefont {Parandekar}\ and\ \citenamefont
  {Tully}(2005)}]{Parandekar2005}%
  \BibitemOpen
  \bibfield  {author} {\bibinfo {author} {\bibfnamefont {P.~V.}\ \bibnamefont
  {Parandekar}}\ and\ \bibinfo {author} {\bibfnamefont {J.~C.}\ \bibnamefont
  {Tully}},\ }\href {\doibase 10.1063/1.1856460} {\bibfield  {journal}
  {\bibinfo  {journal} {J. Chem. Phys.}\ }\textbf {\bibinfo {volume} {122}},\
  \bibinfo {pages} {094102} (\bibinfo {year} {2005})}\BibitemShut {NoStop}%
\bibitem [{\citenamefont {Schmidt}, \citenamefont {Parandekar},\ and\
  \citenamefont {Tully}(2008)}]{Schmidt2008}%
  \BibitemOpen
  \bibfield  {author} {\bibinfo {author} {\bibfnamefont {J.~R.}\ \bibnamefont
  {Schmidt}}, \bibinfo {author} {\bibfnamefont {P.~V.}\ \bibnamefont
  {Parandekar}}, \ and\ \bibinfo {author} {\bibfnamefont {J.~C.}\ \bibnamefont
  {Tully}},\ }\href {\doibase 10.1063/1.2955564} {\bibfield  {journal}
  {\bibinfo  {journal} {J. Chem. Phys.}\ }\textbf {\bibinfo {volume} {129}},\
  \bibinfo {pages} {044104} (\bibinfo {year} {2008})}\BibitemShut {NoStop}%
\bibitem [{\citenamefont {Carof}, \citenamefont {Giannini},\ and\ \citenamefont
  {Blumberger}(2019)}]{Carof2019}%
  \BibitemOpen
  \bibfield  {author} {\bibinfo {author} {\bibfnamefont {A.}~\bibnamefont
  {Carof}}, \bibinfo {author} {\bibfnamefont {S.}~\bibnamefont {Giannini}}, \
  and\ \bibinfo {author} {\bibfnamefont {J.}~\bibnamefont {Blumberger}},\
  }\href {\doibase 10.1039/C9CP04770K} {\bibfield  {journal} {\bibinfo
  {journal} {Phys. Chem. Chem. Phys.}\ }\textbf {\bibinfo {volume} {21}},\
  \bibinfo {pages} {26368} (\bibinfo {year} {2019})}\BibitemShut {NoStop}%
\bibitem [{\citenamefont {Crespo-Otero}\ and\ \citenamefont
  {Barbatti}(2018)}]{Crespo-Otero2018}%
  \BibitemOpen
  \bibfield  {author} {\bibinfo {author} {\bibfnamefont {R.}~\bibnamefont
  {Crespo-Otero}}\ and\ \bibinfo {author} {\bibfnamefont {M.}~\bibnamefont
  {Barbatti}},\ }\href {\doibase 10.1021/acs.chemrev.7b00577} {\bibfield
  {journal} {\bibinfo  {journal} {Chem. Rev.}\ }\textbf {\bibinfo {volume}
  {118}},\ \bibinfo {pages} {7026} (\bibinfo {year} {2018})}\BibitemShut
  {NoStop}%
\bibitem [{\citenamefont {Schwartz}\ \emph {et~al.}(1996)\citenamefont
  {Schwartz}, \citenamefont {Bittner}, \citenamefont {Prezhdo},\ and\
  \citenamefont {Rossky}}]{Schwartz1996}%
  \BibitemOpen
  \bibfield  {author} {\bibinfo {author} {\bibfnamefont {B.~J.}\ \bibnamefont
  {Schwartz}}, \bibinfo {author} {\bibfnamefont {E.~R.}\ \bibnamefont
  {Bittner}}, \bibinfo {author} {\bibfnamefont {O.~V.}\ \bibnamefont
  {Prezhdo}}, \ and\ \bibinfo {author} {\bibfnamefont {P.~J.}\ \bibnamefont
  {Rossky}},\ }\href {\doibase 10.1063/1.471326} {\bibfield  {journal}
  {\bibinfo  {journal} {J. Chem. Phys.}\ }\textbf {\bibinfo {volume} {104}},\
  \bibinfo {pages} {5942} (\bibinfo {year} {1996})}\BibitemShut {NoStop}%
\bibitem [{\citenamefont {Bittner}\ and\ \citenamefont
  {Rossky}(1995)}]{Bittner1995}%
  \BibitemOpen
  \bibfield  {author} {\bibinfo {author} {\bibfnamefont {E.~R.}\ \bibnamefont
  {Bittner}}\ and\ \bibinfo {author} {\bibfnamefont {P.~J.}\ \bibnamefont
  {Rossky}},\ }\href {\doibase 10.1063/1.470177} {\bibfield  {journal}
  {\bibinfo  {journal} {J. Chem. Phys.}\ }\textbf {\bibinfo {volume} {103}},\
  \bibinfo {pages} {8130} (\bibinfo {year} {1995})}\BibitemShut {NoStop}%
\bibitem [{\citenamefont {Heller}(1975)}]{Heller1975}%
  \BibitemOpen
  \bibfield  {author} {\bibinfo {author} {\bibfnamefont {E.~J.}\ \bibnamefont
  {Heller}},\ }\href {\doibase 10.1063/1.430620} {\bibfield  {journal}
  {\bibinfo  {journal} {J. Chem. Phys.}\ }\textbf {\bibinfo {volume} {62}},\
  \bibinfo {pages} {1544} (\bibinfo {year} {1975})}\BibitemShut {NoStop}%
\bibitem [{\citenamefont {Neria}\ and\ \citenamefont
  {Nitzan}(1993)}]{Neria1993}%
  \BibitemOpen
  \bibfield  {author} {\bibinfo {author} {\bibfnamefont {E.}~\bibnamefont
  {Neria}}\ and\ \bibinfo {author} {\bibfnamefont {A.}~\bibnamefont {Nitzan}},\
  }\href {\doibase 10.1063/1.465409} {\bibfield  {journal} {\bibinfo  {journal}
  {J. Chem. Phys.}\ }\textbf {\bibinfo {volume} {99}},\ \bibinfo {pages} {1109}
  (\bibinfo {year} {1993})}\BibitemShut {NoStop}%
\bibitem [{\citenamefont {Granucci}, \citenamefont {Persico},\ and\
  \citenamefont {Toniolo}(2001)}]{Granucci2001}%
  \BibitemOpen
  \bibfield  {author} {\bibinfo {author} {\bibfnamefont {G.}~\bibnamefont
  {Granucci}}, \bibinfo {author} {\bibfnamefont {M.}~\bibnamefont {Persico}}, \
  and\ \bibinfo {author} {\bibfnamefont {A.}~\bibnamefont {Toniolo}},\ }\href
  {\doibase 10.1063/1.1376633} {\bibfield  {journal} {\bibinfo  {journal} {J.
  Chem. Phys.}\ }\textbf {\bibinfo {volume} {114}},\ \bibinfo {pages} {10608}
  (\bibinfo {year} {2001})}\BibitemShut {NoStop}%
\bibitem [{\citenamefont {Fernandez-Alberti}\ \emph {et~al.}(2012)\citenamefont
  {Fernandez-Alberti}, \citenamefont {Roitberg}, \citenamefont {Nelson},\ and\
  \citenamefont {Tretiak}}]{Fernandez-Alberti2012}%
  \BibitemOpen
  \bibfield  {author} {\bibinfo {author} {\bibfnamefont {S.}~\bibnamefont
  {Fernandez-Alberti}}, \bibinfo {author} {\bibfnamefont {A.~E.}\ \bibnamefont
  {Roitberg}}, \bibinfo {author} {\bibfnamefont {T.}~\bibnamefont {Nelson}}, \
  and\ \bibinfo {author} {\bibfnamefont {S.}~\bibnamefont {Tretiak}},\ }\href
  {\doibase 10.1063/1.4732536} {\bibfield  {journal} {\bibinfo  {journal} {J.
  Chem. Phys.}\ }\textbf {\bibinfo {volume} {137}},\ \bibinfo {pages} {014512}
  (\bibinfo {year} {2012})}\BibitemShut {NoStop}%
\bibitem [{\citenamefont {Wang}\ and\ \citenamefont
  {Prezhdo}(2014)}]{Wang2014}%
  \BibitemOpen
  \bibfield  {author} {\bibinfo {author} {\bibfnamefont {L.}~\bibnamefont
  {Wang}}\ and\ \bibinfo {author} {\bibfnamefont {O.~V.}\ \bibnamefont
  {Prezhdo}},\ }\href {\doibase 10.1021/jz500025c} {\bibfield  {journal}
  {\bibinfo  {journal} {J. Phys. Chem. Lett.}\ }\textbf {\bibinfo {volume}
  {5}},\ \bibinfo {pages} {713} (\bibinfo {year} {2014})}\BibitemShut {NoStop}%
\bibitem [{\citenamefont {Mai}, \citenamefont {Marquetand},\ and\ \citenamefont
  {Gonz\'alez}(2020)}]{Mai2020}%
  \BibitemOpen
  \bibfield  {author} {\bibinfo {author} {\bibfnamefont {S.}~\bibnamefont
  {Mai}}, \bibinfo {author} {\bibfnamefont {P.}~\bibnamefont {Marquetand}}, \
  and\ \bibinfo {author} {\bibfnamefont {L.}~\bibnamefont {Gonz\'alez}},\
  }\enquote {\bibinfo {title} {Surface {Hopping} {Molecular} {Dynamics}},}\ in\
  \href {\doibase 10.1002/9781119417774.ch16} {\emph {\bibinfo {booktitle}
  {Quantum {Chemistry} and {Dynamics} of {Excited} {States}}}},\ \bibinfo
  {editor} {edited by\ \bibinfo {editor} {\bibfnamefont {L.}~\bibnamefont
  {Gonz\'alez}}\ and\ \bibinfo {editor} {\bibfnamefont {R.}~\bibnamefont
  {Lindh}}}\ (\bibinfo  {publisher} {John Wiley \& Sons, Ltd},\ \bibinfo {year}
  {2020})\ pp.\ \bibinfo {pages} {499--530}\BibitemShut {NoStop}%
\bibitem [{\citenamefont {Mai}, \citenamefont {Marquetand},\ and\ \citenamefont
  {Gonz\'alez}(2015)}]{Mai2015}%
  \BibitemOpen
  \bibfield  {author} {\bibinfo {author} {\bibfnamefont {S.}~\bibnamefont
  {Mai}}, \bibinfo {author} {\bibfnamefont {P.}~\bibnamefont {Marquetand}}, \
  and\ \bibinfo {author} {\bibfnamefont {L.}~\bibnamefont {Gonz\'alez}},\
  }\href {\doibase 10.1002/qua.24891} {\bibfield  {journal} {\bibinfo
  {journal} {Int. J. Quantum Chem.}\ }\textbf {\bibinfo {volume} {115}},\
  \bibinfo {pages} {1215} (\bibinfo {year} {2015})}\BibitemShut {NoStop}%
\bibitem [{\citenamefont {Qiu}, \citenamefont {Bai},\ and\ \citenamefont
  {Wang}(2018)}]{Qiu2018}%
  \BibitemOpen
  \bibfield  {author} {\bibinfo {author} {\bibfnamefont {J.}~\bibnamefont
  {Qiu}}, \bibinfo {author} {\bibfnamefont {X.}~\bibnamefont {Bai}}, \ and\
  \bibinfo {author} {\bibfnamefont {L.}~\bibnamefont {Wang}},\ }\href {\doibase
  10.1021/acs.jpclett.8b01902} {\bibfield  {journal} {\bibinfo  {journal} {J.
  Phys. Chem. Lett.}\ }\textbf {\bibinfo {volume} {9}},\ \bibinfo {pages}
  {4319} (\bibinfo {year} {2018})}\BibitemShut {NoStop}%
\bibitem [{\citenamefont {Bai}, \citenamefont {Qiu},\ and\ \citenamefont
  {Wang}(2018)}]{Bai2018}%
  \BibitemOpen
  \bibfield  {author} {\bibinfo {author} {\bibfnamefont {X.}~\bibnamefont
  {Bai}}, \bibinfo {author} {\bibfnamefont {J.}~\bibnamefont {Qiu}}, \ and\
  \bibinfo {author} {\bibfnamefont {L.}~\bibnamefont {Wang}},\ }\href {\doibase
  10.1063/1.5020693} {\bibfield  {journal} {\bibinfo  {journal} {J. Chem.
  Phys.}\ }\textbf {\bibinfo {volume} {148}},\ \bibinfo {pages} {104106}
  (\bibinfo {year} {2018})}\BibitemShut {NoStop}%
\bibitem [{\citenamefont {Giannini}, \citenamefont {Carof},\ and\ \citenamefont
  {Blumberger}(2018)}]{Giannini2018}%
  \BibitemOpen
  \bibfield  {author} {\bibinfo {author} {\bibfnamefont {S.}~\bibnamefont
  {Giannini}}, \bibinfo {author} {\bibfnamefont {A.}~\bibnamefont {Carof}}, \
  and\ \bibinfo {author} {\bibfnamefont {J.}~\bibnamefont {Blumberger}},\
  }\href {\doibase 10.1021/acs.jpclett.8b01112} {\bibfield  {journal} {\bibinfo
   {journal} {J. Phys. Chem. Lett.}\ }\textbf {\bibinfo {volume} {9}},\
  \bibinfo {pages} {3116} (\bibinfo {year} {2018})}\BibitemShut {NoStop}%
\bibitem [{\citenamefont {Qiu}, \citenamefont {Bai},\ and\ \citenamefont
  {Wang}(2019)}]{Qiu2019}%
  \BibitemOpen
  \bibfield  {author} {\bibinfo {author} {\bibfnamefont {J.}~\bibnamefont
  {Qiu}}, \bibinfo {author} {\bibfnamefont {X.}~\bibnamefont {Bai}}, \ and\
  \bibinfo {author} {\bibfnamefont {L.}~\bibnamefont {Wang}},\ }\href {\doibase
  10.1021/acs.jpclett.8b03763} {\bibfield  {journal} {\bibinfo  {journal} {J.
  Phys. Chem. Lett.}\ }\textbf {\bibinfo {volume} {10}},\ \bibinfo {pages}
  {637} (\bibinfo {year} {2019})}\BibitemShut {NoStop}%
\bibitem [{\citenamefont {Wang}\ and\ \citenamefont
  {Beljonne}(2013)}]{Wang2013}%
  \BibitemOpen
  \bibfield  {author} {\bibinfo {author} {\bibfnamefont {L.}~\bibnamefont
  {Wang}}\ and\ \bibinfo {author} {\bibfnamefont {D.}~\bibnamefont
  {Beljonne}},\ }\href {\doibase 10.1021/jz400871j} {\bibfield  {journal}
  {\bibinfo  {journal} {J. Phys. Chem. Lett.}\ }\textbf {\bibinfo {volume}
  {4}},\ \bibinfo {pages} {1888} (\bibinfo {year} {2013})}\BibitemShut
  {NoStop}%
\bibitem [{\citenamefont {Qiu}, \citenamefont {Lu},\ and\ \citenamefont
  {Wang}(2022)}]{Qiu2022}%
  \BibitemOpen
  \bibfield  {author} {\bibinfo {author} {\bibfnamefont {J.}~\bibnamefont
  {Qiu}}, \bibinfo {author} {\bibfnamefont {Y.}~\bibnamefont {Lu}}, \ and\
  \bibinfo {author} {\bibfnamefont {L.}~\bibnamefont {Wang}},\ }\href {\doibase
  10.1021/acs.jctc.2c00130} {\bibfield  {journal} {\bibinfo  {journal} {J.
  Chem. Theory Comput.}\ }\textbf {\bibinfo {volume} {18}},\ \bibinfo {pages}
  {2803} (\bibinfo {year} {2022})}\BibitemShut {NoStop}%
\bibitem [{\citenamefont {Burghardt}, \citenamefont {Meyer},\ and\
  \citenamefont {Cederbaum}(1999)}]{Burghardt99}%
  \BibitemOpen
  \bibfield  {author} {\bibinfo {author} {\bibfnamefont {I.}~\bibnamefont
  {Burghardt}}, \bibinfo {author} {\bibfnamefont {H.-D.}\ \bibnamefont
  {Meyer}}, \ and\ \bibinfo {author} {\bibfnamefont {L.~S.}\ \bibnamefont
  {Cederbaum}},\ }\href {\doibase 10.1063/1.479574} {\bibfield  {journal}
  {\bibinfo  {journal} {J. Chem. Phys.}\ }\textbf {\bibinfo {volume} {111}},\
  \bibinfo {pages} {2927} (\bibinfo {year} {1999})}\BibitemShut {NoStop}%
\bibitem [{\citenamefont {Burghardt}, \citenamefont {Nest},\ and\ \citenamefont
  {Worth}(2003)}]{Burghardt03}%
  \BibitemOpen
  \bibfield  {author} {\bibinfo {author} {\bibfnamefont {I.}~\bibnamefont
  {Burghardt}}, \bibinfo {author} {\bibfnamefont {M.}~\bibnamefont {Nest}}, \
  and\ \bibinfo {author} {\bibfnamefont {G.~A.}\ \bibnamefont {Worth}},\ }\href
  {\doibase 10.1063/1.1599275} {\bibfield  {journal} {\bibinfo  {journal} {J.
  Chem. Phys.}\ }\textbf {\bibinfo {volume} {119}},\ \bibinfo {pages} {5364}
  (\bibinfo {year} {2003})}\BibitemShut {NoStop}%
\bibitem [{\citenamefont {Burghardt}, \citenamefont {Giri},\ and\ \citenamefont
  {Worth}(2008)}]{Burghardt08}%
  \BibitemOpen
  \bibfield  {author} {\bibinfo {author} {\bibfnamefont {I.}~\bibnamefont
  {Burghardt}}, \bibinfo {author} {\bibfnamefont {K.}~\bibnamefont {Giri}}, \
  and\ \bibinfo {author} {\bibfnamefont {G.~A.}\ \bibnamefont {Worth}},\ }\href
  {\doibase 10.1063/1.2996349} {\bibfield  {journal} {\bibinfo  {journal} {J.
  Chem. Phys.}\ }\textbf {\bibinfo {volume} {129}},\ \bibinfo {pages} {174104}
  (\bibinfo {year} {2008})}\BibitemShut {NoStop}%
\bibitem [{\citenamefont {Worth}, \citenamefont {Robb},\ and\ \citenamefont
  {Burghardt}(2004)}]{Worth04}%
  \BibitemOpen
  \bibfield  {author} {\bibinfo {author} {\bibfnamefont {G.~A.}\ \bibnamefont
  {Worth}}, \bibinfo {author} {\bibfnamefont {M.~A.}\ \bibnamefont {Robb}}, \
  and\ \bibinfo {author} {\bibfnamefont {I.}~\bibnamefont {Burghardt}},\ }\href
  {\doibase 10.1039/B314253A} {\bibfield  {journal} {\bibinfo  {journal}
  {Faraday Discuss.}\ }\textbf {\bibinfo {volume} {127}},\ \bibinfo {pages}
  {307} (\bibinfo {year} {2004})}\BibitemShut {NoStop}%
\bibitem [{\citenamefont {Worth}\ and\ \citenamefont
  {Burghardt}(2003)}]{Worth03}%
  \BibitemOpen
  \bibfield  {author} {\bibinfo {author} {\bibfnamefont {G.~A.}\ \bibnamefont
  {Worth}}\ and\ \bibinfo {author} {\bibfnamefont {I.}~\bibnamefont
  {Burghardt}},\ }\href {\doibase
  https://doi.org/10.1016/S0009-2614(02)01920-6} {\bibfield  {journal}
  {\bibinfo  {journal} {Chem. Phys. Lett.}\ }\textbf {\bibinfo {volume}
  {368}},\ \bibinfo {pages} {502} (\bibinfo {year} {2003})}\BibitemShut
  {NoStop}%
\bibitem [{\citenamefont {Richings}\ \emph {et~al.}(2015)\citenamefont
  {Richings}, \citenamefont {Polyak}, \citenamefont {Spinlove}, \citenamefont
  {Worth}, \citenamefont {Burghardt},\ and\ \citenamefont
  {Lasorne}}]{Richings15}%
  \BibitemOpen
  \bibfield  {author} {\bibinfo {author} {\bibfnamefont {G.~W.}\ \bibnamefont
  {Richings}}, \bibinfo {author} {\bibfnamefont {I.}~\bibnamefont {Polyak}},
  \bibinfo {author} {\bibfnamefont {K.~E.}\ \bibnamefont {Spinlove}}, \bibinfo
  {author} {\bibfnamefont {G.~A.}\ \bibnamefont {Worth}}, \bibinfo {author}
  {\bibfnamefont {I.}~\bibnamefont {Burghardt}}, \ and\ \bibinfo {author}
  {\bibfnamefont {B.}~\bibnamefont {Lasorne}},\ }\href {\doibase
  10.1080/0144235X.2015.1051354} {\bibfield  {journal} {\bibinfo  {journal}
  {Int. Rev. Phys. Chem.}\ }\textbf {\bibinfo {volume} {34}},\ \bibinfo {pages}
  {269} (\bibinfo {year} {2015})}\BibitemShut {NoStop}%
\bibitem [{\citenamefont {Davydov}(1982)}]{Davydov_1982}%
  \BibitemOpen
  \bibfield  {author} {\bibinfo {author} {\bibfnamefont {A.~S.}\ \bibnamefont
  {Davydov}},\ }\href {\doibase 10.1070/pu1982v025n12abeh005012} {\bibfield
  {journal} {\bibinfo  {journal} {Sov. Phys. Usp.}\ }\textbf {\bibinfo {volume}
  {25}},\ \bibinfo {pages} {898} (\bibinfo {year} {1982})}\BibitemShut
  {NoStop}%
\bibitem [{\citenamefont {Cruzeiro-Hansson}(1994)}]{Cruzeiro-Hansson94}%
  \BibitemOpen
  \bibfield  {author} {\bibinfo {author} {\bibfnamefont {L.}~\bibnamefont
  {Cruzeiro-Hansson}},\ }\href {\doibase 10.1103/PhysRevLett.73.2927}
  {\bibfield  {journal} {\bibinfo  {journal} {Phys. Rev. Lett.}\ }\textbf
  {\bibinfo {volume} {73}},\ \bibinfo {pages} {2927} (\bibinfo {year}
  {1994})}\BibitemShut {NoStop}%
\bibitem [{\citenamefont {Zhao}\ \emph {et~al.}(2012)\citenamefont {Zhao},
  \citenamefont {Luo}, \citenamefont {Zhang},\ and\ \citenamefont
  {Ye}}]{Zhao2012}%
  \BibitemOpen
  \bibfield  {author} {\bibinfo {author} {\bibfnamefont {Y.}~\bibnamefont
  {Zhao}}, \bibinfo {author} {\bibfnamefont {B.}~\bibnamefont {Luo}}, \bibinfo
  {author} {\bibfnamefont {Y.}~\bibnamefont {Zhang}}, \ and\ \bibinfo {author}
  {\bibfnamefont {J.}~\bibnamefont {Ye}},\ }\href {\doibase 10.1063/1.4748140}
  {\bibfield  {journal} {\bibinfo  {journal} {J. Chem. Phys.}\ }\textbf
  {\bibinfo {volume} {137}},\ \bibinfo {pages} {084113} (\bibinfo {year}
  {2012})}\BibitemShut {NoStop}%
\bibitem [{\citenamefont {Ma}\ \emph {et~al.}(2018)\citenamefont {Ma},
  \citenamefont {Bonfanti}, \citenamefont {Eisenbrandt}, \citenamefont
  {Martinazzo},\ and\ \citenamefont {Burghardt}}]{Ma18}%
  \BibitemOpen
  \bibfield  {author} {\bibinfo {author} {\bibfnamefont {T.}~\bibnamefont
  {Ma}}, \bibinfo {author} {\bibfnamefont {M.}~\bibnamefont {Bonfanti}},
  \bibinfo {author} {\bibfnamefont {P.}~\bibnamefont {Eisenbrandt}}, \bibinfo
  {author} {\bibfnamefont {R.}~\bibnamefont {Martinazzo}}, \ and\ \bibinfo
  {author} {\bibfnamefont {I.}~\bibnamefont {Burghardt}},\ }\href {\doibase
  10.1063/1.5062608} {\bibfield  {journal} {\bibinfo  {journal} {J. Chem.
  Phys.}\ }\textbf {\bibinfo {volume} {149}},\ \bibinfo {pages} {244107}
  (\bibinfo {year} {2018})}\BibitemShut {NoStop}%
\bibitem [{\citenamefont {Symonds}, \citenamefont {Kattirtzi},\ and\
  \citenamefont {Shalashilin}(2018)}]{Symonds2018}%
  \BibitemOpen
  \bibfield  {author} {\bibinfo {author} {\bibfnamefont {C.}~\bibnamefont
  {Symonds}}, \bibinfo {author} {\bibfnamefont {J.~A.}\ \bibnamefont
  {Kattirtzi}}, \ and\ \bibinfo {author} {\bibfnamefont {D.~V.}\ \bibnamefont
  {Shalashilin}},\ }\href {\doibase 10.1063/1.5020567} {\bibfield  {journal}
  {\bibinfo  {journal} {J. Chem. Phys.}\ }\textbf {\bibinfo {volume} {148}},\
  \bibinfo {pages} {184113} (\bibinfo {year} {2018})}\BibitemShut {NoStop}%
\bibitem [{\citenamefont {Bramley}, \citenamefont {Symonds},\ and\
  \citenamefont {Shalashilin}(2019)}]{Bramley2019}%
  \BibitemOpen
  \bibfield  {author} {\bibinfo {author} {\bibfnamefont {O.}~\bibnamefont
  {Bramley}}, \bibinfo {author} {\bibfnamefont {C.}~\bibnamefont {Symonds}}, \
  and\ \bibinfo {author} {\bibfnamefont {D.~V.}\ \bibnamefont {Shalashilin}},\
  }\href {\doibase 10.1063/1.5100145} {\bibfield  {journal} {\bibinfo
  {journal} {J. Chem. Phys.}\ }\textbf {\bibinfo {volume} {151}},\ \bibinfo
  {pages} {064103} (\bibinfo {year} {2019})}\BibitemShut {NoStop}%
\bibitem [{\citenamefont {Shalashilin}\ and\ \citenamefont
  {Child}(2008)}]{Shalashilin2008}%
  \BibitemOpen
  \bibfield  {author} {\bibinfo {author} {\bibfnamefont {D.~V.}\ \bibnamefont
  {Shalashilin}}\ and\ \bibinfo {author} {\bibfnamefont {M.~S.}\ \bibnamefont
  {Child}},\ }\href {\doibase 10.1063/1.2828509} {\bibfield  {journal}
  {\bibinfo  {journal} {J. Chem. Phys.}\ }\textbf {\bibinfo {volume} {128}},\
  \bibinfo {pages} {054102} (\bibinfo {year} {2008})}\BibitemShut {NoStop}%
\bibitem [{\citenamefont {Makhov}\ \emph {et~al.}(2017)\citenamefont {Makhov},
  \citenamefont {Symonds}, \citenamefont {Fernandez-Alberti},\ and\
  \citenamefont {Shalashilin}}]{Makhov2017}%
  \BibitemOpen
  \bibfield  {author} {\bibinfo {author} {\bibfnamefont {D.~V.}\ \bibnamefont
  {Makhov}}, \bibinfo {author} {\bibfnamefont {C.}~\bibnamefont {Symonds}},
  \bibinfo {author} {\bibfnamefont {S.}~\bibnamefont {Fernandez-Alberti}}, \
  and\ \bibinfo {author} {\bibfnamefont {D.~V.}\ \bibnamefont {Shalashilin}},\
  }\href {\doibase 10.1016/j.chemphys.2017.04.003} {\bibfield  {journal}
  {\bibinfo  {journal} {Chem. Phys.}\ }\textbf {\bibinfo {volume} {493}},\
  \bibinfo {pages} {200} (\bibinfo {year} {2017})}\BibitemShut {NoStop}%
\bibitem [{\citenamefont {Ronto}\ and\ \citenamefont
  {Shalashilin}(2013)}]{Ronto2013}%
  \BibitemOpen
  \bibfield  {author} {\bibinfo {author} {\bibfnamefont {M.}~\bibnamefont
  {Ronto}}\ and\ \bibinfo {author} {\bibfnamefont {D.~V.}\ \bibnamefont
  {Shalashilin}},\ }\href {\doibase 10.1021/jp310976d} {\bibfield  {journal}
  {\bibinfo  {journal} {J. Phys. Chem. A}\ }\textbf {\bibinfo {volume} {117}},\
  \bibinfo {pages} {6948} (\bibinfo {year} {2013})}\BibitemShut {NoStop}%
\bibitem [{\citenamefont {Itzykson}\ and\ \citenamefont
  {Zuber}(2012)}]{Itzykson2012}%
  \BibitemOpen
  \bibfield  {author} {\bibinfo {author} {\bibfnamefont {C.}~\bibnamefont
  {Itzykson}}\ and\ \bibinfo {author} {\bibfnamefont {J.}~\bibnamefont
  {Zuber}},\ }\href@noop {} {\emph {\bibinfo {title} {Quantum Field Theory}}}\
  (\bibinfo  {publisher} {Courier Corporation},\ \bibinfo {year}
  {2012})\BibitemShut {NoStop}%
\bibitem [{\citenamefont {Makhov}\ \emph {et~al.}(2014)\citenamefont {Makhov},
  \citenamefont {Glover}, \citenamefont {Martinez},\ and\ \citenamefont
  {Shalashilin}}]{Makhov2014}%
  \BibitemOpen
  \bibfield  {author} {\bibinfo {author} {\bibfnamefont {D.~V.}\ \bibnamefont
  {Makhov}}, \bibinfo {author} {\bibfnamefont {W.~J.}\ \bibnamefont {Glover}},
  \bibinfo {author} {\bibfnamefont {T.~J.}\ \bibnamefont {Martinez}}, \ and\
  \bibinfo {author} {\bibfnamefont {D.~V.}\ \bibnamefont {Shalashilin}},\
  }\href {\doibase 10.1063/1.4891530} {\bibfield  {journal} {\bibinfo
  {journal} {J. Chem. Phys.}\ }\textbf {\bibinfo {volume} {141}},\ \bibinfo
  {pages} {054110} (\bibinfo {year} {2014})}\BibitemShut {NoStop}%
\bibitem [{\citenamefont {Makhov}, \citenamefont {Martinez},\ and\
  \citenamefont {Shalashilin}(2016)}]{Makhov2016}%
  \BibitemOpen
  \bibfield  {author} {\bibinfo {author} {\bibfnamefont {D.~V.}\ \bibnamefont
  {Makhov}}, \bibinfo {author} {\bibfnamefont {T.~J.}\ \bibnamefont
  {Martinez}}, \ and\ \bibinfo {author} {\bibfnamefont {D.~V.}\ \bibnamefont
  {Shalashilin}},\ }\href {\doibase 10.1039/C6FD00073H} {\bibfield  {journal}
  {\bibinfo  {journal} {Faraday Discuss.}\ }\textbf {\bibinfo {volume} {194}},\
  \bibinfo {pages} {81} (\bibinfo {year} {2016})}\BibitemShut {NoStop}%
\bibitem [{\citenamefont {Zheng}\ \emph {et~al.}(2014)\citenamefont {Zheng},
  \citenamefont {Xu}, \citenamefont {Meana-Pa\~{n}eda},\ and\ \citenamefont
  {G.~Truhlar}}]{Zheng2014}%
  \BibitemOpen
  \bibfield  {author} {\bibinfo {author} {\bibfnamefont {J.}~\bibnamefont
  {Zheng}}, \bibinfo {author} {\bibfnamefont {X.}~\bibnamefont {Xu}}, \bibinfo
  {author} {\bibfnamefont {R.}~\bibnamefont {Meana-Pa\~{n}eda}}, \ and\
  \bibinfo {author} {\bibfnamefont {D.}~\bibnamefont {G.~Truhlar}},\ }\href
  {\doibase 10.1039/C3SC53290A} {\bibfield  {journal} {\bibinfo  {journal}
  {Chem. Sci.}\ }\textbf {\bibinfo {volume} {5}},\ \bibinfo {pages} {2091}
  (\bibinfo {year} {2014})}\BibitemShut {NoStop}%
\bibitem [{\citenamefont {Zheng}, \citenamefont {Meana-Pa\~{n}eda},\ and\
  \citenamefont {Truhlar}(2014)}]{Zheng2014b}%
  \BibitemOpen
  \bibfield  {author} {\bibinfo {author} {\bibfnamefont {J.}~\bibnamefont
  {Zheng}}, \bibinfo {author} {\bibfnamefont {R.}~\bibnamefont
  {Meana-Pa\~{n}eda}}, \ and\ \bibinfo {author} {\bibfnamefont {D.~G.}\
  \bibnamefont {Truhlar}},\ }\href {\doibase 10.1021/jz500653m} {\bibfield
  {journal} {\bibinfo  {journal} {J. Phys. Chem. Lett.}\ }\textbf {\bibinfo
  {volume} {5}},\ \bibinfo {pages} {2039} (\bibinfo {year} {2014})}\BibitemShut
  {NoStop}%
\bibitem [{\citenamefont {Jasper}, \citenamefont {Hack},\ and\ \citenamefont
  {Truhlar}(2001)}]{Jasper2001}%
  \BibitemOpen
  \bibfield  {author} {\bibinfo {author} {\bibfnamefont {A.~W.}\ \bibnamefont
  {Jasper}}, \bibinfo {author} {\bibfnamefont {M.~D.}\ \bibnamefont {Hack}}, \
  and\ \bibinfo {author} {\bibfnamefont {D.~G.}\ \bibnamefont {Truhlar}},\
  }\href {\doibase 10.1063/1.1377891} {\bibfield  {journal} {\bibinfo
  {journal} {J. Chem. Phys.}\ }\textbf {\bibinfo {volume} {115}},\ \bibinfo
  {pages} {1804} (\bibinfo {year} {2001})}\BibitemShut {NoStop}%
\bibitem [{\citenamefont {Bertini}\ \emph {et~al.}(2021)\citenamefont
  {Bertini}, \citenamefont {Heidrich-Meisner}, \citenamefont {Karrasch},
  \citenamefont {Prosen}, \citenamefont {Steinigeweg},\ and\ \citenamefont {{\v
  Z}nidari{\v c}}}]{Bertini2021}%
  \BibitemOpen
  \bibfield  {author} {\bibinfo {author} {\bibfnamefont {B.}~\bibnamefont
  {Bertini}}, \bibinfo {author} {\bibfnamefont {F.}~\bibnamefont
  {Heidrich-Meisner}}, \bibinfo {author} {\bibfnamefont {C.}~\bibnamefont
  {Karrasch}}, \bibinfo {author} {\bibfnamefont {T.}~\bibnamefont {Prosen}},
  \bibinfo {author} {\bibfnamefont {R.}~\bibnamefont {Steinigeweg}}, \ and\
  \bibinfo {author} {\bibfnamefont {M.}~\bibnamefont {{\v Z}nidari{\v c}}},\
  }\href {\doibase 10.1103/RevModPhys.93.025003} {\bibfield  {journal}
  {\bibinfo  {journal} {Rev. Mod. Phys.}\ }\textbf {\bibinfo {volume} {93}},\
  \bibinfo {pages} {025003} (\bibinfo {year} {2021})}\BibitemShut {NoStop}%
\bibitem [{\citenamefont {Zhu}\ \emph {et~al.}(2004)\citenamefont {Zhu},
  \citenamefont {Nangia}, \citenamefont {Jasper},\ and\ \citenamefont
  {Truhlar}}]{Zhu2004b}%
  \BibitemOpen
  \bibfield  {author} {\bibinfo {author} {\bibfnamefont {C.}~\bibnamefont
  {Zhu}}, \bibinfo {author} {\bibfnamefont {S.}~\bibnamefont {Nangia}},
  \bibinfo {author} {\bibfnamefont {A.~W.}\ \bibnamefont {Jasper}}, \ and\
  \bibinfo {author} {\bibfnamefont {D.~G.}\ \bibnamefont {Truhlar}},\ }\href
  {\doibase 10.1063/1.1793991} {\bibfield  {journal} {\bibinfo  {journal} {J.
  Chem. Phys.}\ }\textbf {\bibinfo {volume} {121}},\ \bibinfo {pages} {7658}
  (\bibinfo {year} {2004})}\BibitemShut {NoStop}%
\bibitem [{\citenamefont {Hashimoto}\ and\ \citenamefont
  {Ishihara}(2017)}]{hashimoto_ishihara_17}%
  \BibitemOpen
  \bibfield  {author} {\bibinfo {author} {\bibfnamefont {H.}~\bibnamefont
  {Hashimoto}}\ and\ \bibinfo {author} {\bibfnamefont {S.}~\bibnamefont
  {Ishihara}},\ }\href {\doibase 10.1103/PhysRevB.96.035154} {\bibfield
  {journal} {\bibinfo  {journal} {Phys. Rev. B}\ }\textbf {\bibinfo {volume}
  {96}},\ \bibinfo {pages} {035154} (\bibinfo {year} {2017})}\BibitemShut
  {NoStop}%
\bibitem [{\citenamefont {Zhu}, \citenamefont {Jasper},\ and\ \citenamefont
  {Truhlar}(2004)}]{Zhu2004}%
  \BibitemOpen
  \bibfield  {author} {\bibinfo {author} {\bibfnamefont {C.}~\bibnamefont
  {Zhu}}, \bibinfo {author} {\bibfnamefont {A.~W.}\ \bibnamefont {Jasper}}, \
  and\ \bibinfo {author} {\bibfnamefont {D.~G.}\ \bibnamefont {Truhlar}},\
  }\href {\doibase 10.1063/1.1648306} {\bibfield  {journal} {\bibinfo
  {journal} {J. Chem. Phys.}\ }\textbf {\bibinfo {volume} {120}},\ \bibinfo
  {pages} {5543} (\bibinfo {year} {2004})}\BibitemShut {NoStop}%
\bibitem [{\citenamefont {Shenvi}, \citenamefont {Roy},\ and\ \citenamefont
  {Tully}(2009)}]{Shenvi2009}%
  \BibitemOpen
  \bibfield  {author} {\bibinfo {author} {\bibfnamefont {N.}~\bibnamefont
  {Shenvi}}, \bibinfo {author} {\bibfnamefont {S.}~\bibnamefont {Roy}}, \ and\
  \bibinfo {author} {\bibfnamefont {J.~C.}\ \bibnamefont {Tully}},\ }\href
  {\doibase 10.1063/1.3125436} {\bibfield  {journal} {\bibinfo  {journal} {J.
  Chem. Phys.}\ }\textbf {\bibinfo {volume} {130}},\ \bibinfo {pages} {174107}
  (\bibinfo {year} {2009})}\BibitemShut {NoStop}%
\bibitem [{\citenamefont {Makhov}\ \emph {et~al.}(2015)\citenamefont {Makhov},
  \citenamefont {Saita}, \citenamefont {Martinez},\ and\ \citenamefont
  {Shalashilin}}]{Makhov2015}%
  \BibitemOpen
  \bibfield  {author} {\bibinfo {author} {\bibfnamefont {D.~V.}\ \bibnamefont
  {Makhov}}, \bibinfo {author} {\bibfnamefont {K.}~\bibnamefont {Saita}},
  \bibinfo {author} {\bibfnamefont {T.~J.}\ \bibnamefont {Martinez}}, \ and\
  \bibinfo {author} {\bibfnamefont {D.~V.}\ \bibnamefont {Shalashilin}},\
  }\href {\doibase 10.1039/C4CP04571H} {\bibfield  {journal} {\bibinfo
  {journal} {Phys. Chem. Chem. Phys.}\ }\textbf {\bibinfo {volume} {17}},\
  \bibinfo {pages} {3316} (\bibinfo {year} {2015})}\BibitemShut {NoStop}%
\bibitem [{\citenamefont {Martinez}, \citenamefont {Ben-Nun},\ and\
  \citenamefont {Levine}(1996)}]{Martinez1996}%
  \BibitemOpen
  \bibfield  {author} {\bibinfo {author} {\bibfnamefont {T.~J.}\ \bibnamefont
  {Martinez}}, \bibinfo {author} {\bibfnamefont {M.}~\bibnamefont {Ben-Nun}}, \
  and\ \bibinfo {author} {\bibfnamefont {R.~D.}\ \bibnamefont {Levine}},\
  }\href {\doibase 10.1021/jp953105a} {\bibfield  {journal} {\bibinfo
  {journal} {J. Phys. Chem.}\ }\textbf {\bibinfo {volume} {100}},\ \bibinfo
  {pages} {7884} (\bibinfo {year} {1996})}\BibitemShut {NoStop}%
\bibitem [{\citenamefont {Ben-Nun}, \citenamefont {Quenneville},\ and\
  \citenamefont {Mart\'inez}(2000)}]{Ben-nun2000}%
  \BibitemOpen
  \bibfield  {author} {\bibinfo {author} {\bibfnamefont {M.}~\bibnamefont
  {Ben-Nun}}, \bibinfo {author} {\bibfnamefont {J.}~\bibnamefont
  {Quenneville}}, \ and\ \bibinfo {author} {\bibfnamefont {T.~J.}\ \bibnamefont
  {Mart\'inez}},\ }\href {\doibase 10.1021/jp994174i} {\bibfield  {journal}
  {\bibinfo  {journal} {J. Phys. Chem. A}\ }\textbf {\bibinfo {volume} {104}},\
  \bibinfo {pages} {5161} (\bibinfo {year} {2000})}\BibitemShut {NoStop}%
\bibitem [{\citenamefont {Wang}(2015)}]{Wang2015}%
  \BibitemOpen
  \bibfield  {author} {\bibinfo {author} {\bibfnamefont {H.}~\bibnamefont
  {Wang}},\ }\href {\doibase 10.1021/acs.jpca.5b03256} {\bibfield  {journal}
  {\bibinfo  {journal} {J. Phys. Chem. A}\ }\textbf {\bibinfo {volume} {119}},\
  \bibinfo {pages} {7951} (\bibinfo {year} {2015})}\BibitemShut {NoStop}%
\end{thebibliography}%

	\clearpage

\end{document}